%% file: Main_arxiv_v3.tex
\documentclass{article}

\input{Packages}
\input{Commandes}

\title{\bf{Tensor invariants for multipartite entanglement classification}}
\author[1]{Sylvain Carrozza\thanks{sylvain.carrozza@ube.fr}}
\author[1,2]{Johann Chevrier\thanks{johann.chevrier@ube.fr}}
\author[2]{Luca Lionni\thanks{luca.lionni@ens-lyon.fr}}
\affil[1]{\normalsize{Universit\'{e} Bourgogne Europe, CNRS, IMB UMR 5584, 21000 Dijon, France}}
\affil[2]{\normalsize{CNRS, ENS de Lyon, LPENSL, UMR 5672, 69342 Lyon cedex 07, France}}

\date{\today}

\begin{document}

\maketitle

\begin{abstract}
    \noindent Organising the space of entanglement structures of a multipartite quantum system is a much more challenging task than its bipartite version: while the local unitary ($\LU$) orbit of a bipartite pure state can be conveniently characterized by its entanglement spectrum, invariants of multipartite entanglement structures are comparatively difficult to define and work with. The root cause of this difference is that the bipartite problem can be reduced to the analysis of matrix invariants, while its multipartite version is governed by a much richer space of tensor invariants. The present work explores the latter through the lens of so-called trace-invariants, which are in one-to-one correspondence with combinatorial objects known as colored graphs. We first explain why trace-invariant evaluations can serve as labels of $\LU$-orbits of multipartite pure states, how this strategy extends to random states, and how the effect of local operations ($\LO$) can be analyzed through such data. We then focus on entanglement classification within an (infinite-dimensional) subspace of reference states, whose basic building blocks are GHZ states of various dimensions. We show that relatively simple subclasses of trace-invariants are sufficient to separate the $\LU$-orbits of reference states, and enable a complete (resp. an incomplete) characterization of their relations in the $\LO$ (resp. $\LOCC$) resource theory of entanglement. Finally, we investigate how a (still infinite) subclass of reference states of local dimension $N$ can be efficiently distinguished at leading and subleading orders in an asymptotic large-$N$ expansion (among themselves, or from Haar-random states). This analysis relies crucially on combinatorial quantities associated to colored graphs, some of which have already played instrumental roles in the recent literature on random tensors. Results of broader relevance are reported along the way.
\end{abstract}

\setcounter{tocdepth}{2}
\tableofcontents

\section{Introduction and summary of our main results}\label{sec:intro}

If, historically, the notion of entanglement was first articulated in thought-experiments to address potential limitations of the quantum formalism \cite{PhysRev.47.777}, almost a century later, it is recognized as an unavoidable ingredient of quantum theory \cite{PhysicsPhysiqueFizika.1.195, PhysRevLett.23.880, PhysRevLett.49.1804}, that is also of great practical relevance: indeed, it is arguably the main resource powering the on-going development of quantum technologies. 

On the mathematical side, a large body of works has been dedicated to the precise definition and characterization of entanglement, both in the bipartite and multipartite settings (see \eg \cite{Horodecki:2009zz, Walter:2016lgl} and references therein). However, it is fair to say that the latter is comparatively much less understood than the former, for reasons that can all be traced back to the following observation: while a bipartite pure state can be naturally represented by a \emph{matrix}, whose \emph{spectrum} (understood as its set of singular values) characterizes its entanglement structure, a multipartite pure state is naturally encoded into a higher order \emph{tensor}, whose space of invariants turns out to be dramatically more complicated than that of a matrix.\footnote{For the same reason, characterizing the entanglement structure of a bipartite \emph{mixed} state, which can be represented by a \emph{tripartite} pure state, is challenging. In this respect, we can consider the theory of bipartite entanglement for mixed states as a branch of the theory of multipartite entanglement.} In particular, no simple and universal notion of spectrum is available to characterize the algebra of invariants of a tensor. As a result, one should expect the development of a complete theory of multipartite entanglement to be intrinsically more challenging than that of bipartite entanglement theory, and to heavily rely on tensors rather than matrices. The present manuscript makes a step in that direction by taking advantage of a combinatorial toolbox that was primarily developed to represent and manipulate tensor invariants in the theory of random tensors (see \eg \cite{gurau_random_2017, Gurau:2011xp} and references therein). As this work will illustrate, this toolbox is perfectly suited to the definition and analysis of tensor invariants in the theory of multipartite entanglement. 

\medskip

The fundamental classification problem of entanglement theory can be formulated as follows. Given a pure state $\ket{\psi}\in \H_1 \otimes \cdots \otimes \H_D$ in a global Hilbert space that has been partitioned into $D$ subsystems (represented by the tensor factors $\{\H_c\}$), the goal is to determine which states in $\H_1 \otimes \cdots \otimes \H_D$ merely differ from $\ket{\psi}$ by the combined action of unitary operators that are local to the subsystems $\{\H_c\}$ (or equivalently, that differ by a change of orthonormal basis in each tensor factor $\H_c$). Any such operation is called a \emph{Local Unitary} ($\LU$) transformation, and any two states related by a $\LU$-transformation are deemed to have equivalent $\LU$-entanglement structures (relative to the $D$-partitioning $\H_1 \otimes \cdots \otimes \H_D$ of the global Hilbert space). Ideally, one would like to be able to distinguish any two inequivalent $\LU$-entanglement structures or, in mathematical terms, to separate any two orbits of the local unitary group. Furthermore, in the context of quantum information \emph{processing}, one is interested in understanding how distinct entanglement structures may or may not be transformed into one another given a set of quantum channels that are operationally available. This type of question gives rise to operational \emph{resource theories} \cite{Chitambar:2018rnj, Coecke:2014svf} of entanglement such as the widely used framework of Local Operations and Classical Communication ($\LOCC$) \cite{PhysRevA.59.1070, PhysRevLett.83.436, PhysRevLett.83.3566, PhysRevLett.127.150503, Chitambar2014}. The latter formalizes the experimental situation in which $D$ agents are allowed to perform any quantum operation they like in their respective quantum laboratories ($\LO$), and can also communicate with each other by classical means \eg they can phone each other ($\mathsf{CC}$). From a mathematical point of view, such a resource theory defines a \emph{preorder} on the set of $D$-partite states which, ideally, one would like to completely characterize. Given that this preorder is compatible with the underlying $\LU$ structure, a natural strategy is to construct a sufficiently rich family of real $\LU$-invariants which vary monotonically under any quantum transformation allowed by the given resource theory (\eg a $\LOCC$ transformation). 

\medskip

Before listing the specific questions we would like to address in the multipartite setting, let us briefly recall a number of relevant results in the bipartite setting (\ie $D=2$). For a bipartite pure state $\ket{\psi} \in \H_1 \otimes \H_2$, by far the most studied quantitative measure of entanglement is its \textit{entanglement entropy}, which is nothing but the \textit{von Neumann entropy} of its reduced density matrix on either subsystem: that is, letting \eg $\rho_{1} \eqdef \tr_{\H_2} \ket{\psi}\bra{\psi}$ denote the reduced density matrix on $\H_1$, the entanglement entropy $\S\left( \ket{\psi}\right)$ is defined as
\begin{equation} \label{eq:vN}
    \S(\ket\psi) \eqdef \S_\rm{vN}(\rho_{1}) \eqdef - \tr \left( \rho_{1} \ln \rho_{1} \right)\,.
\end{equation}
This quantity not only quantifies entanglement for bipartite pure states, but also has operational interpretations in quantum information theory, for instance: it determines the optimal rate of entanglement distillation and formation under $\LOCC$ \cite{PhysRevA.53.2046,PhysRevA.54.3824}, and it provides bounds on the resources required for quantum communication protocols such as teleportation and dense coding (see Refs.~\cite{PhysRevA.51.2738,PhysRevA.53.2046,PhysRevA.56.R3319,Horodecki2006}). However, $\S(\ket\psi)$ being non-polynomial in $\ket\psi$ (or in the eigenvalues of $\rho_1$), evaluating it explicitly can be challenging. 

Another widely used family of entanglement measures are the \textit{entanglement Rényi entropies} $\{\S_{\rm{R}}^{(k)}\}_{ k \in \mathbb{R}_+^* \setminus \{1\}}$, which are defined as follows:
\begin{equation} \label{eq:Renyi}
    \forall k \in \mathbb{R}_+^* \setminus \{1\}\,, \qquad \S_{\rm{R}}^{(k)}(\ket\psi) \eqdef \frac{1}{1-k} \ln \tr \rho_1^k\,.
\end{equation}
In the limit $k\to1$, $\S_{\rm{R}}^{(k)}(\ket\psi)$ converges to the entanglement entropy of Eq.~\eqref{eq:vN}. Because they involve powers of the density matrix, entanglement Rényi entropies are particularly suited to analytic and practical computations, for example, in many-body systems where $S_\rm{R}^{(2)}$ can be evaluated more easily than the von Neumann entropy (see \eg Refs.~\cite{Calabrese2004,Islam2015}). More broadly, for integer $k$, $\S_R^{(k)}(\ket\psi)$ contains the same information as the \emph{polynomial} $\tr\rho_1^k$, which is in general much easier to evaluate explictly than $\S(\ket\psi)$. Moreover, $\tr\rho_1^k$ is simply equal to the power-sum $\displaystyle\sum_{\lambda \in \spec(\rho_1)} \lambda^k$. Knowing $\tr\rho_1^k$ for sufficiently many values of $k$ is therefore sufficient to reconstruct the full spectrum of $\rho_1$, which is also known as the \emph{entanglement spectrum} of $\ket{\psi}$ \cite{PhysRevLett.101.010504}, and is equivalent to the collection of its \emph{Schmidt coefficients}. From elementary linear algebra, we know that the entanglement spectrum of $\ket{\psi}$ completely characterizes its $\LU$-orbit, hence, so does the set $\{\tr\rho_1^k\}_{k\geq2}$ (or, equivalently, $\{\S_R^{(k)}(\ket\psi)\}_{k\geq2}$): such data is therefore sufficient to separate inequivalent bipartite entanglement structures. What's more, the notion of entanglement spectrum is also convenient to characterize order relations in resource theories of entanglement: for instance, the $\LOCC$ preorder on bipartite pure states has been fully characterized in terms of \emph{majorization} conditions on entanglement spectra~\cite{PhysRevLett.83.436}. 

In addition to their practical use in computations, entanglement R\'enyi entropies have direct operational relevance in certain contexts. They are for instance useful in deriving entanglement inequalities, as \eg the second entanglement Rényi entropy provides strong entanglement criteria that can outperform standard Bell inequalities (see Refs.~\cite{Horodecki1996, Horodecki1996_2,Santos2004}). They are also instrumental in the resource theory of \emph{catalytic} $\LOCC$ \cite{Turgut_2007, Klimesh:2007vhg}, where majorization conditions which can be directly expressed in terms of entanglement Rényi entropies relax and replace the majorization conditions of Ref.~\cite{PhysRevLett.83.436}; hence, in that particular context, the values of entanglement R\'{e}nyi entropies are more directly operationally relevant than any individual eigenvalue of the reduced density matrix.

Finally, if the bipartite state $\ket{\psi}$ is sampled from a probability distribution, one is often interested in determining its \emph{typical} entanglement structure in a regime of large dimension, where measure concentration phenomena are expected to be at play. For definiteness, let the Hilbert spaces $\H_1$ and $\H_2$ have the same dimension $N$, and let $\ket{\psi}$ be drawn uniformly at random in the set of pure states on $\H_1 \otimes \H_2$. Then, the following asymptotic expression holds:
\begin{equation} \label{eq:DiffBipartite}
    \mean{\S_{\rm{R}}^{(k)}(\ket{\psi})} \underset{N \to \infty}{\sim} \ln(N)+ \frac{1}{1-k}\ln \rm{Cat}_k\,,
\end{equation}
where $\rm{Cat}_k$ denotes the $k$-th Catalan number. In the $k\to 1$ regime, this can be used to reproduce a celebrated result by Page (see Ref.~\cite{Page:1993df}), which states that $\mean{\S(\ket\psi)} = \ln(N)- 1/2+o(1)$. Given that $\ln(N)$ is the entanglement entropy (resp. entanglement R\'{e}nyi-$k$ entropy) of a (maximally entangled) Bell state of dimension $N$, those relations tell us that the uniform random state is undistinguishable from that deterministic state at leading order in $N$, but not so at the next-to-leading order. Relations such as Eq.~\eqref{eq:DiffBipartite} have been generalized beyond the uniform distribution in the context of \emph{random tensor networks} \cite{Hayden2016}, where it was shown that the typical entanglement entropy of a bipartition is governed by an equation whose holographic features are reminiscent of the Ryu-Takayanagi formula. 

\medskip

Let us come back to the multipartite setting ($D\geq 3$). Our key entry-point into this  world is that, even though no simple generalization of the notion of entanglement spectrum (or, even, of entanglement entropy) is currently available in that regime, the polynomials $\{ \tr \rho_1^k\}_{k \in \mathbb{N}^*}$ appearing in entanglement Rényi entropies (with integer $k$) \emph{do} admit natural multipartite generalizations: namely, through the notion of \textit{trace-invariant} previously introduced in the theory of random tensors (see Ref.~\cite{gurau_random_2017} and references therein). Furthermore, trace-invaritants of a $D$-partite system are in one-to-one correspondence with \emph{$D$-colored graphs}, and are thus intrinsically combinatorial objects. Our aim in the present article (and in follow-up works), is to systematically develop a quantitative theory of multipartite entanglement based on trace-invariants and their underlying combinatorial structure. In particular, we will address the following questions:
\begin{enumerate}[label=(\roman*)]
    \item\label{Q:LU}  How can one efficiently distinguish $\LU$-equivalence classes by means of trace-invariants in a multipartite system?
    \item\label{Q:monotones} How should one combine trace-invariants to obtain valuable monotones in resource theories of entanglement such as $\LO$, $\LOCC$ or $\LOSR$? Can the underlying preorder be efficiently characterized by means of trace-invariants?
    \item\label{Q:typical} How can the typical entanglement structure of a large random state be characterized in terms of correlation functions of trace-invariants?
\end{enumerate}
Furthermore, in connection to the third question, we will investigate the implications of a recently discovered property of random tensors, that is somewhat surprising and subtle. As it turns out, a hallmark of random matrix theory that is often expected (and sometimes assumed) for random tensor distributions -- the large-$N$ factorization property of correlation functions -- is \emph{not} generally valid for random tensors \cite{Gurau2025, Facto2}. We will see that this result has direct implications for question \ref{Q:typical}. Besides, while investigating them, we were naturally led to address a question that we expect to be of independent interest to random tensor enthusiasts: 
\begin{enumerate}[label=(\roman*)]
\setcounter{enumi}{3}
\item\label{Q:factorization} Can one generate richer infinite families of trace-invariants than the ones currently available, whose Gaussian (or Haar) moments can be proven to factorize over their connected components at leading order in the large-$N$ limit?
\end{enumerate}

It goes without saying that the present work is not the first to explore questions \ref{Q:LU}-\ref{Q:typical}. For instance, to mention just a selection of relevant contributions to the literature on this subject: the role of trace-invariants for entanglement classification was recognized in \eg Refs.~\cite{Vrana:2011ehx, Turner2017} (under a different name), multipartite entanglement properties of random tensor networks have been explored by means of observables which are indeed trace-invariants \cite{Dong2021, KudlerFlam2022, Akers2022, Penington2023}, and recent efforts have been initiated to construct valuable entanglement monotones from trace-invariants \cite{Gadde2022, Gadde2023, Gadde2025}. More broadly, multipartite aspects of entanglement theory seem to be catching the attention of a growing number of researchers, from various corners of physics and mathematics. Our aim with the present manuscript is to initiate a \emph{systematic} assessment of the role of trace-invariants in multipartite entanglement classification problems, while taking advantage of a combinatorial toolbox that remains largely untapped in that context, and that we anticipate to be of great practical value to streamline and generalize constructions that have already appeared in the literature. We also made an effort to approach this problem in a largely self-contained manner, that will hopefully speak to a diverse community of researchers.\footnote{This explains in part the significant page count of the present contribution.}  

\

Let us conclude this introduction by an informal summary of our main results, which may also serve as a reader's guide.

\medskip

In {\bf Sec.~\ref{sec:LUinvariants}}, we start out by recalling some basic definitions and properties of local unitary symmetry, and reviewing the main combinatorial properties of trace-invariants that are relevant for the rest of the paper. In Prop.~\ref{prop:charac_LU-1}, we recall in particular that trace-invariants do separate $\LU$-orbits, a result which in principle provides a complete answer to question \ref{Q:LU}. However, in a multipartite quantum state space of even moderately large dimension, the number of independent trace-invariants is expected to be impractically large: on the one hand, the best known bounds on the number of invariants in a generating set grow extremely rapidly with the dimension (as we recall in Prop.~\ref{prop:bound_degree}, which quotes a result of Ref.~\cite{Turner2017}); on the other hand, the number of connected trace-invariants of a given order $k$ grows super-exponentially with $k$ in the multipartite setting \cite{BenGeloun:2013lim}, which starkly contrasts with the bipartite setting (where there exists a \emph{single} connected invariant of a given order). In Sec.~\ref{sec:LU_random_asymptotic}, we introduce definitions and concepts that allow to precisely formalize question \ref{Q:typical}: we first extend the definition of $\LU$-equivalence to random states (Def.~\ref{def:LU-random}), and then to sequences of (random or deterministic) states indexed by a dimensional parameter $N \in \mathbb{N}^*$ (Def.~\ref{def:LargeN_LU_eq}). Def.~\ref{def:LargeN_LU_eq} also introduces two notions of large-$N$ asymptotic $\LU$-equivalence (one weaker than the other), that provide coarser notions of entanglement structure classification than the exact Def.~\ref{def:LU-random}, and that are relevant for generalized Page curve calculations. In Sec.~\ref{sec:separable}, we return to the classification problem of deterministic multipartite pure states, demonstrating in Thm.~\ref{thm:charac_separable} that a subset of trace-invariants -- which we refer to as \emph{genuinely} $D$-partite -- are sufficient to characterize separable $D$-partite states. Sec.~\ref{sec:generalities-on-coarse-graining} explores the notion of \emph{coarse-graining} of a $D$-partite state space (\ie of turning it into a $D'$-partite state space with $D' < D$) in connection to trace-invariants, their graph-theoretical representations, and coarse-graining maps that can be directly implemented at the graphical level. In particular, Prop.~\ref{prop:charac_genuinely_Dpartite} characterizes states that are \emph{genuinely entangled} on a given $D$-partite state space, while Prop.~\ref{prop:partial_sep} characterizes properties related to the notion of \emph{partial separability} directly in terms of trace-invariants. In Sec.~\ref{ss:ent-monotones-trace-inv}, we investigate question \ref{Q:monotones} in the simple context of $\LO$ transformations, which form an integral part of any resource theory of entanglement (such as $\LOCC$). Prop.~\ref{prop:charac_LO} provides a characterization of the $\LO$ preorder in terms trace-invariants, which specializes to a well-known result in the bipartite setting. While this characterization is not constructive and therefore not necessarily easy to apply in practice, its main corollary (Cor.~\ref{cor:LO_monotones}) is: it provides an infinite family of inequalities that are necessarily obeyed by any $\LO$ transformation. This corollary allows introducing multipartite generalizations of entanglement R\'{e}nyi-$k$ entropies for integer $k$ that are labeled by a $D$-colored graph rather than an integer (see Ex.~\ref{ex:renyi_higher-D}): in particular, such quantities are $\LO$-monotones. A key difference with the bipartite setting at this level is that trace-invariants are not in general positive (or even real), which results into the fact that higher order entanglement R\'{e}nyi entropies may take arbitrarily large values, or be strictly infinite (see Ex.~\ref{ex:vanishing_K33}).

\medskip

The dramatic proliferation of the number of independent invariants in the multipartite entanglement classification problem (a manifestation of the fact that this is a \emph{tensor} rather than a matrix classification problem) makes it unreasonable to look for complete characterizations of $\LU$-classes when the dimension of the state space is even moderately large. On the other hand, such a feat is probably not required for practical applications where, resources being necessarily finite, it may be satisfactory enough to be able to classify a limited template of $\LU$-equivalence classes. In turn, inequivalent entanglement stuctures drawn from such a limited template might be perfectly distinguishable with the help of a limited number of trace-invariants. In this spirit, {\bf{Sec.~\ref{sec:Trace_Lit_Ref_States}}} introduces infinite families of invariants, and  infinite families of states, that are structured enough to evade the fundamental challenge of the proliferation of multipartite $\LU$-classes. The infinite families of trace-invariants introduced in Sec.~\ref{sub:inv-from-lit} are in part imported from the existing literature, and in part new. In both situations, we provide graph-theoretic representations (as $D$-colored graphs) of such invariants that prove highly valuable to build-up intuition and reason about them. We encourage the reader to skip this section on a first reading, and only refer to it as needed when going through the rest of the paper. Sec.~\ref{sub:intro-of-ref-states} introduces the limited template of states investigated in later sections. We first define an infinite family of deterministic states of reference, that we refer to as \emph{hypergraph-tensor states} (abbreviated along the text as HT states), whose basic buidling blocks are GHZ states of arbitrary dimension shared by an arbitrary number of subsystems. These states have been for instance considered in Refs.~\cite{Buhrman:2016tif,Vrana:2016edr,Christandl:2018cfb,Christandl:2019zrq,brand2026bilinearcomplexityworksbreaks}. Their structure can be represented graphically in terms of hypergraphs, thus the name. We discuss the effect of coarse-graining in this combinatorial encoding. Second, we recall how the Haar-random state on a $D$-partite state space of local dimension $N$ can be defined, how it can be analyzed with the help of Weingarten calculus, and how its properties compare to those of a Gaussian random tensor (see Sec.~\ref{subsub:Def-of-Haar} and App.~\ref{A:Wick}). This is the only random state we will explicitly consider in the present work, even though some of the results from Sec.~\ref{s:tree} will apply to more general distributions. In Sec.~\ref{sub:LU-inequivalent-reference-states}, we provide a complete classification of the $\LU$-orbits of HT states with a help of a limited number of trace-invariants: Thm.~\ref{th:RefState_Equiv} shows that $D$-partite HT states generate a $(2^D-D-1)$-parameter family of $\LU$-orbits, which can be perfectly distinguished by the joint data of multi-entropies and reflected multi-entropies (whose respective definitions are recalled in Sec.~\ref{sub:inv-from-lit}). In Sec.~\ref{sec:examples_asymptotic_rel}, we illustrate the two asymptotic notions of $\LU$-equivalence introduced in Sec.~\ref{sec:examples_asymptotic_rel}. In particular, we invoke the failure of large-$N$ factorization for Gaussian random tensors to prove that, unlike in $D=2$, in $D\geq 3$ it is \emph{rigorously impossible} to reproduce all the features of the leading-order entanglement structure of a Haar-random state with a deterministic state (see Prop.~\ref{prop:Haar_not_approx_deterministic}). 

\medskip

In {\bf Sec.~\ref{sec:LU-and-ref-states}}, we restrict our attention further to an infinite subclass of HT states: those that can be defined on a $D$-partite Hilbert space with uniform local dimension $N$, and obey an additional condition on the allowed dimension of any GHZ building block. As shown in Refs.~\cite{Looi:2011jrm,Nezami:2016zni}, pure tripartite stabilizer states are $\LU$-equivalent to a subset of these HT states, for some specific choices of parameters, and many results of this paper thereby apply to these states as well. This set-up allows formulating and investigating an asymptotic version of question \ref{Q:LU}: how to efficiently distinguish $\LU$-equivalence classes of HT states \emph{at leading order in the large-$N$ limit}? In other words, our goal in that section is to find a way to distinguish $\LU$-classes by the sole knowledge of the leading-order exponents of trace-invariant evaluations in an asympotic $1/N$ expansion. In Sec.~\ref{ss:discrimination}, we first show that such scaling parameters can be conveniently expressed in terms of combinatorial quantities attached to the representation of trace-invariants by $D$-colored graphs, which have already played a major role in the theory of random tensors, such as: the number of faces of a $D$-colored graph, its genus when $D=3$, and its so-called \emph{Gurau degree} for arbitrary $D\geq 3$. In addition, we introduce the new notion of \emph{$p$-complete degree} that naturally generalizes the Gurau degree (see App.~\ref{A:GurauDeg}). A comparison of the scaling exponents associated to such HT states and those associated to the Haar-random state singles out a combinatorial quantity known as the \textit{degree of compatibility}, which was recently introduced in Ref.~\cite{Collins2025}. Since the degree of compatibility always vanishes in the bipartite case and does not contribute to the leading-order exponents of HT states, its study is particularly relevant to question \ref{Q:typical} (and the related question \ref{Q:factorization}): it is the combinatorial quantity that allows distinguishing the \textit{typical} multipartite entanglement structure of the uniform random state from that of any HT state. However, the numerical search for invariants capable of distinguishing the subfamily of HT states associated with combinatorial quantities quickly becomes intractable, both due to the rapid growth of the quantities involved and the computational difficulty of efficiently evaluating certain quantities, particularly the degree of compatibility (which is defined implicitely via a $\min$, see Eq.~\eqref{eq:Delta}). This motivates a deeper investigation of these combinatorial quantities, as developed in Sec.~\ref{subsec:GraphStructure}. Such an analysis highlights a number of characteristic properties (melonic, planar, compatible, etc.) associated with the invariants typically considered in the literature. As a result, in Sec~\ref{sub:dist-power-inv-from-lit} we classify the invariants introduced in Sec.~\ref{sub:inv-from-lit} based on the values of the combinatorial quantities. In particular, we prove the compatibility or incompatibility of certain invariants (see also App.~\ref{sec:table-in-appendix}) and, building on numerical computations, conjecture the exact value of the degree of compatibility for specific families, a proof of which remains elusive to date. Lastly, we compute the entropies associated with the families of trace invariants introduced in Sec.~\ref{sub:inv-from-lit} for both HT and random states. This provides a preliminary approach to analyzing and understanding the meaning of these proposed multipartite entanglement entropies. Notably, multi-entropies \cite{Gadde2022,Penington2023} and reflected multi-entropies \cite{76vs-rxcs,Iizuka:2025elr} stand out as particularly insightful quantities: despite some limitations, they remain among the few entropies capable of distinguishing several HT states, thus highlighting their potential for characterizing and comparing entanglement patterns in multipartite systems. 

\

{\bf{Sec.~\ref{s:LO_LOCC_invTr}}} is where we return to question \ref{Q:monotones}, now specialized to HT states. In Sec.~\ref{subsec:combLO}, we first invoke necessary conditions previously derived in Sec.~\ref{ss:ent-monotones-trace-inv} to establish or exclude a number of $\LO$ relations among HT states. In particular, we use HT states to prove that the sufficient conditions on $\LO$ relations established in Cor.~\ref{cor:LO_monotones} are not necessary. This motivates a new approach in Sec.~\ref{sec:complete_LO_ref_states}, which relies on the exact (but somewhat implicit) characterization of the $\LO$ preorder from Prop.~\ref{prop:charac_LO}. Applying the conditions provided by this result to certain infinite families of $D$-colored graphs (that have been specifically introduced in Sec.~\ref{sec:Trace_Lit_Ref_States} for that purpose), we arrive at Thm.~\ref{th:LOD}. The proof of this theorem showcases the value of the combinatorial approach to trace-invariants we are following, which allows us to efficiently design and manipulate polynomial invariants of large degree. It results in a complete characterization of the $\LO$ preorder on HT states, that is both explicit and rather simple to state: the $\LU$-orbit of a HT state $\ket{\psi_\alpha}$ can be labeled by an integer-valued function $\alpha$, and given a second HT state $\ket{\psi_\beta}$, we have $\ket{\psi_\alpha}\toLO \ket{\psi_{\beta}}$ if and only if $\alpha$ divides $\beta$. In Sec.~\ref{subsubsec:FlowLOCC}, we turn to a similar analysis of the $\LOCC$ preorder on HT states. Even though we were not able to fully characterize this preorder at this stage, Props.~\ref{prop:move_LOCC-0}, \ref{prop:LOCC_second_order-rel} and \ref{prop:LOCC_QT} establish a number of sufficient conditions for two HT states to be related by $\LOCC$ (and, needless to say, all three results go beyond the $\LO$ relations established in Sec.~\ref{sec:complete_LO_ref_states}). Those partial results rely on a relatively simple subset of $\LOCC$ protocols, all based on measurements followed by at most one round of one-way communication; this includes in particular a quantum teleportation protocol in Prop.~\ref{prop:LOCC_QT}. We leave the question of a more complete characterization of the $\LOCC$ preorder on HT states open for future work.

\

In {\bf{Sec.~\ref{s:tree}}}, we explore combinatorial strategies allowing to \emph{design} trace-invariants with specified entanglement distinction properties (question \ref{Q:LU}). We focus again on the question of how best to distinguish HT states among themselves, or to distinguish a Haar-distributed random state from a deterministic HT state. To this effect, we start out by analysing the behavior of the combinatorial quantities used in the $\LU$-orbit classification  of Sec.~\ref{sec:LU-and-ref-states}, under the action of three types of binary operations: unions, flips, and vertex contractions of pairs of $D$-colored graphs (see Def.~\ref{def:operations}). Of particular interest are the results of Thm.~\ref{th:TreeDegComp} which, under favorable conditions, provide a recursive approach to the computation of the degree of compatibility of a graph. Without going into too much detail, one of the favorable conditions entailing this result is that the graph under consideration only admits \emph{tree-like} contributions to the leading order in the Haar-random state (see Def.~\ref{def:tree-like}, App.~\ref{A:DeltaAB}, as well as our companion paper \cite{Factorization2026}). Another case covered by Thm.~\ref{th:TreeDegComp} is when the graph under consideration is the result of a binary operation on two graphs of sufficiently small degree of compatibility. Altogether, Thm.~\ref{th:TreeDegComp} can be applied to a significant fraction of the invariants introduced in Sec.~\ref{sub:inv-from-lit}. The particularly interesting case of so-called \emph{maximally single-trace} invariants \cite{Ferrari2019} is discussed in and around Prop.~\ref{prop:MST}: while they all appear to be compatible for small enough values of $D$ and small enough number of vertices, recent results on the large-$N$ factorization problem in random tensor theory \cite{Facto2} suggest that incompatible maximally single-trace graphs should exist. Indeed, an example of such graph is provided in Fig.~\ref{fig:MST_incomp}. Some applications of binary operations and Thm.~\ref{th:TreeDegComp} are finally explored in Sec.~\ref{sec:applications-binary}. To begin with, Sec.~\ref{subsec:averageApprox} is devoted to the important question of the large-$N$ asymptotic evaluation of generalized Rényi entropies in the Haar-distributed random state of local dimension $N$. In contrast to the bipartite setting, taking the average of a trace-invariant and its logarithm are operations that can fail to approximately commute in the large-$N$ regime, precisely due to the non-factorization phenomenon uncovered in Ref.~\cite{Gurau2025}. Moreover, the non-positive character of certain multipartite trace-invariants implies that we cannot always bound a generalized Rényi entropy by a constant times $\ln(N)$, which also creates difficulties. Nonetheless, Props.~\ref{prop:concentration}, which is proven in full in our companion article (Ref.~\cite{Factorization2026}), establishes rigorous conditions on a $D$-colored graph and a random state that allow to determine the typical value of the associated generalized Rényi entropy in that state. Similarly, with an additional assumption on the invariant, Prop.~\ref{prop:asymptotic_R_G} (also proven in Ref.~\cite{Factorization2026}) allows computing the leading and subleading contributions to the expectation value of the generalized Rényi entropy at large $N$, in very much the same way as Eq.~\eqref{eq:DiffBipartite} can be derived in a bipartite setting. We apply these results to specific families of $D$-colored graphs in various corollaries. Finally, we conclude by providing concrete illustrations of tree-based designs of trace-invariants in Secs.~\ref{sec:appli_LU_refs}, \ref{subsubsec:DistinctionTree} and \ref{subsubsec:symmViaTree}.

\

Question \ref{Q:factorization} and its implications are more thoroughly explored in a companion publication, Ref.~\cite{Factorization2026}. Moreover, in a third upcoming publication, we will return to question \ref{Q:monotones} by introducing and studying the properties of richer families of entanglement monotones than the ones we restricted our attention to in the present work. 

\paragraph{Authors contributions.} All authors contributed equally to this work.

\paragraph{Acknowledgements.} We thank Austin Conner, Frédéric Holweck and Michael Walter for insightful discussions.
This work was supported by the  ANR JCJC project ``RTFPQuEnt" (ANR-25-CE40-5465). The IMB receives support from the EIPHI Graduate School (contract ANR-17-EURE-0002).  L.L. also acknowledges support  from the  ANR PRC project  ``TAGADA" (ANR-25-CE40-5672).

\section{Local unitary symmetry and its invariants}\label{sec:LUinvariants}

\subsection{Local unitary groups and local unitary equivalence}
\label{ss:LU-groups_LU-eq}

A $D$-partite quantum state space is a Hilbert space $\H$ together with a partition into $D$ local subsystems, represented by Hilbert spaces $\H_1, \ldots , \H_D$, so that $\H = \H_1 \otimes \cdots \otimes \H_D$. We will assume each local subsystem to be finite-dimensional, and denote by $N_c \eqdef \dim (\H_c)$ the dimension of the $c$-th subsystem. For notational convenience, we label such a $D$-partite state space by the $D$-tuple $\sF=(\H_1, \ldots , \H_D)$ of its local tensor factors and set $\H^\sF \eqdef \H_1 \otimes \cdots \otimes \H_D$. In the present paper, we focus exclusively on the entanglement structure of $D$-partite pure states, which can be represented by unit vectors in $\H^\sF$. We denote by $S(\H^\sF)$ the unit sphere in $\H^\sF$. Recall that a pure state $\ket{\psi}\in S(\H^\sF)$ is said to be \emph{separable} if it can be written in the form $\ket{\psi}= \ket{v_1} \otimes \cdots \otimes \ket{v_D}$ with $\ket{v_c} \in S(\H_c)$ for any $c\in \{1 \,, \ldots \,, D\}$; when this is not possible, $\ket{\psi}$ is said to be \emph{entangled}.

The group of \emph{local unitary transformations} of $\H^\sF$ is the group
\begin{equation}
\LU^{\sF} \eqdef \U(\H_1) \times \cdots \times \U(\H_D)\,,    
\end{equation}
where, for any Hilbert space $\H$, $\U(\H)$ denotes its unitary group. $\LU^{\sF}$ acts in an obvious way on $S(\H^\sF)$ and partitions it into orbits, the set of which we denote by
\begin{equation}\label{eq:space_of_orbits}
\cO^\sF \eqdef \bigl\{\LU^\sF \ket{\psi} \,\vert \, \ket{\psi} \in S(\H^\sF)\bigr\} \,.   
\end{equation}
The \emph{$\LU$-entanglement structure} of a pure state $\ket{\psi} \in \H^\sF$ is defined to be its orbit $\LU^\sF \ket{\psi}$. In more physical (but somewhat imprecise) terms, the $\LU$-entanglement structure of $\ket{\psi}$ captures any property of that state that is invariant under changes of orthonormal bases in the local subsystems $\H_1\, , \ldots \,, \H_D$. Accordingly, we will say that two states $\ket{\psi}, \ket{\phi}\in S(\H^\sF)$ are \emph{$\LU^\sF$-equivalent} -- noted $\ket{\psi} \underset{\LU^\sF}{\sim} \ket{\phi}$ -- if they lie in the same $\LU^\sF$ orbit; in other words:
\begin{equation}
    \ket{\psi} \underset{\LU^\sF}{\sim} \ket{\phi} \quad \Leftrightarrow \quad \LU^\sF \ket{\psi} =  \LU^\sF \ket{\phi} \quad \Leftrightarrow \quad \exists (U_1, \ldots , U_D) \in \U(\H_1) \times \ldots \times \U(\H_D)\,, \; \ket{\psi} = (U_1 \otimes \cdots \otimes U_D) \ket{\phi}\,.
\end{equation}
As an example, any two separable states in $\H^\sF$ are $\LU^\sF$-equivalent since, for any $c \in \{1, \ldots , D\}$, $\U(\H_c)$ acts transitively on $S(\H_c)$. By contrast, two entangled states are not necessarily in the same orbit: the partition of the quantum state space into local unitary orbits provides a much finer classification of entanglement than the simple dichotomy between separable and entangled states. The main purpose of the present article is to investigate the properties of particular families of polynomial invariants allowing to distinguish distinct $\LU$-entanglement structures, or equivalently, to separate points in the set of orbits $\cO^\sF$. Before turning to this question, let us introduce a slightly more general equivalence relation, which allows comparing $D$-partite states that do not live in the same $D$-partite state space. For this purpose, we will say that a map $V: \H^\sF \to \H^{\sF'}$ is a \emph{local isometry} between the $D$-partite state spaces $\sF=(\H_1, \ldots, \H_D)$ and $\sF'=(\H_1', \ldots, \H_D')$ if it takes the form of a tensor product $V= V_1 \otimes \cdots \otimes V_D$ where, for any $c \in \{1, \ldots , D\}$, $V_c : \H_c \to \H'_c$ is an isometry. We can then make the following observation: two states in $S(\H^\sF)$ are $\LU^\sF$-equivalent if and only if they are $\LU^{\sF'}$-equivalent in any $D$-partite state space $\H^{\sF'}$ in which they can be embedded via local isometries. 
\begin{lem}\label{lem:equivalence_under_isom}
Let $\sF= (\H_1 , \ldots , \H_D)$ be a $D$-partite state space and $\ket{\psi}, \ket{\phi}\in S(\H^\sF)$. Let $\sF'= (\H_1' , \ldots , \H_D')$ be a second $D$-partite state space such that: for any $c \in \{1, \ldots , D\}$, $\dim(\H_c) \leq \dim(\H_c')$. For any local isometries $V: \H^\sF \to \H^{\sF'}$ and $W: \H^\sF \to \H^{\sF'}$, we have
\begin{equation}
    \ket{\psi} \underset{\LU^\sF}{\sim}  \ket{\phi} \quad \Longleftrightarrow \quad  V\ket{\psi} \underset{\LU^{\sF'}}{\sim} W \ket{\phi}\,. 
\end{equation}
\end{lem}
\begin{proof}
The restrictions on the local dimensions of $\{\H'_c\}_{1 \leq c \leq D}$ ensure that local isometries from $\H^\sF$ to $\H^{\sF'}$ exist. 

    Let us assume that $\ket{\psi}$ and $\ket{\phi}$ are $\LU^\sF$-equivalent, and let $V_1 \otimes \cdots \otimes V_D: \H^\sF \to \H^{\sF'}$ and $W_1 \otimes \cdots \otimes W_D: \H^\sF \to \H^{\sF'}$ be local isometries. Since $\ket{\psi} \underset{\LU^\sF}{\sim} \ket{\phi}$, there is a local unitary $U_1 \otimes \cdots \otimes U_D$ on $\H^\sF$ such that
    \begin{equation}
    \ket{\psi} = (U_1 \otimes \cdots \otimes U_D) \ket{\phi}\,.    
    \end{equation}
    For any $c \in \{1, \ldots , D\}$, we can then find a unitary map $U_c': \H'_c \to \H'_c$ such that
    \begin{equation}
        V_c U_c = U'_c  W_c\,. 
    \end{equation}
    Indeed, we can define $U'_c \eqdef \rho_c \oplus \gamma_c$ as a map from $\H'_c\simeq \im(W_c) \oplus \im(W_c)^\perp$ to $\H'_c\simeq \im(V_c) \oplus \im(V_c)^\perp$, where: $\rho_c: \im(W_c) \to \im(V_c)$ is the isometry that maps any vector of the form $W_c (v)$ with $v \in \H_c$ to $V_c U_c (v)$; and $\gamma_c: \im(W_c)^\perp \to \im(V_c)^\perp$ is an arbitrary isometry. It follows that
    \begin{equation}
    (V_1 \otimes \cdots \otimes V_D) \ket{\psi} = (V_1 \otimes \cdots \otimes V_D) (U_1 \otimes \cdots \otimes U_D) \ket{\phi}= (U'_1 \otimes \cdots \otimes U'_D)  (W_1 \otimes \cdots \otimes W_D) \ket{\phi}  \,,    
    \end{equation}
    hence $(V_1 \otimes \cdots \otimes V_D)\ket{\psi} \underset{\LU^{\sF'}}{\sim} (W_1 \otimes \cdots \otimes W_D)\ket{\phi}$.
    
Reciprocally, let us assume that $(V_1 \otimes \cdots \otimes V_D)\ket{\psi} \underset{\LU^{\sF'}}{\sim} (W_1 \otimes \cdots \otimes W_D)\ket{\phi}$. We then have a local unitary $U'_1 \otimes \cdots \otimes U'_D$ on $\H^{\sF'}$ such that
\begin{equation}\label{eq:equiv_F'-1}
    (V_1 \otimes \cdots \otimes V_D) \ket{\psi} = (U'_1 \otimes \cdots \otimes U'_D)  (W_1 \otimes \cdots \otimes W_D) \ket{\phi}\,.
\end{equation}
Choosing a local orthonormal basis $\{\ket{i_c}_c\}_{1 \leq i_c \leq N_{c}}$ in each tensor factor $\H_c$, we can write the state $\ket{\phi}$ uniquely as
\begin{equation}
    \ket{\phi} = \sum_{i_1=1}^{N_1} \cdots \sum_{i_D=1}^{N_D} \phi_{i_1 \dots i_D} \ket{i_1}_1 \otimes \cdots \otimes \ket{i_D}_D\,,
\end{equation}
where $\{\phi_{i_1 \dots i_D}\}$ are complex coefficients. Let $I = I_1 \times \cdots \times I_D \subset \{1, \ldots , N_1\}\times \cdots \times \{1, \ldots , N_D\}$ denote the minimal rectangular support of the coefficients $\{\phi_{i_1 \dots i_D}\}$, that is to say: $I_1 \subset \{1, \ldots, N_1\}\,, \ldots \, , I_D \subset \{1, \ldots, N_D\}$ are the smallest subsets of indices  such that 
\begin{equation}\label{eq:equiv_F'-2}
    \ket{\phi} = \sum_{i_1\in I_1} \cdots \sum_{i_D\in I_D} \phi_{i_1 \dots i_D} \ket{i_1}_1 \otimes \cdots \otimes \ket{i_D}_D\,.
\end{equation}
For any $c \in \{1, \ldots, D\}$ and any $i \in I_c$, Eqs.~\eqref{eq:equiv_F'-1} and \eqref{eq:equiv_F'-2} implies that $U_c' W_c \ket{i}_c\in \im(V_c)$, and since $V_c$ is injective, there is a unique $\ket{e_i}_c \in \H_c$ such that $V_c \ket{e_i}_c = U'_c W_c \ket{i}_c$. Given that $U'_c$, $V_c$ and $W_c$ are all isometries, the family of vectors $\{\ket{e_i}_c\}_{i \in I_c}$ is orthonormal. We complete it into an orthonormal basis $\{\ket{e_i}_c\}_{1 \leq i \leq N_c}$ of $\H_c$, and define $U_c : \H_c \to \H_c$ as the unique unitary map that sends $\left(\ket{i}_c\right)_{1 \leq i \leq N_c}$ to $\left(\ket{e_i}_c\right)_{1 \leq i \leq N_c}$.  We then have
\begin{align}
(U_1 \otimes \cdots \otimes U_D) \ket{\phi} &= \sum_{i_1\in I_1} \cdots \sum_{i_D\in I_D} \phi_{i_1 \dots i_D} \ket{e_{i_1}}_1 \otimes \cdots \otimes \ket{e_{i_D}}_D \,,\\
&= \sum_{i_1\in I_1} \cdots \sum_{i_D\in I_D} \phi_{i_1 \dots i_D} (V_1)^\dagger U'_1 W_1\ket{i_1}_1 \otimes \cdots \otimes (V_D)^\dagger U'_D W_D\ket{i_D}_D \,, \\
&= ((V_1)^\dagger \otimes \cdots \otimes (V_D)^\dagger)(U'_1 \otimes \cdots \otimes U'_D)(W_1 \otimes \cdots \otimes W_D) \ket{\phi} \,,\\
&= \ket{\psi} \,.
\end{align}
As a result, $\ket{\psi} \underset{\LU^\sF}{\sim}\ket{\phi}$.
\end{proof}

We can now extend the notion of local unitary equivalence.
\begin{defi}
Let $\sF= (\H_1 , \ldots , \H_D)$ and $\sF'= (\H'_1 , \ldots , \H'_D)$ be $D$-partite state spaces and $\ket{\psi} \in S(\H^\sF)$, $\ket{\phi}\in S(\H^{\sF'})$. We say that $\ket{\psi}$ and $\ket{\phi}$ are $\LU$-equivalent -- noted $\ket{\psi} \underset{\LU}{\sim} \ket{\phi}$ -- if there exists a $D$-partite state space $\sF'' = (\H''_1, \ldots , \H''_D)$ together with local isometries $V: \H^\sF \to \H^{\sF''}$ and $W: \H^{\sF'} \to \H^{\sF''}$, such that
\begin{equation}
    V\ket{\psi} \underset{\LU^{\sF''}}{\sim} W\ket{\phi}\,.
\end{equation}
\end{defi}
\noindent As a direct consequence of Lem.~\ref{lem:equivalence_under_isom}, we can replace ``there exists a'' by ``for any'' in the previous definition; furthermore, the generalized notion of local unitary equivalence thus defined is compatible with the previously introduced one, in the sense of the following corollary.
\begin{cor}
 Let $\sF= (\H_1 , \ldots , \H_D)$ be a $D$-partite state space and $\ket{\psi}, \ket{\phi} \in S(\H^\sF)$. Then:
 \begin{equation}
     \ket{\psi} \underset{\LU}{\sim} \ket{\phi} \quad \Longleftrightarrow \quad \ket{\psi} \underset{\LU^\sF}{\sim} \ket{\phi}\,.
 \end{equation}
\end{cor}
The equivalence relation $\underset{\LU}{\sim}$ allows partitioning the set of all $D$-partite finite-dimensional quantum states into equivalence classes. The $\LU$-entanglement structure of a finite-dimensional $D$-partite state $\ket{\psi}$ (as previously defined) is uniquely characterized by the $\LU$-equivalence class of $\ket{\psi}$ in this larger space.  

\subsection{Definition and basic properties of trace-invariants}
\label{sub:trace-invariants}

Let us fix a $D$-partite state space $\sF=(\H_1 , \ldots , \H_D)$, and choose a reference orthonormal basis $\{\ket{i}_c\}_{1 \leq i \leq N_c}$ in each $\H_c$ ($1\leq c \leq D$). For any state $\ket{\psi}\in S(\H^{\sF})$, we denote by $\{\psi_{i_1 \dots i_D}\}$ its complex coordinates relative to this choice of local bases, namely
\begin{equation}\label{eq:tensor_components}
\ket{\psi} = \sum_{i_1=1}^{N_1} \cdots \sum_{i_D=1}^{N_D} \psi_{i_1 \dots i_D} \ket{i_1}_1 \otimes \cdots \otimes \ket{i_D}_D\,.
\end{equation}
A \emph{complex polynomial} on $S(\H^\sF)$ is a map $P:S(\H^\sF) \to \mathbb{C}$ that can be expressed as a complex polynomial function of the tensor components and their conjugates $\{\psi_{i_1 \dots i_D}\,, \bar\psi_{i_1 \dots i_D}\}$. We denote by $\cA^\sF$ the algebra of such polynomials, and by $\cA^\sF_{\inv}$ its subalgebra of \emph{$\LU^\sF$-invariant} polynomials: hence, $P\in \cA_{\inv}^\sF$ whenever, for any $U\in \LU^\sF$, $P\circ U = P$. We have the isomorphism $\cA^\sF \simeq \bb{R}[\H^\sF]\oplus i \bb{R}[\H^\sF]$, where $\bb{R}[\H^\sF]$ denotes the algebra of \emph{real polynomials} on $S(\H^\sF)$: that is, a $P\in \bb{R}[\H^\sF]$ is a map $P:S(\H^\sF) \to \bb{R}$ that can be expressed as a real polynomial function of the real tensor components $\{{\mathrm{Re}}(\psi_{i_1 \dots i_D})\,, {\mathrm{Im}}(\psi_{i_1 \dots i_D})\}$. We then have $\cA_{\inv}^\sF \simeq \bb{R}[\H^{\sF}]^{\LU^{\sF}} \oplus i \bb{R}[\H^{\sF}]^{\LU^{\sF}}$, where $\bb{R}[\H^{\sF}]^{\LU^{\sF}}$ denotes the subalgebra of $\LU^\sF$-invariant real polynomials. We can equivalently understand $\cA_{\inv}^\sF$ as an algebra of complex functions on the set of orbits $\cO^\sF$: for any $P \in \cA_{\inv}^\sF$ and $\ket{\psi}\in S(\H^\sF)$, we simply define $P(\LU^\sF \ket{\psi})\eqdef P(\ket{\psi})$.

Trace-invariants are particular homogeneous polynomials from $\cA_{\inv}^\sF$ that can be labeled by $D$-colored bipartite graphs. As we will explain here and in the next subsections, they are particularly convenient to work with, thanks to the following facts: a) the simple way in which they are defined; b) they generate the full invariant algebra $\cA_{\inv}^\sF$; c) they are \emph{universal}, in the sense that they can be defined on any $D$-partite state space. In fact, we will think of the collection of all trace-invariants as providing a generalization to $D\geq 3$ of the notion of entanglement spectrum of a bipartite pure state (in $D=2$, trace-invariants can be expressed in terms of elementary power sums of elements from this spectrum).

\medskip

In terms of the tensor components defined in Eq.~\eqref{eq:tensor_components}, trace-invariants are the class of homogeneous $\LU$-invariant polynomials obtained by contraction of $k\geq 1$ copies of the tensor $\psi=\{\psi_{i_1 \dots i_D}\}$ with $k$ copies of the conjugate tensor $\bar\psi=\{\bar\psi_{i_1 \dots i_D}\}$, with the constraint that an index of $\psi$ in position $c \in\{1, \ldots , D\}$ need to be summed over with an index of $\bar\psi$ in the same position (see also Refs.~\cite{817508,PhysRevA.58.1833,Hero2011,Vrana:2011ehx,Vrana:2011dka,BenGeloun:2013lim,gurau_random_2017,10.1063/5.0028856}). The label $c$ is often referred to as the \emph{color} of the index, and we will make use of this nomenclature in the rest of the paper. More concretely, we will make use of two complementary representations of trace-invariants, which we now recall. 

\begin{figure}[!ht]
    \centering
    \includegraphics[height = 1.5cm]{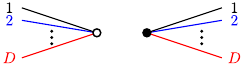}
    \caption{Graphical representation of $\psi$ and $\bar{\psi}$. 
    }
    \label{fig:NotationPSI}
\end{figure}

\paragraph{Trace-invariants and colored graphs.} Trace-invariants can be encoded into certain edge-decorated multigraphs, which are known in the random tensor literature as \emph{colored graphs} (see Refs.~\cite{Gurau:2011xp, Bonzom:2012hw, gurau_random_2017}).\footnote{Independently, the same structure has been described under the name of \emph{graph coverings} in Ref.~\cite{Hero2011,Vrana:2011ehx}. We will adopt the random tensor nomenclature throughout the present paper.}
We represent the tensor $\psi=\{\psi_{i_1 \dots i_D}\}$ and its complex conjugate $\bar \psi = \{\bar{\psi}_{i_1 \dots i_D}\}$ as $D$-valent white and black vertices, respectively (see Fig.~\ref{fig:NotationPSI}).
Each half-edge carries a \textit{color} corresponding to the position  $c \in \paa{1,\dots,D}$ of the index $i_c$. The summation of the indices in position $c$ (that is, of color $c$) between a $\bar \psi$ and a $\psi$ is represented graphically as an edge of color $c$ connecting the corresponding black and white vertices. With these graphical rules, a choice of invariant contraction pattern of all the indices of $k$ copies of $\psi$ and $k$ copies of $\bar{\psi}$ can be represented by a decorated graph with $2k$ vertices which is \emph{bipartite} (any edge connects a black vertex to a white vertex) and \emph{$D$-edge-colored} (each vertex has exactly one incident edge of color $c$ for every color $c \in \{1, \ldots, D\}$). We let $\cG_D$ (resp.~$\cG_D^{\conn}$) denote the set of bipartite $D$-edge-colored graphs (resp.~connected bipartite $D$-edge-colored graphs), and use the notation $\tr_{G}(\cdot)$ for the trace-invariant represented by a graph $G\in \cG_D$. Given $G \in \cG_D$, $k(G) \in \bb{N}^*$ will denote the number of white (resp.~black) vertices of $G$ (equivalently, $k(G)$ is the degree in the coordinates $\{\psi_{i_1 \dots i_D}\}$ of the homogeneous polynomial $\tr_G$). 
In the following, we simply refer to such graphs as \emph{$D$-colored graphs} (or simply as \emph{colored graphs}).  

\begin{figure}[!ht]
    \centering
    \includegraphics[height = 3cm]{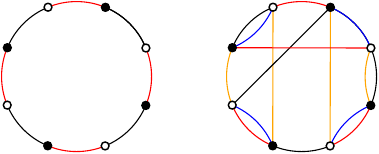}
    \caption{Left: a graph in $\cG_2^{\conn}$. Right: a graph in $\cG_4^{\conn}$.}
    \label{fig:BipAndNew}
\end{figure}

For $D=2$ and $\sF= (\H_1 , \H_2)$, a colored graph is a collection of cycles whose edges are alternatively of colors 1 and 2. We let $\rho_1=\tr_{\H_2}(\ket \psi\bra \psi) \in \End(\H_1)$ and $\rho_2 = \tr_{\H_1}(\ket \psi\bra \psi)\in \End(\H_2)$. Given a choice of local bases as in Eq.~\eqref{eq:tensor_components}, $\ket{\psi}$ is represented by an $N_1 \times N_2$ complex matrix $\psi = (\psi_{i_1,i_2})$, and $\rho_1$ (resp. $\rho_2$) is therefore represented by the matrix $\psi\psi^\dagger$ (resp. $\psi^\dagger \psi$). The trace-invariant associated with a colored graph $C_k$ consisting of a single cycle with $k$ white vertices (see the left of Fig.~\ref{fig:BipAndNew}) is then
\begin{align}
\label{eq:equiv-Ck-entanglement-Renyi-k}
    \tr_{C_k}(\ket{\psi}) &= \tr_{\H_1}( \rho_1^{k})= \tr\left( (\psi \psi^\dagger)^k\right) \\
    &=  \tr_{\H_2}( \rho_2^{k})= \tr\left( (\psi^\dagger \psi)^k\right)\,.\nonumber
\end{align} 
An arbitrary trace-invariant $\tr_G$ with $D=2$ decomposes as a product of such cyclic invariants: 
$\tr_G(\ket{\psi}) = \tr_{C_{k_1}}(\ket{\psi}) \cdots \tr_{C_{k_p}}(\ket{\psi})$, for some $p\ge 1 $ and $k_1\ge \cdots \ge k_p> 0$. 

\begin{figure}[!ht]
    \centering
    \includegraphics[height = 2cm]{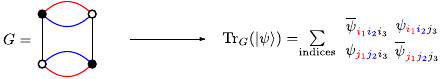}
    \caption{A graph in $\cG_3^{\conn}$ and corresponding trace-invariant $\tr_G$ evaluated on a state $\ket{\psi}$. }
    \label{fig:ExQuartic}
\end{figure}

An example of a connected colored graph with $D=4$ is shown on the right-hand side of Fig.~\ref{fig:BipAndNew}. Another example with $D=3$ is shown in Fig.~\ref{fig:ExQuartic}, together with the explicit expression of its associated trace-invariant. 

\paragraph{Trace-invariants and permutations.} Equivalently, it is possible to encode a \emph{labeled} colored graph into a list of permutations (see Ref.~\cite{BenGeloun:2013lim}). By ``labeling of a graph'' $G\in \cG_D$ with $k(G)=k$, we mean  a labeling of both its black and white vertices by integers from 1 to $k$. The data of $G$ together with a labeling is equivalent to $D$ permutations $\sigma_1, \ldots, \sigma_D\in S_k$, via the following rule: for any $s\in \{1,\ldots, k\}$ and any $c \in \{1, \ldots, D\}$, there is an edge of color $c$ connecting the white vertex with label $s$ to the black vertex with label $\sigma_c(s)$. This establishes a one-to-one correspondence between labeled $D$-colored graphs and $D$-tuple of permutations. For any $D$-tuple of permutations $\vec \sigma=(\sigma_1, \ldots, \sigma_D)\in S_k^D$, we may define:
\begin{equation} \label{eq:LUinv}
    \tr_{\vec \sigma}(\ket{\psi}) = \sum_{\rm{indices}} \pac{\prod_{c = 1}^D \prod_{b=1}^k \delta_{i^{(b)}_c \, j^{(\sigma_c(b))}_c} \pa{\prod_{a=1}^k \psi_{i^{(a)}_1 \dots i^{(a)}_D} \bar{\psi}_{j^{(a)}_1 \dots j^{(a)}_D}}} \,.
\end{equation}
Equivalently (see Refs.~\cite{Penington2023, Gadde2025}), we can express $\tr_{\vec \sigma}$ in terms of unitary representations of $S_k$, as
\begin{equation}\label{eq:trace_inv_permutation_rep}
    \tr_{\vec \sigma}(\ket{\psi}) = \bra{\psi}^{\otimes k} \rho_1(\sigma_1) \otimes \cdots \otimes \rho_D(\sigma_D) \ket{\psi}^{\otimes k}\,,
\end{equation}
where, for any $c \in \{1, \ldots, D\}$ and $\sigma \in S_k$, $\rho_c (\sigma)$ is the linear (and unitary) operator defined by
\begin{equation}\label{eq:permutation_reps}
    \rho_c(\sigma) : \; \H_c^{\otimes k} \ni \ket{v_1} \otimes \cdots \otimes \ket{v_k} \mapsto \ket{v_{\sigma^{-1}(1)}} \otimes \cdots \otimes \ket{v_{\sigma^{-1}(k)}} \,.
\end{equation}

Non-labeled graphs $G\in \cG_D$ are in bijection with $D$-tuples of permutations, up to relabeling of the vertices. More precisely, defining the equivalence relation
\begin{equation}\label{eq:equivalence_permutations}
\vec \sigma \sim \vec \tau \qquad \Longleftrightarrow \qquad \exists\, \eta, \nu \in S_k,\ \ \forall\, c,\ \sigma_c = \eta \tau_c \nu\,,
\end{equation}
each graph $G\in \cG_D$ with $k$ white vertices coincides with an element of the quotient ${S_k^D}{/\sim}$. We denote by $S_k^D(G)$ the subset of permutations in $S_k^D$ which project to $G$ under this quotient: equivalently, $S_k^D(G)$ can be identified with the set of labelings of the unlabeled graph $G$. Note that the definition of $\tr_{\vec\sigma}$ from Eq.~\eqref{eq:LUinv} does not depend on the labeling: it is a class function for $\sim$, meaning that it only depends on the underlying colored graph $G$ represented by $\vec\sigma$. The common value on the equivalence class is given by $\tr_G$:
\begin{equation} \label{eq:LUinv2}
   \forall\, \vec \sigma\in  S_k^D(G)\,, \quad \tr_G = \tr_{\vec \sigma} \,.
\end{equation}

For $D=2$, one has that 
$$
(\sigma_1, \sigma_2) \sim (\tau_1, \tau_2)  \qquad \Leftrightarrow \qquad \exists\, \eta \in S_k,\ \ \sigma_1\sigma_2^{-1} = \eta \tau_1\tau_2^{-1}\eta^{-1}\,,
$$
in which case $\sigma_1\sigma_2^{-1}$ and $\tau_1\tau_2^{-1}$ have the same cycle structure. As stated above, a (non-labeled) graph $G$ is a collection of cycles $C_{k_1}, \ldots, C_{k_p}$, $p\ge 1 $ and $k_1\ge \cdots \ge k_p> 0$. The different labelings of such a $G$ -- the pairs of permutations $(\sigma_1, \sigma_2)$ which are elements of $S_k^D(G)$ -- are such that the cycle structure of $\sigma_1\sigma_2^{-1}$ is $k_1\ge \cdots \ge k_p$. 

\paragraph{Trivial trace-invariants.}  Given $G \in \cG_D$, we will say that $G$ (resp.~$\tr_G$) is \emph{trivial} if it can be represented by a $D$-tuple of permutations $\sigma = (\sigma, \ldots , \sigma )$ with $\sigma \in S_{k(G)}$ (or, equivalently, if it can be represented by the trivial $D$-tuple $(\mathrm{id}, \ldots , \mathrm{id})$). In graphical terms, such a $G$ has $k(G)$ connected components, each of which is the unique connected $D$-colored graph with two vertices. We will denote by $\cG_D^{\triv}$ the subset of trivial graphs in $\cG_D$. Trivial invariants carry no useful information, in the sense that: for any $D$-partite pure state $\ket{\psi}$ (which is a unit vector)
\begin{equation}
    \forall G \in \cG_D^{\triv},\qquad \tr_G (\ket\psi ) = \left\langle \psi | \psi \right\rangle^{k(G)} = 1\,.
\end{equation}

\paragraph{Properties of trace-invariants.} For later purposes, let us collect a few useful properties of trace-invariants. We start with the following simple lemma, which follows straightforwardly from the definitions, and which we therefore recall without proof.
\begin{lem}\label{lem:tr}
    Let $D\geq 2$ and $G \in \cG_D$.
    \begin{enumerate}
        \item The trace-invariant $\tr_G$ is multiplicative under tensor products, namely: for any finite-dimensional $D$-partite pure states $\ket{\psi}$ and $\ket{\varphi}$, we have 
        \begin{equation} \label{eq:tensorprod}
            \tr_G(\ket{\psi} \otimes \ket{\phi}) = \tr_G(\ket\psi) \cdot \tr_G(\ket\phi)\,.
        \end{equation}
        \item The map $\cG_D \ni H \mapsto \tr_H$ is multiplicative under disjoint unions of graphs, namely: for any $H_1, H_2 \in \cG_D$, we have
        \begin{equation} \label{eq:prodUnion}
            \tr_{H_1 \sqcup H_2} =\tr_{H_1}\tr_{H_2}\,. 
        \end{equation}
        In particular, if $G_1,\dots,G_{\kappa(G)}$ (with $\kappa(G)\in \mathbb{N}^*$) denote the connected components of $G$, then
        \begin{equation} \label{eq:nonconn}
            \tr_G = \prod_{i = 1}^{\kappa(G)} \tr_{G_i} \,.
        \end{equation} 
    \end{enumerate}
\end{lem}

Next, we have the following useful bound on the modulus of a trace-invariant.
\begin{lem}\label{lem:modulus_invariants}
    Let $\sF= (\H_1 , \ldots , \H_D)$ be a finite-dimensional $D$-partite state space, $\ket{\psi}\in S(\H^\sF)$, $G \in \cG_D$, and 
    $\vec\sigma = (\sigma_1, \ldots , \sigma_D)\in  S_{k(G)}^D(G)$ (so that $\tr_G = \tr_{\vec\sigma}$).
    The following properties hold.
    \begin{enumerate}
        \item $|\tr_G (\ket{\psi})|\leq1$.
        \item $|\tr_G (\ket{\psi})|=1$ if and only if $\ket{\psi}^{\otimes k(G)}$ is an eigenstate of $\rho_1(\sigma_1)\otimes \cdots \otimes \rho_D(\sigma_D)$.
    \end{enumerate}
\end{lem}
\begin{proof}
Let $k=k(G)$. The operator $\rho_1(\sigma_1) \otimes \cdots \otimes \rho_D(\sigma_D)$ is unitary on $(\H^\sF)^{\otimes k}$ (since $\rho_c (\sigma_c)$ is unitary on $\H_c^{\otimes k}$, for any $c$). Hence, invoking the Cauchy-Schwarz inequality and Eq.~\eqref{eq:trace_inv_permutation_rep} (as in Ref.~\cite{Gadde2025}), we have:
\begin{equation}    
|\tr_G (\ket{\psi})| = |\tr_{\vec \sigma}(\ket{\psi})| = |\bra{\psi}^{\otimes k} \rho_1(\sigma_1) \otimes \cdots \otimes \rho_D(\sigma_D) \ket{\psi}^{\otimes k}| \leq \| \ket{\psi}^{\otimes k}\| \,  \| \rho_1(\sigma_1) \otimes \cdots \otimes \rho_D(\sigma_D) \ket{\psi}^{\otimes k}\| = 1 \,,
\end{equation}
which yields the first claim. Furthermore, the previous inequality is saturated if and only if the vectors $\ket{\psi}^{\otimes k}$ and $\rho_1(\sigma_1) \otimes \cdots \otimes \rho_D(\sigma_D) \ket{\psi}^{\otimes k}$ are linearly-dependent, or in other words, if and only if $\rho_1(\sigma_1) \otimes \cdots \otimes \rho_D(\sigma_D) \ket{\psi}^{\otimes k}$ is proportional to $\ket{\psi}^{\otimes k}$. This yields the second claim.
\end{proof}

\subsection{Separation of \texorpdfstring{$\LU$}{LU} orbits by trace-invariants}

It is a well-known fact that the algebra $\cA_{\inv}^{\sF}$ separates the orbits $\cO^\sF$ presented in Eq.~\eqref{eq:space_of_orbits}; we provide a proof here for completeness, which closely follows Ref.~\cite{meyer2002invariants}. 
\begin{prop}
    $\cA_{\inv}^{\sF}$ separates points in $\cO^\sF$.
\end{prop}
\begin{proof}
    Let us show that $\bb{R}[\H^ \sF]^{\LU^\sF}$ separates orbits; the proposition will follow since $\cA_{\inv}^\sF \simeq \bb{R}[\H^{\sF}]^{\LU^{\sF}} \oplus i \bb{R}[\H^{\sF}]^{\LU^{\sF}}$. Let $\ket{\psi}, \ket{\phi}\in S(\H^\sF)$ be such that $\ket{\psi} \underset{\LU^\sF}{\nsim} \ket{\phi}$. Since $\LU^\sF$ is a compact group, the orbits $\LU^\sF \ket{\psi}$ and $\LU^\sF \ket{\phi}$ are both compact in $\H^\sF$ (seen as a $\bb{R}$-vector space). They can thus be separated by the topology of $\H^\sF$. By Uryshon's lemma, there is thus a continuous function $f:S(\H^\sF)\to \bb{R}$ such that:
    \begin{equation}
        \restr{f}{\LU^\sF\ket{\psi}} = 0 \,, \qquad \restr{f}{\LU^\sF\ket{\phi}} = 1\,. 
    \end{equation}
    Define $\tilde{f}:S(\H^\sF)\to \bb{R}$ by group-averaging as
    \begin{equation}
        \tilde{f}(\cdot)= \int_{\LU^\sF} f(U \cdot ) \, \extd U\,,
    \end{equation}
    where $\extd U$ denotes the normalized Haar measure on $\LU^\sF$. $\tilde{f}$ is a $\LU^\sF$-invariant function such that $\restr{f}{\LU^\sF\ket{\psi}} = 0$ and $\restr{f}{\LU^\sF\ket{\phi}} = 1$. By the Stone-Weierstrass theorem, it is possible to approximate $\tilde{f}$ by a polynomial: more precisely, we can find $P\in \bb{R}[\H^\sF]$ such that, for any $\ket{v}\in S(\H^\sF)$, $\left| \tilde{f}(\ket{v})- P(\ket{v})\right| \leq \frac{1}{4}$. Defining $\tilde{P}(\cdot)\eqdef \int_{\LU^\sF} P(U \cdot ) \, \extd U$, we have $\tilde{P}\in \bb{R}[\H^\sF]^{\LU^\sF}$ and: for any $\ket{v}\in \H^\sF$,
    \begin{align}
        \left| \tilde{f}(\ket{v}) - \tilde{P}(\ket{v}) \right| = \left| \int_{\LU^\sF} \left( \tilde{f}(U \ket{v}) - P(U \ket{v}) \right) \extd U \right| \leq  \int_{\LU^\sF} \left| \tilde{f}(U \ket{v}) - P(U \ket{v}) \right| \extd U \leq \frac{1}{4}\,.
        \end{align}
As a result, $\tilde{P}(\ket{\psi})\leq \frac{1}{4}$ and $\tilde{P}(\ket{\phi})\geq \frac{3}{4}$, so $\tilde{P}$ is an invariant polynomial that separates the orbits of $\ket{\psi}$ and $\ket{\phi}$.
\end{proof}

The connected trace-invariants introduced in the previous subsection are particularly useful because they generate the full algebra of polynomial invariants $\cA_{\inv}^\sF$.
\begin{prop}
    $\left\{ \tr_G \, |\, G \in \cG_D^{\conn} \right\}$ is a generating set of $\cA^\sF_{\inv}$.
\end{prop}
\begin{proof}
    $\cA^\sF$ decomposes as a direct sum of subalgebras of degree $(k,\bar k)\in \bb{N}^2$, where $k$ denotes the degree in the tensor entries $\{\psi_{i_1 \dots i_D}\}$ and $\bar k$ denotes the degree in the complex conjugated variables $\{\bar\psi_{i_1 \dots i_D}\}$. The action of $\LU^\sF$ on $\cA^\sF$ preserves this degree, hence $\cA^\sF_{\inv}$ is generated by homogeneous polynomials of fixed degree. Let $P\in \cA_{\inv}^\sF$ of degree $(k,\bar k)\in \bb{N}^2$. For any phase $\eta \in \U(1) \subset \LU^\sF$, we have $P(\cdot) = P(\eta \cdot )=\eta^k \bar\eta^{\bar k}P(\cdot)$, which imposes that $k = \bar k$. It follows that there exists some linear operator $A \in \End((\H^\sF)^{\otimes k})$ such that
    \begin{equation}\label{eq:inv_pol-1}
        \forall \ket{\psi}\in S(\H^\sF)\,, \qquad P(\ket{\psi}) = \bra{\psi}^{\otimes k} A\ket{\psi}^{\otimes k} \,.
    \end{equation}
Since $\End((\H^\sF)^{\otimes k})\simeq \End(\H_1^{\otimes k}) \otimes \cdots \otimes \End(\H_D^{\otimes k})$, one can decompose $A$ as
\begin{equation}
    A = \sum_{\ell = 1}^p A_{1}^{(\ell)} \otimes \cdots \otimes A_{D}^{(\ell)}\,,
\end{equation}
where $p \in \bb{N}^*$ and: for any $\ell\in \{1,\ldots , p\}$ and any $c \in \{1, \ldots, D\}$, $A_{c}^{(\ell)} \in \End(\H_c^{\otimes k})$. We can now average Eq.~\eqref{eq:inv_pol-1} over $\LU^\sF$, to obtain
\begin{equation}\label{eq:inv_pol-2}
        \forall \ket{\psi}\in S(\H^\sF)\,, \qquad P(\ket{\psi}) = \sum_{\ell=1}^p \bra{\psi}^{\otimes k} B_1^{(\ell)} \otimes \cdots \otimes B_D^{(\ell)} \ket{\psi}^{\otimes k}\,,
        \end{equation}
where, for any $\ell\in \{1,\ldots , p\}$ and any $c \in \{1, \ldots, D\}$,
\begin{equation}
    B_c^{(\ell)} \eqdef \int_{\U(\H_c)} U^{\otimes k}  A_c^{(\ell)}(U^\dagger)^{\otimes k} \, \extd U
\end{equation}
is invariant under the diagonal adjoint action of $\U(\H_c)$ on $\End(\H_c^{\otimes k})$. In other words, $B_c^{(\ell)}$ is in the commutant of the subalgebra $\Span\left\{ U^{\otimes k}\, \vert \, U \in \U(\H_c) \right\}$ of $\End(\H_c^{\otimes k})$. It is straightforward to check that, for any permutation $\sigma \in S_k$, this commutant contains the linear operator $\rho_c (\sigma)$ introduced in Eq.~\eqref{eq:permutation_reps}. Furthermore, a main result of Schur-Weyl duality asserts that the full commutant is actually spanned by such permutation operators. It follows that
\begin{equation}
    B_c^{(\ell)} \in \Span \left\{ \rho_c(\sigma)\, | \, \sigma\in S_k \right\}\,,
\end{equation}
and therefore that
\begin{equation}
 \sum_{\ell =1}^p B_1^{(\ell)} \otimes \cdots \otimes B_D^{(\ell)} \in \Span \left\{ \rho_1(\sigma_1)\otimes \cdots \otimes\rho_D(\sigma_D) \, | \, \sigma_1 \,, \ldots \,, \sigma_D \in S_k \right\}\,.
\end{equation}
We conclude by observing that, for any $\sigma_1 \,, \ldots \,, \sigma_D \in S_k$, the map 
\begin{equation}
    \H^\sF \ni \ket{\psi} \mapsto \bra{\psi}^{\otimes k} \rho_1(\sigma_1) \otimes \cdots \otimes \rho_D(\sigma_D) \ket{\psi}^{\otimes k} \in \bb{C}
\end{equation}
is a trace-invariant expressed in the form of Eq.~\eqref{eq:trace_inv_permutation_rep}. As we have seen, the latter decomposes as a product of connected trace-invariants, which concludes the proof.
\end{proof}
\noindent As a result, we obtain the following characterization of $\LU$-equivalence.
\begin{prop}\label{prop:charac_LU-1}
    Let $\ket{\psi}$ and $\ket{\phi}$ be two finite-dimensional $D$-partite states. Then:
    \begin{equation}
        \ket{\psi} \underset{\LU}{\sim} \ket{\phi} \qquad \Longleftrightarrow \qquad \forall G \in \cG^{\conn}_D\,, \quad \tr_G (\ket{\psi}) = \tr_G (\ket{\phi})\,. 
    \end{equation}
\end{prop}
\noindent Furthermore, it was shown in \eg Refs.~\cite{Vrana:2011ehx, Collins2025} that the previous characterization cannot be improved if no supplementary information on the states $\ket{\psi}$ and $\ket{\phi}$ is provided: indeed, connected trace-invariants of degree less or equal to $k \in \mathbb{N}^*$ are \emph{independent} on any $D$-partite Hilbert space $\H_1 \otimes \cdots \otimes \H_D$ with $\dim(\H_1), \ldots , \dim(\H_D)$ large enough (see Ref.~\cite{Collins2025} for a more precise statement). However, given some restrictions on the dimensions of the $D$-partite state spaces into which $\ket{\psi}$ and $\ket{\phi}$ can be embedded, one can ask whether $\LU$-equivalence can be asserted by evaluating a finite number of trace-invariants. The answer is positive, and well-known in the bipartite case $(D=2)$.
\begin{prop}
    Let $\sF=(\H_1 , \H_2)$ be a bipartite state space, and $N= \min\left(\dim(\H_1),\dim(\H_2)\right)$. For any $\ket{\psi}, \ket{\phi}\in S(\H^\sF)$, we have
    \begin{equation}
        \ket{\psi} \underset{\LU}{\sim} \ket{\phi} \qquad   \Longleftrightarrow \qquad \forall G\in \cG_2^{\conn}\,, \quad k(G) \leq N \;\Rightarrow\; \tr_G (\ket{\psi}) = \tr_G (\ket{\phi})\,.  
    \end{equation}
\end{prop}
\begin{proof}
The direct implication ($\Rightarrow$) is immediate given Prop.~\ref{prop:charac_LU-1}. For the reverse direction, let us assume, for definiteness, that $N=\dim(\H_1)$. Suppose that $\tr_G (\ket{\psi}) = \tr_G (\ket{\phi})$ for any $G \in \cG_2^{\conn}$ with $k(G)\leq N$. We can write $\ket{\psi}$ (resp.~$\ket{\phi}$) in a local unitary basis in terms of matrix components $\{\psi_{ij}\}$ (resp.~$\{\phi_{ij}\}$). The $N \times N$ matrices $\psi\psi^\dagger$ and $\phi\phi^\dagger$ are Hermitian, positive, and of unit trace: indeed, they are the reduced density matrices associated with $\ket{\psi}$ and $\ket{\phi}$ in subsystem $\H_1$. Let us denote by $\pmb{p}_\psi = ( p_1, \ldots , p_N )$ the vector of eigenvalues of $\psi \psi^\dagger$ (with, say, $p_1 \geq \ldots \geq p_N$) and by $\pmb{p}_\phi =(p'_1 , \ldots , p'_{N})$ the vector of eigenvalues of $\phi \phi^\dagger$ (with $p'_1 \geq \ldots \geq p'_N$). We can rewrite the assumed condition on $\ket{\psi}$ and $\ket{\phi}$ in terms of power sums of these eigenvalues:
\begin{equation}
    \forall k \in \{1, \ldots , N\}\,, \qquad \sum_{i=1}^N (p_i)^k = \sum_{i=1}^N (p'_i)^k\,.
\end{equation}
Since the first $N$ power sums in $\{ p_1 , \ldots , p_N \}$ (resp.~$\{p'_1 , \ldots , p'_{N}\}$) generate the whole ring of symmetric polynomials in $\{ p_1 , \ldots , p_N\}$ (resp.~$\{ p'_1 , \ldots , p'_N\}$), we conclude that 
\begin{equation}
    \sum_{i=1}^{N} (p_i)^k = \sum_{i=1}^{N} (p'_i)^k
\end{equation}
for any $k \in \bb{N}^*$. Equivalently, we have
\begin{equation}
    \forall G \in \cG_2^{\conn}\,, \quad \tr_G (\ket{\psi}) = \tr_G (\ket{\phi})\,,
\end{equation}
and therefore, by Prop.~\ref{prop:charac_LU-1}, we conclude that $\ket{\psi}$ and $\ket{\phi}$ are $\LU$-equivalent.
\end{proof}
\begin{rem}
    It is more typical in the literature to express the previous characterization directly in terms of the eigenvalues, that is, one usually writes: $\ket{\psi} \underset{\LU}{\sim} \ket{\phi}$ if and only if $(p_1 , \ldots , p_N) = (p'_1, \ldots , p'_N)$. The list of eigenvalues $(p_1 , \ldots , p_N)$ is referred to in this context as the \emph{entanglement spectrum} of $\ket{\psi}$. The two formulations are, of course, equivalent; the reason why we prefer to work with power sums of the eigenvalues is that they are straightforwardly interpreted as connected trace-invariants, which continue to provide a natural generating set of invariants when $D\geq 3$.
\end{rem}
In the multipartite setting, finding generating and independent sets of trace-invariants is much less straightforward. However, it is known that the algebra $\cA_{\inv}^\sF$ of a $D$-partite state space $\sF$ is finitely generated also when $D\geq 3$, and explicit degree bounds on generators have been computed in Refs.~\cite{derksen2015computational, Turner2017}. Those results imply the following proposition, which we state without proof.\footnote{The proofs of such results we are aware of start by recasting the orbit problem for the local unitary group into an analogous problem for a group that is a product of general linear groups. Contrary to the former, the latter is a reductive group, which allows us to exploit algebraic geometry tools that are not directly available otherwise. Reviewing this approach further would take us too far from the core of the present contribution.}
\begin{prop}\label{prop:bound_degree}
    Let $D\geq 3$, $\sF = (\H_1 , \ldots , \H_D)$ a $D$-partite state space, and $\ket{\psi}, \ket{\phi} \in S(\H^\sF)$. There exists $k^\sF_{\max} \in \bb{N}^*$ such that:
     \begin{equation}
        \ket{\psi} \underset{\LU}{\sim} \ket{\psi} \qquad \Longleftrightarrow \qquad \forall G \in \cG^{\conn}_D\,, \quad k(G) \leq k_{\max}^\sF \;\Rightarrow\;  \tr_G (\ket{\psi}) = \tr_G (\ket{\phi})\,. 
    \end{equation}
Moreover, the previous equivalence holds with (see Corollary 4.11 of Ref.~\cite{Turner2017}) 
\begin{equation}
    k_{\max}^\sF = \max\left( 2, \frac{3}{8} \max_{1 \leq c \leq D}\{ \dim(\H_c)\} \left(\prod_{c=1}^D \dim(\H_c)\right)^4 \left(2D\right)^{2\delta} \right)\,,
\end{equation}
where $\displaystyle\delta = \sum_{c=1}^D (\dim(\H_c) - 1)$. 
\end{prop}

Our goal in the following will be to compare the entanglement structures of various $D$-partite states by evaluating trace-invariants on them. That is, we will consider two $D$-partite entanglement structures to be ``close'' whenever the evaluation of any trace-invariant on them gives ``close'' results. At the most basic level, this idea can be implemented by equipping the space of orbits $\cO^\sF$ of a $D$-partite state space $\sF$ with a topology that makes all trace-invariants continuous (or, equivalently, that makes any $f \in \cA_{\inv}^\sF$ continuous). The coarsest such topology is the Zariski topology, namely, the topology on $\cO^\sF$ whose closed sets are of the form $V(\cI)\eqdef\{O \in \cO^\sF \, | \, \forall P \in \cI \,, P(O) = 0\}$, where $\cI$ is a subset of $\cA_{\inv}^\sF$.\footnote{One can check that, so defined, arbitrary intersections (resp.~finite unions) of Zariski-closed sets are Zariski-closed. Furthermore, let $T$ denote another topology on $\cO^\sF$ that makes any element of $\cA_{\inv}^\sF$ continuous. To show that $T$ is finer than the Zariski topology, we can consider a Zariski-closed set $V(\cI)$ with $\cI \subset \cA_{\inv}^\sF$, and explain why $V(\cI)$ is also a closed subset of $T$. By definition, $V(\cI)=\underset{P \in \cI}{\bigcap} P^{-1}(\{0\})$, and since any $P\in \cI$ is continuous with respect to $T$, $P^{-1}(\{0\})$ must be closed in $T$ ($\{0\}$ being closed in $\bb{C}$, considered with its standard topology). $V(\cI)$ is thus closed in $T$ as an intersection of closed subsets.} Any quantitative approach to entanglement based on trace-invariants can only be consistent if $\cO^\sF$ is equipped with a topology that is finer than the Zariski topology. We will work under this very mild assumption in the rest of the paper, without committing to any particular such choice. In future works, it will be particularly interesting to investigate whether $\cO^\sF$ can be equipped with an \emph{operationally-meaningful} metric structure that can be directly expressed in terms of trace-invariants (note the emphasis on ``operationally-meaningful'').

\subsection{\texorpdfstring{$\LU$}{LU}-equivalence of random states, and asymptotic \texorpdfstring{$\LU$}{LU}-equivalence of sequences of states}\label{sec:LU_random_asymptotic}

To be able to discuss properties of random entangled states, let us now generalize the notion of $\LU$-equivalence slightly. We will say that $\ket{\psi}$ is a \emph{random $D$-partite state} if $\ket{\psi}$ is a random variable on $S(\H^\sF)$ for some $D$-partite state space $\sF$, such that the expectation value of any trace-invariant of $\ket{\psi}$ is well-defined and finite: 
\begin{equation}
    \forall G \in \cG_D\,, \quad  \mean{\tr_G(\ket{\psi})} \in \bb{C}\,.
\end{equation}
There is then a natural generalization of the notion of $\LU$-equivalence that applies to such random states while remaining consistent with the previously described deterministic setting.
\begin{defi}\label{def:LU-random}
Let $\ket{\psi}$ and $\ket{\phi}$ be two $D$-partite random states. We say that $\ket{\psi}$ is $\LU$-equivalent to $\ket{\phi}$ -- noted $\ket{\psi} \underset{\LU}{\sim} \ket{\phi}$ -- if:
\begin{equation}
    \forall G \in \cG_D\,, \qquad \mean{\tr_G(\ket{\psi})} = \mean{\tr_G(\ket{\phi})}\,.
\end{equation}
\end{defi}

\begin{rem}
    Note that, crucially, $G$ is not restricted to be connected in the previous definition, unlike in Prop.~\ref{prop:charac_LU-1}. However, for deterministic states (\ie random states whose distributions are Dirac delta functions), $\mean{\tr_G(\ket{\psi})}$ and $\mean{\tr_G(\ket{\phi})}$ factorize over the connected components of $G$, and one recovers the characterization of $\LU$-equivalence given in Prop.~\ref{prop:charac_LU-1}. 
\end{rem}

In the present paper, we are particularly interested in distinguishing entangled states in the asymptotic limit of $N\to +\infty$, where $N$ is a dimension parameter. We formalize two weaker notions of $\LU$-equivalence that are sufficient for such distinctions: asymptotic $\LU$-equivalence on the one hand, and $\LU$-equivalence in scaling on the other hand.
\begin{defi} \label{def:LargeN_LU_eq}
    Let $\psi=(\ket{\psi_N})_{N\in \bb{N}^*}$ and $\phi=(\ket{\phi_N})_{N\in \bb{N}^*}$ be two sequences of $D$-partite deterministic or random states. 
    \begin{enumerate}
        \item For any $G \in \cG_D$, we say that the pair $(G, \psi)$ -- or, by slight abuse of notation, $\mean{\tr_G \left( \ket{\psi_N}\right)}$ -- obeys the \emph{large-$N$ Ansatz} if one can find $\mu_G(\psi)\in \mathbb{C}^*$ and $s_G (\psi) \in \mathbb{R}$ such that:\footnote{Note that, owing to the fact that $|\tr_G (\cdot )|$ is bounded by $1$, $s_G(\psi)$ is necessarily non-positive.}
        \begin{equation}\label{eq:Ansatz_large-N}
            \mean{\tr_G(\ket{\psi_N})} \underset{N \to \infty}{=} \mu_G(\psi) N^{s_G(\psi)} \left( 1 +  O\pa{1/N}\right) \,.
        \end{equation}
        \item We say that $\psi$ and $\phi$ are \emph{asymptotically $\LU$-equivalent} -- noted $\ket{\psi_N} \underset{N\to\infty}{\sim}\ket{\phi_N}$ -- if:
        \begin{equation}
        \label{eq:asympt-LU-eq-random}
            \forall G \in \cG_D\,,  \qquad \mean{\tr_G(\ket{\psi_N})} \underset{N\to\infty}{\sim} \mean{\tr_G(\ket{\phi_N})}\,.
        \end{equation}
        \item We say that $\psi$ and $\phi$ are \emph{$\LU$-equivalent in scaling} -- noted $\ket{\psi_N} \underset{N\to\infty}{\approx}\ket{\phi_N}$ -- if: for any $G \in \cG_D$, there exists $\lambda_G\in\bb{C}\setminus \{0\}$ such that 
        \begin{equation}
             \mean{\tr_G(\ket{\psi_N})} \underset{N\to\infty}{\sim} \lambda_G \mean{\tr_G(\ket{\phi_N})}\,.
        \end{equation}
        \item Let us assume that the pairs $\{ (G, \psi), (G, \phi) \}_{G \in \cG_D}$ all obey the large-$N$ Ansatz, meaning that: for any $G\in \cG_D$, we can find $s_G(\psi)\in \bb{R}$ (resp.~$s_G(\phi) \in \bb{R}$) and $\mu_G(\psi)\in \bb{C}^*$ (resp.~$\mu_G(\phi)\in \bb{C}^*$) such that
        \begin{equation}
            \mean{\tr_G(\ket{\psi_N})} \underset{N \to \infty}{=} \mu_G(\psi) N^{s_G(\psi)} +  O\pa{N^{s_G(\psi)-1}} \,, \quad \mean{\tr_G(\ket{\phi_N})} \underset{N \to \infty}{=} \mu_G(\phi) N^{s_G(\phi)} +  O\pa{N^{s_G(\phi)-1}}\,.
        \end{equation}
        We then have:
        \begin{equation}
            \ket{\psi_N} \underset{N\to\infty}{\sim}\ket{\phi_N} \qquad \Longleftrightarrow \qquad \forall G \in \cG_D\,, \quad \begin{cases}s_G (\psi)= s_G(\phi)\\
                \mu_G (\psi)= \mu_G(\phi) 
                \end{cases}
        \end{equation}
        and
        \begin{equation}
            \ket{\psi_N} \underset{N\to\infty}{\approx}\ket{\phi_N} \qquad \Longleftrightarrow \qquad \forall G \in \cG_D\,, \quad s_G (\psi) = s_G(\phi)\,.
        \end{equation}
    \end{enumerate}
\end{defi}
\begin{rem}\label{rem:barG_1}
Given $G \in \cG_D$, it is convenient to denote by $\bar G$ the colored graph obtained from $G$ by flipping the colors of all its vertices. One then has $\tr_{\bar{G}}(\cdot)= \overline{\tr_G (\cdot)}$. According to the first item in the previous list of definitions, one trivially has that: $(G, \psi)$ obeys the large-$N$ Ansatz if and only if $(\bar G , \psi )$ does. Furthermore, in that case: $s_{\bar G} (\psi) = s_{ G} (\psi)$ and $\mu_{\bar{G}} (\psi) = \overline{\mu_{ G} (\psi)}$.
\end{rem}
\begin{rem}
    When $\psi$ and $\phi$ are both deterministic sequences, the previous definitions can be simplified: expectation values can be omitted, and since trace-invariants factorize over their connected components, it suffices to compare connected invariants to assess equivalence. Furthermore, for $\psi$ and $\phi$ deterministic, $s_G$ is clearly additive  under decomposition into connected components, while $\mu_G$ is multiplicative, or in other words: for any $G_1, G_2 \in \cG_D$, $s_{G_1 \sqcup G_2}=s_{G_1}+ s_{G_2}$ and $\mu_{G_1 \sqcup G_2}=\mu_{G_1} \cdot \mu_{G_2}$. This is not always true for random states, see Refs.~\cite{Gurau2025, Facto2} and Sec.~\ref{subsec:averageApprox}. 
\end{rem}
It is clear that $\LU$-equivalence implies asymptotic $\LU$-equivalence, which itself implies $\LU$-equivalence in scaling; moreover, none of the reverse implications hold.

\subsection{Characterizing separable states by means of trace-invariants}\label{sec:separable}

We have seen in Prop.~\ref{prop:charac_LU-1} that the entanglement structure of a (deterministic) $D$-partite pure state $\ket{\psi} \in \H_1 \otimes \cdots \otimes \H_D$ is fully characterized by the trace-invariant evaluations $\{\tr_G(\ket{\psi})\}_{G \in \cG_D^{\conn}}$. In practice, one is often interested in answering much coarsest classification questions, such as: is $\ket{\psi}$ separable or entangled? As we will see shortly, such a question can be answered by evaluating only \emph{one} suitably chosen trace-invariant. In the well-known bipartite context ($D=2$), it suffices to measure \eg the $2$-purity of the reduced density matrix on one of the subsystems: $\ket{\psi}$ is then separable if and only if the $2$-purity is unity.\footnote{Recall that, for any integer $k\geq 3$, the $k$-purity of a mixed state $\rho$ is defined to be $\tr(\rho^k)$, and that $\rho$ is pure (\ie rank one) if and only if $\tr(\rho^k)=1$. Hence, if $\rho$ is the reduced density matrix of a bipartite state $\ket\psi$, then $\ket\psi$ is separable if and only if $\tr(\rho^k)=1$.} It turns out that such simple tests of separability can also be obtained when $D\geq 3$. To describe them, we introduce the following subclass of trace-invariants. 
\begin{defi}\label{def:genuinely_D-partite_graph}
    Let $D \geq 2$, $G \in \cG_D$, and $c_1 , c_2 \in \{1, \ldots ,D\}$.  

  \noindent  We will say that the colors $c_1, c_2$ are \emph{parallel} in $G$ if the graph obtained by removing all the edges of color $c\notin \{c_1 , c_2\}$ from $G$ is trivial.

    \noindent We will say that $G$ (resp.~$\tr_G$) is \emph{genuinely $D$-partite} whenever no two colors are  parallel in $G$. Equivalently, $G$ (resp.~$\tr_G$) is genuinely $D$-partite whenever, for some $\vec{\sigma}=(\sigma_1 , \ldots , \sigma_D) \in S_{k(G)}^D(G)$, the permutations $\sigma_1 , \ldots , \sigma_D$ are pairwise disjoint.
\end{defi}
\noindent Clearly, the second formulation of the second definition does not depend on the choice of representative $\vec{\sigma}\in S_{k(G)}^D(G)$ and is therefore consistent. Note also that, when $D=2$, $G$ is genuinely $D$-partite if and only if it is a union of cyclic graphs $C_{k_1}, \ldots , C_{k_p}$, with at least one $k_i$ larger or equal to $2$. 

According to the following theorem, any genuinely $D$-partite trace-invariant can be used to characterize separable states: more precisely, the modulus of such an invariant is maximized precisely on the orbit of separable states.\footnote{A similar result was reported in \cite{Ma:2026wpv}, shortly after the present work was first submitted to arXiv.}
\newpage
\begin{theo}\label{thm:charac_separable}
    Let $\sF= (\H_1 , \ldots , \H_D)$ be a finite-dimensional $D$-partite state space, $\ket{\psi}\in S(\H^\sF)$, and $G \in \cG_D$. If $G$ is genuinely $D$-partite, then:  $\lvert\tr_G (\ket{\psi})\rvert=1$ if and only if $\ket{\psi}$ is separable in $\H^\sF$. 
\end{theo}
\begin{proof}
    We proceed by induction on $D \geq 2$. 
    
    Suppose that $D=2$ and $G \in \cG_2$ is genuinely $2$-partite. Then $G$ can be represented by $\vec\sigma = (\sigma_1 , \sigma_2)$ with $\sigma_1 \neq \sigma_2$. In other words, $\sigma_1 \sigma_2^{-1}$ admits at least one cycle of length at least 2. It follows that $G$ can be decomposed as a union of cyclic graphs $C_{k_1}\,, \ldots , C_{k_p}$ with $k_1 \geq 2$. We then have 
    \begin{equation}
        |\tr_G (\ket\psi )| = \prod_{i =1}^p \tr_{C_{k_i}}(\ket{\psi}) = \prod_{i =1}^p \tr\left((\psi\psi^\dagger)^{k_i}\right)\,. 
    \end{equation}
    This evaluates to $1$ if and only if 
    \begin{equation}
        \tr\left((\psi\psi^\dagger)^{k_i}\right) = 1
    \end{equation}
    for any $i\in\{1,\ldots, p\}$. Because $k_1 \geq 2$, the latter condition is in turn equivalent to the reduced density matrix $\psi\psi^\dagger$ being of rank one, or in other words, to $\ket{\psi}$ being separable in $\H_1 \otimes \H_2$.
    
    Suppose now that $D \geq 3$ and that the proposition holds in any $D'$-partite state space with $D' \leq D-1$. Let $G \in \cG_D$ be a genuinely $D$-partite trace-invariant and $k = k(G)$. $G$ can therefore be represented by a $D$-tuple $\vec\sigma = (\sigma_1, \ldots , \sigma_D) \in S_k^D(G)$ of pairwise disjoint permutations. In particular, one has $\sigma_1 \neq \sigma_2$, meaning that there exists some $k_0 \in \{ 1, \ldots , k\}$ such that $\sigma_1 (k_0) \neq \sigma_2 (k_0)$. Making use of the relabeling symmetry (or the equivalence relation \eqref{eq:equivalence_permutations}), we can choose a labeling such that $\sigma_1 = \mathrm{id}$ and $k_0 = 1$. Introducing
    \begin{equation}
        B_1 = \{ c \in \{1, \ldots , D\}\, | \, \sigma_c(1) = 1\}\qquad \mathrm{and} \qquad B_2 = \{ c \in \{1, \ldots , D\}\, | \, \sigma_c(1) \neq 1\}\,,
    \end{equation}
    we have the partition
    \begin{equation}
        \{1, \ldots , D\} = B_1 \sqcup B_2 
    \end{equation}
    with $B_1 \neq \emptyset$ (since $1\in B_1$) and $B_2 \neq \emptyset$ (since $2 \in B_2$). Suppose that $\ket{\psi} \in S(\H^\sF)$ is such that $|\tr_G(\ket\psi)|=1$. By Lem.~\ref{lem:modulus_invariants}, we must have
    \begin{equation}\label{eq:proof_sep_ev-eq}
        \left( \mathrm{id}_{\H_1} \otimes \rho_2(\sigma_2) \otimes \cdots \otimes \rho_D(\sigma_D) \right) \ket\psi^{\otimes k} = \eta \ket\psi^{\otimes k}\,,
    \end{equation}
    where $\eta$ is a unit complex number. The right-hand side is manifestly separable in the bipartition $(\H^{\sF})^{\otimes k} = \H^{\sF} \otimes (\H^{\sF})^{\otimes k-1}$, hence the left-hand side must be too. A necessary condition for this to hold is that $\ket{\psi}$ must be separable as a bipartite state in $\H^\sF \simeq \H_{B_1} \otimes \H_{B_2}$. Indeed, any entanglement for $\ket\psi$ across the bipartition $\H_{B_1}\otimes \H_{B_2}$ would generate entanglement for $\left( \mathrm{id}_{\H_1} \otimes \rho_2(\sigma_2) \otimes \cdots \otimes \rho_D(\sigma_D) \right) \ket\psi^{\otimes k}$ across the bipartition $\H^{\sF} \otimes (\H^{\sF})^{\otimes k-1}$, as is made intuitively clear by Fig.~\ref{fig:proof_separable} (and can be checked analytically by computing $2$-purities).
    \begin{figure}
        \centering
        \includegraphics[scale=.7]{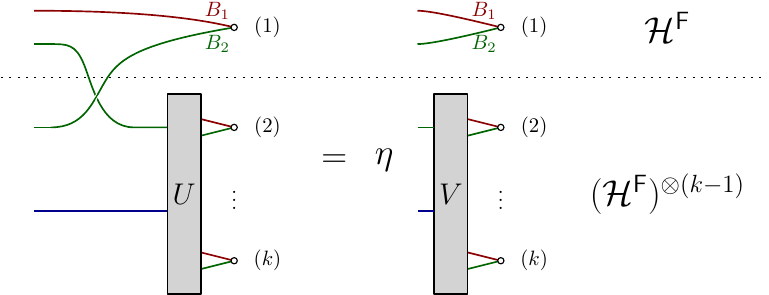}
        \caption{
        Graphical representation of equation \eqref{eq:proof_sep_ev-eq}. Multi-indices associated with the color set $B_1$ are represented in red, while those associated with $B_2$ are represented in green. Both $U$ and $V$ are unitary operators from $(\H^\sF)^{\otimes (k-1)}$ to $\H_{B_2} \otimes \left( \H_{B_1}\otimes (\H^{\sF})^{\otimes (k-2)}\right)$, and multi-indices associated with the tensor factor $\H_{B_1}\otimes (\H^{\sF})^{\otimes (k-2)}$ are represented in blue. Note that $U$ depends on the permutations $\sigma_2,\ldots, \sigma_D$, while $V$ does not. Any entanglement of $\ket{\psi}$ relative to the bipartition $\H^\sF\simeq \H_{B_1}\otimes \H_{B_2}$ necessarily generates entanglement of the state appearing on the left-hand side relative to the bipartition $\H^\sF \otimes (\H^\sF)^{\otimes(k-1)}$ (which is represented by the dotted line).
        }
        \label{fig:proof_separable}
    \end{figure}
    Hence, we can find states $\ket\phi \in S(\H_{B_1})$ and $\ket\eta \in S(\H_{B_2})$ such that $\ket{\psi} = \ket\phi \otimes \ket\eta$. We then have
    \begin{equation}
    \tr_G(\ket{\psi}) = \tr_{H_1}(\ket\eta) \tr_{H_2}(\ket\phi)\,,
    \end{equation}
    where $H_1 \in \cG_{|B_1|}$ is genuinely $|B_1|$-partite and $H_2 \in \cG_{|B_2|}$ is genuinely $|B_2|$-partite. More precisely, $H_1$ is the $|B_1|$-colored graph represented by the $|B_1|$-tuple of permutations $(\sigma_c)_{c \in B_1}$ (which are pairwise distinct); and, similarly, $H_2$ is the $|B_2|$-colored graph encoded by $(\sigma_c)_{c \in B_2}$. By Lem.~\ref{lem:tr}, $|\tr_G (\ket{\psi})|$, $|\tr_{H_1} (\ket{\eta})|$ and $|\tr_{H_2} (\ket{\phi})|$ are all bounded from above by $1$. It results that $|\tr_G (\ket{\psi})|=1$ if and only if: 
    \begin{equation}
        |\tr_{H_1} (\ket{\eta})| = 1 \quad \mathrm{and}\quad |\tr_{H_2} (\ket{\phi})| = 1\,.
    \end{equation} 
    By the induction hypothesis, these conditions hold if and only if $\ket\eta$ is separable in $\H_{B_1}$ and $\ket\phi$ is separable in $\H_{B_2}$, which, together, are equivalent to $\ket\psi$ being separable in $\H^\sF = \H_{B_1}\otimes \H_{B_2}$. This concludes the proof.
\end{proof}
\begin{rem}
\label{remark:sep-and-all-invariants}
    Note that, if $\ket{\psi}$ is a separable $D$-partite state, then $\tr_H (\ket\psi)=1$ for any $H \in \cG_D$. Hence, with the hypotheses of the previous proposition, we have:
    \begin{equation}
        |\tr_G(\ket\psi)|= 1 \qquad \Rightarrow \qquad \forall H \in \cG_D\,, \quad \tr_H(\ket\psi)=1\,.
    \end{equation}
\end{rem}

\begin{ex}
    Let us illustrate the previous result in $D=3$. It is possible to assess whether a tripartite state $\ket{\psi} \in \H_1 \otimes \H_2 \otimes \H_3$ is separable by measuring the $2$-purity across at least two bipartitions: for instance, the bipartitions $\H_1 \otimes (\H_2 \otimes \H_3)$ and $(\H_1 \otimes \H_2) \otimes \H_3$. Denoting by $\rho_1$ (resp. $\rho_3$) the reduced density matrix to subsystem $1$ (resp. subsystem $3$), one has:
    $$
    \ket\psi \; \mathrm{separable} \quad \Leftrightarrow \quad  \begin{cases}\tr(\rho_1^2) = 1 \\
    \tr(\rho_3^2) = 1
    \end{cases} \quad \Leftrightarrow \quad  
     \tr(\rho_1^2)\tr(\rho_3^2) =1 \quad \Leftrightarrow  \quad \tr_G (\ket\psi) = 1 \quad \Leftrightarrow  \quad |\tr_G (\ket\psi)| = 1\,,
     $$
where $G= \vcenter{\hbox{\includegraphics[scale=.5]{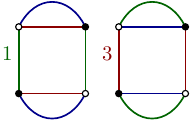}}}$. The graph $G$ is genuinely $3$-partite, and provides a test of separability of order $4$ (\ie with $k(G)=4$). According to Thm.~\ref{thm:charac_separable}, any other genuinely $3$-partite trace-invariant can be used as a separability test, and it turns out that four of those invariants have degree $3$ (which is the lowest degree of a genuinely $3$-partite trace-invariant in $D=3$): for any $c \in \{1,2,3\}$, $\mathrm{M}_{6}^{(c)} = \vcenter{\hbox{\includegraphics[scale=.5]{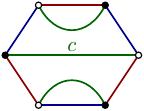}}}$ is genuinely $3$-partite (this is an example of melon graph, see Sec.~\ref{sss:Melo}), and so is $\PT_3 = \vcenter{\hbox{\includegraphics[scale=.5]{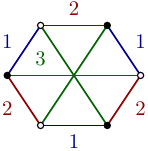}}}$ (this graph encodes the third moment of the partial transpose of the reduced density matrix to any of the three subsystems, see Sec.~\ref{subsubsec:PT}). 
\end{ex}
For arbitrary integers $D\geq 2$ and $k \in \mathbb{N}^*$, it is clear that one can find a genuinely $D$-partite graph $G \in \cG_D$ with $k(G)=k$ if and only if $|S_k|=k! \geq D$. In particular, one can always take $k=D$, as in the $D=3$ example just given. However, when $D>3$, this is not the optimal choice: in the asymptotic limit $D \to \infty$, the minimal degree $k_{\mathrm{min}}(D)$ of a genuinely $D$-partite trace-invariant scales like the functional inverse of the $\Gamma$ function. The approximation scheme laid out in Ref.~\cite{borwein2018gamma} implies that
\begin{equation}
k_{\mathrm{min}}(D) \underset{D \to \infty}{\sim} \frac{1}{2} +\frac{\ln\left( \frac{D}{\sqrt{2\pi}}\right)}{W\left( \frac{1}{e} \ln\left( \frac{D}{\sqrt{2\pi}}\right)\right)}\,,
\end{equation}
where $W$ is Lambert's $W$ function. This scales much more slowly than $D$, and dramatically more slowly than the degree of a trace-invariant assessing separability by means of bipartite $2$-purity tests: for instance, taking the product of the $2$-purities associated with all bipartitions of the subsystem, one obtains a genuinely $D$-partite invariant of degree $k=2(2^D-1)$, which grows exponentially with $D$.

\subsection{Coarse-graining of \texorpdfstring{$D$}{D}-partite state spaces}
\label{sec:generalities-on-coarse-graining}

In this subsection, we introduce natural coarse-graining maps relating $D$-partite state spaces to $D'$-partite state-spaces, with $D' \leq D$. When $D'<D$, this will allow us to distinguish multipartite entanglement properties which can be captured by effectively $D'$-partite invariants from those that can only be identified by genuinely $D$-partite invariants.  

Let $\sF = (\H_1 , \ldots , \H_D)$ be a finite-dimensional $D$-partite state space. Given a non-empty subset $B \subset \{1 , \ldots, D\}$, we will denote
\begin{equation}
    \H_B \eqdef \bigotimes_{c \in B} \H_{c}\,.
\end{equation}
Let $D' \leq D$ and $\sF' = (\H'_1, \ldots , \H'_{D'})$ a $D'$-partite state space. We say that $\sF'$ is \emph{coarser} than $\sF$ (or that $\sF'$ is a \emph{coarse-graining} of $\sF$, or that $\sF$ is a \emph{fine-graining} of $\sF'$) --  noted $\sF \preceq \sF'$ -- if there exists a partition $\zeta=\{B_1 , \ldots , B_{D'}\}$ of $\{1, \ldots , D\}$,\footnote{That is, the subsets $B_1, \ldots , B_{D'}$ are all non-empty and obey the condition 
$
\displaystyle\{1\,, \ldots \,, D\} = \bigsqcup_{c=1}^{D'} B_c\,.
$
The subsets $B_1 \,, \ldots \,, B_{D'}$ are called the \emph{blocks} of the partition $\zeta$, and $D'=\#(\zeta)$ its \emph{length}.} such that 
\begin{equation}
        \forall c' \in \{1, \ldots , D'\}\,, \quad \H'_{c'} = \H_{B_{c'}}\,.
\end{equation}
If $\sF \preceq \sF'$, $\LU^\sF$ is a subgroup of $\LU^{\sF'}$ ($\LU^\sF \preceq \LU^{\sF'}$). We therefore have a natural surjective map between the space of orbits
\begin{align}
    \pi_{\sF \to \sF'}:\qquad  \cO^\sF &\to \cO^{\sF'} \\
    \LU^\sF \ket{\psi} &\mapsto \LU^{\sF'} \ket{\psi} \nonumber
\end{align}
and an injective algebra homomorphism
\begin{align}
    \iota_{\sF' \to \sF}:\qquad  \cA_{\inv}^{\sF'} &\to \cA_{\inv}^{\sF}  \label{eq:inj_homo}\\
    P &\mapsto P \circ \pi_{\sF \to \sF'} \nonumber
\end{align}
Furthermore, the projection map $\pi_{\sF \to \sF'}$ is \emph{continuous} with respect to the Zariski topology,\footnote{Indeed, let $V(S)$ be a Zariski-closed set in $\cO^{\sF'}$, with $S\subset \cA_{\inv}^{\sF'}$. Then:
$$
\pi_{\sF \to \sF'}^{-1}\left( V(S) \right) = \big\{O \in \cO^\sF \, \vert \, \forall P \in S\,, \; P\circ \pi_{\sF \to \sF'} (O)  = 0 \big\} = V(\iota_{\sF' \to \sF}(S)) 
$$
is Zariski-closed in $\cO^{\sF}$.} and therefore, with respect to any consistent refinement of this topology in $\sF$ and $\sF'$. 

It is clear from the previous considerations that the set of coarse-grainings of $\sF$ is in one-to-one correspondence with the set of partitions of $\{1, \ldots, D\}$; let us denote by $\sF_\zeta$ the coarse-graining  associated with a given partition $\zeta$. The map $\zeta \mapsto \sF_\zeta$ is then order-preserving, in the sense that:\footnote{Here, $\leq$ denotes the usual partial order on partitions of $\{1, \ldots , D\}$, $\zeta \leq \zeta'$ meaning that $\zeta$ is finer than $\zeta'$. For example, $\zeta_{\mathrm{ min}}\eqdef\{\{1\}, \{2\}, \ldots, \{D\}\}$ is the finest partition of $\{1, \ldots, D\}$, and $\sF_{\zeta_{\mathrm{ min}}}=\sF$.}
\begin{equation}\label{eq:order_preserving_cg}
    \zeta \leq \zeta' \qquad \Rightarrow \qquad \sF_\zeta \preceq \sF_{\zeta'}\,. 
\end{equation}

\paragraph{Trace-invariants and coarse-graining.} 
Coarse-graining can be understood at the level of trace-invariants and their underlying combinatorics. For any partition $\zeta$ of $\{1, \ldots , D\}$, we denote by $\mathcal{G}_{D, \zeta}$ the set of $D$-colored graphs $G$ for which all  the colors in the same block $B\in \zeta$ are parallel in $G$ (see Def.~\ref{def:genuinely_D-partite_graph}).\footnote{Equivalently, if $\vec\sigma= (\sigma_1, \ldots, \sigma_D) \in S_{k(G)}^D(G)$, then: for any $B \in \zeta$ and any $c, c' \in B$, $\sigma_c = \sigma_{c'}$.} Clearly, if $\zeta'\ge \zeta$, then $\mathcal{G}_{D, \zeta'}\subset \mathcal{G}_{D, \zeta}$ (which we recognize as the dual of property \eqref{eq:order_preserving_cg}). We let $\mathcal{G}_{D, \zeta}^\neq$ be the subset of $\mathcal{G}_{D, \zeta}$ comprised of graphs which do not belong to $\mathcal{G}_{D, \zeta'}$ for any $\zeta'>\zeta$. In terms of permutations, a graph $G$ belongs to $\mathcal{G}_{D, \zeta}^\neq$ if it can be represented by a $D$-tuple $\vec\sigma = (\sigma_1 , \ldots, \sigma_D)$ with the following property: there exists a $D'$-tuple $\vec\tau= (\tau_{B})_{B \in \zeta}$ of \emph{pairwise distinct} permutations such that, for any $B \in \zeta$ and any $c \in B$, $\sigma_c = \tau_B$. In particular, if $\zeta_{\mathrm{min}}$ denotes the finest partition, then $\cG_{D, \zeta_{\mathrm{min}}}^{\neq}$ is nothing but the set of genuinely $D$-partite graphs, as introduced in Def.~\ref{def:genuinely_D-partite_graph}. Finally, if $\zeta_{\mathrm{max}}= \{\{ 1, \ldots , D\}\}$ denotes the coarsest partition of $\{1, \ldots , D\}$, we note that $\cG_{D}^{\triv}= \cG_{D, \zeta_{\mathrm{max}}}$ and therefore  $\cG_{D}^{\triv} \subset \cG_{D,\zeta}$ for any partition $\zeta$.

Having fixed the partition $\zeta= (B_1, \ldots, B_{D'})$, the $D'$-partite invariant algebra $\cA_{\inv}^{\sF_\zeta}$ is naturally generated by the set of trace-invariants $\{\tr_{G'} \, \vert \, G' \in \cG_{D'}\}$. Given $G' \in \cG_{D'}$, $\tr_{G'}$ is mapped by the injection $\iota_{\sF^\zeta\to\sF}$ to some $D$-partite trace-invariant $\tr_G \in \cA_{\inv}^\sF$, where $G \in \cG_{D, \zeta}$. By construction, we then have
\begin{equation}
    \forall \ket{\psi} \in S(\H^{\sF})\simeq S(\H^{\sF_\zeta}) \,,\quad  \tr_{G'}(\ket{\psi}) = \tr_{G}(\ket{\psi})\,.
\end{equation}
In more detail, $G$ is the $D$-colored graph of $\mathcal{G}_{D, \zeta}$ obtained from $G'$ by the following substitution rule: each edge of color $c'\in\{1, \ldots, D'\}$ in $G'$ is replaced by $|B_{c'}|$ parallel edges whose colors are the elements of $B_{c'}$. Moreover, we see that $G$ belongs to $\cG_{D, \zeta}^\neq$ if and only if $G'$ is genuinely $D'$-partite (in the sense of Def.~\ref{def:genuinely_D-partite_graph}). The map $\cG_{D'}\ni G' \mapsto G \in \cG_{D, \zeta}$ is a reexpression, at the level of generators and in purely combinatorial terms, of the injective homomorphism $\iota_{\sF^\zeta\to\sF}$ from equation \eqref{eq:inj_homo}. 

\medskip

\paragraph{Genuinely $D$-partite entangled states.} In Sec.~\ref{sec:separable}, we have introduced the notion of genuinely $D$-partite invariants, which allow testing the separability of a state. In the existing literature, the notion of genuinely $D$-partite entangled \emph{state} was also introduced, which we now recall (see \eg Ref.~\cite{Guhne:2008qic}).
\begin{defi}
    Let $D \geq 2$ and $\sF$ a $D$-partite state space. A state $\ket{\psi}\in S(\H^\sF)$ is said to be \emph{genuinely $D$-partite entangled} -- or \emph{genuinely entangled} on $\sF$ -- if, for any coarse-graining $\sF' \succeq \sF$ with $|\sF'|=2$, $\ket\psi$ is entangled in $\H^{\sF'}$. 
\end{defi}
\noindent Genuinely $D$-partite states admit a useful characterization in terms of trace-invariants.
\begin{prop}\label{prop:charac_genuinely_Dpartite}
        Let $D \geq 2$, $\sF=(\H_1 , \ldots , \H_D)$ a $D$-partite state space, and $\ket{\psi}\in S(\H^\sF)$. $\ket\psi$ is \emph{genuinely entangled on $\sF$} if and only if:
        \begin{equation}
            \forall G \in \cG_D\setminus \cG_D^{\triv}, \quad |\tr_G(\ket\psi)| < 1\,.
        \end{equation} 
\end{prop}
\begin{proof}
    Let us proceed by contraposition, and assume that $\ket\psi$ is \emph{not} genuinely entangled. Then we have a bipartition $\zeta=(B_1 , B_2)$ such that $\ket\psi$ is separable as a bipartite state on $\H^{\sF_\zeta}= \H_{B_1} \otimes \H_{B_2}$. The $2$-purity of the reduced density matrix on $B_1$ is then equal $1$, which provides a graph $G \in \cG_D \setminus \cG_{D}^{\triv}$ such that $|\tr_G (\ket\psi ) |= 1$.

     Reciprocally, suppose we are given a graph $G \in \cG_D \setminus \cG_{D}^{\triv}$ such that $|\tr_G (\ket\psi ) |= 1$. Then there exists a coarse-graining to a $D'$-partite state space with $2\leq D'\leq D$, and a genuinely $D'$-partite graph $G' \in \cG_{D'}$ such that $1=\lvert\tr_G(\ket\psi)\rvert = \lvert\tr_{G'}(\ket\psi)\rvert$. Indeed, such a $G'$ can be obtained from $G$ by merging any maximal block of parallel colors in $G$ into a single color. We are necessarily left with $D'\geq 2$ coarse-grained color labels after this is done, otherwise $G$ would have to be in $\cG_D^{\triv}$. By construction, the resulting graph $G'$ has no parallel colors left \ie it is genuinely $D'$-partite. From Thm.~\ref{thm:charac_separable}, it follows that $\ket\psi$ is separable on this coarse-graining, therefore it is not genuinely entangled on $\sF$.
\end{proof}

\medskip

\paragraph{Partially separable states.} A $D$-partite state is said to be \emph{partially separable}  (see \eg Refs.~\cite{Seevinck:2008ztx, Szalay2015, Szalay:2012wxa, Szalay:2018xvw}) if it is separable relative to some coarse-graining of its $D$-partite state space (hence, as a $D'$-partite state with $D' \leq D$). In more detail, consider a pure state $\ket\psi$ on $\sF = (\H_1, \ldots, \H_D)$, and let $\zeta$ be a partition of $\{1,\ldots, D\}$. The state $\ket\psi$ is said to be \emph{$\zeta$-separable} if it is separable on the coarse-graining $\sF_\zeta$ of $\sF$. We let $\xi$ be the \emph{finest} partition such that $\ket{\psi}$ is $\xi$-separable. We then have $\ket{\psi}=\displaystyle \bigotimes_{B \in \xi} \ket\psi_B$, with $\ket\psi_B \in \H_B$ for any $B \in \xi$. Furthermore, the (pure) state $\ket{\psi}_B$ induced on each block $B$ of $\xi$ is \emph{genuinely $\lvert B\rvert$-partite entangled}. The partition $\xi$ will be important in later sections. 

We also recall the following related concepts from the existing literature (see \eg Refs.~\cite{Szalay:2019, Toth2020stretchinglimitsof, Hong:2025ewo}). One says that $\ket\psi$ is \emph{$r$-separable} (resp.~\emph{$h$-producible}) if there exists a partition $\zeta$ with $\#(\zeta)=r$ (resp.~$\displaystyle\max_{B \in \zeta}|B| = h$) such that $\ket{\psi}$ is $\zeta$-separable. The smallest $m$ such that $\ket\psi$ is $m$-producible is called the \emph{entanglement depth} of $\ket\psi$. Equivalently, $m$ is the size of the largest block of the partition $\xi$ introduced above. Similarly, the number of blocks of $\xi$, denoted $\#(\xi)$, is the largest $r$ such that $\ket\psi$ is $r$-separable. 
Finally, note that a state $\ket\psi$ is genuinely entangled  on $\H^\sF = \H_1\otimes \cdots \otimes \H_D$ if and only if it is $1$-separable (if and only if it is $D$-producible, in which case $m=D$). 

These concepts can all be nicely reformulated in terms of trace-invariants. 
\begin{prop}\label{prop:partial_sep}
    Let $\sF$ be a $D$-partite state space and $\ket\psi \in S(\H^\sF)$. We denote by $\xi$ the finest partition of $\{1, \ldots, D\}$ such that $\ket\psi$ is $\xi$-separable, and by $m= \displaystyle\max_{B \in \xi}|B|$ the entanglement depth of $\ket\psi$. The following  holds.
    \begin{enumerate}
    \item Let $\zeta$ be a partition of $\{1,\ldots, D\}$. The following statements are equivalent:
    \begin{enumerate}
        \item $\ket\psi$ is $\zeta$-separable;
        \item there exists a graph $G\in \mathcal{G}_{D, \zeta}^\neq$ such that $\abs{\tr_G(\ket\psi)}=1$;
        \item for any $G\in \mathcal{G}_{D, \zeta}$, $\abs{\tr_G(\ket\psi)}=1$.
    \end{enumerate}  
    \item For any partition $\zeta$ of $\{1, \ldots, D\}$, $\ket\psi$ is genuinely entangled on $\sF_\zeta$  if and only if:  for any $G\in \cG_{D, \zeta}\setminus \cG_{D}^{\triv}$, $\abs{\tr_G(\ket\psi)}<1$.
    \item $\xi$ is the finest partition of $\{1,\ldots, D\}$ such that there exists a graph $G\in \mathcal{G}_{D, \xi}^\neq$ obeying $\abs{\tr_G(\ket\psi)}=1$: for any $\zeta\ngeq \xi$, and any $G\in \mathcal{G}_{D, \zeta}^\neq$, one has $\abs{\tr_G(\ket\psi)}<1$. 
    \item $\ket\psi$ is $h$-producible if and only if there exist a partition $\zeta$ with $\displaystyle\max_{B \in \zeta} |B| = h$ and $G\in \mathcal{G}_{D, \zeta}^\neq$ such that $\abs{\tr_G(\ket\psi)}=1$.
    \item For any partition $\zeta$ such that $\displaystyle\max_{B \in \zeta} |B| \le m-1$, and any $G\in \mathcal{G}_{D, \zeta}^\neq$, one has $\abs{\tr_G(\ket\psi)}<1$. 
    \end{enumerate}
\end{prop}
\begin{proof}
Property $1$ follows straightforwardly from Thm.~\ref{thm:charac_separable}, Rk.~\ref{remark:sep-and-all-invariants}, together with the definitions. 

Property $2$ is a direct consequence of Prop.~\ref{prop:charac_genuinely_Dpartite} upon coarse-graining to $\sF_\zeta$.

Let us turn to Property $3$. The definition of $\xi$ implies that we can write $\ket\psi$ in the product form
\begin{equation}
    \ket\psi = \bigotimes_{B \in \xi} \ket{\psi}_B\,,
\end{equation}
where, for any $B \in \xi$, $\ket{\psi}_B$ is genuinely entangled on $\H_B$. Suppose that $\zeta\ngeq \xi$, and let $G \in \cG_{D, \zeta}^\neq$. We have pairwise distinct permutations $(\tau_{B'})_{B' \in \zeta}$ such that $G$ can be represented by the $D$-tuple $\vec\sigma= (\sigma_1 , \ldots , \sigma_D)$, defined by: for any $B' \in \xi$ and any $c' \in B'$, $\sigma_{c'}= \tau_{B'}$. We then have
\begin{equation}
    \tr_G (\ket\psi)= \prod_{B \in \xi} \tr_{G_B} (\ket{\psi}_B)
\end{equation}
where, for any $B \in \xi$, $G_B \in \cG_{|B|}$ is the colored graph represented by the $|B|$-tuple $(\sigma_c)_{c\in B}$. Given that $\zeta\ngeq \xi$, we can find a block $B_0$ of $\xi$ that intersects two (or more) blocks of $\zeta$, say $B'_1$ and $B'_2$. Hence, the graph $G_{B_0}$ is represented by a list of permutations that contain both $\tau_{B'_1}$ and $\tau_{B'_2}$, which are distinct. As a result, $G_{B_0}$ belongs to $\cG_{|B_0|}\setminus \cG_{|B_0|}^{\triv}$. It follows that
\begin{equation}
    |\tr_G (\ket\psi)|= |\prod_{B \in \xi} \tr_{G_B} (\ket{\psi}_B)| \leq |\tr_{G_{B_0}} (\ket{\psi}_{B_0})| < 1\,,
\end{equation}
where the first inequality follows from Lem.~\ref{lem:modulus_invariants} and the second from Prop.~\ref{prop:charac_genuinely_Dpartite}. This concludes the proof of Property $3$.

Suppose that $\ket\psi$ is $h$-producible. By definition, we can find a partition $\zeta$ with $\displaystyle \max_{B\in \zeta} |B|=h$ such that $\ket\psi$ is $\zeta$-separable. But the latter property is equivalent, thanks to Property $1$, to the existence of some $G\in \cG_{D, \zeta}^\neq$ such that $|\tr_G(\ket\psi)|=1$. This establishes Property~$4$.

By definition of the entanglement depth, one has $m = \displaystyle \max_{B\in \xi} |B|$. For any partition $\zeta$ such that $\displaystyle \max_{B\in \zeta} |B| leq m-1$, one necessarily has $\zeta\ngeq \xi$. Hence, by Property~$3$, for any $G \in \cG_{D, \zeta}^\neq$, one has $|\tr_G (\ket\psi)|<1$. This establishes Property~$4$.
\end{proof}

\begin{figure}[ht]
    \centering
    \includegraphics[height = 3.2cm]{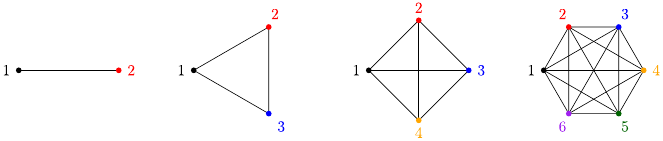}
    \caption{On the left, a visual representation of a Bell pair shared between parts $1$ and $2$. On the middle left, the so-called \textit{triangle state} or \textit{matrix multiplication tensor} (see \eg Ref.~\cite{Buhrman:2016tif,Zou:2020bly,Iizuka:2025caq}) for a tripartite system $1,2$ and $3$. On the middle right and right, a generalization of this state to a quantum system made up of $4$ and $6$ subsystems (see \eg Refs.~\cite{Christandl:2018cfb,Christandl:2019zrq}).}
    \label{fig:shareBP}
\end{figure}

While coarse-graining can render a partially separable state fully separable (it is precisely the definition of partial separability), it can also diminish the number of blocks of $\xi$ (as introduced above): this happens when the coarse-graining identifies two colors in different blocks of $\xi$. This is illustrated for instance by the examples of Fig.~\ref{fig:shareBP} (which are all tensor products of Bell states): any of these four states is genuinely entangled on its native $D$-partite state space (with $D= 2, 3, 4$ and $6$, respectively), but it is not necessarily so on a fine-graining of this state space. In particular, for each of those examples, we can actually find a fine-graining of the state space relative to which the entanglement depth of the state is $2$.\footnote{For instance, in the second example (the ``triangle state'', with $D=3$), this is obtained by further partitioning each subsystem in two, and thereby viewing the triangle state as a $6$-partite state.} This reflects the fact that, in all those examples, entanglement is mediated by Bell states, which, in a natural sense, only contain bipartite entanglement. In certain situations, one is rather interested in characterizing a stronger notion of ``genuine $D$-partite entanglement'', that is, one that cannot be reduced to entanglement between $D'$ parties with $D' < D$ upon fine-graining. This can motivate the definition of new entanglement measures, such as the genuine multi-entropy introduced in Ref.~\cite{Iizuka:2025ioc}, which have the property to vanish on products of Bell states. In a forthcoming paper, we will investigate further how this stronger notion of ``genuine $D$-partite entanglement'' can be quantified by means of trace-invariants. 

\medskip

The notion of coarse-graining defined above naturally leads to the introduction of several equivalence relations that are coarser than full $\LU$-equivalence. For instance, we can decide to analyse multipartite entanglement through the (limited) lens of bipartite entanglement, by comparing multipartite states in terms of the entanglement properties of all their bipartitions.  
\begin{defi}\label{def:2partite_eq}
    Let $\sF$ be a $D$-partite state space, and $\ket{\psi}, \ket{\phi}$ two deterministic or random states on $S(\H^\sF)$. We say that $\ket{\psi}$ and $\ket{\phi}$ have \emph{equivalent bipartite entanglement structures relative to $\sF$} -- noted $\ket{\psi}\underset{(2, \sF)}{\sim} \ket{\phi}$ -- if: for any $\sF' \succeq \sF$ such that $|\sF'|=2$, and for any $P \in \cA_{\inv}^{\sF'}$, we have
    \begin{equation}
        \mean{P(\ket{\psi})} = \mean{P(\ket{\phi})}\,.
    \end{equation}
\end{defi}
\noindent According to this definition, we have $\ket{\psi}\underset{(2, \sF)}{\sim} \ket{\phi}$ whenever, relative to any bipartite coarse-graining of $\sF$, the distributions followed by the entanglement spectra of the two states $\ket{\psi}$ and $\ket{\phi}$ are equivalent in the sense of moments. This characterization is easily transcribed in terms of trace-invariants. For this purpose, let us introduce the notion of \emph{cyclic graph}. Let $G \in \cG_D$. We say that $G$ is \emph{cyclic} if there exists a bipartition $\zeta$ of $\{1, \ldots, D\}$ such that $G \in \cG_{D,\zeta}$. Equivalently, $G$ is cyclic if there is a permutation $\tau \in S_{k(G)}$ such that $G$ can be represented by a $D$-tuple of permutations $(\sigma_1 , \ldots , \sigma_D)$ obeying: for any $c = \{1, \ldots , D\}$, $\sigma_c \in \{ \mathrm{id} , \tau \}$. Trivial graphs (\ie graphs from $\cG_D^{\triv}$) are cyclic. Any other cyclic graph is one for which the color set $\{1, \ldots , D\}$ can be partitioned into two non-trivial subsets; any two edges incident to the same vertex in $G$ are then required to be parallel if and only if their color belong to the same subset of the color partition. It is not difficult to see that a $G \in \cG_D$ is a \emph{connected} cyclic graph if and only if it can be represented as above with $\tau = (1 \, 2\,  \cdots \,D)$ a full cycle. This includes so-called \emph{necklace} graphs (see Sec.~\ref{sss:necklace}). We then have the following characterization.
\begin{prop}\label{prop:2partite_carac}
     Let $\sF$ be a $D$-partite state space, and $\ket{\psi}, \ket{\phi}$ two deterministic or random states on $S(\H^\sF)$. We have $\ket{\psi}\underset{(2, \sF)}{\sim} \ket{\phi}$ if and only if: for any cyclic graph $G \in \cG_D$,
    \begin{equation}
        \mean{\tr_G(\ket{\psi})} = \mean{\tr_G(\ket{\phi})}\,.
    \end{equation}
\end{prop}
\begin{proof}
    For any $\sF'=(\H_{B_1} , \H_{B_2}) \succeq \sF$, $\cA_{\inv}^{\sF'}$ is spanned by cyclic invariants associated with $D$-tuples of permutations $(\sigma_1 , \ldots , \sigma_D)$ with: $\sigma_c = \mathrm{id}$ for any $c \in B_1$, and $\sigma_c = \tau$ for any $c \in B_2$, where $\tau$ is some permutation. The result immediately follows. 
\end{proof}
Furthermore, one can straightforwardly iterate this construction. Given $D' \in \{1, \ldots , D \}$, we will say that $G \in \cG_D$ is \emph{$D'$-partite} if there exists a partition $\zeta$ of $\{1, \ldots , D\}$ with $\#(\zeta) = D'$ such that $G \in \cG_{D, \zeta}$. In other words, $G$ is $D'$-partite if one can find $D'$ permutations $\tau_1, \ldots , \tau_{D'} \in S_{k(G)}$ and a $D$-tuple $(\sigma_1 , \ldots , \sigma_D)$ representing $G$ such that: for any $c \in \{1, \ldots , D\}$, $\sigma_c \in \{\tau_1, \ldots , \tau_{D'}\}$. This definition generalizes the notion of being cyclic, in the sense that $G$ is cyclic if and only if it is $2$-partite. Moreover, it is also clear that: (1) $G \in \cG_D$ is $1$-partite if and only if $G\in \cG_D^{\triv}$; (2) at the opposite end, any $G\in \cG_D$ is $D$-partite. If $1 \leq D'' \leq D' \leq D$, we also have that: (3) if $G$ is $D''$-partite, then $G$ is also $D'$-partite. Finally, we will say that a $D'$-partite graph $G \in \cG_D$ is \emph{genuinely $D'$-partite} if it is $D'$-partite but \emph{not} $(D'-1)$-partite. This consistently generalizes Def.~\ref{def:genuinely_D-partite_graph}.  
\begin{defi}\label{def:D'partite_eq}
        Let $\sF$ be a $D$-partite state space, $D' \in \{ 2, \ldots , D-1\}$, and $\ket{\psi}, \ket{\phi}$ two deterministic or random states on $S(\H^\sF)$. We say that $\ket{\psi}$ and $\ket{\phi}$ are \emph{$D'$-partite equivalent relative to $\sF$} -- noted $\ket{\psi}\underset{(D', \sF)}{\sim} \ket{\phi}$ -- if: for any $\sF' \succeq \sF$ such that $|\sF'|=D'$, and for any $P \in \cA_{\inv}^{\sF'}$, we have
    \begin{equation}
        \mean{P(\ket{\psi})} = \mean{P(\ket{\phi})}\,.
    \end{equation}
\end{defi}
\noindent Clearly, this definition is consistent with Def.~\ref{def:2partite_eq}. Prop.~\ref{prop:2partite_carac} then generalizes in a straightforward way to the following observation.
\begin{prop}\label{prop:D'partite_carac}
     Let $\sF$ be a $D$-partite state space, $D'\in \{2, \ldots, D-1\}$, and $\ket{\psi}, \ket{\phi}$ two deterministic or random states on $S(\H^\sF)$. We have $\ket{\psi}\underset{(D', \sF)}{\sim} \ket{\phi}$ if and only if: for any $D'$-partite graph $G \in G_D$,
    \begin{equation}
        \mean{\tr_G(\ket{\psi})} = \mean{\tr_G(\ket{\phi})}\,.
    \end{equation}
\end{prop}

\subsection{Entanglement monotones from trace-invariants}
\label{ss:ent-monotones-trace-inv}

We will now illustrate how trace-invariants can be used to investigate order relations in operational resource theories of entanglement such as $\LO$ (Local Operations), $\LOCC$ (Local Operations and Classical Communication), or $\LOSR$ (Local Operations and Shared Randomness). 
We will focus on the common layer of such operational theories, namely $\LO$, and only discuss relations between deterministic pure states. Furthermore, we will limit our analysis to the most elementary family of $\LO$ monotones one can construct from trace-invariants, which we may interpret as multipartite generalizations of R\'{e}nyi entanglement entropies (see Ex.~\ref{ex:renyi_higher-D} below). Other interesting families of monotones will be introduced and investigated in a forthcoming paper by the same authors.   

The orbit of separable states, which we can interpret as the orbit of \emph{least} entangled states in $\LO$ (and in any other operational resource theory of entanglement) has already been characterized by means of trace-invariants: by Thm.~\ref{thm:charac_separable}, a $D$-partite state $\ket{\psi}$ is separable if and only if $|\tr_G(\ket\psi)|=1$ for any $G \in \cG_D^{\conn}$. This raises the following natural question: are there restrictions on how the quantities $\{|\tr_G (\ket{\psi})|\}_{G \in \cG_D^{\conn}}$ can change if one transforms $\ket\psi$ into another pure state via a $\LO$ transformation? Let us find out.

\medskip

Given two finite-dimensional $D$-partite state spaces $\sF= (\H_1 , \ldots , \H_D)$ and $\sF'= (\H'_1 , \ldots , \H'_D)$, we will denote by $\LO^{\sF\to \sF'}$ the set of \emph{local operations} mapping states in $\sF$ to states in $\sF'$. Recall that a local operation $A \in \LO^{\sF\to \sF'}$ can be represented by a linear operator $A \in \cL(\H^\sF, \H^{\sF'})$ of the form $A= A_1 \otimes \cdots \otimes A_D$ such that, for any $c \in \{1, \ldots , D\}$, $A_c \in \cL(\H_c, \H'_c)$ is completely positive and trace-preserving (in particular, it sends any density matrix on $\H_c$ to a density matrix on $\H'_c$). Given two density matrices $\psi \in \cL(\H^\sF)$ and $\phi \in \cL(\H^{\sF'})$, we will write
\begin{equation}
    \psi \toLO \phi
\end{equation}
whenever there exists $A \in \LO^{\sF \to \sF'}$ such that $A (\psi) = \phi$. By extension, given two (deterministic) pure states $\ket{\psi} \in \H^\sF$ and $\ket{\phi} \in \H^\sF$, we will write
\begin{equation}
    \ket{\psi} \toLO \ket{\phi}
\end{equation}
whenever $\ket{\psi}\bra{\psi} \toLO \ket{\phi}\bra{\phi}$. The binary relation $\toLO$ is reflexive and transitive, so it defines a \emph{preorder} on the set of all finite-dimensional $D$-partite pure states. Local unitary transformations are reversible local operations, which makes the $\LO$ preorder compatible with $\LU$-equivalence:
\begin{equation}
    \ket{\psi} \simLU \ket{\phi} \quad \Longrightarrow \quad \ket{\psi} \toLO \ket{\phi} \; \mathrm{and} \; \ket{\phi} \toLO \ket{\psi}\,. 
\end{equation}
It turns out that the converse implication also holds (see Refs.~\cite{Schmid:2020pgv, PhysRevA.53.2046}), and we will rederive this result using trace-invariants below (see Cor.~\ref{cor:LO_classes}). Moreover, as far as pure states are concerned, the preorder induced by $\LO$ operations is the same as the preorder induced by $\LOSR$ operations (see Ref.~\cite{Schmid:2020pgv}). The two resource theories do differ for mixed states. 

Thanks to Stinespring's theorem, a quantum operation can always be implemented by a unitary transformation on a larger system including ancilla degrees of freedom, followed by a trace over the ancilla Hilbert space. Combined with Prop.~\ref{prop:charac_LU-1}, this allows us to characterize the $\LO$ preorder in terms of trace-invariants.
\begin{prop}\label{prop:charac_LO}
    Let $\sF= (\H_1 , \ldots , \H_D)$ and $\sF'=(\H'_1, \ldots , \H'_D)$ be two finite-dimensional $D$-partite state spaces, $\ket{\psi}\in S(\H^\sF)$, and $\ket{\phi}\in S(\H^{\sF'})$. We have $\ket{\psi} \toLO \ket{\phi}$ if and only if there exists a $D$-partite state space $\sF''=(\H''_1, \ldots , \H''_D)$ and a state $\ket{\eta} \in S(\H^{\sF''})$ such that:\footnote{Obviously, the proposition remains true if one replaces $\cG_D^{\conn}$ by $\cG_D$ in Eq.~\eqref{eq:equivalence_LO}.}
    \begin{equation}\label{eq:equivalence_LO}
        \forall G \in \cG_D^{\conn}\,, \qquad \tr_G(\ket{\psi}) = \tr_G(\ket{\phi}) \tr_G(\ket{\eta})\,. 
    \end{equation}
\end{prop}
\begin{proof}
    By Stinespring's theorem, $\ket{\psi}\toLO \ket{\phi}$ if and only if there exist two auxillary $D$-partite state spaces $\sF^{\aux}=(\H^{\aux}_1, \ldots , \H^{\aux}_D)$ and $\sF'=(\H'_1, \ldots , \H'_D)$, as well as unitary maps $\{U_c: \H_c \otimes \H_c^{\aux} \to \H_c' \otimes \H_c''\}_{1\leq c \leq D}$ and ancillary states $\{\ket{v_c} \in S(\H_c^{\aux})\}_{1\leq c \leq D}$,  such that
    \begin{equation}
        \ket{\phi}\bra{\phi} = \tr_{\H^{\sF''}}\left( (U_1 \otimes \cdots \otimes U_D)(\ket{\psi}\bra{\psi} \otimes \bigotimes_{c=1}^D \ket{v_c}\bra{v_c} ) (U_1 \otimes \cdots \otimes U_D)^\dagger \right) \,.
    \end{equation}
$\phi\eqdef\ket{\phi}\bra{\phi}$ is the reduced density matrix to subsystem $\H^{\sF'}$ of the bipartite pure state
\begin{equation}
    (U_1 \otimes \cdots \otimes U_D)(\ket{\psi} \otimes \bigotimes_{c=1}^D \ket{v_c}) \in \H^{\sF'}\otimes\H^{\sF''}\,. 
\end{equation}
Since $\phi$ is pure (\ie rank one), this bipartite pure state must be separable, hence there exists a state $\ket{\eta} \in S(\H^{\sF''})$ such that
\begin{equation}
    (U_1 \otimes \cdots \otimes U_D)(\ket{\psi} \otimes \bigotimes_{c=1}^D \ket{v_c})  = \ket{\phi} \otimes \ket{\eta}\,.
\end{equation}
In other words, we have 
\begin{equation}
    \ket{\psi} \otimes \bigotimes_{c=1}^D \ket{v_c} \simLU \ket{\phi} \otimes \ket{\eta} \,.
\end{equation}
According to Prop.~\ref{prop:charac_LU-1}, this equivalence relation holds if and only if: for any $G \in \cG_D^{\conn}$,
\begin{equation}
    \tr_G \left(\ket{\psi} \otimes \bigotimes_{c=1}^D \ket{v_c}\right) = \left( \prod_{c=1}^D \langle v_c \vert v_c \rangle^{k(G)} \right)   \tr_G (\ket{\psi}) =  \tr_G (\ket{\psi})
\end{equation}
is equal to 
\begin{equation}
    \tr_G \left( \ket{\phi} \otimes \ket{\eta} \right) = \tr_G (\ket{\phi}) \tr_G (\ket{\eta})\,.
\end{equation}
This concludes the proof.
\end{proof}
\begin{rem}
    According to Stinespring's theorem, we can take each $\H^{\aux}_c$ in the preceding proof to have dimension lower or equal to $\dim(\H_c)^2$. Combined with the degree bound from Prop.~\ref{prop:bound_degree}, this leads to a refinement of the previous proposition, in which only trace-invariants with degree smaller than some finite maximal degree need to be examined. Since it will be of little use for us in the present paper, we leave the determination of the exact value of this bound as an exercise for the interested reader.
\end{rem} 
In the special case of bipartite systems, we recover a previously known characterization of the $\LO$ order; see \eg Corollary 7 of Ref.~\cite{Schmid:2020pgv}.
\begin{cor}\label{cor:LO_bipartite}
    Let $\ket{\psi}$ and $\ket{\phi}$ be two bipartite pure states, and let $\pmb{p}_\psi$ (resp.~$\pmb{p}_\phi$) denote the entanglement spectrum of $\ket{\psi}$ (resp.~$\ket{\phi}$).\footnote{Previously, we defined $\pmb{p}_\psi= (p_1, \ldots , p_N)$ as the vector of eigenvalues of the reduced density matrix $\psi\psi^\dagger$, with the convention $p_1\geq \ldots \geq p_N$. We make the further assumption that only nonzero eigenvalues are recorded in $\pmb{p}_\psi$ (so $p_N >0$), and similarly for $\pmb{p}_\phi$ and $\pmb{p}_\eta$.} We have $\ket{\psi}\toLO \ket{\phi}$ if and only if there exists a bipartite pure state $\ket{\eta}$ whose entanglement spectrum $\pmb{p}_\eta$ obeys
    \begin{equation}
        \pmb{p}_\psi = \left( \pmb{p}_\phi \otimes \pmb{p}_\eta \right)^\downarrow\,.
    \end{equation}
In this equation, the symbol $\downarrow$ indicates that the vector components on the right-hand side are ordered from highest to lowest in magnitude.  
\end{cor}
\begin{proof}
  Thanks to Prop.~\ref{prop:charac_LO}, we have $\ket{\psi}\toLO \ket{\phi}$ if and only if there exists a state $\ket{\eta}$ such that:
  \begin{equation}
      \forall k \in \mathbb{N}^*\,, \qquad \tr\left( (\psi\psi^\dagger)^k\right) = \tr\left( (\phi\phi^\dagger)^k\right) \tr\left( (\eta\eta^\dagger)^k\right)\,. 
  \end{equation}
If $\pmb{p}_\psi = (p_1 , \ldots , p_N)$, $\pmb{p}_\phi = (p_1' , \ldots , p'_{N'})$ and $\pmb{p}_\eta = (p''_1 , \ldots , p''_{N''})$, these conditions are equivalent to 
\begin{equation}
      \forall k \in \mathbb{N}^*\,, \qquad \sum_{l=1}^N (p_i)^k = \sum_{m=1}^{N'} \sum_{n=1}^{N''} (p'_m p''_n)^k\,.
\end{equation}
Since power sums generate the algebra of symmetric polynomials, this is equivalent to the condition that the eigenvalues $(p_i)_{1 \leq i \leq N}$ be identical to $(p'_m p''_n)_{\substack{1\leq m \leq N'\\ 1\leq n \leq N''}}$, up to permutation. In other words, $\pmb{p}_\psi = \left( \pmb{p}_\phi \otimes \pmb{p}_\eta \right)^\downarrow$.
\end{proof}
\begin{rem}
    In Ref.~\cite{Schmid:2020pgv} (Corollary 8), a necessary (but insufficient) condition for $\ket{\psi} \toLO \ket{\phi}$ was provided in the multipartite setting. In the language of Prop.~\ref{prop:charac_LO}, that condition is equivalent to the existence of a $D$-partite state $\ket{\eta}$ such that, for any \emph{bipartite} connected graph $G$, $\tr_G(\ket{\psi})= \tr_G (\ket{\phi}) \tr_G (\ket{\eta})$. Clearly, this condition can be equivalently expressed as relations involving the entanglement spectra of arbitrary bipartite coarse-grainings of the states $\ket{\psi}$, $\ket{\phi}$, and $\ket{\eta}$. Such a formulation is analogous to that of Cor.~\ref{cor:LO_bipartite}, and it is in this form that this necessary condition was originally stated in Ref.~\cite{Schmid:2020pgv}. A clear benefit of our reliance on trace-invariants instead of eigenvalues is that it allows for a complete characterization of the $\LO$ preorder, in the bipartite and multipartite settings alike. 
\end{rem}

The next observation provides a sufficient condition for two $D$-partite states to be incomparable under the $\LO$ preorder.
\begin{prop}\label{prop:states_not_LO}
    Let $D \geq 3$, $G\in \cG_D$, and $\ket{\psi}$, $\ket{\phi}$ two finite-dimensional (and deterministic) $D$-partite pure states. Suppose that:
    \begin{enumerate}
        \item $G$ is genuinely $D$-partite and $\tr_G(\ket{\psi})= \tr_G (\ket{\phi})$;
        \item $\ket{\psi}\underset{\LU}{\not\sim}\ket{\phi}$.
    \end{enumerate}
    Then, $\ket{\psi}$ and $\ket{\phi}$ are unrelated under $\LO$.
\end{prop}
\begin{proof}
Let us proceed by contraposition, by assuming that $\ket{\psi}\toLO\ket{\phi}$ or $\ket{\phi}\toLO\ket{\psi}$. According to Prop.~\ref{prop:charac_LO}, this implies that one can find a state $\ket{\eta}$ such that:
    \begin{equation}\label{eq:proof_notLO_1}
        \forall H \in \cG_D^{\conn}\,, \quad \tr_H (\ket{\psi})= \tr_H(\ket\eta) \tr_H(\ket\phi)\,,
    \end{equation}
    or 
    \begin{equation}\label{eq:proof_notLO_2}
        \forall H \in \cG_D^{\conn}\,, \quad \tr_H (\ket{\phi})= \tr_H(\ket\eta) \tr_H(\ket\psi)\,.
    \end{equation}
Under these conditions, let us show that property \emph{1} implies the negation of property \emph{2}. Assuming that $G$ is genuinely $D$-partite and $\tr_G(\ket\psi)=\tr_G(\ket\phi)$ (property \emph{1.}), and taking $H=G$ in \eqref{eq:proof_notLO_1} and \eqref{eq:proof_notLO_2}, we find that $\tr_H (\ket\eta) = 1$. Given that $G$ is genuinely $D$-partite, it follows from Thm.~\ref{thm:charac_separable} that $\ket\eta$ is separable. But then $\tr_H(\ket\eta)=1$ for any $H \in \cG_D$, and thus: 
\begin{equation}
        \forall H \in \cG_D^{\conn}\,, \quad \tr_H (\ket{\psi})= \tr_H(\ket\phi)\,.
    \end{equation}
By Prop.~\ref{prop:charac_LU-1}, it follows that $\ket{\psi}\simLU \ket{\phi}$ (\ie property \emph{2} does not hold), which concludes the proof.
\end{proof}

As an immediate consequence of Prop.~\ref{prop:charac_LO}, we can construct an infinite family of $\LO$-entanglement monotones from trace-invariants.

\begin{cor}
    \label{cor:LO_monotones}
     Let $\ket{\psi}$ and $\ket{\phi}$ be two finite-dimensional (and deterministic) $D$-partite pure states. \begin{enumerate}
         \item If $\ket{\psi} \toLO \ket{\phi}$, then
    \begin{equation}
        \forall G \in \cG_D\,, \qquad |\tr_G(\ket{\psi})| \leq |\tr_G(\ket{\phi})|\,.
    \end{equation}
    \item Moreover, if $\ket{\psi} \toLO \ket{\phi}$ and $\ket{\psi}\underset{\LU}{\not\sim}\ket{\phi}$, then: for any genuinely $D$-partite graph $G$,
    \begin{equation}
        |\tr_G(\ket{\psi})| < |\tr_G(\ket{\phi})|\,.
    \end{equation}
     \end{enumerate}
\end{cor}
\begin{proof}
The first item follows directly from Prop.~\ref{prop:charac_LO}, together with the fact that $|\tr_G (\ket{\eta})|\leq1$ for any $G$ and $\ket{\eta}$ (Lem.~\ref{lem:modulus_invariants}). The second item relies in addition on Prop.~\ref{prop:states_not_LO}.
\end{proof}
\begin{ex} \label{ex:Rényi}
For any integer $k\geq 2$, the \emph{entanglement Rényi-$k$ entropy} of a bipartite state $\ket{\psi}$ takes the form
\begin{equation}
    S_R^{(k)}(\ket{\psi}) = \frac{1}{1-k} \ln\left( \tr_{C_k} (\ket{\psi}) \right) \in \mathbb{R}_+^*\,,
\end{equation}
where $C_k \in \cG_2^{\conn}$ denotes the unique $2$-colored connected graph with $2k$ vertices (see Eq.~\eqref{eq:equiv-Ck-entanglement-Renyi-k}). As is well-known, $S_R^{(k)}$ defines a $\LO$-monotone (it is non-increasing under any $\LO$ transformation), and this is indeed implied by Cor.~\ref{cor:LO_monotones}. Furthermore, $S_R^{(k)}(\ket\psi)$ vanishes if and only if $\ket\psi$ is separable (as is well-known, and also implied by Thm.~\ref{thm:charac_separable}).
\end{ex}
\begin{ex}\label{ex:renyi_higher-D}
Cor.~\ref{cor:LO_monotones} allows us to generalize the construction recalled in the previous example to the multipartite context. Suppose $D \geq 3$, and let $G \in \cG_D$. For any $D$-partite state $\ket{\psi}$, we can introduce the quantity $R_G (\ket{\psi})\in [0, +\infty]$ defined by
\begin{equation}
    R_G (\ket{\psi})\eqdef \begin{cases} - \ln\abs{\tr_G(\ket{\psi})} &\; \mathrm{if}\; \tr_G(\ket{\psi}) \neq 0 \\ 
    +\infty &\; \mathrm{otherwise}\end{cases} 
\end{equation}
$R_G$ defines a $\LO$-monotone (it is non-decreasing under any $\LO$ transformation). If one makes the further assumption that $G$ is genuinely $D$-partite, Thm.~\ref{thm:charac_separable} ensures that: $\ket\psi$ is separable if and only if $R_G(\ket\psi) =0$. In other words, $R_G$ enjoys the same basic properties as an entanglement R\'{e}nyi entropy; hence, at least in this somewhat limited sense, it provides a  multipartite generalization of this concept. 
\end{ex}

\

Let us now explain how we can recover the following known fact from our analysis: any two states that can be related by $\LO$ transformations in both directions are $\LU$-equivalent.
\begin{cor}\label{cor:LO_classes}
    Let $\ket{\psi}$ and $\ket{\phi}$ be two finite-dimensional (and deterministic) $D$-partite pure states. Then:
    \begin{equation}
            \ket{\psi} \simLU \ket{\phi} \quad \Longleftrightarrow \quad \ket{\psi} \toLO \ket{\phi} \; \mathrm{and} \; \ket{\phi} \toLO \ket{\psi}\,.
    \end{equation}
\end{cor}
\begin{proof}
    The direct implication ($\Rightarrow$) is trivial since $\LU$ transformations are both invertible and in $\LO$. In the reverse direction, let us assume that $\ket{\psi} \toLO \ket{\phi}$ and $\ket{\phi} \toLO \ket{\psi}$. By Cor.~\ref{cor:LO_monotones}, we have:
    \begin{equation}
    \forall G \in \cG_D^{\conn}\,, \qquad |\tr_G(\ket{\psi})|= |\tr_G(\ket{\phi})|\,.
    \end{equation}
    By Prop.~\ref{prop:charac_LO}, we can also find a $D$-partite state space $\H^{\sF''}$ and a pure state $\ket{\eta}\in S(\H^{\sF''})$ such that: 
    \begin{equation}
        \forall G \in \cG_D^{\conn}\,, \qquad \tr_G(\ket{\psi}) = \tr_G(\ket{\phi}) \tr_G(\ket{\eta})\,.
    \end{equation}
In particular, if $G$ is a cyclic connected graph, then: $\tr_G (\ket{\psi}) \in \bb{R}_+^*$, $\tr_G (\ket{\phi}) \in \bb{R}_+^*$, and therefore $\tr_G (\ket{\eta})=1$. This implies that $\ket{\eta}$ is separable with respect to any bipartite coarse-graining of $\H^{\sF''}$. But this can only be the case if $\ket{\eta}$ is separable in the $D$-partite state space $\H^{\sF''}$. Hence $\ket{\eta}= \ket{\eta_1}\otimes \cdots \otimes \ket{\eta_D}$ with $\ket{\eta_c}\in S(\H_c)$ for any $c \in \{1,\ldots, D\}$. It follows that
\begin{equation}
    \tr_G(\ket{\psi}) = \tr_G(\ket{\phi}) \tr_G(\ket{\eta}) = \tr_G(\ket{\phi}) \prod_{c=1}^D \langle \eta_c \vert \eta_c \rangle^{k(G)} = \tr_G(\ket{\phi})
\end{equation}
for any $G\in \cG_D^{\conn}$. By Prop.~\ref{prop:charac_LU-1}, we conclude that $\ket{\psi} \simLU \ket{\phi}$.
\end{proof}

A straightforward but interesting consequence of Cor.~\ref{cor:LO_monotones} and Cor.~\ref{cor:LO_classes} is the following observation.
\begin{cor}\label{cor:vanishing_invariant}
Let $G \in \cG_D^{\conn}$ and $V(G)= \{\ket{\psi} \;D\mathrm{-partite}\;\mathrm{pure}\;\mathrm{state}\,|\, \tr_G (\ket{\psi})=0 \}$. For any $D$-partite pure state $\ket{\phi}$, if there exists a $D$-partite state $\ket{\psi} \in V(G)$ such that $\ket{\phi}\toLO \ket{\psi}$, then $\ket{\phi} \in V(G)$. In other words, there is no incoming $\LO$ arrow into $V(G)$.  
\end{cor}
\begin{proof}
    Let $G\in \cG_D^{\conn}$, and $V(G)$ as defined above. If $\ket{\phi}$ and $\ket{\psi}$ are two $D$-partite states such that $\ket{\psi} \in V(G)$ and $\ket{\phi}\toLO\ket{\psi}$, then: $|\tr_G(\ket{\phi})|\leq |\tr_G(\ket{\psi})|$ by Cor.~\ref{cor:LO_monotones}, which implies that $\tr_G(\ket{\phi}) = 0$ since $\tr_G(\ket{\psi}) = 0$. Hence, $\ket{\phi}\in V(G)$.
\end{proof}
\begin{ex}\label{ex:vanishing_K33} Consider the graph $\PT_3=\vcenter{\hbox{\includegraphics[scale=.5]{images/K_33.pdf}}}$ (\ie the complete $3$-colored bipartite graph, also discussed in and around Fig.~\ref{fig:PT_EX}). It is clear that $\tr_{\PT_3}$ takes value in $\bb{R}$ (since exchanging white and black vertices in $\PT_3$ leaves the structure of the graph invariant), and that $\tr_{\PT_3}(\ket{\psi})=1$ for any separable state. It is also known that $\tr_{\PT_3}$ evaluates to negative values for some states.\footnote{This was first checked numerically by Austin Conner and Johann Chevrier (unpublished).} By continuity, we conclude that $V(\PT_3)$ is a non-empty union of $\LU$-classes that has only outgoing arrows in the $\LO$ resource theory of entanglement. More concretely, taking $\sF= \left( \mathbb{C}^3 , \mathbb{C}^3 , \mathbb{C}^3\right)$ and denoting by $(\ket{0},\ket{1},\ket{2})$ the canonical basis of $\mathbb{C}^3$, one can verify that
\begin{equation}
    \ket{\psi_\rm{ex}} \eqdef \frac{1}{\sqrt{3}} \pa{\ket{0}\ot \frac{\ket{01} - \ket{10}}{\sqrt{2}} + \ket{1}\ot \frac{\ket{02} - \ket{20}}{\sqrt{2}} + \ket{2}\ot \frac{\ket{12} - \ket{21}}{\sqrt{2}}} 
\end{equation}
does obey $\tr_{\PT_3}(\ket{\psi_{\rm{ex}}})=0$. This establishes that, indeed, $V(\PT_3)\neq \emptyset$. 
\end{ex}

It would be interesting to investigate the nature of the subvariety defined by the equation $\tr_{\PT_3} (\ket{\psi})=0$ in concrete, low-dimensional examples of $D$-partite Hilbert spaces (such as the three-qutrit state space of $\ket{\psi_{\rm{ex}}}$), building on insightful connections between algebraic geometry and the problem of entanglement classification (see \eg \cite{holweck:tel-02913351} for a review of this approach). More broadly, Cor.~\ref{cor:vanishing_invariant} calls for a combinatorial characterization of trace-invariants that admit zeroes, and for an analysis of the subvarieties formed by these zeroes in low dimensions.

\begin{rem}
    The situation described by Cor.~\ref{cor:vanishing_invariant} and illustrated in the context of a tripartite system in Ex.~\ref{ex:vanishing_K33} cannot be realized in the bipartite setting. Indeed, given that reduced density matrices are positive definite, any bipartite trace-invariant is valued in $\mathbb{R}^*_+$ and thus never vanishes. Hence, Cor.~\ref{cor:vanishing_invariant} captures a genuinely multipartite phenomenon.
\end{rem}

To conclude this section, we establish a useful connection between the concept of partial separability and the $\LO$ preorder.
\begin{prop}
    Let $\sF$ be a $D$-partite state space and $\ket\psi, \ket{\psi'} \in S(\H^\sF)$. We denote by $\xi$ (resp.~$\xi'$) the finest partition of $\{1, \ldots, D\}$ such that $\ket\psi$ (resp.~$\ket{\psi'}$) is $\xi$-separable (resp.~$\xi'$-separable). If $\ket\psi \toLO \ket{\psi'}$, then $\xi \leq \xi'$.
\end{prop}
\begin{proof}
    By contraposition, suppose that $\xi \ngeq \xi'$. Property~$3$ from Prop.~\ref{prop:partial_sep} implies that there exists a graph $G \in \cG_{D, \xi}^\neq \subset \cG_D$ such that $|\tr_{G}(\ket{\psi'})| < 1$. But given that $\ket{\psi}$ is $\xi$-separable, Thm.~\ref{thm:charac_separable} implies that $ |\tr_{G}(\ket{\psi})|=1$.  Hence, we have $|\tr_{G}(\ket{\psi'})| < |\tr_{G}(\ket{\psi})|$. By Cor.~\ref{cor:LO_monotones}, we conclude that one cannot have $\ket{\psi} \toLO \ket{\psi'}$, which concludes the proof.  
\end{proof}

\section{Families of trace-invariants, hypergraph-tensor states, Haar-random states} \label{sec:Trace_Lit_Ref_States} 

In this section, we start by describing and reviewing the properties of particular families of trace-invariants which have appeared in the recent literature, emphasizing their combinatorial description in terms of colored graphs (which, despite its convenience, is not always explicit in the literature). We then introduce the particular families of states of reference on which we will focus in later sections: the hypergraph-tensor states (abbreviated to HT states) on one hand, and the Haar-random state (the only non-deterministic state we will investigate in detail in the present contribution) on the other hand.

\subsection{Families of trace-invariants under consideration}\label{sub:inv-from-lit}

\subsubsection{Entanglement Rényi entropies and cyclic graphs} \label{sss:necklace}

\begin{figure}[!ht]
    \centering
    \includegraphics[height = 1.8cm]{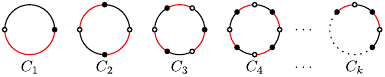}
    \caption{Family of two-colored graph $C_k$ associated with $k$-purities of reduced density matrices in the bipartite setting.} 
    \label{fig:bip}
\end{figure}

Let $k > 1$ be an integer. Given a pure state $\ket{\psi}$ on a bipartite Hilbert space $\H_{1} \ot \H_{2}$, the \emph{entanglement Rényi-$k$ entropy} of $\ket{\psi}$ is the Rényi-$k$ entropy of the state $\rho_{1}$ induced on $\H_{1}$ (resp. of $\rho_{2}$ induced on  $\H_{2}$). As already stated in Sec.~\ref{sec:separable}, the associated trace-invariant - the $k$-\textit{purity} of $\rho_{1}$ (resp. of $\rho_{2}$) - corresponds to a cyclic graph having $2k$ vertices, denoted by $C_k$ and represented in Fig.~\ref{fig:bip}:
\begin{equation}
    \tr_{C_k}(\ket{\psi}) = \tr \rho_{1}^k = \tr \rho_{2}^k\,.
\end{equation}
While the entanglement Rényi-$k$ entropies are defined for $k \geq 2$, we remark that $C_k$ and the associated trace-invariants are defined for $k \geq 1$. As already mentioned in the introduction and in Sec.~\ref{sec:LUinvariants}, the knowledge of this quantity for arbitrary integer values of $k$ characterizes fully the entanglement spectrum of $\ket{\psi}$, or equivalently, its entanglement structure. 

\begin{figure}[!ht]
    \centering
    \includegraphics[height = 2.3cm]{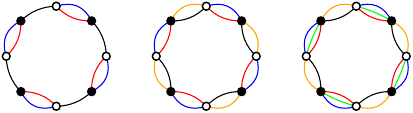}
    \caption{On the left, a cyclic graph corresponding to a bipartition of $\H$ of the form $\H_c \ot \H_{\bar{c}}$ where $c$ is the black color. In the middle, a cyclic graph with 4 colors. On the right, a cyclic graph with 5 colors.}
    \label{fig:Necklace}
\end{figure}

More generally, starting from a $D$-partite state space $\sF=(\H_1 , \ldots , \H_D)$, with $D\geq 3$, one can coarse-grain its Hilbert space $\H^\sF = \H_1 \otimes \cdots \otimes \H_D$ into a $2$-partite Hilbert space (see Sec.~\ref{sec:generalities-on-coarse-graining}): choosing $B \subsetneq \{ 1, \ldots , D\}$ with $B \neq \emptyset$, the partition $\zeta= \{ B , \bar B\}$ (where $\bar B$ denotes the complement of $B$) produces a bipartite coarse-graining $\sF_\zeta \succeq \sF$ represented by the Hilbert space $\H^{\sF_\zeta} = \H_B \otimes \H_{\bar B}$. As explained in Sec.~\ref{sec:generalities-on-coarse-graining}, the set of cyclic $2$-colored graphs characterizing bipartite entanglement in $\H^{\sF_\zeta}$ can be identified to the subset of $D$-colored graphs $\cG_{D, \zeta} \subset \cG_D$. Accordingly, a $D$-colored graph $G \in \cG_D$ was thus defined to be \emph{cyclic}\footnote{In the literature on random tensor models, such a graph is often referred to as a \emph{necklace} (see \eg Refs.~\cite{Bonzom:2013lda,BONZOM2015161}).} whenever it belongs to $\cG_{D, \zeta}$ for some bipartition $\zeta$. Examples of (connected) cyclic graphs are represented in Fig.~\ref{fig:Necklace}.

In the following, the entanglement Rényi-$k$ entropy of $\ket{\psi} \in \H^\sF$, seen as a bipartite state in $\H^{\sF_\zeta} = \H_B \ot \H_{\bar{B}}$, will be denoted $\S_B^{(k)}(\ket\psi)$ (note that $\S_{B}^{(k)}(\ket\psi) = \S_{\bar{B}}^{(k)}(\ket\psi)$).

\subsubsection{Melonic invariants} \label{sss:Melo}

The melonic family plays a central role in the study of random tensor models, particularly in the context of their $1/N$ expansion. In these models, the leading order of the expansion is dominated by melonic diagrams, which are specific Feynman graphs characterized by their tree-like series-parallel structure. This dominance has been proven in various settings, including colored tensor models (see Ref.~\cite{Gurau2011_1}), multi-orientable models (see Ref.~\cite{Dartois2014}), and uncolored tensor models (see Ref.~\cite{PhysRevD.95.046004}). The melonic family is significant due to its connection with the \textit{Sachdev–Ye–Kitaev} (SYK) model (see Refs.~\cite{Sachdev:1992fk,Kitaev2015}), which has motivated their study as toy models for quantum gravity and holography (see Ref.~\cite{10.1063/1.4983562}). 

\begin{figure}[!ht]
    \centering
    \includegraphics[height = 1.5cm]{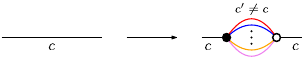}
\caption{Graph operation involved in the recursive definition of melonic graphs, called melonic insertion.}
    \label{fig:MelonConst}
\end{figure}

\begin{figure}[!ht]
    \centering
    \includegraphics[height = 3cm]{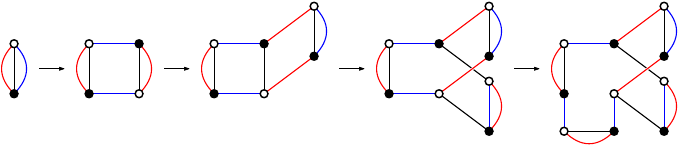}
    \caption{Recursive construction of a melonic 3-colored graph starting from a $2$-vertex graph (on the left).}
    \label{fig:ExMelo}
\end{figure}

Melonic graphs can be defined recursively. The only graph with $k=1$  -- called the \textit{$2$-vertex graph} and which represents the square norm of a tensor -- is melonic. 
Starting from a $2$-vertex graph, one constructs all connected melonic graphs by inductively performing the graph operation of Fig.~\ref{fig:MelonConst} (referred to as a \textit{melonic insertion}). Non-connected melonic graphs have connected components constructed this way.  An example of the construction of connected melonic graphs is depicted in Fig.~\ref{fig:ExMelo}. See Ref.~\cite{gurau_random_2017} for more details on melonic graphs. 

\

Finally, for $c_0\in\{1,\ldots, D\}$ and $B\subset \{1,\ldots, D\}$, we call \emph{$c_0$-melonic graph} (resp.~$B$-melonic graph) a melonic graph for which the move of Fig.~\ref{fig:MelonConst} is never performed on an edge of color $c_0$ (resp.~on any edge of color $c\in B$). Equivalently, a melonic graph $G$ is $c_0$-melonic (resp.~$B$-melonic) if and only if the graph $G^{\widehat {c_0}}$ obtained by removing all the edges of color $c_0$ (resp.~the graph $G\lvert_{\bar B}$) is connected.

The following properties - which follow from the definition - will be useful later on: if $G$ is $c_0$-melonic, then $G^{\widehat {c_0}}$ is melonic, and if $G$ is $B$-melonic, it is $B'$-melonic for any $B'\subset B$. If a connected graph $G$ is melonic but not $c$-melonic for any $c$ (resp.~$B$-melonic but not $B'$ melonic for any $B'\nsubseteq B$), then for any $c$ and any $B'\nsubseteq B$,  $G^{\hat c}$ (resp.~$G\lvert_{\bar {B'}}$) is disconnected and melonic.   Furthermore, adding an extra color $D+1$ to a connected and melonic $G$  to obtain $G'\in\cG_{D+1}^{\conn}$,  then $G'$ is melonic if and only if it is $(D+1)$-melonic. 

For any color $c \in \{1, \ldots,  D\}$, we can consider the bipartition $\zeta= \{ \{c\}, \bar{\{c\}}\}$. Any element $G \in \cG_{D, \zeta}$ is then $\bar{\{c\}}$-melonic, which we will abridge to $\bar{c}$-melonic throughout the paper (see \eg the leftmost graph of Fig.~\ref{fig:Necklace}). By contrast, a cyclic graph associated with any other bipartition is not melonic.

\subsubsection{Planar invariants} \label{sss:planar}

Another kind of colored graphs (and corresponding trace-invariants) of interest are those with a planar \emph{regular embedding} (see Ref.~\cite{gagliardi1981regular}) or \emph{jacket} (see Refs.~\cite{BenGeloun:2010wbk, Ryan:2011qm, Gurau:2011aq}).  

Given a connected graph, one may view it as embedded into a surface by choosing a cyclic ordering of its edges around each vertex; that is, a graph together with this additional structure can be seen as drawn on a two-dimensional oriented surface (up to isomorphisms): the cyclic ordering corresponds to the list of edges encountered going around each vertex on the surface. Such an embedded graph may equivalently be interpreted as a \emph{ribbon graph} or a \emph{combinatorial map} (see \eg~\cite{lando2004graphs, ellis2013graphs}). In particular, given a connected $D$-colored graph $G\in \cG_D^{\conn}$ and a cyclic ordering of the color-set $\{1, \ldots, D\}$, which we prescribe in the form of a permutation $\tau\in S_D$ comprised of a single cycle of length $D$, we can cyclically orient the edges around the black (resp.~white) vertices according to $\tau$ (resp.~$\tau^{-1}$). The embedded graphs obtained from $G$ for different choices of $\tau$ are its \emph{regular embeddings} $\{G_\tau\}$ (see Ref.~\cite{gagliardi1981regular}), also called \emph{jackets} (see Refs.~\cite{BenGeloun:2010wbk, Ryan:2011qm, Gurau:2011aq}) of $G$. Note also that $G_\tau = G_{\tau^{-1}}$ for any $\tau$, so that $G$ admits $\frac{(d-1)!}{2}$ regular embeddings (there exist other embeddings of $G$, which are obtained by allowing vertex-dependent color orderings).

The genus $g(G_\tau) \in \mathbb{N}$ is the number of holes of the underlying orientable surface. We introduce the notation
\begin{equation}
    \label{def:g-tau}
    g_\tau(G)=g(G_\tau)
\end{equation}
Remark that $g_\tau(G) = g_{\tau^{-1}}(G)$ and that if $g_\tau(G)=0$ for some $\tau$, the corresponding embedding of $G$ is planar (the underlying surface is a sphere). Moreover, a graph is melonic if and only if all its regular embeddings are planar (see Refs.~\cite{Gurau:2011aq, Gurau:2011xp, gurau_random_2017}). Trace-invariants associated with colored graphs with one or more planar jackets will be relevant in Sec.~\ref{sec:LU-and-ref-states}. 

\subsubsection{Maximally single-trace invariants} \label{sss:MST}

Maximally single-trace colored graphs were introduced as a distinguished class of interactions in tensorial quantum field theories, motivated in particular by strongly coupled models such as the SYK model and other related holographic models  (see Refs.~\cite{Ferrari2019,Valette:2019nzp}). A \emph{maximally single-trace} $D$-colored graph $G$ is connected and satisfies the properties that for any pair of colors $(i,j) \in \paa{1,\dots,D}^2$, the restriction of the graph $G$ to edges labeled by colors $i$, and $j$ has only one connected component. Said differently, this means that removing any subset of colors $C \subset \paa{1,\dots,D}$, with $0 \leq  \abs{C}\leq D-2$, from a maximally single-trace invariant leaves the graph connected. 

\begin{figure}[H]
    \centering
    \includegraphics[width = \textwidth]{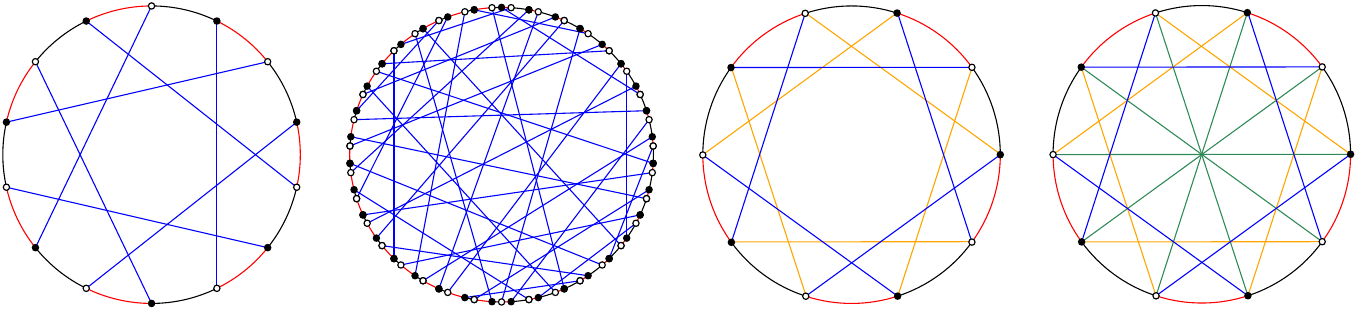}
    \caption{Examples of maximally single-trace invariants. From left to right: an edge-coloring of the Heawood graph which is maximally single-trace; an edge-coloring of the Balaban 10-cage graph which is maximally single-trace; a $D=4$ example of a maximally single-trace invariant; and the $5$-colored complete bipartite graph $\K_{5,5}$.}
    \label{fig:MST}
\end{figure}

While maximally single-trace graphs do not exist for arbitrary order $k$,\footnote{An exception stands for $D=2$ since all connected graphs are maximally single-trace.} it is known that for any number $D$ of colors there exist maximally single-trace $D$-colored graphs (see proof in Ref.~\cite{Ferrari2019}). In Fig.~\ref{fig:MST}, we show particular examples of maximally single-trace colored graphs with a varying number of colors. As a remark, we observe that certain edge-colorings of some \textit{cage graphs} such as the \textit{Heawood graph} (see the leftmost graph of Fig.~\ref{fig:MST} and Ref.~\cite{heawood_map-colour_1949}) and the \textit{Balaban 10-cage graph} (see the middle-left graph of Fig.~\ref{fig:MST} and Ref.~\cite{Balaban1972}) are maximally single-trace.

In the following subsection, we present a distinguished family of trace-invariants arising in quantum information theory: the moments of the partial transpose, whose odd moments are maximally single-trace.

\subsubsection{Moments of the partial transpose and negativities}  \label{subsubsec:PT}

\begin{figure}[H]
    \centering
    \includegraphics[height = 1.3cm]{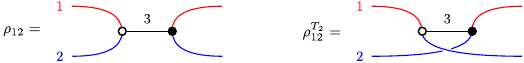}
    \caption{Illustration of the partial transpose.}
    \label{fig:PT}
\end{figure}

In the tripartite setting, the moments of the partial transpose ($\PT$) are used to compute the trace norm of the partial transpose of the reduced density matrix. For $\ket{\psi} \in \H_1 \ot \H_2 \ot \H_3$ a tripartite pure state, consider, for instance, the reduced density matrix $\rho_{12} \eqdef \tr_{\H_3} \ket{\psi} \bra{\psi}$  induced on $\H_1\ot \H_2$. As illustrated in Fig.~\ref{fig:PT}, the partial transpose of $\rho_{12}$ has components 
\begin{equation}
(\rho_{12}^{T_{2}})_{i_1 i_2 ; j_1 j_2} = (\rho_{12})_{i_1 j_2 ; j_1 i_2}\,,
\end{equation}
where we remark that $\rho_{12}^{T_2} = \pa{\rho_{12}^{T_1}}^T$ where $T$ is the total transpose.
    
\begin{figure}[H]
    \centering
    \includegraphics[height = 3.5cm]{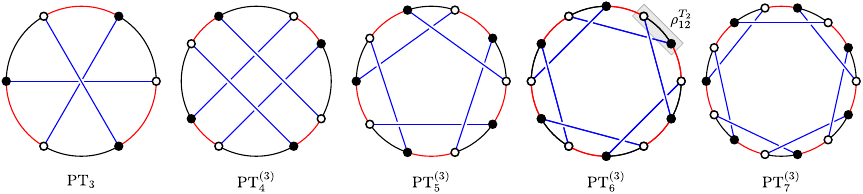}
    \caption{Colored graphs $\PT_3$, $\PT_4^{(3)}$, $\PT_5^{(3)}$, $\PT_6^{(3)}$ and $\PT_7^{(3)}$ corresponding to the partial transpose moments. The black edge denotes the part $3$ that has been traced out.}
    \label{fig:PT_EX}
\end{figure}

The traces of powers of the partial transpose of $\rho_{12}$ are called \textit{moments of the partial transpose} of $\rho_{12}$, and the graphs  associated with these trace-invariants:
\begin{equation}
    \tr_{\PT_k^{(3)}}(\ket{\psi}) \eqdef \tr \pac{\pa{\rho^{T_{2}}_{12}}^k} = \tr \pac{\pa{\rho^{T_{1}}_{12}}^k} \,,
\end{equation} 
will be denoted by $\PT_k^{(3)}$, with $k \geq 1$ an integer. Examples for the even and odd partial transpose moments of $\rho_{12}$ are shown in Fig.~\ref{fig:PT_EX}. Note that $\PT_{1}^{(c)}$ and $\PT_{2}^{(c)}$ are melonic, and that $\{\PT_{3}^{(c)}\}_{1 \leq c\leq 3}$ define a unique graph, which we denote simply as $\PT_{3}$ \footnote{If $\tau=(12\ldots k)$, then one can represent: $\PT_k^{(1)}$ by $(\mathrm{id},\tau, \tau^{-1})\sim (\mathrm{id},\tau^{-1}, \tau)$; $\PT_k^{(2)}$ by $(\tau, \mathrm{id}, \tau^{-1})\sim (\tau^{-1}, \mathrm{id} , \tau)$; $\PT_k^{(3)}$ by $(\tau,  \tau^{-1} , \mathrm{id})\sim (\tau^{-1}, \tau, \mathrm{id})$. We then have $\PT_k^{(1)}= \PT_k^{(2)}$ if and only if $(\mathrm{id},\tau, \tau^{-1})\sim (\tau^{-1}, \mathrm{id} , \tau)$, which is equivalent to $(\mathrm{id},\tau, \tau^{-1})\sim (\mathrm{id}, \tau, \tau^{2})$. Moreover, the latter condition holds if and only if one can find $\sigma\in S_k$ such that: $\sigma \tau \sigma^{-1} = \tau$ and $\sigma \tau^2 \sigma^{-1} = \tau^{-1}$. Taking the product of those two equations, one obtains the necessary condition $\sigma \tau^3 \sigma^{-1} = \mathrm{id}$, which is equivalent to $\tau^3 = \mathrm{id}$. This holds if and only if $k = 3$ and, similarly, one can show that $\PT_k^{(2)}= \PT_k^{(3)}$ implies $k=3$. Hence, for any $k \neq 3$, the graphs $(\PT_k^{(c)})_{1\leq c \leq 3}$ are pairwise distinct.}.  

The moments of the partial transpose have been extensively studied in the literature, as they provide information on the bipartite entanglement properties of mixed states. Examples include inequalities satisfied only for entangled mixed states (see Refs.~\cite{10.1063/1.4799440,Yu:2021uzg,Neven:2021igr,Carrasco:2022rcy,Tarabunga:2025znd}).

Partial transpose moments provide a way to compute the trace norm of the partial transpose\footnote{The operator $\rho_{12}^{T_{2}}$ being Hermitian (as is clear from its graphical representation in Fig.~\ref{fig:PT}), it is diagonalizable, and one has $\sqrt{\rho_{12}^{T_{2}} \,^\dagger \rho_{12}^{T_{2}}}= \sqrt{\rho_{12}^{T_{2}}  \rho_{12}^{T_{2}}} = \abs{\rho_{12}^{T_{2}}}$ in the sense of spectral calculus.} 
\begin{equation}
    \norm{\rho_{12}^{T_{2}}}_1 \eqdef \tr \sqrt{\rho_{12}^{T_{2}} \,^\dagger \rho_{12}^{T_{2}}} = \tr\abs{\rho_{12}^{T_{2}}} = \lim_{n \to 1/2} \tr_{\PT_{2n}^{(3)}}(\ket{\psi}) \,.
\end{equation}
A rigorous treatment of this limit would require an analytic continuation in $n$. However, carrying out this analytic continuation lies beyond the scope of the present work. \\
The even moments of the partial transpose are sometimes called \textit{even Rényi negativities}, see Ref.~\cite{Dong2021}.  From the trace norm one derives the \textit{entanglement negativity} (EN) and the \textit{logarithmic negativity} (LN). They are entanglement monotones\footnote{They constitute entanglement monotones under $\LOCC$ for bipartite mixed states (see Refs.~\cite{Vidal2002, Plenio:2005cwa, Eisert2006}).} defined as: 
\begin{equation}
    \S_\rm{EN}(\ket{\psi}) \eqdef \frac{1}{2}\pa{\norm{\rho_{12}^{T_{2}}}_1 - 1} \quad \rm{ and } \quad \S_\LN(\ket{\psi}) \eqdef \ln \norm{\rho_{12}^{T_{2}}}_1 \,.
\end{equation}
These quantities have emerged as central probes of mixed-state entanglement, capturing nontrivial correlations beyond those accessible through the entanglement entropy (see Refs.~\cite{Schneeloch2023,PhysRevA.80.012325,Castelnovo2013}). Moreover, they are closely tied to the \textit{positive partial transpose} (PPT) or \textit{Peres-Horodecki} criterion, which states that 
\begin{equation} 
    \norm{\rho^{T_{2}}_{12}}_1 > 1 \quad \Rightarrow \quad \rho_{1 2} \rm{ entangled as a bipartite mixed state}\,.
\end{equation}

\

On the other hand, the odd moments of the partial transpose, sometimes referred to as \textit{odd Rényi negativities}, allow computing the so-called \textit{partially transposed entropy} (PTE),\footnote{We adopt the terminology of Refs.~\cite{Vidal2002,Dong2021}; in the literature, EN is often referred to simply as Negativity, while PTE is also known as the odd entanglement entropy.} defined as 
\begin{equation}
    \S_\PTE(\ket{\psi}) \eqdef - \frac{1}{2} \cdot \lim_{n \to 1} \partial_n \ln \tr_{\PT_{2n-1}^{(3)}}(\ket{\psi})\,.
\end{equation}
As for the trace norm of the partial transpose, this limit requires, strictly speaking, an analytic continuation in $n$. The definition of the PTE mirrors that of the von Neumann entropy, but applied to the eigenvalues of the partially transposed density matrix $\rho^{T_{2}}_{12}$ rather than those of $\rho_{12}$. Explicitly, one has (see \eg Ref.~\cite{Dong2021}):
\begin{equation}
    \S_\PTE(\ket{\psi}) = - \sum_{i=1}^N \lambda_i \ln \abs{\lambda_i} \quad \rm{ with } \quad \lambda_i \in \spec \pa{\rho^{T_{2}}_{12}}\,.
\end{equation}

\begin{figure}[H]
    \centering
    \includegraphics[height = 4cm]{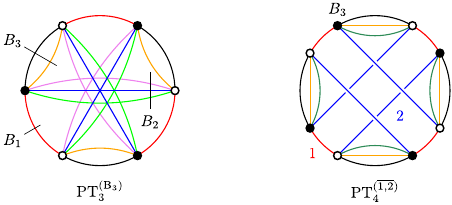}
    \caption{Tripartite block versions of the moments of the partial transpose. On the right, we highlighted the fact that blocks $B_1$ and $B_2$ are singlets by the use of $\bar{1,2}$ instead of $B_3$.}
    \label{fig:genPT}
\end{figure}

In a $D$-partite Hilbert space $\H^\sF$ with $D > 3$, a tripartite block version of the moments of the partial transpose can be achieved by considering a tripartition $\zeta = \{ B_1 , B_2 , B_3\}$ of $\paa{1,\dots,D}$, and the coarse-grained Hilbert space $\H^{\sF_\zeta}= \H_{B_1} \otimes \H_{B_2} \otimes \H_{B_3}$. Given $\rho_{B_1 B_2} \eqdef \tr_{\H_{B_3}} \ket{\psi} \bra{\psi}$, the reduced density matrix induced by $\ket{\psi}\in \H^\sF \simeq \H^{\sF_\zeta}$ on $\H_{B_1}\otimes \H_{B_2}$, the $k$-th moment of the partial transpose is given by 
\begin{equation}
    \tr_{\PT_k^{(B_3)}} (\ket{\psi}) \eqdef  \tr \pac{(\rho_{B_1 B_2}^{T_{B_2}})^k} \,,
\end{equation}
and is associated with the colored graph $\PT_k^{(B_3)}$. Fig.~\ref{fig:genPT} shows different tripartite block versions of the moments of the partial transpose.

\subsubsection{Realignment moments} \label{sss:RM_JRM}

\begin{figure}[H]
    \centering
    \includegraphics[height = 1.5cm]{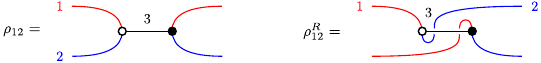}
    \caption{Illustration of the realignment operation.}
    \label{fig:R}
\end{figure}

Let $\ket{\psi}$ denote a $3$-partite pure state, and $\rho = \ket{\psi}\bra{\psi}$. The \textit{realignment} (or \textit{reshuffling}) operation (see \eg Ref.~\cite{aubrun_realigning_2012}), depicted in Fig.~\ref{fig:R}, acts on the matrix elements of the reduced density matrix $\rho_{12}$ according to: 
\begin{equation}
    (\rho_{12}^{R})_{i_1 i_2 ; j_1 j_2} = (\rho_{12})_{i_1 j_1 ; i_2 j_2}\,.
\end{equation}

\begin{figure}[H]
    \centering
    \includegraphics[height = 4cm]{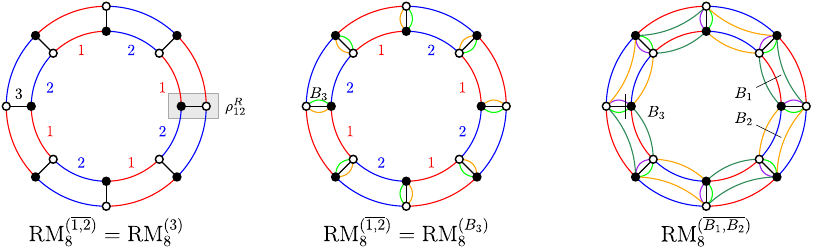}
    \caption{Left: an example for $k=8$ white vertices of colored graph $\RM^{(3)}_k$ associated with the realignment moments (the color 3 is represented in black). Middle and right: examples of multipartite generalizations of the realignment moments.}
    \label{fig:RMgen}
\end{figure}

We use the notation $\rho_{21}^{R} = {\rho_{12}^{R}}^\dagger$. The traces of powers of $\rho_{12}^R {\rho_{12}^R}^\dagger = \rho_{12}^R \rho_{21}^{R} $ are called \textit{realignment moments}. They are associated with colored graphs $\RM^{(3)}_{2n}$ with $k=2n$ ($n \in \mathbb{N}^*$) white vertices:
\begin{equation}
    \tr_{\RM_{2n}^{(3)}}(\ket{\psi}) \eqdef \tr \pac{\pa{ \rho_{12}^R \rho_{21}^{R}}^n}\,.
\end{equation}
The exponent $(3)$ refers to the fact that $\rho_{12}$ is obtained from $\ket \psi$ by  partial trace over $\H_3$. The graph $\RM_{8}^{(3)}$ is presented in the left of Fig.~\ref{fig:RMgen}. Note that $\RM_{2}^{(c)}$ is the same graph as  $\PT_{2}^{(c)}$, and that $\RM_{4}^{(c)}$ are the same graphs for all $c\in \{1,2,3\}$, denoted by $\RM_{4}$. The graph $\RM_{4}$ is the same as $\ME_2^3$ introduced for multi-entropies (see Ref.~\cite{Penington2023}).

There are also inequalities based on realigned moments, whose violation guarantees the entanglement of the density matrix, see \eg Ref.~\cite{Aggarwal:2023qbj}. 

The trace norm of the realigned density matrix can be computed from the realignment moments as previously done for the partial transpose: 
\begin{equation}
    \norm{\rho^{R}_{12}}_1 = \lim_{n \to 1/2} \tr \pac{\pa{\rho_{12}^R {\rho_{21}^R} }^{n} } = \lim_{n \to 1/2} \tr_{\RM^{(3)}_{2n}}(\ket{\psi}) \,,
\end{equation}
where $\rho_{12}^R$ is no longer Hermitian. To make this limit well defined, one should, in principle, perform an analytic continuation in the parameter $n$. The realignment transformation also has a corresponding criterion which is independent of the PPT criterion and can detect certain PPT entangled states, called the \textit{computable cross-norm} or \textit{realignment} criterion: 
\begin{equation}
    \norm{\rho^{R}_{1 2}}_1 > 1 \quad\Rightarrow\quad \rho_{12} \rm{ is entangled as a bipartite mixed state}\,.
\end{equation}

If $\ket\psi \in \H^\sF$ is a $D$-partite pure state with $D>3$, tripartite block version can be provided using a tripartition $\zeta=\{B_1 , B_2 , B_3\}$ of $\paa{1,\dots,D}$. The corresponding realignment moment of $\rho_{B_1 B_2} = \tr_{\H_{B_3}} \ket{\psi} \bra{\psi}$ is given by the trace-invariant 
\begin{equation}
    \tr_{\RM_{2n}^{(B_3)}}(\ket{\psi}) \eqdef \tr \pac{\pa{\rho_{B_1 B_2}^R \rho_{B_2 B_1}^R}^n} \,,
\end{equation}
whose associated colored graph is denoted by $\RM_{2n}^{(B_3)}$. We give explicit examples of tripartite block versions of the realignment moments in the middle and right-hand side of Fig.~\ref{fig:RMgen}.

\

\begin{figure}[ht]
    \centering
    \includegraphics[height = 4cm]{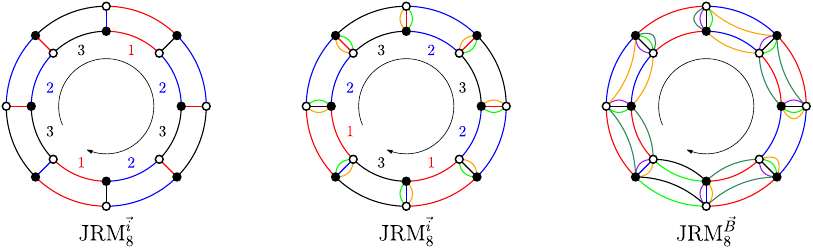}
    \caption{Left: the joint realignment moment and its corresponding colored graph. The sequence of colors is $(3,2,3,1,2,3,2,1)$. Middle: a tripartite block version of the joint realignment moment with a sequence of color $(1,2,3,2,3,2,1,3)$. Right: a tripartite block version of the joint realignment moment with a sequence of blocks.}
    \label{fig:JRM}
\end{figure}

One can more generally define \emph{joint realignment moments}, of the form:
\begin{equation}
\label{eq:joint-realignment-moment}
\tr_{\JRM_k^{\vec{i}}} (\ket{\psi}) \eqdef  \tr \pac{ \rho_{12}^R  \rho_{23}^R \rho_{31}^R \rho_{13}^R   \rho_{32}^R  \cdots}\,
\end{equation}
which correspond to graphs with a similar cyclic structure as $\RM^{(3)}_{k}$ on the left of Fig.~\ref{fig:RMgen}, but for which the colors when going around the cycle are given by a cyclic sequence of colors $\vv{i} = (1,2,3,1,3,2,\ldots)$ instead of $(1,2,1,2,\ldots)$.  The joint moment $\tr \pac{ \rho_{12}^R  \rho_{23}^R \rho_{31}^R}$ corresponds for instance to the colored graph $\PT_{3}$, while the graph $\PT^{(3)}_{4}$ corresponds to $\tr \pac{ \rho_{32}^R  \rho_{23}^R \rho_{31}^R\rho_{13}^R}$ (see the first two examples on the left of Fig.~\ref{fig:PT_EX}). Another example is shown in the left of Fig.~\ref{fig:JRM} and represents
\begin{equation}
    \tr_{\JRM_8^{(3,2,3,1,2,3,2,1)}} (\ket{\psi}) = \tr \pac{\rho_{32}^R \rho_{23}^R \rho_{31}^R \rho_{12}^R \rho_{23}^R \rho_{32}^R \rho_{21}^R \rho_{13}^R} \,.
\end{equation}

In a $D$-partite state space with $D>3$, tripartite block versions can again be straightforwardly introduced: given a fixed tripartition $\{ B_1 , B_2 , B_3\}$ of $\{1, \ldots , D\}$, a block version of joint realignment moments is defined by a cyclic sequence of blocks of colors $\vec{B} = (B_1, B_2, B_3, \dots)$ as illustrated in the middle and right-hand side of Fig.~\ref{fig:JRM}.

\subsubsection{Lattice extension of the joint realignment moments} \label{ss:latticeJRM}

In this paper, we introduce a class of colored graphs defined for $D>3$ which can be viewed as an extension of the joint realignment moments to a two-parameter-dependent lattice, denoted by $L_{m,n}^{\vec i}$ with $m,n\geq 1$. The construction proceeds as follows.

\begin{figure}[ht]
    \centering
    \includegraphics[width = \textwidth]{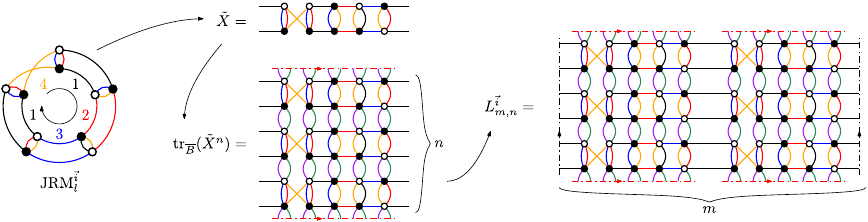}
    \caption{Principle of the construction of the lattice extension of the joint realignment moment. The construction is explicit for the $6$-colored lattice extension with parameters $m=2$, $n=3$ and cyclic sequence $(4,\, 1,\, 2,\, 3,\, 1)$. Remark that $B = \paa{1,2,3,4}$ and $\bar{B} = \paa{5,6}$. The dashed-dotted lines on the rightmost colored graph represent an identification of the edges.}
    \label{fig:JRMLattice}
\end{figure}

Let $B \subset \paa{1,\dots,D}$ be a subset of colors with $\abs{B} > 2$. We first introduce the $\abs{B}$-colored graph $\JRM_{l}^{\vec i}$, where $\vec i$ is a sequence of colors containing each element of $B$ at least once,\footnote{Although this requirement is not essential, it will play an important role in subsequent developments.} implying $l = \lvert \vec i \rvert \geq \abs{B}$. Next, we cut two edges of the same color $c \in \vec i$ resulting in a matrix $\tilde{X} \in \End(\H_c \ot \H_c^*)$ which can be interpreted as an ``open ladder'' (see $\tilde{X}$ in Fig.~\ref{fig:JRMLattice}).

We then construct an $l \times n$ lattice by duplicating $n$ times the matrix $\tilde{X}$ and tracing over the remaining $D-\abs{B}$ colors. This procedure yields a matrix in $\End\pac{(\H_c \ot \H_c^*)^n}$, explicitly given by
\begin{equation}
    \tr_{\bar{B}} \pa{\tilde{X}^n} \,,
\end{equation}
and illustrated in Fig.~\ref{fig:JRMLattice}. Finally, we take $m$ copies of this matrix and perform a trace over the color $c$. The resulting object (represented as a lattice with periodic boundary conditions, and hence that can visually be represented on a torus) is the lattice extension of the joint realignment moment considered in this work, namely
\begin{equation}
    L_{m,n}^{\vec i} \eqdef \tr \pac{ \pa{\tr_{\bar{B}} \pa{\tilde{X}^n}}^m} \qquad \rm{with} \qquad k(L_{m,n}^{\vec i}) = \|\vec i \| mn\,.
\end{equation}
A colored graph representation of $L_{m,n}^{\vec i}$ and its construction is presented Fig.~\ref{fig:JRMLattice}.

\subsubsection{Multi-entropies} \label{ss:ME}

\begin{figure}[ht]
    \centering
    \includegraphics[height = 6cm]{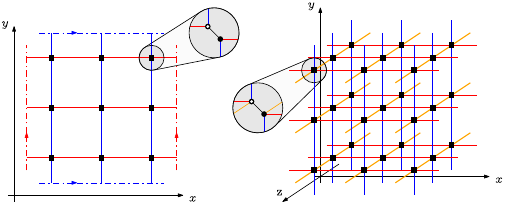}
    \caption{Colored graphs $\ME_n^D$ associated with multi-entropy trace-invariants for $D=3$, $n=3$ (left) and $D=4$, $n=3$ (right). To improve readability, black edges have been contracted and represented as black boxes, and coordinate axes have been added. The underlying topology of the invariant is a $(D-1)$-dimensional torus. For clarity, boundary identifications are not shown in the $D=4$ case, but, for example, the top blue edges are identified with the bottom ones.}
    \label{fig:ME2}
\end{figure}

Multi-entropies have garnered considerable attention in recent years due to their holographic interpretation (see \eg Refs.~\cite{Gadde2022,Penington2023,Gadde2023,Harper2024,Iizuka:2025ioc,Iizuka2025BH,Iizuka:2025caq}), as well as their intrinsic quantum-theoretic properties (see Refs.~\cite{Gadde2024,Iizuka:2025caq}). The construction relies on $\LU$-invariants called \textit{Rényi multi-entropies}: for any integer $n \geq 2$, and any $D$-partite state $\ket\psi$, one considers 
\begin{equation} \label{eq:RényiME}
    \S_{\ME_n^D}(\ket{\psi}) =  \frac{1}{(1-n)n^{D-2}} \ln \pac{\frac{\tr_{\ME_n^D}(\ket{\psi})}{\pa{\tr_{\ME_1^D}(\ket{\psi})}^{n^{D-1}}}} \,,
\end{equation}
associated with the $D$-colored graph denoted by $\ME_n^D$. Unlike Rényi multi-entropies, $\ME_n^D$ is still defined for $n=1$. Furthermore, the colored graph $\ME_n^D$ respects a lattice symmetry with identified boundaries, as illustrated in Fig.~\ref{fig:ME2}. One observes that $\ME_n^D$ contains $k(\ME_n^D) = n^{D-1}$ copies of the reduced density matrix $\rho_{\bar D}$ induced by $\ket\psi$ on $\H_1 \otimes \cdots \otimes \H_{D-1}$, and can be represented by a $D$-tuple of permutations of the form $\vec{\sigma}=(\sigma_1 , \ldots , \sigma_{D-1}, \mathrm{id})$, where each $\sigma_i$ consists in $n^{D-2}$ cycles of size $n$.

For any $D$, one defines from such trace-invariants the $D$-partite \textit{multi-entropy} as the $n \to 1$ limit of Rényi multi-entropies with proper normalization (see Refs.~\cite{Penington2023,Iizuka:2025caq}). Indeed, from the literature, we define\footnote{Some authors omit the $n^{D-2}$ normalization factor (see Refs.~\cite{Gadde2022,Iizuka2025BH}).}
\begin{equation}
    \S_\ME(\ket{\psi}) = \lim_{n \to 1}  \S_{\ME_n^D}(\ket{\psi})\,,
\end{equation}
where this limit formally requires an analytic continuation in $n$. Let us emphasize that for $D=2$, one recovers the cycle graph shown in Fig.~\ref{fig:bip}, \ie $\ME_n^2 = C_n$. Therefore, the multi-entropy of a bipartite system coincides with the usual entanglement entropy of a bipartite system.

\subsubsection{Reflected entropy and multipartite generalizations}\label{ss:RE}

The \textit{reflected entropy} (RE) has recently attracted significant attention (see Refs.~\cite{Akers2022,Akers2023,Akers2024}), both as a refined probe of entanglement for mixed states and as a quantity with direct holographic interpretation: it captures the entanglement wedge cross-section.

\begin{figure}[ht]
    \centering
    \includegraphics[height = 5cm]{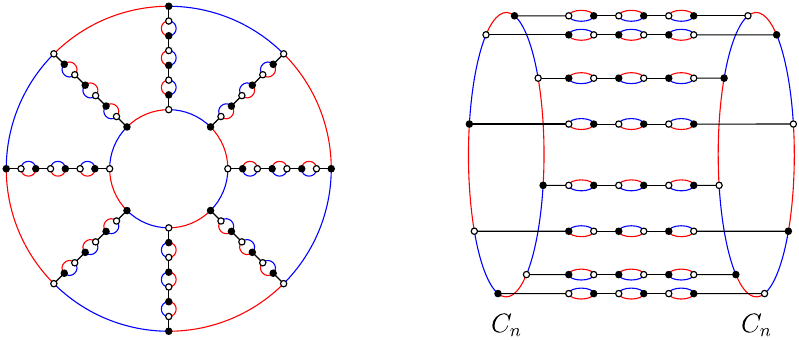}
    \caption{Colored graph $\RE_{8,4}^{(3)}$ associated with the construction of the Rényi reflected entropy, obtained from $\RM_{8}^{(3)}$ by inserting a sequence of 
    3 melons on each color 3 (black) edge.}
    \label{fig:RE_mn}
\end{figure}

Consider the mixed state $\rho_{1 2} = \tr_{\H_3} \ket{\psi} \bra{\psi} \in \End(\H_1\otimes\H_2)$ associated with a given tripartite pure state $\ket{\psi} \in \H_1 \otimes \H_2 \otimes \H_3$. The reflected entropy of $\ket{\psi}$ is constructed via the so-called \textit{Rényi reflected entropies} of $\rho_{1 2}$, as: 
\begin{equation}
    \S_{\RE_{m,n}^{(3)}}(\ket{\psi}) = \frac{1}{1-n} \ln \pac{\frac{\tr_{\RE_{m,n}^{(3)}}(\ket{\psi})}{\pa{\tr_{\RE_{m,1}^{(3)}}(\ket{\psi})}^n}} \,,
\end{equation}
where $n > 1$ is an integer, and $m \in 2 \mathbb{N}^*$. These quantities are related to a $2$-parameter family of colored graphs $\RE_{m,n}^{(3)}$ defined for $n\geq 1$ and $m\in 2 \bb{N}^*$, an example of which is represented in Fig.~\ref{fig:RE_mn}, with $m=8$ and $n=4$. As seen from the left-hand side of Fig.~\ref{fig:RE_mn}, the colored graph $\RE_{m,n}^{(3)}$ is obtained from the realignment moment graph $\RM_{2n}^{(3)}$ (Fig.~\ref{fig:RMgen}) by inserting $(m-2)/2$ melons (operation of Fig.~\ref{fig:MelonConst}) on each black edge (the traced out color, here color 3). Said differently, the colored graph $\RE_{m,n}^{(3)}$ is obtained by taking two copies of $C_n$ (see Fig.~\ref{fig:bip}) joined by edges of color $3$, each of which is decorated by $(m-2)/2$ melons as illustrated on the right-hand side of Fig.~\ref{fig:RE_mn}.

\begin{figure}[ht]
    \centering
    \includegraphics[height = 2.5cm]{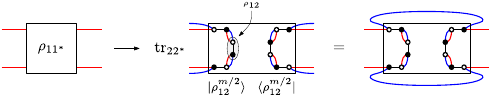}
    \caption{Construction of $\rho_{11^*}$ using the replica trick on even powers of $\sqrt{\rho_{12}}$.}
    \label{fig:RE_const}
\end{figure}

This construction is commonly implemented to obtain the reflected entropy via a two-step replica trick: 
\begin{equation}
    \S_{\RE^{(3)}}(\ket{\psi}) \eqdef \S_\rm{ent}(\ket{\sqrt{\rho_{12}}}) = \lim_{n \to 1} \lim_{m \to 1} \S_{\RE_{m,n}^{(3)}}(\ket{\psi}) \,,
\end{equation}
where $\ket{\sqrt{\rho_{12}}} \in \H_{\{1,2\}}\otimes \H^*_{\{1,2\}}$ with $\H^*_{\{1,2\}}\simeq \H_{\{1,2\}}$ denotes the canonical purification of $\rho_{12}$. Again, a proper definition of this limit necessitates an analytic continuation in the replica numbers $m$ and $n$.
Since $\S_{\RE^{(3)}}(\psi) = \S_\rm{vN}(\rho_{1 1^*})$, where $\rho_{11^*}$  is defined as 
\begin{equation}
    \rho_{1 1^*} \eqdef \tr_{\H_2 \ot \H_2^*} \ket{\sqrt{\rho_{1 2}}} \bra{\sqrt{\rho_{1 2}}} \,,
\end{equation}
the replica index $n$ is used to compute the von Neumann entropy of $\rho_{11^*}$, and the index $m$ (taken to be even) is used to define powers of $\sqrt{\rho_{12}}$. Fig.~\ref{fig:RE_const} illustrates this double-replica construction.

\ 

Beyond the tripartite setting, that is for arbitrary $D \geq 3$, the \textit{reflected multi-entropies} and their Rényi generalizations provide $D$-partite extensions of the reflected entropies, which are built from the canonical purification of a mixed quantum state $\rho_{\bar{D}} = \tr_D \ket{\psi} \bra{\psi}$ (see Refs.~\cite{76vs-rxcs,Iizuka:2025elr}). They reduce to the standard reflected entropy when $D=3$. 

\begin{figure}[!ht]
    \centering
    \includegraphics[width = \textwidth]{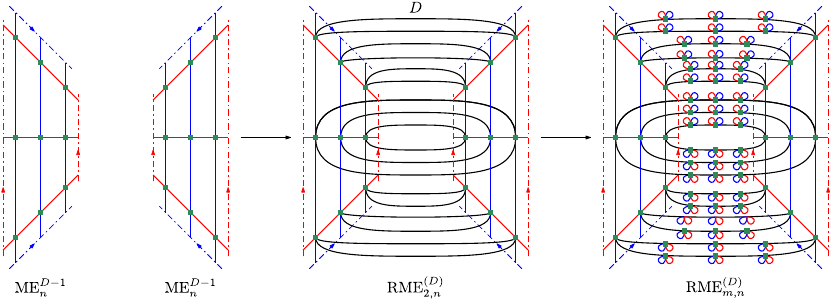}
    \caption{Colored graph $\RME_{m,n}^{(D)}$ constructed step by step. On the left, we start with $2$ copies of $\ME_n^{D-1}$ represented using boxes as in Fig.~\ref{fig:ME2}. Then, in the middle, we connect the vertices with the color that is traced out (here the color $D$). On the right, we finish by adding melons. Here, the illustration is given for $D=4$, $n=3$, and $m=8$.}
    \label{fig:RME}
\end{figure}

The \textit{Rényi reflected multi-entropies} of $\rho_{\bar{D}}$ are defined in terms of $\ket{\psi}$ for $n >1$ and $m \in 2 \bb{N}^*$ as: 
\begin{equation}
    \S_{\RME_{m,n}^{(D)}}(\ket{\psi}) = \frac{1}{(1-n)n^{D-3}} \ln \pac{\frac{\tr_{\RME_{m,n}^{(D)}}(\ket{\psi})}{\pa{\tr_{\RME_{m,1}^{(D)}}(\ket{\psi})}^{n^{D-2}}}} \,,
\end{equation}
and are associated with the colored graphs denoted by  $\RME_{m,n}^{(D)}$ and defined for $n\geq 1$ and $m \in 2 \bb{N}^*$, an example of which is shown in  Fig.~\ref{fig:RME} for $D=4$, $m=8$ and $n=3$. The construction is analogous to the $D=3$ case (see on the right of Fig.~\ref{fig:RE_mn}), namely one takes two copies of $\ME_n^{D-1}$ and connects them using edges of color $D$. On each edge of color $D$, one then inserts $(m-2)/2$ melons. The resulting colored graph $\RME_{m,n}^{(D)}$ has  $m n^{D-2}$ white vertices. 

By the use of the usual two-step replica trick and taking care of the analytical continuation in $m$ and $n$, one then defines the \textit{reflected multi-entropy}:
\begin{equation}
    \S_{\RME^{(D)}}(\ket{\psi}) \eqdef \S_\ME(\ket{\sqrt{\rho_{\bar{D}}}}) = \lim_{n \to 1} \lim_{m \to 1} \S_{\RME_{m,n}^{(D)}}(\ket{\psi}) \,,
\end{equation}
where $\ket{\sqrt{\rho_{\bar{D}}}}$ denotes the canonical purification of $\rho_{\bar{D}}$. The replica index $n$ is used for the multi-entropy of the canonical purification, while the (even) index $m$ is used to compute the square root.

\ 

As a remark, we stress that the reflected entropy without melons, namely when $m = 2$, is a copy of the colored graph used for the computation of the realignment moments, \ie $\RE_{2,n}^{(c)} = \RM_{2n}^{(c)}$. 

Furthermore, the definition of the colored graphs $\RME_{m,n}^{(c)}$ admits a natural generalization. Indeed, instead of tracing over a single color $c$ and defining $\RME_{m,n}^{(c)}$, one may trace over an arbitrary subset of colors $\bar{B} \subset \paa{1,\dots,D}$ and define the corresponding colored graph $\RME_{m,n}^{(\bar{B})}$. The latter being constructed as two copies of $\ME_n^{\abs{B}}$, \ie the colored graphs associated with the multi-entropies restricted to the set of colors $B$. The trace is then performed over the colors in $\bar{B}$. Finally, melonic insertions are added along the edges of colors in $\bar{B}$, treating the entire subset $\bar{B}$ effectively as a single color.

In particular, if we trace over a block $\bar{B}$ of size $D-2$, one recovers the tripartite version of the realignment moment presented in the middle of Fig.~\ref{fig:RMgen}. Indeed, if $\abs{\bar{B}} = D-2$ we have the equality $\RME_{2,n}^{(\bar{B})} = \RE_{2,n}^{(\bar{B})} = \RM_{2n}^{(\bar{B})}$. 

\begin{figure}[!ht]
    \centering
    \includegraphics[height = 3.5cm]{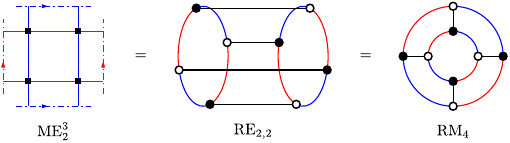}
    \caption{Equality between the colored graphs corresponding to $\ME_2^3$, $\RE_{2,2}^{(c)}$ and $\RM_4$.}
    \label{fig:ME_RE_RM}
\end{figure}

What is more, due to the equality between $\ME_2^3$, $\RM_4$ and $\RE_{2,2}^{(c)}$ for any $c$ (as exemplified in Fig.~\ref{fig:ME_RE_RM}), the supscript $(c)$ can be removed from $\RE_{2,2}^{(c)}$. The same goes for $\RME_{2,2}^{(c)}$ since we have the equality $\RME_{2,2}^{(c)} = \ME_2^D$.

Furthermore, the reflected entropy as well as its multipartite generalization can be related to some invariants from the melonic family. Indeed, take $n=1$, the colored graphs $\RME_{m,1}^{(c)}$ for a given $c$ (or for $D=3$, $\RE_{m,1}^{(c)}$) is the melonic graph corresponding to a bipartition of $\H$ of the form $\H_c \ot \H_{\bar{c}}$. More broadly, the colored graph $\RME_{m,1}^{(\bar{B})}$ is also a cyclic graph associated with a bipartition of $\H$ of the form $\H_B \ot \H_{\bar{B}}$.

\subsection{Quantum states under consideration}
\label{sub:intro-of-ref-states}

In this section, we introduce an infinite family of deterministic states of reference, which have a modular structure that can be conveniently encoded into graphs, and that we will refer to as \emph{hypergraph-tensor states} (or HT states), see \eg Refs.~\cite{Buhrman:2016tif,Vrana:2016edr,Christandl:2018cfb,Christandl:2019zrq,brand2026bilinearcomplexityworksbreaks} and references therein.\footnote{These states are sometimes called hypergraph tensors, or graph tensors in certain simpler cases. We thank Péter Vrana for mentionning these references to us.} We also recall the definition and main properties of a random Haar-distributed state. Both types of states will serve as benchmarks to assess the distinguishing power of various families of trace-invariants in later sections.

\subsubsection{Hypergraph-tensor states}\label{sec:ref_states}
 
In this subsection, we fix $D\geq 2$, as well as a $D$-partite Hilbert space $\H_1 \otimes \cdots \otimes \H_D$, with $\H_c$ of dimension $N_c \in \mathbb{N}^*$ for any $c \in \{ 1, \ldots , D\}$.
 
\paragraph{GHZ states.} The basic building blocks of our construction are GHZ states shared between any number of subsystems. For any $p \in \paa{1,\dots,D}$, any distinct colors $c_1,\dots,c_p \in \paa{1,\dots,D}$, let $B = \paa{c_1,\dots,c_p}$, and for any $N \in \mathbb{N}^*$, we let $\ket{\GHZ}_{B, N}$ denote a $\GHZ$ state of dimension $N$ on $B$, namely:
\begin{equation}
\label{eq:def-GHZ}
    \ket{\GHZ}_{B,N} \eqdef\frac{1}{\sqrt{N}} \sum_{i=1}^{N} \ket{i}_{c_1} \ot \cdots \ot \ket{i}_{c_p} \in \H_{B}\,,
\end{equation}
where, for any $c \in B $, $\{\ket{i}_{c}\}_{1\le i\le N}$, is an orthonormal basis on $\H_{c}$ (and, in particular, we must have $N\leq N_{c}$).  
For this state, the tensor $\psi_{i_{c_1} \dots i_{c_p}}$
in Eq.~\eqref{eq:tensor_components} reads
\begin{equation}
\label{eq:comp-ghz}
    \psi_{i_{c_1} \dots i_{c_p}} = \frac{1}{\sqrt{N}}\sum_{i=1}^N \delta_{i \, i_{c_1}} \cdots \delta_{i \, i_{c_p}} \,.
\end{equation}
$\ket{\GHZ}_{B,N} \in \H_B$ is naturally interpreted as a $|B|$-partite state; for any connected $G\in \cG_{|B|}^{\conn}$, we have
\begin{equation} \label{eq:GHZ}
    \tr_G(\ket{\GHZ}_{B,N}) = \pa{\frac{1}{\sqrt{N}}}^{2k(G)} \cdot \sum_{i=1}^{N} 1 = N^{1-k(G)} \,,
\end{equation}
as can be seen by inserting Eq.~\eqref{eq:comp-ghz} in Eq.~\eqref{eq:LUinv}. Note that, given that we have left the choice of local bases $\{\ket{i}_c\}_{1\le i\le N}$ arbitrary, our notation $\ket{\GHZ}_{B,N}$ actually labels a $\LU$-equivalent class of states in the $|B|$-partite state space $\H_B$. With this in mind, we also note that
\begin{equation}\label{eq:equiv_product_GHZ}
    \ket{\GHZ}_{B,N}\otimes \ket{\GHZ}_{B,N'} \simLU \ket{\GHZ}_{B,NN'}
\end{equation}
for any $B\subset\{1, \ldots , D\}$ and $N,N' \in \mathbb{N}^*$, which follows immediately from \eqref{eq:GHZ} and Prop.~\ref{prop:charac_LU-1} (in this equation, $\LU$ refers to local unitary equivalence of $|B|$-partite states).\footnote{In particular, taking $N=1$ in the previous equation, we see that any factor of the form $\ket{\GHZ}_{B,1}$ can be omitted if we are only interested in the $\LU$-class of the state, which is to be expected since $\ket{\GHZ}_{B,1}$ is separable in $\H_B$.} Finally, by convention, we will denote by $\ket{\GHZ}_{N}\eqdef\ket{\GHZ}_{\{1, \ldots , D\}, N}$ the GHZ state of dimension $N$ on $\H_1 \otimes \cdots \otimes \H_D$. 

\medskip

\begin{figure}[!ht]
    \centering
    \includegraphics[width = \textwidth]{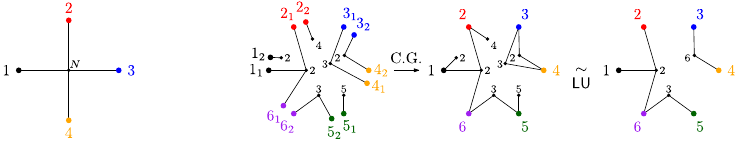}
    \caption{
    Left: the GHZ state $\ket{\GHZ}_N \in \H_1 \ot \H_2 \ot \H_3\ot  \H_4$ (associated with the weighted partition $\{ \{1, 2, 3, 4\}_N\}$). Right: state corresponding to the weighted partition $(\pi_0 , w_0) = \paa{\paa{1_1,2_1,6_1}_2 ,\paa{5_2,6_2}_3,\paa{3_1,4_1}_3,\paa{3_2,4_2}_2,\paa{1_2}_2,\paa{2_2}_4,\paa{5_1}_5}$ (and to the multiset $\sM_0 = \{\{1,2,6\}_2,\{5,6\}_3,\{3,4\}_3,\{3,4\}_2,\paa{1}_2,\paa{2}_4,\paa{5}_5,\} \sim \{\{1,2,6\}_2,\{5,6\}_3,\{3,4\}_6\}$). 
    }
    \label{fig:phi_p}
\end{figure}

The family of HT states is introduced in two steps: we define them first at the level of a fine-graining of the original $D$-partite state space, and then investigate their property under coarse-graining. In each case, we will introduce a graphical representation that captures invariant $\LU$ properties of the states and explain how they are related to one another under coarse-graining. 
\paragraph{Fine-grained definition of HT states.} Let us start out from a fine-graining of the $D$-partite state space $(\H_1 , \ldots , \H_D)$. That is, we assume that for any color $c$, the subsystem $\H_c$ is subdivided into $I_c$ subsystems (with $I_c \in \mathbb{N}^*$), so that   
\begin{equation}
    \H_c = \bigotimes_{1\le i \le I_c} \H_{(c,i)}\,,
\end{equation}
where, for every pair $(c,i)$, $\H_{(c,i)}$ is a Hilbert space of dimension $N_{(c,i)}$ (and we therefore have $\displaystyle\prod_{i=1}^{I_c} N_{(c,i)} = N_c$). Denoting by $\fS= \{ (c,i) \, | \, 1 \leq c \leq D \,, 1 \leq i \leq I_c\}$ the set labeling the elementary subsystems, $\sF = (\H_{(c,i)})_{(c,i) \in \fS}$ defines a $|\fS|$-partite state space $\H^\sF$. Furthermore, introducing the partition $\zeta= \{ \{ (c,i)\}_{1 \leq i \leq I_c} \}_{1\leq c\leq D}$ of $\fS$, we have:
\begin{equation}
    \sF_{\zeta} = (\H_1 , \ldots , \H_D) \,, \qquad \H^{\sF_\zeta} = \H_1 \otimes \cdots \otimes \H_D \,,
\end{equation}
\ie our original $D$-partite state space is the $\zeta$-coarse-graining of $\sF$. Consider now the class of states that can be represented by products of GHZ states in the fine-grained Hilbert space $\H^\sF$. Any such state can be labeled by a \emph{weighted partition} of $\fS$, by which we mean: a couple $(\pi, w)$, where $\pi$ is a partition of $\fS$, and $w:\pi \to \mathbb{N}^*$ is a map assigning a dimension $w(A)$ to any subset $A \in \pi$ appearing in the partition. One then defines
\begin{equation}
    \ket{\Psi_{(\pi,w)}} \eqdef \bigotimes_{A \in \pi} \ket{\GHZ}_{A, w(A)}\,.
\end{equation}
Again, given that we have left the choice of local bases in the fine-grained state space implicit, what $\ket{\Psi_\pi}$ actually represents is an infinite family of states that are all $\LU$-equivalent in $\sF$; as such, they are also $\LU$-equivalent in $\sF_\zeta$, so we are not losing any relevant information with this combinatorial encoding. Before providing some examples, let us introduce two convenient shorthand notations: 1) we will use the notation $c_i \eqdef (c,i)$ to label subsystems in the fine-grained state space, and 2) a weighted partition $(\pi, w)$ will be represented by its collection of blocks, with their weights indicated in subscript, \ie $(\pi, w) \eqdef\{ A_{w(A)}\}_{A \in \pi}$. As a simple example, the $\GHZ$ state $\ket{\GHZ}_N \in \H_1 \otimes \cdots \H_D$ can be represented by the weighted partition $\{ \{ 1, \ldots , D\}_N\}$, indicating that the partition has one block $\{ 1, \ldots , D\}$ of weight $w(\{1, \ldots, D\})=N$.\footnote{This particularly simple example does not require the introduction of a fine-graining, meaning that one can take $\sF = \sF_\zeta$.} More involved, the notation 
\begin{equation}\label{eq:ex_weighted_partition}
    (\pi_0 , w_0) = \paa{\paa{1_1,2_1,6_1}_2 ,\paa{5_2,6_2}_3,\paa{3_1,4_1}_3,\paa{3_2,4_2}_2,\paa{1_2}_2,\paa{2_2}_4,\paa{5_1}_5}
\end{equation}
indicates that $\pi_0 = \paa{\paa{1_1,2_1,6_1} ,\paa{5_2,6_2},\paa{3_1,4_1},\paa{3_2,4_2},\paa{1_2},\paa{2_2},\paa{5_1}}$ and $w(\paa{1_1,2_1,6_1}) =2$, $w(\paa{5_2,6_2})=3$, \etc We can give a graphical interpretation of such a piece of data by encoding it into a \emph{decorated hypergraph}, as follows: a) any subsystem $c_i \in \fS$ is represented by a vertex labeled by $c_i$; b) any element $A \in \pi$ is represented by an hyper-edge connecting the vertices associated with the subsystems comprising $A$, and decorated by its weight $w(A)$. Given the way in which we introduced HT states in the fine-grained picture, the hypergraph representing $\ket{\Psi_{(\pi, w)}}$ has exactly $|\pi|$ connected components, which is also the number of elementary $\GHZ$ states making up $\ket{\Psi_{(\pi , w)}}$. Furthermore, each of its vertices belongs to exactly one hyper-edge. See Fig.~\ref{fig:phi_p} for graphical representation of a single $4$-partite $\GHZ$ state, as well as of the state $\ket{\Psi_{(\pi_0 , w_0)}}$ defined by \eqref{eq:ex_weighted_partition}.

\paragraph{Coarse-grained picture of HT states.} Two HT states $\ket{\Psi_{(\pi , w)}}$ and $\ket{\Psi_{(\pi' , w')}}$ which are $\LU$-inequivalent in the fine-grained multipartite state space $\sF$ may well have the same $D$-partite entanglement structure in the coarse-grained state space $\sF_\zeta$. Indeed, we can easily identify structures in the weighted partitions $(\pi , w)$ and $(\pi' , w')$ that become irrelevant after coarse-graining. To begin with, if $\ket{\Psi_\pi}$ and $\ket{\Psi_{\pi'}}$ only differ by a relabeling of the Hilbert spaces $\H_{(c,i)}$ that subdivide the same $\H_c$, they will have the same $D$-partite entanglement properties in $\mathcal{H}^{\sF_\zeta}=\mathcal{H}_1\otimes \cdots \ot \mathcal{H}_D$. They indeed belong to the same $\LU^{\sF_{\zeta}}$-class, since a permutation of several subsystems carrying the same color $c \in \{1, \ldots, D\}$ can be implemented by a local isometry in $\H^{\sF_\zeta}$.\footnote{Up to a local isometry, we can assume all $I_c$ Hilbert spaces $\{\H_{(c,i)}\}$ comprising $\H_c$ to have the same dimension (per Prop.~\ref{lem:equivalence_under_isom}), and apply a unitary map of the form $\rho_c(\sigma)$ with $\sigma \in S_{I_c}$ on $\H_c$, as introduced in Eq.~\eqref{eq:trace_inv_permutation_rep}.} This leads to a first simplification of the combinatorial data defining an HT state: up to $\LU^{\sF_\zeta}$-equivalence, one only needs to specify a \emph{multiset} of weighted subsets of $\{1,\ldots, D\}$ induced by $\pi$; this is obtained by omitting the sub-index $i$ of any $c_i$: for any weighted block $A_{w} =\{{c^{(1)}}_{i_1}, \ldots, {c^{(s)}}_{i_s} \}_w$ of $\pi$ (with $c^{(j)}_{i_j}\in \fS$ for any $j\in \{1, \ldots , s\}$), we will only specify the colors $C_w=\{c^{(1)}, \ldots, c^{(s)}\}_w$ appearing in that block (with repetitions allowed). In the previous expression, $C$ should be interpreted as a \emph{multiset of colors}, and we therefore refer to $C_w$ as a \emph{weighted multiset of colors}. After this procedure has been implemented to any weighted block of $(\pi , w)$, one is left with a multiset of weighted multisets (since each weighted multiset $C_w$ is allowed to be repeated, and it should be counted with multiplicity). This leads to a first coarse-grained description of HT states in terms of \emph{multisets} (of weighted multisets) rather than weighted partitions. 
As an example, the state represented by the weighted partition $\{ \{1_1,1_2,2_1\}_4 , \{2_2\}_3 \}$ can be written as 
\begin{equation} \label{eq:relab-subcolors}
\frac 1 {\sqrt{4 \cdot 3}}\sum_{i=1}^{4} \ket{i}_{1_1}\otimes \ket{i}_{1_2} \otimes \ket{i}_{2_1} \bigotimes \sum_{j=1}^{3} \ket{j}_{2_2}\,,
\end{equation}
and is associated with the multiset $\sM = \{\{1,1,2\}_4,\{2\}_3\}$ at the coarse-grained level. The same multiset $\sM$ will be associated with the state obtained by exchanging the labels $2_1$ and $2_2$ in Eq.~\eqref{eq:relab-subcolors}. As a second example, the weighted partition $(\pi_0, w_0)$ introduced in Eq.~\eqref{eq:ex_weighted_partition} is associated with the multi-set $\sM_0$ given by $\sM_0 = \{\{1,2,6\}_2,\{5,6\}_3, \{3,4\}_3, \{3,4\}_2, \paa{1}_2, \paa{2}_4, \paa{5}_5\}$. Graphically, such a multiset can be represented by a graph with $D$ vertices, one for each subsystem of the coarse-grained state space $\sF_\zeta$, obtained as follows. Starting from the fine-grained graphical representation of a weighted partition $(\pi, w)$, and for any $c \in \{ 1, \ldots , D\}$, one simply merges all the vertices labeled by $c_i$ (with $1 \leq i \leq I_c$) into a single vertex labeled by their common color $c$. This results in a decorated hypergraph in which multi-hyper-edges are allowed. See Fig.~\ref{fig:phi_p} for an illustration of this construction for the weighted partition defined in \eqref{eq:ex_weighted_partition}.

\begin{figure}[!ht]
    \centering
    \includegraphics[height = 2.5cm]{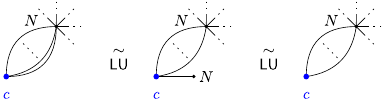}
    \caption{Graphical illustration of Eq.~\eqref{eq:equiv_ref_3}: adding or removing connections between a hyper-edge and a colored vertex without changing the support of that hyper-edge preserves the $\LU$-class of an HT state.}
    \label{fig:Move}
\end{figure}

\begin{figure}[!ht]
    \centering
    \includegraphics[height = 2.8cm]{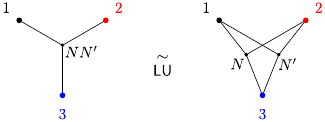}
    \caption{Graphical illustration of the equivalence relation introduced in Eq.~\eqref{eq:equiv_ref_4} for $B \subset \paa{1,\dots,D}$ with $\abs{B} = 3$.}
    \label{fig:LUeq_Ref2}
\end{figure}

\paragraph{Equivalence relations on HT states.} The weighted multiset $\sM$ associated with an HT state $\ket{\Psi_{(\pi , w)}}$ still carries redundant information if one only cares about the $\LU^{\sF_\zeta}$-orbit of $\ket{\Psi_{(\pi , w)}}$. It is convenient for this purpose to make use of two standard notions from the formalism of multisets. First, there is the notion of \emph{sum of multisets}: denoted $+$, it is the symmetric binary operation which, to any multisets $\sM = \{x_1 , \ldots , x_p\} , \sM'= \{y_1 , \ldots, y_q\}$, associates the multiset $\sM+ \sM'= \{ x_1 , \ldots, x_p , y_1 , \ldots, y_q\}$. In other words, $+$ is a kind of multiset version of the notion of union of sets, with respect to which the multiplicity of any element is additive. We can also define the \emph{support of a multiset}: if $C$ is a multiset, we will denote by $\red(C)$ the ordinary set consisting of all the color labels $c$ that occur at least once in $C$. We will make use of this concept in relation to multisets of colors; for instance, if $C= \{1,1,1,3,3,5\}$, then $\red(C)= \{1,3,5\}$. With these notions at hand, we can introduce four independent equivalence conditions on the multisets representing HT states. First of all, any separable state appearing as a tensor factor of $\ket{\Psi_{(\pi , w)}}$ can be omitted. In other words, if we can find a weighted multiset $\sM'$, a color $c \in \{ 1, \ldots , D\}$ and some integer $N \in \mathbb{N}^*$ such that $\sM = \sM' + \{ c\}_N$, then $\sM$ and $\sM'$ will represent the same $D$-partite $\LU$ orbit in $\sF_\zeta$; hence, we will write:
\begin{equation}\label{eq:equiv_ref_1}
    \sM = \sM' + \{ c\}_N \quad \Rightarrow \quad \sM \sim \sM'\,.
\end{equation}
For instance, for the HT state introduced in \eqref{eq:relab-subcolors}, we have $\{\{ 1, 1, 2\}_4,  \{ 2\}_3 \}\sim \{\{ 1, 1, 2\}_4\}$. In graphical terms, any univalent hyper-edge can be deleted. Furthermore, for any $B \subset \{ 1, \ldots , D\}$ with $|B| \geq 2$, a tensor factor of the form $\ket{\psi}_{B, 1}$ is itself a separable state, which leads to a second equivalence condition:     \begin{equation}\label{eq:equiv_ref_2}
    \sM = \sM' +B_1 \quad \Rightarrow \quad \sM \sim \sM'\,.
\end{equation}
In graphical terms, any hyper-edge with weight one can be omitted. Third, for any $(c,i)\in \fS$, we can choose a local orthonormal basis of the form $\{ \ket{i_1}_{c_{1}} \otimes \cdots \otimes \ket{i_{I_c}}_{c_{I_c}} \}_{1 \leq i_k \leq N_{(c_k,i_k)}} $ in $\H_{(c,i)}$. It is then clear that any $\GHZ$ state involving $p$ subsystems $\{ \H_{(c, i_1)}, \ldots , \H_{(c, i_p)} \}$ from $\H_c$ (with $1 \leq i_1 <\cdots < i_p \leq I_c$) can be equivalently understood as a GHZ state involving a single subsystem $H_{(c, i_1)}\otimes \ldots \otimes \H_{(c, i_p)}$ of $\H_c$ (\ie defined relative to a coarser fine-graining of the original $D$-partite state space). This leads to a third equivalence condition: any multiset of colors $C$ (with repetitions of colors allowed)  can be equivalently replaced by its support $\red(C)$ (with no repetition). Said differently, for any multiset $\sM$, any multisets of colors $C$ and $C'$, and any $N \in \mathbb{N}^*$, we have
\begin{equation}\label{eq:equiv_ref_3}
    \red(C) = \red(C') \quad \Rightarrow \quad \sM + C_N \sim \sM + C'_N \,.
\end{equation}
For example, applying this condition to the HT state introduced in \eqref{eq:relab-subcolors}, we have $\{\{ 1, 1, 2\}_4, \{ 2\}_3\} \sim \{\{ 1, 1, 2\}_4\} \sim \{\{ 1 , 2 \}_4\}$. Graphically, this is saying that two half-edges connecting the same vertex $v$ to the same $\GHZ$ block $C$ are equivalent to a single half-edge. This is illustrated in Fig.~\ref{fig:Move}.
Finally, the $\LU$-equivalence relation \eqref{eq:equiv_product_GHZ} yields a fourth equivalence condition on multisets: for any multiset $\sM$, any $B \subset \{ 1, \ldots , D\}$ with $|B|\geq 1$, and any $N, N' \in \mathbb{N}^*$, we have
\begin{equation}\label{eq:equiv_ref_4}
    \sM + B_{N} + B_{N'} \sim \sM + B_{NN'}\,.
\end{equation}
In graphical terms, this is saying that two hyper-edges sharing the same vertices can be merged into a single hyper-edge connecting those vertices, and that weights are multiplicative under such a move. For instance, we have already seen that the HT state of \eqref{eq:relab-subcolors} can be represented by the multiset $\{\{ 1 , 2 \}_4\}$; equivalently, it can also be represented by the multiset $\{\{ 1 , 2 \}_2,\{ 1 , 2 \}_2\}$. Another example is provided in Fig.~\ref{fig:LUeq_Ref2}.

\paragraph{Canonical representation of HT states.} Equations \eqref{eq:equiv_ref_1}, \eqref{eq:equiv_ref_2}, \eqref{eq:equiv_ref_3} and \eqref{eq:equiv_ref_4} define an equivalence relation $\sim$ on the set of multisets representing HT states (or, equivalently, on their associated graphs). Combining these relations, it is then clear than, given an HT state $\ket{\Psi_{(\pi , w)}}$, one can always find a multiset $\sM$ representing it that obeys the following additional conditions: for any block $C_w\subset\{ 1, \ldots , D\}$ appearing in $\sM$, one must have 1) $|C|\geq 2$; 2) $w\geq 2$; 3) $C= \red(C)$ (\ie $C$ does not contain repeated colors); 4) there is no other block of the form $C_{w'}$ with $w' \in \mathbb{N}^*$ appearing in $\sM$. In graphical terms, such a multiset can be represented by a hypergraph having: 1) no hyper-edge connecting a single vertex; 2) no hyper-edge of weight $1$; 3) no repeated half-edge in any hyper-edge; 4) no repeated hyper-edge. It is also clear that such a multiset (resp.~graph) is unique, and we will refer to it as the \emph{canonical multiset} (resp.~\emph{canonical hypergraph}) representing $\ket{\Psi_{(\pi , w)}}$ (resp. $(\pi , w)$). For example, the canonical weighted multiset representing the weighted partition $(\pi_0 , w_0)$ introduced in \eqref{eq:ex_weighted_partition} is
\begin{equation}\label{eq:ex_canonical_weighted_multiset}
    \{\{1,2,6\}_2,\{5,6\}_3,\{3,4\}_6\}\,,
\end{equation}
and its associated canonical graph is represented in Fig.~\ref{fig:phi_p}. The data of a canonical multiset $\sM$ can be equivalently (and more conveniently) encoded into a \emph{weight function} $\alpha_{\sM}: \{ B \subset \{ 1, \ldots , D\} \,|\, |B| \geq 2\} \to \mathbb{N}^*$, defined as follows. For any block $B\subset\{1, \ldots, D\}$ involving at least two subsystems, $\alpha_\sM (B)$ denotes the weight assigned to $B$ if $B$ appears in $\sM$ (in which case $\alpha_\sM (B) \geq 2$), and is set to $\alpha(B)=1$ whenever $B$ does not occur in $\sM$. For any weight function $\alpha: \{ B \subset \{ 1, \ldots , D\} \,|\, |B| \geq 2\} \to \mathbb{N}^*$, let us define the $D$-partite state
\begin{equation}\label{eq:def_alpha_states}
\ket{\psi_\alpha} \eqdef \bigotimes_{B \subset \{ 1, \ldots , D\}, \, |B|\geq 2} \ket{\GHZ}_{B, \alpha(B)}\,,
\end{equation}
with the convention that $\ket{\GHZ}_{B, 1}\eqdef 1$ for any $B$. More precisely, $\ket{\psi_\alpha}$ represents an $\LU$-equivalence class of $D$-partite pure states and, from the preceding discussion, any $D$-partite HT state is $\LU$-equivalent to a state of the form $\ket{\psi_\alpha}$. In addition, we will prove in Sec.~\ref{sub:LU-inequivalent-reference-states} that two HT states are $\LU$-equivalent if and only if they have the same weight function. In other words, the equivalence relation $\sim$ we have introduced on the space of multisets $\{\sM\}$ coincides with the equivalent relation $\simLU$ in the space of HT states. Note that an element of the family $\{\ket{\psi_\alpha}\}$ is specified by $2^{D}-(D+1)$ integer parameters (\ie the number of $B \subset\{1, \ldots , D\}$ such that $|B|\geq 2$), which is therefore equal to the number of independent parameters defining $D$-partite HT states (up to $\LU$-equivalence).

\begin{ex}[Pure tripartite stabilizer states] \label{ex:stab_states}
    As stated in \cite{Looi:2011jrm, Nezami:2016zni}, any pure stabilizer state $\ket{S}$ on a tripartite space $(\H_1 , \H_2 , \H_3)$ is equivalent to an HT state $\ket{\Psi_{(\pi , w)}}$ of the following form. For any such state $\ket{S}$, there exists a prime integer $p\in \mathbb{N}^*$ and some non-negative integers $m_{123}, m_{12}, m_{13}, m_{23}, m_1, m_2, m_3$, such that each $\H_c$, $1\le c \le 3$ decomposes as $\H_c = \bigotimes_{1\le i \le I^S_c} \H_{(c,i)},$ where $I^S_1=m_{123} + m_{12}+ m_{13} + m_1$ and similarly for $I^S_2$ and $I^S_3$, and where $\dim(\H_{(c, i)}) = p$. One then has $\ket{S}\simLU \ket{\Psi_{(\pi , w)}}$ where $(\pi, w)$ is a weighted partition inducing the multiset 
    \begin{equation}
    \label{eq:tripartite-stabilizer}
        \sM_S=\{\{1,2,3\}_p^{m_{123}},\{1,2\}_p^{m_{12}},\{1,3\}_p^{m_{13}},\{2,3\}_p^{m_{23}},\{1\}_p^{m_1},\{2\}_p^{m_2},\{3\}_p^{m_3}\}\;,
    \end{equation} 
    where $\{\cdot\}_p^a$ means that the block $\{\cdot\}_p$ appears $a$ times in $\sM_S$. States of this kind will be consider in Sec.~\ref{sec:LU-and-ref-states} with the notations $N_s=p$ and $N=p^{I_c}$ (more generally without the requirement that $p$ is a prime number). 
    Such states have a canonical representation whose multiset is given by $\{\{1,2,3\}_{p^{m_{123}}},\{1,2\}_{p^{m_{12}}},\{1,3\}_{p^{m_{13}}},\{2,3\}_{p^{m_{23}}}\}$, corresponding to the weight function $\alpha_S: B \mapsto p^{m_B}$. 
    
    Due to this fact, our results concerning the $\LU$-equivalence and $\LO$ or $\LOCC$ comparison of HT states in Sec.~\ref{sub:LU-inequivalent-reference-states}, Sec.~\ref{sec:LU-and-ref-states} and Sec.~\ref{s:LO_LOCC_invTr} apply to tripartite pure stabilizer states as well.
\end{ex}

\paragraph{Evaluation of trace-invariants on HT states.} Trace-invariants can easily be evaluated on an arbitrary HT state $\ket{\Psi_{(\pi, w)}}$. If $\sM$ is a multiset representing $(\pi, w)$, then, thanks to the factorization Eqs.~\eqref{eq:tensorprod}, \eqref{eq:nonconn} and~\eqref{eq:GHZ}, we have: 
\begin{equation} \label{eq:tr_psi_pi}
    \forall G \in \cG_D\,,\qquad  \tr_G \left( \ket{\Psi_{(\pi, w)}}\right) = \prod_{C_w \in \sM} w^{\kappa(G\vert_{\red(C)}) - k(G)}\,,
\end{equation}
where the product over elements of $\sM$ is understood with multiplicity, and $\kappa(G\vert_{\red(B)})$ denotes the number of connected components of the colored subgraph $G\vert_{\red(B)}$ (\ie the restriction of the graph $G$ to its set of edges labeled by colors that appear in $C$). Combined with the characterization of $\LU$-equivalence given in Prop.~\ref{prop:charac_LU-1}, Eq.~\eqref{eq:tr_psi_pi} can be used to recover the four elementary equivalence conditions listed in the previous paragraph. Eq.~\eqref{eq:tr_psi_pi} takes a particularly simple form when applied to canonical multiset descriptions of HT states. If $\alpha$ is a weight function, we find that:
\begin{equation}\label{eq:trace-inv_alpha-states}
    \forall G \in \cG_D, \qquad \tr_G \left( \ket{\psi_\alpha}\right) = \prod_{B \subset \{1, \ldots , D\}\,,\, |B|\geq 2} \alpha(B)^{\kappa(G\vert_{B}) - k(G)}\,.
\end{equation}
As an example, for the weighted partition $(\pi_0 , w_0)$ introduced in \eqref{eq:ex_weighted_partition} (and represented in Fig.~\ref{fig:phi_p}), which can be represented by the canonical weighted multiset of Eq.~\eqref{eq:ex_canonical_weighted_multiset}, we have
\begin{equation}
    \forall G \in \cG_D\,, \quad \tr_G(\ket{\Psi_{(\pi_0, w_0)}}) = 2^{\kappa(G\vert_{\paa{1,2,6}}) - k(G)} \cdot 3^{\,\kappa(G\vert_{\paa{5,6}}) - k(G)} \cdot 6^{\,\kappa(G\vert_{\paa{3,4}}) - k(G)}\,.
\end{equation}

\paragraph{Partial separability and related notions.} Consider an HT state $\ket{\Psi_{(\pi, w)}}$ and consider an hypergraph representation of that state obeying the following condition: none of its hyper-edges of valency $2$ or higher have weight $1$ (\ie we forbid the use of hyper-edges of weight $1$ to represent the inclusion of a trivial $\GHZ$ state on two or more subsystems). This condition is in particular obeyed if we choose the canonical hypergraph representation of $\ket{\Psi_{(\pi, w)}}$. Let $\xi$ be the partition of $\{1,\ldots, D\}$ given by the connected components of the hypergraph representing $\ket{\Psi_{(\pi, w)}}$. On each connected component, the state is genuinely entangled: it is the finest $\zeta$ such that $\ket{\Psi_{(\pi, w)}}$ is $\zeta$-entangled (see Sec.~\ref{sec:generalities-on-coarse-graining}).  Then $\#(\xi)$ is the maximum $r$ for which $\ket{\Psi_{(\pi, w)}}$ is $r$-separable. If $m$ is the number of colored vertices in the largest connected component of the hypergraph representing $\ket{\Psi_{(\pi, w)}}$ (that is, $m$ is the maximum size of blocks of $\xi$), then $m$ is the entanglement depth of $\ket{\Psi_{(\pi, w)}}$. 

Graphically, coarse-graining amounts to the following: merging $\H_c$ and $\H_{c'}$ to $\H_{c''}=\H_c\otimes \H_{c'}$ amounts to merging the vertices of color $c$ and $c'$ in the graphical representation of the hyper-edges. Conversely, fine-graining  the system by splitting  $\H_{c''}=\H_{c_1}\otimes \cdots \otimes H_{c_{\cal{I}}}$ into $\H_{c}=\H_{c_1}\otimes \cdots \otimes H_{c_s}$ and $\H_{c'}=\H_{c_{s+1}}\otimes \cdots \otimes H_{c_{\cal{I}}}$ amounts to splitting the colored vertex of color $c''$ in two colored vertices of colors $c'$ and $c''$, keeping the right hyper-edges attached to each one of the two resulting vertices. Note that this last procedure requires specifying the actual weighted partiton $(\pi, w)$ and not just the multisets, which is unavoidable since we are breaking the local unitary invariance to a subgroup in the process. 

\begin{figure}[!ht]
    \centering
    \includegraphics[width = \textwidth]{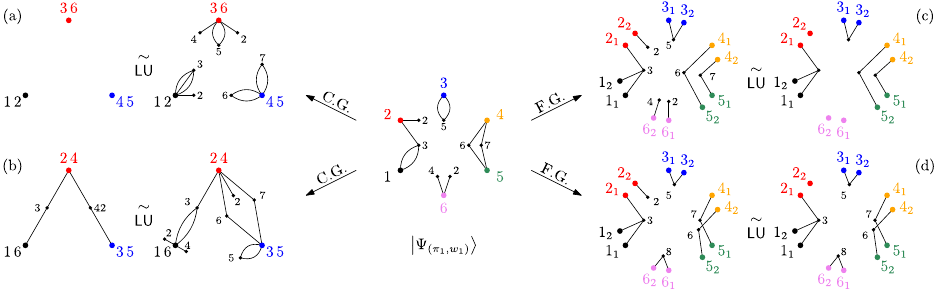}
    \caption{HT states represented by the central hypergraph can, for instance, be coarse-grained to the two examples on the left, or fine-grained to the two examples on the right. }
    \label{fig:CG}
\end{figure}

We recover graphically the facts stated in  Sec.~\ref{sec:generalities-on-coarse-graining}: coarse-graining and fine-graining may respectively diminish and raise the number of connected components of $\xi$, thereby changing the partial separability properties. The states  represented in (a), (b) (resp.~(c), (d)) in Fig.~\ref{fig:CG} are obtained by coarse graining (resp.~fine graining\footnote{To be more precise, the two states (c) and (d) are obtained by fine-graining two $\LU$-equivalent states  $\ket{\Psi_{(\pi_1,w_1)}}$ and $\ket{\Psi_{(\pi_1',w'_1)}}$ that differ only by a relabelling of the $c_j$ for each $c$, and which are therefore represented by the same hypergraph, see the details above Eq.~\eqref{eq:relab-subcolors}.}) of the central state $\ket{\Psi_{(\pi_1,w_1)}}$ associated with the weighted partition 
\begin{equation}
    (\pi_1,w_1) = \{\{1_1,1_2,2_1\}_3,\{2_2\}_2,\{3_1,3_2\}_5,\{4_1,5_2\}_6,\{4_2,5_1\}_7,\{6_1\}_2,\{6_2\}_4\} \,,
\end{equation} 
and whose entanglement depth $m$ and maximal $r$ of $r$-separability respectively take the values 2 and 4. For the other 4 states, these values are respectively: (a) 1 and 3; (b) 3 and 1; (c) and (d) 3 and 7 (but the partition $\xi$ differs).

\subsubsection{Haar-random states}
\label{subsub:Def-of-Haar}

The Haar-random state on a Hilbert space $\mathcal{H}$ of dimension $N^D$ is defined as:
\begin{equation}
\label{eq:def-of-Haar-state}
    \ket{\varphi} = U \ket{\varphi_0},
\end{equation}
where $U$ is a random Haar-distributed $N^D$ by $N^D$ unitary matrix and $\ket{\varphi_0}$ is any pure state of $\mathcal{H}$ ($\ket{\varphi}$ does not depend on $\ket{\varphi_0}$). 
This state can therefore be seen as taken uniformly at random in the space of pure states $S(\H)$.  
Equivalently (see Prop. 2 of Ref.~\cite{Nechita2007}), if we view $\H\simeq (\mathbb{C}^N))^{\otimes D}$ as a $D$-partite state space with local dimension $N$, the components of $\ket{\varphi}$ in an adapted basis can be shown to be normalized complex Gaussian variables 
\begin{equation}
    \varphi_{i_1 \ldots i_D} = \frac{T_{i_1 \dots i_D}}{\norm{T}}\,,
\end{equation}
where $\displaystyle\norm{T}^2 \eqdef \sum_{i_1, \dots, i_D=1}^D T_{i_1 \dots i_D} \bar{T}_{i_1 \dots i_D}$.

As explained in Sec.~\ref{sub:trace-invariants}, for any  $G\in\cG_D$  with $k(G)=k$ and $\vec \sigma\in S_{k}^D(G)$, one has $\tr_G=\tr_{\vec \sigma}$ defined in Eq.~\eqref{eq:LUinv}, where $\vec \sigma$ is a   $D$-tuple of permutations $\vec\sigma=(\sigma_1,\dots,\sigma_D)$. Trace-invariants of Haar-random states can be evaluated using Weingarten calculus (see App.~\ref{A:Wick}): 
\begin{equation} 
\label{eq:meanLUWeing}
    \mean{\tr_{G}(\ket{\varphi})} = f_{k,D,N} \sum_{\nu \in S_k}  N^{- \sum_{c=1}^D d(\sigma_c,\nu)} \,.
\end{equation}
where $f_{k,D,N}=\frac{N^{Dk}\pa{N^D - 1}!}{\pa{N^D - 1 + k}!}$, $\vec \sigma\in S_{k}^D(G)$, and $d$ is the \textit{Cayley distance} on $S_k$:
\begin{equation} \label{eq:Cayley}
    d(\sigma_i,\nu) \eqdef k - \# (\sigma_i \cdot\nu^{-1})\,,
\end{equation}
where $\#(\rho)$ is the number of disjoint cycles of a permutation $\rho$. The permutations $\sigma_i$ encode the edges of color $i$ of $G$. The permutation $\nu$ can be seen as encoding the edges of a $(D+1)$-th color: in the same way as $\vec \sigma$ for $G$, one has 
$$ 
(\sigma_1,\dots,\sigma_D, \nu)\in S_{k(G)}^{D+1}(\widehat G)\,,\qquad \mathrm{with} \quad \widehat G\in \cG_{D+1}\,.
$$
By convention, we assign the color 0 to the edges corresponding to the additional permutation $\nu$.\footnote{This is a common convention in the literature on random tensor models.} The cycles of $\sigma_i \nu^{-1}$ are in bijection with the cycles in the graph $\widehat G$ whose edges are alternatively of color $i$ and 0, which correspond to the connected components of $\widehat G\vert_{0i}$ and are called \emph{faces} of colors $(0,i)$. 
In the same way,  the connected components of $G\vert_{ij}$ are the faces of colors $(i,j)$ of $G$.  We introduce the following notations:
\begin{equation}
    F_{0i}(\widehat G)\eqdef \#(\sigma_i \nu^{-1})\,, \qquad  F_{ij}(G) = \# (\sigma_i \sigma_j^{-1})\, 
\end{equation}
as well as
\begin{equation}
\label{eq:faces}
   F_{c}(G)\eqdef  \sum_{i\neq c} F_{ic}(G)\,\qquad F(G)   \eqdef \sum_{i<j} F_{ij}(G)\,.
\end{equation} 
We let $\cG_{D+1}(G)$ be the subset of $\cG_{D+1}$ (that is, with the $(D+1)$-th color relabeled as 0) of graphs $\widehat G$ satisfying $(\widehat G)^{\hat 0}=G$ (the graph $G$ is recovered when all edges of color 0 are removed for $\widehat G$), we may therefore also formulate Eq.~\eqref{eq:meanLUWeing} without permutations as:
\begin{equation} 
\label{eq:meanLUWeing2}
\mean{\tr_{G}(\ket{\varphi})} = f_{k,D,N} \sum_{\widehat G\in \cG_{D+1}(G)} \Xi(\widehat G) N^{ F_{0}(\widehat G)-Dk} \,.
\end{equation}
where, $\Xi(\widehat G) \eqdef \abs{\paa{\nu \in S_{k(G)} \;\; \rm{s.t.} \;\;\pa{\sigma_1,\dots,\sigma_D,\nu} \in S_{k(G)}^{D+1}(\widehat G)}}$ is a degeneracy factor.
\begin{rem}
    \label{rem:Haar-vs-Gauss}
    The distribution for the coefficients $\varphi_{i_1 \ldots i_D}$ of the Haar-random states can be compared at large $N$ with centered i.i.d.~Gaussian complex variables $X_{i_1, \ldots, i_D}$ with variance $1/N^D$, as explained in App.~\ref{A:Wick}. For this distribution, trace-invariants can be computed using Wick's theorem, leading (with the notations above) to:
    \begin{equation} \label{eq:meanLUinv}
        \mean{\tr_{G}(X)} = \sum_{\nu \in S_k} N^{- \sum_{c=1}^D d(\sigma_c,\nu)} \,.
    \end{equation}
    In particular, the terms of Eq.~\eqref{eq:meanLUWeing} and Eq.~\eqref{eq:meanLUinv} that dominate when $N$ is large are given in both cases by the permutations that minimize  $\sum_c d(\sigma_c,\nu)$, and, fixing $k,D$, one has $\displaystyle\lim_{N\rightarrow \infty} f_{k,D,N}=1$. 
\end{rem}

\subsection{\texorpdfstring{$\LU$}{LU}-classification of hypergraph-tensor states}
\label{sub:LU-inequivalent-reference-states}

Let $D \geq 2$. The goal of this subsection is to establish a property we already alluded to, namely: that the weight function $\alpha$ associated with an HT state $\ket{\phi_\alpha}$ (as defined in \eqref{eq:def_alpha_states}) labels the $\LU$-equivalence classes of HT states. 

A first hint that this could be the case is that this description of HT states is stable under coarse-graining. To see this, let $\sF= (\H_1 , \ldots , \H_D)$ a finite-dimensional $D$-partite state space, and $\zeta$ a partition of $\{ 1, \ldots , D\}$. Given $\alpha$ a weight function and $\ket{\psi_\alpha}\in \H^{\sF}=\H_1 \otimes \cdots \otimes \H_D$, one can introduce the weighted function $\alpha_\zeta$ on the coarse-grained state space $\H^{\sF_\zeta}$, defined as follows:
\begin{equation}
    \forall B' \subset \zeta\,, |B'|\geq 2\,, \quad \alpha_\zeta (B') \eqdef \prod_{\substack{B \subset \{1 , \ldots , D\}\, ,\, |B| \geq 2 \\
    \{A \in \zeta \, | \, A \cap B \neq \emptyset \}= B'}} \alpha(B)\,.
\end{equation}
One can then verify that $\ket{\psi_\alpha}$ and $\ket{\psi_{\alpha_{\zeta}}}$ are $\LU$-equivalent in the coarse-grained state space $\H^{\sF_\zeta}$:
\begin{equation}
    \ket{\psi_\alpha} \underset{\LU^{\sF_\zeta}}{\sim} \ket{\psi_{\alpha_\zeta}} \,.
\end{equation}
In particular, for a bipartition $\zeta = \{A, \bar{A}\}$, $\ket{\psi_{\alpha_\zeta}}$ is fixed by the single coefficient
\begin{equation}\label{eq:bipartite_alpha_state}
    \alpha_\zeta \eqdef a_\zeta (\zeta) = \prod_{\substack{B \subset \{1 , \ldots , D\}\, ,\, |B| \geq 2 \\
    A\cap B \neq \emptyset\,,\, \bar{A}\cap B \neq \emptyset }} \alpha(B)\,.
\end{equation}
The coefficient $a_\zeta$ is a measure of the entanglement spectrum of $\ket{\phi_\alpha}$ across the bipartition $\zeta$, which, given the Ansatz \eqref{eq:def_alpha_states}, is necessarily flat.

In the bipartite setting, HT states are Bell states with flat entanglement spectra, which leads to the following elementary result.
\begin{lem}\label{lem:LU_ref_class}
    Suppose that $D=2$, and let $\ket{\psi_\alpha}$, $\ket{\psi_\beta}$ denote two bipartite HT states. Then, the following statements are equivalent:
    \begin{enumerate}
        \item $\ket{\psi_\alpha} \simLU \ket{\psi_\beta}$;
        \item there exists $k \in \mathbb{N}^*\setminus \{ 1\}$ such that $\tr_{C_k} (\ket{\psi_\alpha}) = \tr_{C_k} (\ket{\psi_\beta})$;
        \item $\alpha = \beta$.
    \end{enumerate}
\end{lem}
\begin{proof}
    The entanglement spectrum of $\ket{\psi_\alpha}$ contains a single eigenvalue $\frac{1}{\alpha}$ of multiplicity $\alpha$, and similarly for $\ket{\psi_\beta}$. Hence, those two states are $\LU$-equivalent if and only if $\alpha = \beta$. In turn, this is equivalent to $\ket{\psi_\alpha}$ and $\ket{\psi_\beta}$ having the same $k$-purity for some fixed $k\geq 2$.
\end{proof}
To generalize this result to $D\geq 3$, we can look for a family of $D$-colored graphs that could play an analogous role to that of $k$-purities in the bipartite setting. We should expect such a family to be rather non-unique, but the next result provides a choice that works in any $D \geq 3$. It relies on the family of graphs $\{ \ME_n^D\}$ associated with multi-entropies and the family of graphs  $\{ \RME_{2,n}^{(D)} \}$ associated with reflected multi-entropies, which have respectively been introduced in  Sec.~\ref{ss:ME} and  Sec.~\ref{ss:RE}. In the latter case, the role of the subsystem $\H_D$ in the construction of $\RME_{2,n}^{(D)}$ in Sec.~\ref{ss:RE} may be played instead by $\H_{\bar B}=\bigotimes_{c\in \bar B} \H_c$, where $B\subset \{1,\ldots, D\}$ and $\bar B$ is the complement of $B$ in $\{1,\ldots, D\}$. 
\begin{theo}\label{th:RefState_Equiv}
    Let $D\geq 3$, $\alpha: \{ B \subset \{ 1, \ldots , D\} \,|\, |B| \geq 2\} \to \mathbb{N}^*$ and $\beta : \{ B \subset \{ 1, \ldots , D\} \,|\, |B| \geq 2\} \to \mathbb{N}^*$ two weight functions. The following statements are equivalent:
    \begin{enumerate}
        \item $\ket{\psi_{\alpha}} \simLU \ket{\psi_{\beta}}$;
        \item for any $n \in \{2, \ldots , D \}$, any $B \subset \{ 1, \ldots , D\}$ such that $2 \leq |B| \leq D-1$, and any $n_B \in \{ 1, \ldots , |B|\}$, one has: $\tr_{\ME_n^D} \left( \ket{\psi_{\alpha}} \right) = \tr_{\ME_n^D} \left( \ket{\psi_{\beta}} \right)$ and $\tr_{\RME_{2,n_B}^{(\bar B)}} \left( \ket{\psi_{\alpha}} \right) = \tr_{\RME_{2,n_B}^{(\bar B)} } \left( \ket{\psi_{\beta}} \right)$;
        \item $\alpha = \beta$.
    \end{enumerate}
\end{theo}
\begin{proof}
    Property $1$ clearly implies Property $2$ by Prop.~\ref{prop:charac_LU-1}. Furthermore, since the evaluation of a trace-invariant on an HT state $\ket{\psi_\alpha}$ only depends on its weight function $\alpha$ (by \eqref{eq:trace-inv_alpha-states}), Prop.~\ref{prop:charac_LU-1} also ensures that: Property $3$ implies Property $1$. 
    
    Let us finally prove that Property $2$ implies Property $3$. For any $G \in \cG_D$, we have $\tr_{G} \left( \ket{\psi_{\alpha}} \right) = \tr_{G} \left( \ket{\psi_{\beta}} \right)$ if and only if
\begin{equation}
\prod_{C \subset \{1, \ldots , D\}\,,\, |C|\geq 2} \alpha (C)^{\kappa(G\vert_{C}) - k(C)} = \prod_{C \subset \{1, \ldots , D\}\,,\, |C|\geq 2} \beta (C)^{\kappa(G\vert_{C}) - k(G)}\,,
\end{equation}
which, upon taking a log, is equivalent to
\begin{equation}
    \label{eq:Syst_Sigma}
\sum_{C \subset \{1, \ldots , D\}\,,\, |C|\geq 2} \left( k(G) - \kappa(G\vert_{C}) \right) x_C\left( \alpha , \beta\right)= 0\,, \quad \mathrm{where} \quad x_C\left( \alpha , \beta\right) \eqdef \ln\left( \frac{\alpha (C)}{\beta (C)}\right)\,.
\end{equation}
Suppose that this equation holds for any $G$ taken in the families $\{ \ME_n^D \, |\, 2 \leq n \leq D\}$ and $\{ \RME_{2, n_B}^{(\bar{B})} \, |\, B \subset \{1, \ldots , D \}\,, 2 \leq |B| \leq D-1 \,, 1 \leq n_B \leq |B| \}$. For the first family, this yields the conditions:
\begin{equation}
    \label{eq:ME-vanish-proof-LU}
    \forall n \in  \{ 2, \ldots , D \}\,, \qquad        \sum_{\substack{{C\subset\{1,\ldots, D\}}\\{\abs{C}>1}}} \pa{n^{D-1} - n^{D-\abs{C}}} x_C (\alpha , \beta) = 0\,.
\end{equation}
The real polynomial
\begin{equation}
P(X) \eqdef \sum_{\substack{{C\subset\{1,\ldots, D\}}\\{\abs{C}>1}}} \pa{X^{D-1} - X^{D-\abs{C}}} x_C (\alpha , \beta)
\end{equation}
is of degree at most $D-1$ and admits at least $D$ roots: any $n\in\{ 2, \ldots , D\}$ is a root by \eqref{eq:ME-vanish-proof-LU} and we also trivially have $P(1)=0$. Hence $P$ must vanish identically. In particular, for any $p \in \{2, \ldots ,D \}$, its coefficient of order $D-p$ must vanish, leading to:  
    \begin{equation}
    \label{eq:cons-ME-proof-LU}
        \forall p \in \paa{2, \dots, D}\,, \qquad \sum_{\substack{{C\subset\{1,\ldots, D\}}\\{\abs{C}=p}}}x_C (\alpha , \beta )  = 0\,.
    \end{equation}
In particular, since there is only one set with $D$ elements, we proved that $x_{\{1,\dots,D\}}(\alpha, \beta) = 0$. 

\medskip
   
Now, let $B\subset \{ 1, \ldots , D\}$ with $|B|\geq 2$, and let us prove that $x_B (\alpha , \beta )=0$. We proceed by downward induction on $p\eqdef |B| \in \{2, \ldots , D\}$. 

If $p =D$, then $B=\{ 1, \ldots , D\}$ and therefore $x_{\{1,\dots,D\}}(\alpha, \beta) = 0$, as was already proven. 

Let us then assume that $p \leq D-1$, and suppose that $x_C (\alpha , \beta)=0$ for any $C \subset \{1, \ldots , D\}$ with $|C|\geq p+1$. We want to evaluate Eq.~\eqref{eq:Syst_Sigma} on the family of graphs $\{ \RME_{2, n}^{(\bar{B})} \, |\,  2 \leq n \leq |B| \}$. For this purpose, note that: for any $n \in   \{1 , \ldots, p\}$ and any $C \subset \{1, \ldots , D\}$ with $2 \leq |C|\leq p$, we can evaluate the number of connected components of $\RME_{2,n}^{(\bar{B})}$ to deduce  
     \begin{equation} 
     \label{eq:coeffRME}
     k(\RME_{2,n}^{(\bar{B})}) - \kappa\bigl(\RME_{2,n}^{(\bar{B})}\vert_{C}\bigr) = \left\{
        \begin{array}{ll}
            2n^{p-1} - 2 \,, \rm{ if } C = B\,, \\
                   2n^{p-1}- 2n^{p - \abs{C}} \,, \rm{ if } C \subsetneq B \,, \\
            2n^{p-1}- n^{p - \abs{C \cap B}} \,, \rm{ if } C\cap  B\neq \emptyset\, \textrm{ and } C\cap \bar B\neq \emptyset, \\
            0 \,, \rm{ otherwise.}
        \end{array}
        \right.
    \end{equation}
Plugging this into Eq.~\eqref{eq:Syst_Sigma} leads to $p$ independent conditions: for any $n \in \{1, \ldots, p\}$,  
     \begin{equation}
     \label{eq:vanish-RME-proof-LU}
         2(n^{p-1} - 1)x_B(\alpha, \beta) + \sum_{\substack{{C \subsetneq B}\\{1<\abs{C}<p}}} 2(n^{p-1}- n^{p - \abs{C}})x_C(\alpha, \beta) +  \sum_{\substack{{C\cap B\neq \emptyset \textrm{ and } C\cap \bar B\neq \emptyset}\\{\abs{C\cap B} < p}}} (2n^{p-1}- n^{p - \abs{C\cap B}})x_C(\alpha, \beta) =0\,.
     \end{equation}
This is a polynomial in $n$ of degree at most $p-1$ with at least $p$ roots, hence it must vanish identically. In particular, its constant coefficient, which is equal to $2 x_B (\alpha , \beta)$, must vanish. 

\medskip

In conclusion, we have shown that $x_B (\alpha , \beta)=0$ for any $B \subset \{1, \ldots , D\}$ with $|B|\geq 2$, which is equivalent to $\alpha = \beta$.
\end{proof}

\begin{figure}[!ht]
    \centering
    \includegraphics[width = \textwidth]{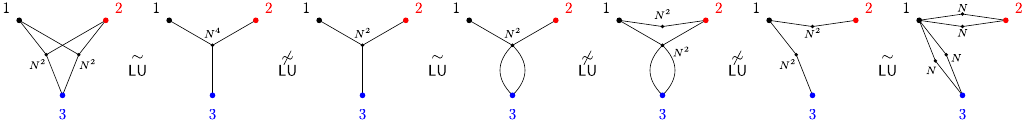}
    \caption{Examples of $\LU$ (in)equivalence relations between HT states. This illustrates both the graphical equivalence relations introduced in Sec.~\ref{sec:ref_states} (Eqs.~\eqref{eq:equiv_ref_1}--\eqref{eq:equiv_ref_4}), and the $\LU$-classification result of Thm.~\ref{th:RefState_Equiv}. 
    }
    \label{fig:LUeq_Ref}
\end{figure}

Thm.~\ref{th:RefState_Equiv} is illustrated in Fig.~\ref{fig:LUeq_Ref} for $3$-partite HT states. This result confirms that weight functions label the $\LU$-classes of HT states. As a result, the four elementary equivalence relations Eqs.~\eqref{eq:equiv_ref_1}--\eqref{eq:equiv_ref_4} introduced in Sec.~\ref{sec:ref_states} allow us to explore all the multisets representing a given $\LU$-class of HT states. Equivalently, the set of decorated hypergraphs representing a given $\LU$-class of HT states can be explored by means of four elementary graphical moves: 1) the addition or removal of an univalent hyper-edge of arbitrary weight; 2) the addition or removal of an hyper-edge of weight $1$; 3) the graphical move illustrated in  Fig.~\ref{fig:Move}; 4) the graphical move illustrated in Fig.~\ref{fig:LUeq_Ref2}.
 
In the following, we will always rely on graphical representations of HT states that do not feature hyper-edges of weight $1$ (since those carry no information and can always be suppressed by means of graphical moves of type 2)). In that case, all the $\LU$-equivalent HT states generated by the graphical moves 1), 3) and 4) induce the same $\xi$, defined as the partition of $\{1, \ldots, D\}$ induced by the connected components of the hypergraph representing the states. The converse is not true, however. 

Point 2 of Thm.~\ref{th:RefState_Equiv} states that it is sufficient to consider the trace-invariants corresponding to $\ME_n^D$ for any $n\in\{2,\ldots, D\}$  and $\RME_{2,n_B}^{(\bar B)}$  for  any $B\subset \{1,\ldots, D\}$ with $2\le \abs{B}\le D-1$ and any $n_B\in \{1,\ldots,\abs{B}\}$ to know the $\LU$-class of a HT state. The proof of the theorem can in fact be adapted to derive explicit expressions relating the (canonical) weight function $\alpha$ of an unknown HT state $\ket{\Psi}$ to the evaluations $\{\ln \tr_G (\ket{\Psi})\}$ with $G$ belonging to one of the two families $\{\ME_n^D\}$ and $\{\RME_{2,n_B}^{(\bar B)}\}$. More concretely, for any $G$ in those families, we have
\begin{equation}\label{eq:linear_sys}
-\ln\tr_G \left( \ket\Psi \right) = \sum_{C\subset \{1, \ldots , D\}\,, \, |C|\geq 2} \left( k(G)-\kappa(G\vert_{C})   \right) P_C^D(\ket\Psi)\,, 
\end{equation}
where, for any $C$, $P_C^D (\ket\Psi)$ denotes the log of the evaluation of the weight function of $\ket\Psi$ on $C$. Namely, for any weight function $\alpha$, we have 
\begin{equation}
    P_C^D(\ket{\psi_\alpha}) \eqdef \ln\left( \alpha(C) \right)\,.
\end{equation}
The linear system \eqref{eq:linear_sys} can be inverted to express the parameters $\{P_C^D(\ket\Psi)\}_C$ in terms of the evaluations $\{ \ln\tr_G \left( \ket\Psi \right)\}_G$ (possibly in a non-unique way since the system is expected to be over-determined in general). For any $C \subset \{ 1, \ldots, D\}$ with $|C|\geq 2$, we can thus derive a linear combination of the form:
\begin{equation}
    P_C^D(\ket\Psi) = \sum_{n=2}^D \lambda_{n}(C) \ln \tr_{\ME_n^D} (\ket{\Psi}) + \sum_{\substack{B, n_B\\2 \leq |B|\leq D-1 \,,\, 1\leq n_B \leq |B|}} \mu_{B,n_B}(C) \ln \tr_{\RME_{2,n_B}^{(\bar B)}} (\ket{\Psi}) \,,
\end{equation}
where $\{ \lambda_{n}(C) , \mu_{B,n_B}(C) \}$ are rational coefficients. Knowing those coefficients for any $C$ allows us to reconstruct the weight function of the unknown HT state $\ket{\Psi}$ from the sole knowledge of the graph evaluations $\{\tr_{\ME_n^D}(\ket{\Psi})\}$ and $\{\tr_{\RME_{2,n_B}^{(\bar B)}} (\ket{\Psi})\}$ (with $n$, $B$ and $n_B$ having a range as defined above). For instance, for $D=3$, the linear system \eqref{eq:linear_sys} takes the form: 
\begin{equation}
    \begin{cases}
    -\ln\tr_{\ME_2^3} &= 2 \left( P_{\{ 1,2\}}^3 + P_{\{ 1,3\}}^3 + P_{\{ 2,3\}}^3\right) + 3 P_{\{ 1, 2, 3\}}^3\\
    -\ln\tr_{\ME_3^3} &= 6 \left( P_{\{ 1,2\}}^3 + P_{\{ 1,3\}}^3 + P_{\{ 2,3\}}^3\right) + 8 P_{\{ 1, 2, 3\}}^3 \\
    -\ln\tr_{\RME_{2,1}^{(k)}} &= P^3_{\{i,k\}} + P^3_{\{j,k\}} + P_{\{ 1, 2, 3\}}^3 \quad \text{for any} \; i,j,k \quad \text{s.t.} \quad \{i,j,k\}= \{1,2,3\}\\
    -\ln\tr_{\RME_{2,2}^{(k)}} &= 2 P^3_{\{1,2\}} + 2 P^3_{\{1,3\}} + 2 P^3_{\{2,3\}} + 3 P_{\{ 1, 2, 3\}}^3 \quad \text{for any} \; k\in \{1,2,3\}
    \end{cases}
\end{equation}
This system can be inverted, \eg in the following form:
\begin{equation}
\begin{cases}
P^{3}_{\{1,2,3\}} &= - 3 \ln \tr_{\ME_2^3} + \ln \tr_{\ME_3^3}\\ 
P^{3}_{\{i,j\}}  &=  \frac 3 2  \ln \tr_{\ME_2^3} - \frac 1 2 \ln \tr_{\ME_2^3} - \frac 1 2   \ln \tr_{\RME_{2,2}^{(k)}} +  \ln \tr_{\RME_{2,1}^{(k)}} \quad \text{for any} \; i,j,k \quad \text{s.t.} \quad \{i,j,k\}= \{1,2,3\}
\end{cases}
\end{equation}
It is also clear in this example that distinct (but ultimately equivalent) expressions of the maps $\{ P_C^3 \}$ in terms of  multi-entropy and reflected multi-entropy invariants can be written down: this is simply a manifestation of the fact that those invariants over-determine the $\LU$-class of an HT state. It would be interesting to investigate whether there exists a smaller (and perhaps simpler) set of trace-invariants that allows one to discriminate any pair of $\LU$-inequivalent HT states. Here are some comments on this problem for small values of $D$. 

\begin{itemize}
    \item For $D=2$, we have seen (in Lem.~\ref{lem:LU_ref_class}) that it is sufficient to consider $C_2$ (see Fig.~\ref{fig:bip}), which is in fact equal to $\ME_2^2$ (so that Lem.~\ref{lem:LU_ref_class} can be understood as a $D=2$ version of Thm.~\ref{th:RefState_Equiv}). 
    \item For $D=3$, Thm.~\ref{th:RefState_Equiv} states that it is sufficient to consider: $\ME_2^3$, which corresponds to $\RM_4$ (see Fig.~\ref{fig:ME_RE_RM}); $\ME_3^3$, shown on the left of Fig.~\ref{fig:ME2}; $\RME_{2,n}^{(\bar B)}$ for $n=1$ and $\abs{B}=2$, which are simply the three connected graphs with $k=2$ (and are in particular melonic); and finally, $\RME_{2,n}^{(\bar B)}$ for $n=2$ and $\abs{B}=2$, but this actually coincides with $\RM_4$ (see Fig.~\ref{fig:ME_RE_RM}). This is too much information: it is actually sufficient to consider $\RM_4$, $\ME_3^3$, together with any two connected graphs with $k=2$. Alternatively, if one wishes to rely on the simplest possible graphs, one may instead consider $\PT_3$ (left of Fig.~\ref{fig:PT_EX}), for which Eq.~\eqref{eq:Syst_Sigma} reads $2(x_{\{1,2,3\}} +x_{\{1,2\}}+x_{\{1,3\}}+x_{\{2,3 \}})=0$, and the three connected graphs with $k=2$, which give: $x_{\{1,2,3\}} +x_{\{1,2\}}+x_{\{2,3\}}=0$ (together with two independent equations obtained by cyclic permutation of the color labels). 
    \item We expect that for $D>3$, there always exists a simpler set of invariants playing the role of the set introduced in Point~2 of Thm.~\ref{th:RefState_Equiv}: indeed, our proof only relies on the cancellation of constant terms in polynomials that may have degree up to $D-1$, and we should therefore not expect it to be optimal (if anything, we might expect it to be less and less optimal as $D$ grows). Note also, for instance, that the value of $\RME_{2,n}^{(\bar B)}$ for $\abs{B}=D-1$ and $n=2$ always concides with $\ME_2^D$ (Sec.~\ref{ss:RE}). 
\end{itemize}

\

The complete $\LU$-classification of HT states obtained in Thm.~\ref{th:RefState_Equiv} allows us to illustrate some of the weaker entanglement classification notions introduced in Sec.~\ref{sec:generalities-on-coarse-graining}. For instance, two states can be $2$-partite equivalent without being $\LU$-equivalent. 
\begin{ex}
    Let $D = 3$, $N \geq 2$, and $\sF= (\H_1 , \H_2 , \H_3)$, with each tensor factor $\H_c$ assumed to be of dimension $N^2$. Let also $\alpha$ and $\beta$ be the two weight functions defined by:
    \begin{align}
        \alpha(\{1,2,3\}) = N^2 \,, \quad \forall 1 \leq i <j \leq 3\,, \;\alpha(\{i,j\})=1\,, \\
        \beta(\{1,2,3\}) = 1 \,, \quad \forall 1 \leq i <j \leq 3\,, \;\beta(\{i,j\})=N\,.
    \end{align}
    The HT states $\ket{\psi_\alpha}$ and $\ket{\psi_\beta}$ can both be realized in the $3$-partite state space $\sF$. Moreover, for any connected cyclic graph $G \in \cG_3$, Eq.~\eqref{eq:trace-inv_alpha-states} yields
    \begin{align}
        \tr_G (\ket{\psi_\alpha}) &= (N^2)^{\kappa(G) - k(G) } = N^{2 - 2k(G)}\,, \\
        \tr_G (\ket{\psi_\beta}) &= \prod_{{1 \leq i < j \leq 3}} N^{ \left( \kappa(G\vert_{\{i,j\}}) - k(G) \right)} = N^{\left( k(G) - k(G) \right) + 2 (1 - k(G))} = N^{2 - 2k(G)} \,.
    \end{align}
        It follows by Prop.~\ref{prop:2partite_carac} that $\ket{\psi_\alpha}$ and $\ket{\psi_\beta}$ are $2$-partite equivalent (recall Def.~\ref{def:2partite_eq}):
    \begin{equation}
        \ket{\psi_\alpha } \underset{(2, \sF)}{\sim} \ket{\psi_\beta }\,.
    \end{equation}
    However, by Thm.~\ref{th:RefState_Equiv}, we also have:  $\ket{\psi_\alpha } \underset{\LU}{\not\sim} \ket{\psi_\beta }$. 
    
    \begin{figure}[!ht]
        \centering
        \includegraphics[height = 3cm]{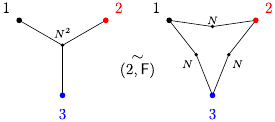}
        \caption{Example of $\LU$-inequivalent states that are $2$-partite equivalent according to Def.~\ref{def:2partite_eq}.}
        \label{fig:bip_eq}
    \end{figure}
\end{ex}

\subsection{Asymptotic \texorpdfstring{$\LU$}{LU}-equivalences and \texorpdfstring{$\LU$}{LU}-equivalences in scaling}\label{sec:examples_asymptotic_rel}

We conclude by discussing a couple of examples of asymptotic $\LU$-relations (as defined in Sec.~\ref{sec:LU_random_asymptotic}) that involve HT states.

\medskip 

To begin with, it was recently established that the correlation functions of Gaussian random tensors of order $D \geq 3$ fail to factorize in the same way as those of Gaussian (or Haar-distributed) random matrices in the large $N$ limit (see Ref.~\cite{Gurau2025}). This result has a straightforward but interesting implication in our context: while a Haar-random bipartite quantum state can be well-approximated by a deterministic state in the large $N$ limit  (namely, a maximally entangled Bell state, with flat entanglement spectrum, as first observed by Page in Ref.~\cite{Page:1993df}), a similar result is provably impossible to achieve in the multipartite setting.
\begin{prop}\label{prop:Haar_not_approx_deterministic}
    Let $D \geq 2$. For  every $N \in \mathbb{N}^*$, let $\ket{\varphi_N}$ denote the Haar-distributed random $D$-partite pure state on $(\mathbb{C}^N)^{\otimes D}$, and let $\ket{\Bell_N}$ denote a maximally entangled bipartite pure state on $(\mathbb{C}^N)^{\otimes 2}$. 
    \begin{enumerate}
        \item If $D=2$, the sequence of random pure states $(\ket{\varphi_N})_{N \in \mathbb{N}^*}$ is $\LU$-equivalent in scaling to the sequence of deterministic pure states $(\ket{\Bell_N})_{N \in \mathbb{N}^*}$:
        \begin{equation}
            \ket{\varphi_N} \underset{N \to \infty}{\approx} \ket{\Bell_N}\,.
        \end{equation}
        \item By contrast, if $D \geq 3$, it is not possible to find a sequence of determinisitc $D$-partite pure states that is $\LU$-equivalent in scaling to $(\ket{\varphi_N})_{N \in \mathbb{N}^*}$ (and, thus, neither is there one that is asymptotically $\LU$-equivalent to it).
    \end{enumerate}
\end{prop}
\begin{proof}
Let us assume that $D=2$. For any $k \geq 2$, a standard application of Wick's theorem allows us to prove that
\begin{equation}
\langle \tr_{C_k}(\ket{\varphi_N}) \rangle    \underset{N\to \infty}{\sim}
\rm{Cat}_k N^{1- k}\,, 
\end{equation}
where $\rm{Cat}_k = \#\{\sigma \in S_k \, |\, \#(\sigma) + \#(\sigma^{-1} \tau)= k+1 \}$ denotes the $k$-th Catalan number and $\tau = (12 \cdots k)$. More generally, arbitrary correlation functions of this random matrix ensemble factorize over their connected components in the large $N$ limit, namely: for any $p \geq 2$ and $k_1, \ldots , k_p \geq 2 $, we have:
\begin{equation}
    \langle \prod_{\ell = 1}^p \tr_{C_{k_\ell}}(\ket{\varphi_N}) \rangle    \underset{N\to \infty}{\sim} \prod_{\ell = 1}^p \langle  \tr_{C_{k_\ell}}(\ket{\varphi_N}) \rangle \underset{N\to \infty}{\sim} \left( \prod_{\ell = 1}^p \rm{Cat}_{k_\ell}\right) N^{p - \sum_{\ell=1}^p k_\ell }\,.
\end{equation}
In other words, for any $G \in \cG_2$, we have:
\begin{equation}
    \langle \tr_G \left( \ket{\varphi_N} \right) \rangle \underset{N\to \infty}{\sim}\lambda_G N^{\kappa(G) - k(G)}\,, 
\end{equation}
where the coefficient $\lambda_G >0$ is a product of Catalan numbers. On the other hand, Eq.~\eqref{eq:trace-inv_alpha-states} implies that
\begin{equation}
    \forall G \in \cG_2\,, \forall N \in \mathbb{N}^*\,, \qquad  \tr_G \left( \ket{\Bell_N} \right) = N^{\kappa(G) - k(G)}\,.
\end{equation}
It follows that $(\ket{\varphi_N})_{N \in \mathbb{N}^*}$ is $\LU$-equivalent in scaling to $(\ket{\Bell_N})_{N \in \mathbb{N}^*}$ (in the sense of Def.~\ref{def:LargeN_LU_eq}).

Suppose next that $D \geq 3$. Let us assume, by contradiction, that there exists a sequence of deterministic pure states $(\ket{\psi_N})_{N \in \mathbb{N}^*}$ that is $\LU$-equivalent in scaling to $(\ket{\varphi_N})_{N \in \mathbb{N}^*}$. According to Ref.~\cite{Gurau2025} (Thm.~1),\footnote{Strictly speaking, the authors of Ref.~\cite{Gurau2025} only considered real Gaussian random tensors in their work, but their proof generalizes straightforwardly to the complex setting, and by extension, to Haar-distributed random quantum states.} there exists a connected graph $G \in \cG_D$ such that 
\begin{equation}\label{eq:proof_no_deterministic_approx}
\frac{    \langle \tr_G \left( \ket{\varphi_N} \right) \tr_G \left( \ket{\varphi_N}\right) \rangle }{ \langle \tr_G \left( \ket{\varphi_N} \right) \rangle^2 } \underset{N \to\infty}{\rightarrow} + \infty\,.
\end{equation}
But by assumption, we also have non-zero constants $\lambda_G$ and $\lambda_{G \sqcup G}$ such that:  
\begin{equation}
    \frac{    \langle \tr_G \left( \ket{\varphi_N} \right) \tr_G \left( \ket{\varphi_N}\right) \rangle }{  \langle \tr_G \left( \ket{\varphi_N} \right) \rangle^2 } \underset{N\to \infty}{\sim}     \frac{    \lambda_{G \sqcup G}\, \langle \tr_G \left( \ket{\psi_N} \right) \tr_G \left( \ket{\psi_N}\right) \rangle }{ (\lambda_G)^2 \, \langle \tr_G \left( \ket{\psi_N} \right) \rangle^2 } = \frac{    \lambda_{G \sqcup G}}{(\lambda_G)^2} \,,
\end{equation}
where in the last equality we have used the fact that $\ket{\psi_N}$ is deterministic. We therefore obtain a contradiction with Eq.~\eqref{eq:proof_no_deterministic_approx}, meaning that the deterministic sequence $(\ket{\psi_N})_{N \in \mathbb{N}^*}$ cannot exist.
\end{proof}

\medskip 

Next, let us provide elementary examples of HT states that are asymptotically $\LU$-equivalent or $\LU$-equivalent in scaling. 
\begin{ex}
    Let $D \geq 2$ and $B \subset \{1, \ldots , D\}$ with $|B|\geq 2$. Recalling the definitions of Def.~\ref{def:LargeN_LU_eq}, one can verify that \eg
    \begin{equation}
        \ket{\GHZ}_{B,N+N^2} \underset{N\to \infty}{\sim} \ket{\GHZ}_{B,N^2}\,,
    \end{equation}
    and \eg
    \begin{equation}
 \ket{\GHZ}_{B,N} \underset{N\to \infty}{\approx} \ket{\GHZ}_{B,2N} \,, \quad \ket{\GHZ}_{B,N} \underset{N\to \infty}{\not\sim} \ket{\GHZ}_{B,2N}\,.
    \end{equation}
Such relations can be composed in a straightforward manner; for instance, if $D=4$, $B_1 = \{1,2,4\}$ and $B_2 = \{1,3,4\}$, one has \eg
\begin{equation}
    \ket{\GHZ}_{B_1, N^3+N^2} \otimes \ket{\GHZ}_{B_2, 3N^4+N} \underset{N\to \infty}{\sim} \ket{\GHZ}_{B_1, N^3} \otimes \ket{\GHZ}_{B_2, 3N^4} \underset{N\to \infty}{\approx} \ket{\GHZ}_{B_1, N^3} \otimes \ket{\GHZ}_{B_2, N^4} \,. 
\end{equation}
\end{ex}
Next, one may be interested in quantum states that are not HT states \emph{per se} but that are asymptotically $\LU$-equivalent (or merely $\LU$-equivalent in scaling) to HT states.  
\begin{ex}\label{ex:asymptotic_eq}
    Let $D \geq 2$ and $B \subset \{1, \ldots , D\}$ with $|B|\geq 2$. For any $N \in \mathbb{N}^*$, we may define
    \begin{align}
        \ket{\rho_1}_{B,N} &\eqdef \frac{1}{\sqrt{N}}\pac{\sum_{i=1}^{N-1}\sqrt{1+\frac{1}{N}}\ket{i}^{\ot \abs{B}} + \frac{1}{\sqrt{N}}\ket{N}^{\ot \abs{B}}} \,, \\
        \ket{\rho_2}_{B,N} &\eqdef \sqrt{\frac{2}{N(N+1)}} \sum_{j=1}^N \sqrt{j} \ket{j}^{\ot \abs{B}}\,.
    \end{align}
    For any $G \in \cG_D^{\conn}$, we then have:
    \begin{align}
        \tr_G\left( \ket{\rho_1}_{B,N} \right)&= N^{-k(G)} \left( \sum_{i=1}^{N-1} \left(1+ \frac{1}{N}\right)^{k(G)} + \left( \frac{1}{N}\right)^{k(G)}\right) \underset{N\to \infty}{\sim} N^{1- k(G)}\,, \\
        \tr_G\left( \ket{\rho_2}_{B,N} \right)&=\left( \frac{2}{N(N+1)}\right)^{k(G)} \sum_{j=1}^N j^{k(G)} \underset{N\to \infty}{\sim} \frac{2^{k(G)}}{1+k(G)} N^{1- k(G)}\,,
    \end{align}
where, in the second line, we have invoked Faulhaber's formula (truncated to its leading order term). It follows that
\begin{equation}
\ket{\rho_1}_{B,N}  \underset{N\to \infty}{\sim} \ket{\GHZ}_{B,N} \qquad \mathrm{and} \qquad  \ket{\rho_2}_{B,N}  \underset{N\to \infty}{\approx} \ket{\GHZ}_{B,N} \qquad (\mathrm{but} \quad \ket{\rho_2}_{B,N}  \underset{N\to \infty}{\not\sim} \ket{\GHZ}_{B,N}) \,. 
\end{equation}
Again, such relations can be composed: for instance, if $D=5$, $B_1 = \{1,2,3,4\}$ and $B_2 = \{ 2,4,5\}$, one has the asymptotic relations
\begin{equation}
    \ket{\rho_1}_{B_1,N}\otimes \ket{\rho_2}_{B_2,N^3} \underset{N\to \infty}{\sim} \ket{\GHZ}_{B_1,N}\otimes \ket{\rho_2}_{B_2,N^3} \underset{N\to \infty}{\approx} \ket{\GHZ}_{B_1,N}\otimes \ket{\GHZ}_{B_2,N^3}\,.
\end{equation}
\end{ex}

\

In the following section, we will focus on specific examples of HT states, defined on $D$-partite state spaces with equally sized parties and equally sized fine-grained building blocks. We will see that the evaluation of a trace-invariant $G$ on such a state is fully encoded into the asymptotic scaling $s_G$ introduced in Sec.~\ref{sec:LU_random_asymptotic}. As a result, we will see that, for such states, the problem of state discrimination can be approached by purely combinatorial methods. More precisely, we will find out that it can be related in a precise manner to properties of certain graph-theoretic combinatorial quantities, some of which have already been extensively studied in the literature (in most cases, for different purposes). In addition, given that our combinatorial analysis will rely solely on the values of asymptotic scaling parameters $\{s_G\}_{G \in \cG_D^{\conn}}$, any statement regarding the exact state discrimination of a pair of HT states implies a statement regarding the asymptotic state discrimination of whole families of states (that are $\LU$-equivalent in scaling to one of those two HT states). For instance, coming back to Ex.~\ref{ex:asymptotic_eq}: if the scaling $s_G$ of a graph $G$ discriminates between $\ket{\GHZ}_{B,N}$ and $\ket{\GHZ}_{B,N^2}$, then the graph $G$ can also be used to discriminate the states $\ket{\rho_1}_{B,N}$ and $\ket{\rho_2}_{B,N^2}$. 

\section{Distinguishing power of trace-invariants for a subset of hypergraph-tensor states and Haar-random states}
\label{sec:LU-and-ref-states}

In this section, we investigate the ability of trace invariants to discriminate between a subset of HT states and Haar-random states (at leading order, in the latter case). Our goal is to characterize the classes of trace-invariants for which such distinctions are effective or, conversely, fail to hold.

\subsection{Some combinatorial quantities responsible for state discrimination} \label{ss:discrimination}

Let us come back to the fine-grained definition of HT states from the beginning of Sec.~\ref{sec:ref_states}. We have a $D$-partite state space $\H_1 \otimes \cdots \otimes \H_D$ (with $D\geq 2$), and each tensor factor $\H_c$ is itself divided into $I_c$ subsystems $\{\H_{(c,i)} \}_{1\leq i \leq I_c}$. From the remainder of the present section, we will make the following simplifying assumptions: a) that all $D$ local subsystems have the same dimension $N \geq 2$, and b) that they are subdivided into equally sized elementary building blocks. Calling $\cI$ the number of elementary building blocks in each local subsystem, we thus have: for any color $c \in \{ 1, \ldots , D\}$,
\begin{equation}
\label{eq:def-ref-states-for-combi}
     I_c = \cI\,, \quad \dim(\H_c) = N\,, \quad \mathrm{and}\qquad  \forall i \in \{ 1, \ldots, \cI\}\,, \quad \dim(\H_{(c,i)})= N^{1/\cI} \eqdef N_s\,. 
\end{equation}
We will fix $\cI$ and consider $N$ as a free parameter, conditioned on $N_s = N^{1/\cI}$ being an integer. In this setting, we now restrict our attention to a subclass of HT states that admit particularly simple combinatorial encodings, namely: we will only consider HT states $\ket{\Psi_{(\pi , w)}}$ where each block of the partition $\pi$ has constant weight $w= N_s$ (in other words, each $\GHZ$ building block of $\ket{\Psi_{(\pi , w)}}$ is assumed to be of maximal dimension). Since the weight of each block is fixed, we can use the short-hand $\ket{\Psi_\pi}\eqdef \ket{\Psi_{(\pi, w)}}$ to label such a state. Likewise, at the graphical level, we can always choose a representation of $\ket{\Psi_\pi}$ by a hypergraph whose hyper-edges all have weight $N^{1/ \cI}$; when such a choice is made, we can keep the weights implicit in our graphical representations of HT states (and we will). 

As detailed in Ex.~\ref{ex:stab_states}, for the particular case where $D=3$ and $N_s=p$ is a prime number, such $\ket{\Psi_\pi}$ are representatives of the $\LU$-equivalence classes of pure stabilizer states.\footnote{The requirement thay all $I_c$ be equal in Eq.~\eqref{eq:def-ref-states-for-combi} is not restrictive, as one can always complement  by some separable states to satisfy this requirement if the $I_c^S$ differ in Ex.~\ref{ex:stab_states}}

For any partition $\pi$ of $\{ c_i \,| \, 1\leq c \leq D\,, 1 \leq i \leq \cI\}$, we then have (as a special case of Eq.~\eqref{eq:tr_psi_pi})
\begin{align}\label{eq:general-scaling-ref-state-recall}
    \forall G \in \cG_D\,, \qquad \tr_G(\ket{\Psi_\pi}) &= \prod_{B \in \pi} N_\rm{s}^{\kappa(G\vert_{\red(B)}) - k(G)} = N^{s_G (\ket{\Psi_\pi}) } \,, \nonumber \\
  \mathrm{with}\quad  &s_G(\ket{\Psi_\pi}) \eqdef \frac 1 {\mathcal{I}}\sum_{B \in \pi} \pac{\kappa(G\vert_{\red(B)}) - k(G)} \,.
\end{align}
If one interprets $\ket{\Psi_\pi}$ as defining a sequence of deterministic states labeled by $N$,\footnote{For consistency, $N$ must take value in $\{ N \geq 2 \, | \, N^{1/\cI}\in \mathbb{N}\}$ rather than $\mathbb{N}^*$.} then the definition of $s_G(\ket{\Psi_\pi})$ in Eq.~\eqref{eq:general-scaling-ref-state-recall} coincides with the definition of asymptotic scaling introduced in Def.~\ref{def:LargeN_LU_eq}.\footnote{To be extra precise, Def.~\ref{def:LargeN_LU_eq} allows understanding the quantity $\cI\times s_G\left(\ket{\Psi_\pi}\right)$ as the asymptotic scaling of the sequence of states $\ket{\Psi_\pi}$ indexed by $N_s\in \mathbb{N}^*$. We introduced a uniform rescaling that slightly deviates from Def.~\ref{def:LargeN_LU_eq} in order to be able to use $1/N$ as our small parameter.} Moreover, the quantity $\mu_G$ introduced generically in Def.~\ref{def:LargeN_LU_eq} evaluates to $1$ for any graph $G$ and any such HT state:
\begin{equation}
    \forall G \in \cG_D\,, \qquad \mu_G(\ket{\Psi_\pi}) = 1\,. 
\end{equation}
Hence, for this subclass of HT states, the problem of $\LU$-orbit discrimination can be analyzed in terms of the asymptotic scalings $\{s_G (\ket{\Psi_{\pi}}) \}_{G \in \cG_D}$ of a state $\ket{\Psi_{\pi}}$ which, in view of Eq.~\eqref{eq:general-scaling-ref-state-recall}, are rational coefficients of combinatorial origin. This allows for a combinatorial analysis of the problem, which is the purpose of the remainder of this section. 

We will now introduce specific notations for a number of HT states that we will focus our attention on in the following. The exact $\LU$ characterization results that will follow in the rest of the section can be straightforwardly translated into weaker \emph{asymptotic} $\LU$ characterization results \emph{in scaling} for a sequence of states that are \emph{asymptotically equivalent in scaling} to those HT states (see Sec.~\ref{sec:LU_random_asymptotic}).

\begin{figure}[!ht]
    \centering
    \includegraphics[width = \textwidth]{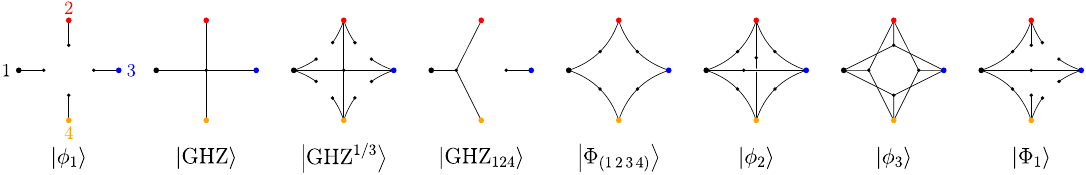}
    \caption{Visual representation of the selected HT states presented in this section by the use of hypergraphs. In this picture, each hyper-edge has (implicit) weight $N_s = N^{1/\cI}$.}
    \label{fig:SelectedRef}
\end{figure}

\paragraph{The GHZ state.}
In Sec.~\ref{sub:intro-of-ref-states}, Eq.~\eqref{eq:GHZ}, we expressed the value of a connected trace invariant evaluated for the GHZ state. It is the second graph from the left in Fig.~\ref{fig:SelectedRef}.
The scaling $s_G$ for a graph $G\in \cG_D$ is given by
\begin{equation}
    s_G(\ket{\GHZ}) = \kappa(G) - k(G) \le 0\,.
\end{equation}
It  vanishes if and only if $G$ is a collection of colored graphs with only one white vertex per connected component ($G$ is the square norm of the state to the power $k$, which is one). More generally, since the scaling of HT states of Eq.~\eqref{eq:general-scaling-ref-state-recall} is a sum of non-negative terms, $G$ allows discrimination with the separable HT state $\ket{\phi_1}$  (left of Fig.~\ref{fig:SelectedRef}) if at least one of the terms is strictly negative.

\paragraph{Lower dimensional GHZ state.}Fixing $\cal{I}$, we call dimension $N^{\frac 1 {\cal{I}}}$ and denote by $\lvert\GHZ^{\frac{1}{\cal{I}}}\rangle$ a $\GHZ$ state shared by $D$ subsystems of dimension $N^{\frac 1 {\cal{I}}}$, supplemented by separable HT states. It is an HT state whose partition $\pi$ induces the multiset of colors $\{1,\ldots, D\} \bigcup_{c=1}^D\{c\}^{\cal{I} - 1}$, meaning that the singlet $\{c\}$ is repeated $\cal{I}-1$ times (Fig.~\ref{fig:SelectedRef}). One has:
\begin{equation}
\label{eq:lower-dim-ghz}    s_G(\lvert\GHZ^{\frac{1}{\cal{I}}}\rangle) = 
     s_G(\ket{\GHZ}) / \cal{I}, \hspace{1.5cm} s_G(\ket{\GHZ}) - s_G(\lvert\GHZ^{\frac{1}{\cal{I}}}\rangle) = \frac{\cal{I} - 1}{\cal{I}}  s_G(\ket{\GHZ})  \,. 
     \end{equation}
Therefore, the conditions that a graph $G$ should satisfy to discriminate between two states among the $\GHZ$ state,  a lower-dimensional $\GHZ$, or a separable state are the same: any graph with $\kappa>k$ works.

\paragraph{GHZ state shared by less parties.}  For $B\subset\{1,\ldots,D\}$, $B=\{c_1,\ldots,c_p\}$, we will use the notation $\ket{\GHZ\vert_{B}}$ for a GHZ state shared by the $p$ subsytems $\H_{c_1}$,\dots, $\H_{c_p}$ and supplemented by separable states $\sum_s\ket{s}$ (appropriately normalized) on the remaining $D-p$ subsystems (fourth from the left in Fig.~\ref{fig:SelectedRef}). 
Its scalings are given by
\begin{equation}
    s_G(\ket{\GHZ\vert_B}) = \kappa(G\vert_B) - k(G)\le 0\,,
\end{equation}
and it vanishes exactly for the graphs for which the edges of color $c_1, \ldots, c_p$ incident to a given white vertex are all incident to the same black vertex.\footnote{See the discussion regarding $\xi$-separability in Sec.~\ref{sec:generalities-on-coarse-graining}.} Any graph that does not satisfy this property  allows discriminating $\ket{\GHZ\vert_B}$ from the HT separable states. The particular case where $p=2$ corresponds to a Bell pair on two subsystems as presented Sec.~\ref{sec:intro}. In that case,  $G\vert_{c_1,c_2}$ consists of the cycles of $G$ alternating the colors $c_1, c_2$, and $\kappa(G\vert_{c_1,c_2})=F_{c_1 c_2}(G)$ (defined in Eq.~\eqref{eq:faces}). 

The scalings of $\ket{\GHZ\vert_B}$ compare to those of a GHZ state shared by all parties as:
\begin{equation}
\label{eq:def-KB}
    s_G(\ket{\GHZ\vert_B}) - s_G(\ket{\GHZ}) = K_B(G),\hspace{1cm}K_B(G)=\kappa(G\vert_B) - \kappa(G) \geq 0\,.
\end{equation}
The non-negativity holds because the number of connected components  of $G$ can only increase when removing some edges.  The invariant $G$ can distinguish $\ket{\GHZ}$ from $\ket{\GHZ\vert_B}$ iff the number of connected components strictly increases when removing from $G$ all the edges whose colors are not in the set $B=\{c_1,\dots,c_p\}$.

In summary, any connected $G$ satisfying the condition
\begin{equation}
    1<  \kappa(G\vert_B) < k(G)\,,
\end{equation}
allows discriminating $\ket{\GHZ\vert_B}$ from both the separable HT state and a $\GHZ$ state shared by all parties.

\paragraph{Cyclic states.} We call cyclic quantum states a family of states labeled by permutations $\tau \in S_D$, each consisting of a single cycle on $D$ elements (also known as \textit{iterated matrix multiplication tensors}, see \eg Refs.~\cite{Buhrman:2016tif,Christandl:2018cfb,Christandl:2019zrq}). They correspond to HT states $\ket{\Phi_\tau} \eqdef \ket{\Psi_{\pi_\tau}}$ for which each $\H_c$ is subdivided in two, and $\pi_\tau$ induces the multiset of colors  $\paa{\paa{i,\tau(i)} \, \vert \, 1\leq i \leq D}$ (fourth from the right in Fig.~\ref{fig:SelectedRef}). Note that two cyclic permutations $\tau$ and $\tau^{-1}$ lead to the same state. 
The scaling of the trace-invariant associated with $G\in \cG_D^{\conn}$ evaluated for a cyclic state is
\begin{equation}
    s_G(\ket{\Phi_\tau}) = \frac 1 2 \sum_{i=1}^D \left[F_{i \tau(i)}(G) - k(G)\right]\,,
\end{equation}
where the $F_{i\tau(i)}$ are defined in \eqref{eq:faces}, and correspond to the number of connected components of the complement of the graph $G_\tau$ on the surface on which it is embedded (the jacket (regular embedding) $G_\tau$ has been introduced in Sec.~\ref{sss:planar}).  The $F_{i\tau(i)}$ are related to the genus of $G_\tau$ by the Euler characteristic formula
\begin{equation}
\label{eq:Euler-jacket}
(2-D)k(G) + \sum_{i=1}^D F_{i \tau(i)}(G) = 2 - 2g_\tau(G)
\end{equation}
Comparing the scalings for cyclic states and the GHZ state on all parties, one therefore gets:
\begin{equation} \label{eq:GHZ_phi_tau}
    s_G(\ket{\Phi_\tau}) - s_G(\ket{\GHZ}) = - g_\tau(G) \leq 0 \,,
\end{equation}
so that a graph allows discriminating between these two states if and only if the corresponding jacket is non-planar.

\paragraph{The $2$-complete state $\ket{\phi_2}$.} For the states $\ket{\phi_2}$ shown in Fig.~\ref{fig:shareBP} and Fig.~\ref{fig:SelectedRef}, each $\H_c$ is subdivided  in $\mathcal{I}=D-1$ subspaces, and $\pi$ are partitions containing all possible pairs of colors $(c,c')$ with $c\neq c'$ (see for instance Ref.~\cite{Christandl:2019zrq}). For a given $D$, the state is represented as a complete graph on $D$ vertices: we will refer to it as the \emph{$2$-complete state}. Its scalings are given by
\begin{equation} 
\label{eq:scaling-of-2complete}
    s_G(\ket{\phi_2}) = 
    \frac 1 {D-1} F(G) - \frac{D}{2} k(G)  \,,
\end{equation}
where $F$ has been defined in  \eqref{eq:faces}. 
Its scalings compare to those of the GHZ state as
\begin{equation} 
\label{eq:GHZ_phi2}
        s_G(\ket{\phi_2}) - s_G(\ket{\GHZ}) = - \frac{1}{D-1} \omega_2(G) \leq 0 \,,
\end{equation}
where $\omega_2(G)$ is a non-negative integer defined for $G\in\cG_D$ by 
\begin{equation} \label{eq:GurauDeg}
    \omega_2(G) \eqdef (D-1)\kappa(G) + \frac{(D-1)(D-2)}{2}k(G) - F(G)\,.
\end{equation}
The trace-invariant $\tr_G$ therefore allows distinguishing the GHZ state  from the state $\ket{\phi_2}$ if and only if $\omega_2(G) > 0$.
For $D=3$, $\ket{\phi_2}$ coincides with $\ket{\Phi_\tau}$ for both $\tau=(123)$ or $\tau^{-1}=(132)$. 
For both choices, Eq.~\eqref{eq:GurauDeg} reduces to Eq.~\eqref{eq:Euler-jacket}, so that:
\begin{equation}
\label{eq:degree-is-genus-for-D=3}
\forall G\in \cG_2^{\conn}, \qquad \omega_2(G) = 2g_\tau(G)\,.
\end{equation}
For $D\ge 4$, the quantity  $\omega_2$ is known as the \textit{Gurau degree} (see Ref.~\cite{gurau_random_2017} and references therein).

\begin{rem}
    The following  relation holds between the state $\ket{\phi_2}$ and the cyclic states introduced above:\footnote{On the right, each state $\phi_\tau$ is therefore obtained twice, once for $\tau$ and once for $\tau^{-1}$.}
    \begin{equation}
        \ket{\phi_2}^{\ot (D-1)!} = \bigotimes_{\tau \rm{ cyclic}} \ket{\Phi_\tau} \,.
    \end{equation}
 This relation between quantum states allows recovering - at the scaling level -  a well-known  identity (Ref.~\cite{gurau_random_2017}) between the degree $\omega_2$ and the genera of the jackets:
    \begin{equation} \label{eq:GurauDegGenus}
        \omega_2(G) = \frac{1}{(D-2)!} \sum_{\tau \rm{ cyclic}} g_\tau(G) \,.
    \end{equation}
\end{rem}

\paragraph{The $p$-complete state $\ket{\phi_p}$.}For $1 \leq p \leq D$, we more generally define the states $\ket{\phi_p}$ that share maximally GHZ states with $p$ parts (see \eg Ref.~\cite{Vrana:2016edr}): each $\H_c$ is subdivided in $\mathcal{I}=\frac p D \binom{D}{p} $ subspaces, and the partitions  $\pi$ contain all possible $p$-tuples of different colors ($\ket{\phi_3}$ is represented in Fig.~\ref{fig:SelectedRef}). Concretely, we have 
\begin{equation} \label{eq:phi_p}
    \ket{\phi_p} \eqdef \bigotimes_{1 \leq c_1 < \cdots < c_p \leq D} \ket{\GHZ }_{c_1 \dots c_p} \,,
\end{equation}
where we recover the definition of $\ket{\phi_2}$ when $p=2$. Note that the state $\ket{\phi_D}$ corresponds to the state $\ket{\GHZ}$, while $\ket{\phi_1}$ is separable. The state $\ket{\phi_p}$ is represented  graphically  by a $p$-complete hypergraph (see the representation of $\ket{\phi_3}$ in Fig.~\ref{fig:phi_p}), which explains the name $p$-complete state. 
For $1 \leq p \leq D$,  its scalings are given by 
\begin{equation}
    s_G(\ket{\phi_p})  =  \frac 1 {\cal{I}_p} \kappa^{(p)}(G) - \frac D p k(G), \qquad  \cal{I}_p = \binom{D-1}{p-1}\,,
    \end{equation}
where  $\kappa^{(p)}$ is the following generalization of $F=\kappa^{(2)}$:
\begin{equation}
\label{eq:def-of-kappa-p}
\kappa^{(p)}(G) \eqdef \sum_{1 \leq c_1 < \cdots < c_p \leq D} \kappa(G\vert_{{c_1,\dots,c_p}})\,.
    \end{equation}

We introduce a generalization of the Gurau degree, the $p$\textit{-complete degree} $\omega_p(G)$, defined for $1 \leq p \leq D-1$ by
\begin{equation} 
\label{eq:GenGurauDeg}
    \omega_p(G) \eqdef 
 \cal{I}_p\kappa(G) + 
    \cal{I}_{p+1} k(G) - \kappa^{(p)}(G)\,.
\end{equation}
We may also set $\omega_D = 0$ identically. In App.~\ref{A:GurauDeg}, we show that for $2 \leq p \leq D-1$, $\omega_p$ are non-negative integers. These quantities play a role analogous to that of $\omega_2$ for $\ket{\phi_2}$, as for any $2 \le p \le D$:
\begin{equation} \label{eq:rem_omega_p}
    s_G(\ket{\phi_p}) - s_G(\ket{\GHZ}) = - \frac{1}{\cal{I}_p} \omega_p(G)\leq 0 \,.
\end{equation}

Comparing the scalings for two complete states $\ket{\phi_p}$ and $\ket{\phi_q}$, with $2 \leq p \leq q \leq D$, leads to another non-negative quantity:
\begin{equation}
    \label{eq:zeta_pq}
     s_G(\ket{\phi_q}) - s_G(\ket{\phi_p}) = \frac 1 {\cal{I}_p} \omega_p(G) - \frac 1 {\cal{I}_q} \omega_q(G) = 
     \frac{1}{\cal{I}_p \binom{D-p}{q-p}}
     \omega_p^{(q)}(G)\ge 0\,,
\end{equation} 
so that $\omega_p^{(D)}= \omega_p$ and $\omega_p^{(p)}$ is zero. One verifies that for  $2\le p \le q\le D$, this is equivalent to the definition
\begin{equation}
\label{eq:def-omega-pq}
    \omega_p^{(q)}(G) \eqdef \binom{q-1}{p-1} \kappa^{(q)}(G) + \binom{q-1}{p} \binom{D}{q} k(G) - \binom{D-p}{q-p} \kappa^{(p)}(G) \,,
\end{equation}
along with $\omega_p^{(p)} = 0$ for $2 \leq p \leq D$. To show that this quantity is non-negative, we prove in Lem.~\ref{lem:omega_p,q} of App.~\ref{A:GurauDeg} that
    \begin{equation}
        \label{eq:toprove-omega-pq}
        \omega_p^{(q)}(G) = \sum_{c_1< \cdots < c_q} \omega_p(G\vert_{c_1,\dots,c_q}) \,.
    \end{equation}

\paragraph{$c$-Star states.} We now consider the HT states $\ket{\Phi_c}$ for which one of the subsystems $\H_c$ shares Bell pairs with all the other subsystems. Each $\H_{c'}$ is subdivided in $D-1$, and $\pi$ induces the multiset of colors $\bigcup_{c'\neq c} \{c,c'\}\cup\{c'\}^{D-2}$, meaning that the singlet $\{c'\}$ is repeated $D-2$ times (rightmost state in  Fig.~\ref{fig:SelectedRef}).
Its scalings are given by
\begin{equation} 
\label{eq:scaling-of-2-star}
    s_G(\ket{\Phi_c}) = \frac 1 {D-1} F_c(G) - k(G)  \,,\qquad F_c(G)=\sum_{c'\neq c} F_{cc'}(G)\,.
\end{equation}
They compare to those of a $\GHZ$ state shared by all parties as
\begin{equation}
\label{eq:phic-vs-cstar}
    s_G(\ket{\Phi_c})  - s_G(\ket{\GHZ}) = \frac{1}{D-1} F_c(G) - \kappa(G) = \frac 1 {D-1} \sum_{i\neq c} K_{ic}(G)\ge 0\,,
\end{equation}
since for each $c'\neq c$, $F_{cc'}\ge \kappa$, 
and they compare to those of a  GHZ state of local dimension $N^{\frac{1}{D-1}}$   as 
\begin{equation}
    s_G(\ket{\Phi_c})  - \frac{1}{D-1} s_G(\ket{\GHZ}) = - \frac{1}{D-1} \Omega_c(G) \leq 0
\end{equation}
where the $c$-degree $\Omega_c$ is defined as: 
\begin{equation} \label{eq:c-Degree}
    \Omega_c(G) \eqdef \kappa(G) + (D-2)k(G) - F_c(G)\,.
\end{equation}
This quantity is known to be a non-negative integer, see \eg Refs.~\cite{Gurau2019,FUSY2020103066,Collins2023}.

More generally, considering $B\subset\{1,\ldots, D\}$ and letting $\lvert\Phi_c^{1/r}\rangle$ be the $c$-star state but for a local dimension $N^{\frac 1 r}$ instead of $N$ (each subsystem is still subdivided in $D-1$ parts), we define  $\ket{\Phi_B} = \bigotimes_{c\in B}\lvert\Phi_c^{1/ {\lvert B \rvert}}\rangle$.
Its scalings compare to those of $\GHZ$ as 
\begin{equation}
    s_G(\ket{\Phi_B}) - s_G(\ket{\GHZ}) =  + \frac{1}{\lvert B \rvert}  \sum_{c\in B} \left(\frac {F_c(G)} {D-1} - \kappa(G)\right)\,,
\end{equation}
and to those of a GHZ state of local dimension $N^{\frac{1}{D-1}}$ as 
\begin{equation}
\label{eq:B-degree}
    s_G(\ket{\Phi_B}) -\frac 1 {D-1} s_G(\ket{\GHZ}) =  - \frac{1}{\lvert B \rvert (D-1)} \Omega_B(G) , \hspace{1.2cm} \Omega_B\eqdef \sum_{c\in B} \Omega_c\,.
\end{equation},

\paragraph{Haar-random state.} The Haar-random state $\ket{\varphi}$ defined in Sec.~\ref{sub:intro-of-ref-states}, Eq.~\eqref{eq:def-of-Haar-state}. The scaling of $\ket{\varphi}$ for a graph $G\in \cG_D$ is now given by the term that dominates the sum in Eq.~\eqref{eq:meanLUWeing} (see Def.~\ref{def:LargeN_LU_eq}):
$$ 
 \mean{\tr_G(\ket{\varphi})} \underset{N \to \infty}{=} \mu_G(\ket{\varphi}) N^{s_G(\ket{\varphi})} +  o\bigl(N^{s_G(\ket{\varphi})}\bigr), 
$$
where using the notations introduced in Sec.~\ref{subsub:Def-of-Haar}, one has for $\vec \sigma\in S_{k(G)}^D(G)$:
\begin{equation} \label{eq:sG-Haar}
    s_G(\ket{\varphi}) = -\min_{\nu\in S_{k(G)}}\sum_{c=1}^D d(\sigma_c,\nu)  = \max_{\widehat G\in \cG_{D+1}(\widehat G)}F_{0}(\widehat G) - D k(G)\,,
\end{equation}
and the factor $\mu_G$ counts the number of extremizers: fixing $\vec\sigma \in S_{k(G)}^D(G)$, we have
\begin{equation} 
    \mu_G(\ket{\varphi}) = \abs{\paa{\nu \in S_{k(G)} \, \rm{ s.t. } \,\sum_{c = 1}^D d(\sigma_c,\nu) = - s_G(\ket{\varphi}) }}\,.
\end{equation}
Equivalently, one can introduce the set \begin{equation}
    \label{eq:MD}
    \MD{}(G) \eqdef \paa{\widehat G \in\cG_{D+1}(G) \;\; \rm{s.t.} \;\; F_{0}(\widehat G) = D k(G) + s_G(\varphi)} \,, 
\end{equation}
to formulate the combinatorial factor without permutations as
\begin{equation} \label{eq:mu_vs_chi}
    \mu_G(\ket{\varphi}) = \sum_{\widehat G \in \MD{}(G)} \Xi(\widehat G) \,,
\end{equation} 
where the degeneracy $\Xi$ was introduced in Eq.~\eqref{eq:meanLUWeing2}.

Comparing the scalings of the Haar-random state with those of $\ket{\phi_2}$ in Eq.~\eqref{eq:scaling-of-2complete}:
\begin{equation}
    s_G(\ket{\varphi}) - s_G(\ket{\phi_2}) = - \frac{2}{D-1} \Delta(G)  \leq 0 \,,
\end{equation}
where the non-negative quantity $\Delta(G)$ is called  \textit{degree of compatibility} (see Ref.~\cite{Collins2025}), defined in terms of the degrees introduced above as:
\begin{equation}
\label{eq:Delta}
  \Delta(G)\eqdef  \min_{\widehat G\in \cG_{D+1}(G)}\Delta_0(\widehat G) \,,  
  \end{equation}
  where:
\begin{equation} \label{eq:Delta_0}
 \Delta_0(\widehat G)\ \eqdef  \  \frac {D-1} 2 \Omega_{0}(\widehat G) -  \frac 1 2 \omega_2(G)\ = \ \frac{D(D-1)}4 k(G) + \frac 1 2 F(G) - \frac {D-1} 2 F_{0}(\widehat G)\,.  
\end{equation}

 A graph $G\in \cG_D$ with $\Delta(G)>0$ allows discriminating $\ket{\phi_2}$ and $\ket \phi$ at the scaling level (the first order in $N$ for the logarithm of the trace-invariant). 
 
 It is useful to reformulate $\Delta$ in terms of permutations. For any $\sigma,\tau,\nu \in S_{k(G)}$, we introduce the following notation for the Gromov product of $\sigma$ and $\tau$ at $\nu$: 
      \begin{equation}
            \GP{\sigma}{\tau}{\nu} \eqdef \frac{1}{2}\pac{d(\sigma,\nu) + d(\nu,\tau) - d(\sigma,\tau)} \,.
        \end{equation}
        
Letting for $(\vec \sigma, \nu)\in S_k^{D+1}$: 
\begin{equation}
\label{eq:Delta-nu-def}
\Delta_\nu(\vec \sigma) \eqdef \sum_{i<j} \GP{\sigma_i}{\sigma_j}{\nu} \,,
\end{equation}
one  has the following identifications for any $G\in\cG_D$, $\widehat G\in \cG_{D+1}(G)$, and $(\vec \sigma, \nu)\in S_{k(G)}^{D+1}(\widehat G)$:
\begin{equation}
 \Delta_\nu(\vec \sigma) = \Delta_0(\widehat G) \,,
 \end{equation}
 and with the same notations, 
     \begin{equation}
\label{eq:Gromov-vs-0-degree-vs-genus}
    \GP{\sigma_i}{\sigma_j}{\nu} \ =\  \frac 1 2 K_{ij}(\widehat G\vert_{0ij}) + \frac 1 2 \Omega_{0}(\widehat G\vert_{0ij})\  =\  K_{ij}(\widehat G\vert_{0ij}) + g( \widehat G\vert_{0ij}) \,.
\end{equation}   
This last equation justifies that $\Delta\in\mathbb{N}$. 
One may also express $\Delta$ for any $\vec \sigma\in S_{k(G)}^D(G)$ as
\begin{equation}
\label{eq:Delta-perm}
        \Delta(G) = \min_{\nu \in S_{k(G)}} \Delta_\nu(\vec \sigma)\,.
\end{equation}

\paragraph{Summary.} The tables below summarize the difference between the scaling for $G\in\cG_D$ of the top state and that of the bottom state (above the condition). For $G$ to discriminate between the two states, the scalings must differ.  
In the following, $p,q\in\{2,\ldots, D-1\}$ with $p < q$, $c, c'\in\{1,\ldots, D\}$, $B,B'\subset \{1,\ldots,D\}$, and  $\tau,\tau'$ are two cyclic permutations of $D$ elements. See the definitions of $K_B$ in Eq.~\eqref{eq:def-KB}, $g_\tau$ in Eq.~\eqref{eq:Euler-jacket}, $\omega_2$ in Eq.~\eqref{eq:GurauDeg}, $\omega_p$ in 
Eq.~\eqref{eq:GenGurauDeg}, $\Omega_c$ in Eq.~\eqref{eq:c-Degree} and $\Omega_B$ in Eq.~\eqref{eq:B-degree}, $\omega_p^{(q)}$ in Eq.~\eqref{eq:def-omega-pq}, and $\Delta$ in Eq.~\eqref{eq:Delta}. All these quantities are non-negative.

\begin{table}[H]
    \begin{center}
        \begin{tabular}{ |c|c|c|c|c|c| } \hline
            \multicolumn{6}{|c|}{$\ket{\GHZ}$} \\ \hline
             $\ket{\phi_1}$ & $\lvert\GHZ^{\frac{1}{\cal{I}}} \rangle$ & $\ket{\GHZ\vert_{B}}$ & $\ket{\Phi_B}$ & $\ket{\Phi_\tau}$ &  $\ket{\phi_p}$   \\ \hline
             $\kappa - k \le 0$ &  $\frac{\cal{I} - 1}{\cal{I}} (\kappa - k) \le 0$ & $- K_B \le 0$ &
             $\frac{1}{\lvert B \rvert } \sum_{c\in B}[\frac{1}{D-1}F_c - \kappa] \ge 0$ & $g_\tau \ge 0$ & $\frac{1}{\cal{I}_p}\omega_p \ge 0$ \\ \hline
        \end{tabular}
        \label{table:Distinction vs GHZ}

\vspace{0.5cm}

        \begin{tabular}{ |c| } \hline
            \multicolumn{1}{|c|}{$\ket{\Phi_B}$} \\ \hline
              $\lvert\GHZ^{\frac{1}{D-1}}\rangle$  \\ \hline
              $-\frac{1}{\lvert B \rvert (D-1)} 
              \Omega_{B} \le 0$   \\ \hline
        \end{tabular} 
        \begin{tabular}{ |c| } \hline
            \multicolumn{1}{|c|}{$\ket{\phi_2}$} \\ \hline
             $\ket{\varphi}$ \\ \hline
             $\frac{2}{D-1} \Delta \ge 0$ \\ \hline
        \end{tabular}
        \label{table:Distinction vs phi_p}
               \begin{tabular}{ |c| } \hline
            \multicolumn{1}{|c|}{$\ket{\phi_p}$} \\ \hline
             $\ket{\phi_q}$\\ \hline
             $\cal{I}_p^{-1} \binom{D-p}{q-p}^{-1} \omega_{p}^{(q)} \ge 0$ \\ \hline
        \end{tabular}
    \end{center}
\end{table}
From the tables, one can compute the difference in scaling between the selected HT states considered in this section. For instance:
\begin{equation}
    s_G(\ket{\GHZ\vert_B}) - s_G(\ket{\Phi_\tau}) = \pac{s_G(\ket{\GHZ\vert_B}) - s_G(\ket{\GHZ})} + \pac{s_G(\ket{\GHZ}) - s_G(\ket{\Phi_\tau})} = K_B(G) + g_\tau(G) \ge 0\,.
\end{equation}
If the terms in the sum have the same sign, it results again in a positivity condition: $G$ distinguishes $\ket{\GHZ\vert_B}$ and  $\ket{\Phi_\tau}$ if and only if either $K_B(G)>0$ or $g_\tau(G)>0$. This is not always the case: for instance, when comparing the states $\ket{\GHZ\vert_B}$ and $\ket{\GHZ\vert_{B'}}$, one may find graphs with  $K_B(G)>K_{B'}(G)$, and conversely. This will have consequences in Sec.~\ref{s:LO_LOCC_invTr}.

\begin{rem}
    Some of these quantities are well-known in the literature on random matrices and tensors. The genus of 3-colored graphs is the quantity involved in the $1/N$ expansion of two-matrix models, such as those describing the Ising model on random surfaces (see \eg Refs.~\cite{kazakov1986ising, DIFRANCESCO1998543,  bernardi2011counting}), or three-matrix models, such as in Ref.~\cite{ gurau_random_2017}. The Gurau degree $\omega_2$ or the degree of compatibility $\Delta$ play the same role regarding the $1/N$ expansion of random tensor models, see Refs.~\cite{Gurau2011_1, Gurau2011_2, Gurau:2011aq, Bonzom:2012hw, gurau_random_2017}. The $c$-degree $\Omega_c$ governs the $1/N$ expansion of the SYK model, see Ref.~\cite{10.1063/1.4983562}. 
    
    In the same way as for the genus, most of the quantities introduced here have a geometric interpretation: $D$-colored graphs are dual to certain $(D-1)$-dimensional triangulations, and the restrictions of the graphs to different subsets of colors count the number of lower-dimensional simplices of different kinds. The jackets are related to so-called Heegaard splittings, and so on. See Refs.~\cite{Ryan:2011qm, Casali:2017tfh} for more details. 

   The $\kappa^{(p)}$, introduced in Eq.~\eqref{eq:def-of-kappa-p}, then count the number $f_{D-p-1}$ of $(D-p-1)$-dimensional simplices of the triangulation (and $f_{-1}$ is the number of connected components). Recall that $x_C(\alpha,\beta)$ was introduced in Eq.~\eqref{eq:Syst_Sigma}. Setting $x_C(\alpha,\beta)=x_p$ for all $C\subset\{1,\ldots, D\}$ satisfying $\abs{C}=p >1$, $N \in \bb{N}^*$, and introducing  $x_0,x_1\in\mathbb{R}$, one can rewrite the scaling difference as 
\begin{align} \label{eq:symmetricCond}
   \pa{s_G(\ket{\psi_\beta}) - s_G(\ket{\psi_{\alpha}})} \cdot \ln N  &= \sum_{p=2}^D \pa{\binom{D}{p} k(G) - \kappa^{(p)}(G)} x_p \,,  \\
   &= k(G)\sum_{p = 0}^{D} \binom{D}{p} x_p   -  \sum_{p = 0}^{D} \kappa^{(p)}(G)x_p + k(G) x_0\,.
\end{align}
For different choices of the $x_p$, the rightmost sum coincides, up to $+ k(G) x_0$, with different geometrical invariants of $(D-1)$-dimensional triangulations. For instance, for the choice $x_p=(-1)^{D-p-1}$, one recovers  $ \sum_{p = 0}^{D} \kappa^{(p)}x_p = \kappa + \chi$, where $\chi = \sum_{i = 0}^{D-1} (-1)^{i}f_i$ is  the Euler characteristics of the triangulation, which  vanishes for odd-dimensional manifolds. Fixing $R\in\{0,\ldots, D\}$ and choosing $x_p=0$ for $p \le D-R-1 $ and $x_p= (-1)^{R-D+p} \binom{p}{R-D+p}$ for $p \in\{D-R, \ldots, D\}$, one obtains $\sum_{p = 0}^{D} \kappa^{(p)}x_p = h_R$, the $R$-th component of the $h$-vector of the triangulation see \eg Refs.~\cite{McMullen1971,billera1980sufficiency,Lutz_Sulanke_Swartz_2009}.
\end{rem}

\paragraph{Producing graphs satisfying these conditions.}

Thm.~\ref{th:RefState_Equiv} provides a set of trace-invariants whose knowledge allows determining the  $\LU$-equivalence class of an HT state. As discussed at the end of Sec.~\ref{sub:LU-inequivalent-reference-states}, for fixed $D$ and for some fixed sub-family of $\LU$-inequivalent HT states, one may wonder whether there exists a simpler set of trace-invariants that indeed take a different set of values for these states. 
Considering for instance the HT states $\ket{\GHZ}$, $\ket{\GHZ\vert_B}$, $\ket{\phi_p}$, $\ket{\Phi_\tau}$ and $\ket{\varphi}$ introduced in this subsection, the question is whether one can produce a simpler set of graphs $G$ satisfying all the conditions listed in the tables above.

\begin{figure}[ht]
    \centering
    \includegraphics[height = 4cm]{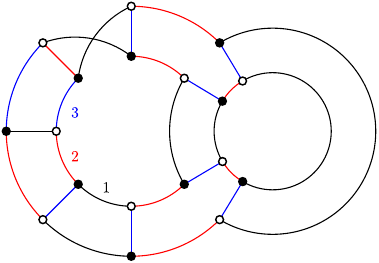}
    \caption{Example of a $3$-colored graph satisfying all the distinction conditions summarized in the tables.}
    \label{fig:ExDistinction}
\end{figure}

In the tripartite case, Fig.~\ref{fig:ExDistinction} shows such a graph\footnote{Note that the first graph fulfilling all these conditions appears at $k(G) = 9$. This is not surprising, since requiring faces of different sizes together with a non-vanishing genus forces $k(G)\geq 9$.} $G\in\cG_3$. Choosing the same labels for the black and white vertices linked by an edge of color 1, the permutations describing this example are
\begin{equation} 
    \sigma_1 = \id \qquad \sigma_2 = (1\, 2\, 3\, 4\, 5\, 6\, 7) (8 \,9) \qquad \sigma_3 = (1 \, 4) (2 \,5\, 9) (3 \,5) (6\, 8) \,.
\end{equation}
Recalling the definition of $\mu_G$ in Eq.~\eqref{eq:MD}, the resulting graph has the following properties: 
\begin{equation}
    k(G) = 9 \,,\quad F_{12}(G) = 2\,,\quad F_{13}(G) = 4\,,\quad F_{23}(G) = 3\,,\quad g(G) = 1 \,,\quad \Delta(G) = 2\,,\quad \mu_G(\ket{\varphi}) = 3 \,.\footnote{For the example, the degeneracy is given by the permutations $\nu_1 = (1 \,3)(2)(4)(5 \,6)(7)(8)(9)$, $\nu_2 = (1\, 3) (2) (4\, 7) (5 \,6) (8\, 9)$ and $\nu_3 = (1 \, 3)(2) (4\, 8) (5\, 6) (7 \,9)$.}
\end{equation}

Already at $D=4$, numerically computing an example of a 4-colored graph satisfying all distinction conditions is out of reach. Indeed, the requirement that all combinatorial quantities be well-defined, together with the condition that the faces and connected components of subgraphs are pairwise disjoint, implies $k(G) \geq 21$. Moreover, excluding the conditions on the Haar-random states, there are $70$ constraints to fulfill. In practice, an exhaustive numerical search would involve verifying more than $2.07 \times 10^{42}$ quadruples of permutations. 
Furthermore, computing $\Delta$ (Eq.~\eqref{eq:Delta-perm}) requires  computing $\Delta_\nu(G)$ for all $\nu$, which entails evaluating approximately $5.11 \times 10^{19}$ possibilities.

For a finite set of $\LU$-inequivalent HT states, we expect that a $D$-colored graph with $k$ white vertices chosen uniformly at random will satisfy the required distinction conditions with high probability when $k$ is sufficiently large. It would be interesting to rigorously prove such expectation.

\subsection{Graph structure of distinguishing invariants} \label{subsec:GraphStructure}

The conditions listed in Sec.~\ref{ss:discrimination} for discriminating  the specific examples of  $\LU$-inequivalent HT states introduced in that section involve several non-negative combinatorial quantities. For a given graph, one can compute these quantities to verify whether the conditions are satisfied, but it is harder to produce a graph satisfying this or that condition, as discussed at the end of Sec.~\ref{ss:discrimination}. 

For some examples of HT states, one actually knows the structure of graphs for which the corresponding conditions are  satisfied or not, and in some cases, one knows the structure of graphs for which the corresponding combinatorial quantities take a given specific value. The present subsection aims to review these situations. 

\ 

As a first example, it is trivial to classify the graphs according to the value taken by $s_G(\ket{\GHZ})=\kappa(G)-k(G)$, which is the non-positive quantity responsible for discrimination between $\ket{\GHZ}$ and a separable state. 

\paragraph{Graphs of fixed $K_B$.} A second trivial example is given by $K_B(G)=\kappa(G\vert_B) - \kappa(G)$, where $B\subset\{1,\ldots,D\}$: the graphs of vanishing $K_B$ - which do not distinguish between  $\ket{\GHZ\vert_B}$ and $\ket{\GHZ}$ on all parties -  are again collections of $2$-vertex graphs. Graphs of positive $K_B$ are relevant for the discrimination of a number of the HT states considered in  Sec.~\ref{ss:discrimination}. To construct all the graphs with a given fixed value of $K_B\in [0,k-1]$, one considers the $\abs{B}$-colored graphs whose edges are colored according to the edges in $B$ and with $\kappa(G\vert_B)$ connected components, and then connect them using edges of the remaining colors to reduce the number of connected components to $\kappa$. 

A graph $G$ for which  $\kappa(G\vert_B)=\kappa(G\vert_{B'})$ (that is, $K_B=K_{B'}$) is unable to distinguish the states $\ket{\GHZ\vert_B}$ and $\ket{\GHZ\vert_{B'}}$: distinguishing $\GHZ$ states shared by different parties requires non-symmetric graphs.

\paragraph{Fine-grained states and fixed scaling.}
Another simple example is given by $s_G(\ket{\GHZ\vert_B})= \kappa(G\vert_B)- k(G)$, the non-positive quantity responsible for discrimination between $\ket{\GHZ\vert_B}$ and a separable state. The graphs $G$ for which it takes a specific value $x\le0$ are those for which $G\vert_B$ has $k+x$ connected components, while the edges of $G$ whose colors are not in $B$ may be in any positions.   More generally, we may consider the ``fine-grained states'' for which $\cal{I}=1$ and $\pi$ is directly a partition of $\{1,\ldots, D\}$ (the subsystems $\H_c$ are not subdivided). Then, the relevant combinatorial quantity for discrimination with a separable state is the scaling 
$$
\sum_{B\in \pi} s_G(\GHZ\vert_B) \le 0
$$ 
itself  (Eq.~\eqref{eq:general-scaling-ref-state-recall}), but where a color $c$ appears in only one  block of $\pi$: the blocks of $\pi$ are ``independent''. This renders the task of constructing graphs of fixed $s_G$ easy: as just explained, for the different blocks $B\subset\{1,\ldots, D\}$ with $\abs{B}>1$, we know how to construct the graphs with $k$ fixed and  whose edges are colored according to $B$,  and for which the  $\{s_G(\ket{\GHZ\vert_B})\}_B$ take some fixed values that sum to $s_G$. One may then choose a labeling of the white vertices from 1 to $k$  for each $B$, and  merge the vertices with the same label. If $B=\{c\}$, the color-$c$ edges may be in any position.  The graphs constructed this way are all the graphs with the desired value $s_G$.  In accordance with the discussion of Sec.~\ref{sec:generalities-on-coarse-graining} on partially separable states, we recover the fact that the fine-grained graphs of vanishing $s_G$ are those for which   $G\in \cG_{D,\pi}$ (since $\xi=\pi$ for these states).
 
\paragraph{Genus of a jacket.}For any $D$ and any cyclic permutation $\tau\in S_D$, it is possible to build (and count) the graphs $G$ for which $g_\tau(G)$ takes some fixed value. It is indeed possible to show\footnote{Let us just sketch this briefly: considering an embedded graph $G_\tau$, its dual $G_\tau^\star$ has a vertex per face $(i,\tau(i))$, and an edge of color $i$ between two vertices if the two corresponding faces in $G$ share an edge of color $i$. One has $g(G_\tau)=g(G_\tau^\star)$. The black and white vertices of $G_\tau$ are mapped to the black and white faces of $G_\tau^\star$.  The colors may be passed to the vertices on the left when going around the boundaries of the black faces, clockwise. The resulting embedded graph with bicolored faces and labeled vertices is a $D$-constellation as defined in Ref.~\cite{BOUSQUETMELOU2000337}, Def.~2.1. Both the black and white faces are bounded by $D$ edges, that is, with the notations of Prop.~2.2 in this article, the resulting $D$-constellations satisfy the additional constraint $\sigma_0=\mathrm{id}$. One may remove one color without changing the genus, but removing the constraint that $\sigma_0=\mathrm{id}$ (using the encoding with permutations of Prop.~2.2 of \cite{BOUSQUETMELOU2000337}, one may just multiply both sides by $\sigma_D^{-1}$ and set $\sigma_0\eqdef \sigma_D^{-1}$). 
Building or counting $D$-colored graphs $G$ with $g_\tau(G)=g$  fixed therefore amounts to doing so for $(D-1)$-constellations of genus $g$. } that $D$-colored graphs $G$ with $g_\tau(G)=g$  fixed are in bijection with so-called $(D-1)$-constellations of same genus $g$. They are specific kinds of graphs drawn on surfaces, introduced for counting branched coverings of the 2-sphere. There is a literature in on how to build and count constellations of fixed genus: see Ref.~\cite{BOUSQUETMELOU2000337} for the planar case, and for instance in Ref.~\cite{CHAPUY_2009} for higher genus. 

\paragraph{Gurau degree $\omega_2$.}For $D=3$, $\omega_2$ is equal to twice the genus of any of the two jackets (Eq.~\eqref{eq:degree-is-genus-for-D=3}): this case is treated in the previous paragraph. 

For $D>3$, graphs of vanishing Gurau degree cannot distinguish $\ket{\GHZ}$ from $\ket{\phi_2}$ (Eq.~\eqref{eq:GHZ_phi2}). It is a well-known result from the literature on tensor models (see Refs.~\cite{Gurau:2011aq,Gurau2011_1,Gurau2011_2})  that 
\begin{equation}
    \omega_2(G) = 0 \qquad \Longleftrightarrow \qquad G\textrm{ is melonic},
\end{equation}
where we recall that the latter have been introduced in  Sec.~\ref{sss:Melo}. We show the following generalization of this result in App.~\ref{A:GurauDeg}.
\begin{theo}
\label{thm:omegap-vs-omega2}
For $D>3$, the following holds for any $2\le p  \le  D-1$:
\begin{equation}
 G\textrm{ is melonic} \quad \Longleftrightarrow \quad \omega_2(G) = 0 \quad \Longleftrightarrow \quad \omega_p(G)=0 \,.
\end{equation}
\end{theo}
As a consequence, $\omega_p^{(q)}(G)=0$ for $G$ melonic.\footnote{\label{footnote:omegapq}The converse is not true as there are non-melonic graphs with vanishing $\omega_p^{(q)}$, for instance, for $p=2$ and $q=D-1$ one can consider joint realignment moments defined in Eq.~\eqref{eq:joint-realignment-moment}, for which all $D$ colors occur in the sequence of colors.} Furthermore, from Eq.~\eqref{eq:GurauDegGenus}  $\omega_2(G)=0$ if and only if $g_\tau(G)=0$ for any $\tau\in S_D$ cyclic.  From Ref.~\cite{Collins2025} lemma 5.4., melonic graphs $G$ satisfy $\Delta(G)=0$. A melonic graph $M$ therefore satisfies for any  $2\le p \le q \le D$ and  any cyclic $\tau\in S_D$:
\begin{equation}
\label{eq:diff-quantities-vanish-for-melo}
    \omega_2(M) =  \omega_p(M)= \Delta(M) = \omega_p^{(q)}(M) = g_\tau (M)= 0\,.
\end{equation}

On the other hand, one can show (see e.g.~Ref.~\cite{Gurau2019,Collins2023}) that a melonic graph can have positive $\Omega_c$ (it occurs if and only if removing the edges of color $c$ disconnects the graph), 
and that for any $B\subset \{1,\ldots, D\}$ and any $K\in[0, k-1]$, there exists a melonic graph $G$ with $K_B(G)=K$.\footnote{
This follows from two standard properties of melonic graphs (see Refs.~\cite{Gurau:2011aq,Gurau2011_1,Gurau2011_2}): Assume that $D\notin B$.  (a) Considering a melonic graph $G_B$ whose edges have colors in $B$ and with $K(G_B)=K$, then one trivially constructs a $B'$-melonic graph $G_{D-1}$ with edges of colors in $\{1, \ldots, D-1\}$ by adding the edges of colors in $B'=\{1, \ldots, D-1\}\setminus B$ between the vertices at each step of a recursive construction of $G_B$.
 (b) one can complete any such $G_{D-1}$  into a connected $D$-colored melonic graph $G$ by adding color $D$ edges. By construction, $G\vert_B = G_B$, so that $K_B(G)=K$.} Combining  this data with  the tables of Sec.~\ref{ss:discrimination}, one sees that melonic graphs can in fact only discriminate a few of the HT states listed in Sec.~\ref{ss:discrimination}: $\ket{\GHZ}$ and $\Phi_c$ and $\ket{\GHZ\vert_B}$, etc (this will be summarized in Sec.~\ref{sub:dist-power-inv-from-lit}).

On the other hand, graphs with $\omega_2>0$ for $D>3$ are useful for distinguishing a number of the HT states in the tables of Sec.~\ref{ss:discrimination}. The structure of graphs of fixed positive $\omega_2$ has been studied in Refs.~\cite{Gurau2016,10.1063/1.4983562,FUSY2020103066}. 

\paragraph{The $p$-complete degrees.} From Thm.~\ref{thm:omegap-vs-omega2}, the graphs of vanishing $\omega_p$ are planar for $D=3$ and melonic for $D>3$. We can generate the graphs of fixed positive $\omega_{D-1}$ using Rem.~\ref{rm:graphs-of-positive-omega-D-1} in App.~\ref{A:GurauDeg}. We have not studied the structure of graphs of fixed positive $\omega_p$ for $2 < p < D-1$. Regarding graphs of vanishing $\omega_p^{(q)}$: from  Eq.~\eqref{eq:toprove-omega-pq} and Thm.~\ref{thm:omegap-vs-omega2}, they are such that $q$-colored subgraphs are melonic. An example is given in Footnote~\ref{footnote:omegapq} for $q=D-1$. This is the only characterization we have at this stage.

\paragraph{Graphs with few faces.}The quantity $F_c - (D-1)\kappa$ arising in Eq.~\eqref{eq:phic-vs-cstar} when comparing the scalings of $\ket{\Phi_c}$ and $\ket{\GHZ}$ vanishes if and only if there is a single face for each pair of color $(c,c')$, $c'\neq c$, and fixed positive values are obtained distributing the numbers of faces. All  moments of the partial transpose satisfy this condition for one color.  Imposing the condition for all $c$, one obtains graphs with minimal $F$, called maximally single trace (see Sec.~\ref{sec:Trace_Lit_Ref_States} and Refs.~\cite{Ferrari2019,Wallis1997}). Complete graphs and bipartite complete graphs are examples of such graphs. Only the partial transpose moments for odd $n$ satisfy this condition. As explained in Ref.~\cite{Ferrari2019}, one can build new examples from known ones by one of the binary  operations that we will study in Sec.~\ref{s:tree}. We do not know of any complete characterization of the graphs with minimal $F_c$ or minimal $F$.

\paragraph{The $c$-degree $\Omega_c$.}This quantity arises in Eq.~\eqref{eq:c-Degree} when comparing $\Phi_c$ and a lower dimensional $\GHZ$ state $\lvert\GHZ^{\frac 1 {D-1}}\rangle$. It is known (see \eg in Refs.~\cite{10.1063/1.4983562,Gurau2019,FUSY2020103066,Collins2023}) that for $G$ connected:
\begin{equation}
\label{eq:Omega_c-vs-cmelonic}
    \Omega_c(G) = 0 \qquad \Longleftrightarrow \qquad G\textrm{ is } c\textrm{-melonic}\,,
\end{equation}
where $c$-melonic graphs have been defined in  Sec.~\ref{sss:Melo}, and therefore for $B\subset\{1,\ldots, D\}$, 
\begin{equation}
\label{eq:Omega_B-vs-Bmelonic}
    \Omega_B(G) = 0 \qquad \Longleftrightarrow \qquad G\textrm{ is } B\textrm{-melonic}\,.
\end{equation}
 Since $B$-melonic graphs are melonic, they satisfy  Eq.~\eqref{eq:diff-quantities-vanish-for-melo}. 

\begin{figure}[ht]
    \centering
    \includegraphics[height = 3cm]{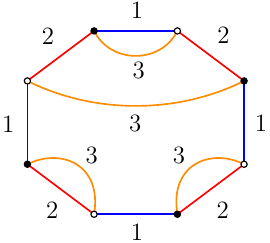}
    \caption{A 3-colored, 3-melonic graph. For $D=3$, $c$-melonic graphs are such that there is a unique cycle for the edges of colors $i,j\neq c$, and the edges of color $c$ form a  non-crossing pairing or  chord diagram on this cycle.}
    \label{fig:NC-pairing}
\end{figure}

For $D=3$ for instance, a $c$-melonic graph is planar ($g=0$) and such that the two colors $i,j\neq c$ form a unique cycle ($K_{ij} = 0$), see Sec.~4.1.1 in Ref.~\cite{Collins2023}. Said otherwise, the edges of color $c$ form a  non-crossing chord diagram on the face of colors $(i,j)$ (see Fig.~\ref{fig:NC-pairing}).  This is coherent with Eq.~\eqref{eq:Omega_c-vs-cmelonic}, since for $G\in\cG_3$ and $i,j,c$ two by two different in $\{1,2,3\}$: 
\begin{equation}
\label{eq:Omegac-D3-genus}
   \Omega_c(G) = K_{ij}(G) +2g(G)\,,
\end{equation}
where $g(G)=g_\tau(G)$ for $\tau=(123)$ and $K_{ij}(G)=F_{ij}(G)-1$ for $G$ connected.
For $D=3$ and $k>1$, one has $\frac 1 2 F_{c} - \kappa >0$. One may have $K_{c i } = 0$  or  $\frac 1 2 F_{i} - \kappa =0$ or   $\Omega_{i} = 0$, but this only occurs if the graph is cyclic.

Graphs of fixed positive $\Omega_c$ have been studied in great detail in Refs.~\cite{10.1063/1.4983562,FUSY2020103066}.

\paragraph{The degree of compatibility.}It is denoted by $\Delta$ and expressed in Eq.~\eqref{eq:Delta-perm} using Gromov products of the form $\GP{\sigma}{\tau}{\nu}$, where $\sigma, \tau, \nu \in S_k$. The latter are non-negative, thanks to the triangular inequality for the Cayley distance, and: 
\begin{equation} \label{eq:GP}
\GP{\sigma}{\tau}{\nu} =0 \qquad \Longleftrightarrow \qquad d(\sigma, \nu) + d(\nu, \tau) = d(\sigma, \tau)\,,
\end{equation}
which occurs when $\nu$ lies on a \textit{geodesic} between $\sigma$ and $\tau$ on the Cayley graph of the symmetric group, with the transpositions as generators. 

For $G\in\cG_D$,  $\widehat G\in\cG_{D+1}(G)$,  and  $(\vec\sigma,\nu)\in S_{k(G)}^{D+1}(\widehat G)$, thanks to Eq.~\eqref{eq:Gromov-vs-0-degree-vs-genus}, $\GP{\sigma_i}{\sigma_j}{\nu}=0$  if and only if the edges of color $0$ (corresponding to the permutation $\nu$) form non-crossing chord diagrams on the faces of colors $i,j$ of $G$ (see Fig.~\ref{fig:NC-pairing}).

A graph $G$ of vanishing $\Delta$ is said to be \textit{compatible}, and otherwise it is \textit{incompatible}.  From  Eq.~\eqref{eq:Delta}, a graph is compatible if and only if 
\begin{equation}
\label{eq:compatible-first-caract}
\min_{\widehat G\in \cG_{D+1}(G)} \Omega_{0}(\widehat G) = \frac 1 {D-1}  \omega_2(G)\,.
\end{equation}
From Eq.~\eqref{eq:Delta-perm}, one also has, with the notations above, the more intuitive characterization
\begin{equation}
    \Delta(G) = 0\qquad \Longleftrightarrow \qquad \forall i<j,\quad \GP{\sigma_i}{\sigma_j}{\nu} = 0 \qquad \Longleftrightarrow \qquad \forall i<j,\quad \Omega_{0}(\widehat G\vert_{0ij})=0\,,
\end{equation}
that is, a graph $G$ is compatible if there is a way to add new edges of color $0$ to it such that these edges form non-crossing chord diagrams on the faces of colors $i,j$ of $G$ for every $i<j$ (see Fig.~\ref{fig:NC-pairing}). On the other hand, incompatibility indicates the impossibility to do so, or equivalently, the fact that no permutation that lies on geodesics between any two permutations $\sigma_i$ and $\sigma_j$, $i<j$.

For $D=3$, most planar graphs are incompatible, as shown in the following lemma.  \begin{lem} \label{lem:non_melo_and_planar}
     A 3-colored graph $G$ is planar and compatible if and only if it is melonic. 
\end{lem}
\begin{proof}
Melonic graphs are planar and compatible. Reciprocally, from the characterization of compatible graphs in Eq.~\eqref{eq:compatible-first-caract} together with Eq.~\eqref{eq:degree-is-genus-for-D=3}, if $G$ is planar and compatible, then  there is a $\widehat G\in \cG_{4}(G)$ such that  $ \Omega_{0}(\widehat G)=0$, so that $\widehat G$ is 0-melonic, and therefore $G$ is melonic.  
\end{proof} 

\subsection{Combinatorial quantities and trace-invariants from the literature}
\label{sub:dist-power-inv-from-lit}
The tables below review whether the combinatorial quantities introduced in Sec.~\ref{ss:discrimination} vanish or not for the trace-invariants presented in  Sec.~\ref{sec:Trace_Lit_Ref_States}.

For the following tables, we will need the values of $\Delta$ for $\RM_{2n}^{(\bar{1,2})}$,  $\JRM_k^{\vec{i}}$ for $\vec i$ a cyclic sequence of colors and $\RE_{m,n}^{(3)}$ which are deduced from Ref.~\cite{Lionni2018},\footnote{See the tables in Ch.~4 of Ref.~\cite{Lionni2018}: one has $2\Delta(G) = (D-1)S_G - \omega_2(G)$, where $S_G$ is denoted in Ref.~\cite{Lionni2018} by $s_G$,  $\omega_2$ by $\delta_\textrm{Gur}$, and $k$ by $V/2$.} see also  Ref.~\cite{10.1063/1.4983562} and   Ref.~\cite{Akers2022,Akers2023,Akers2024}. The compatibility of the odd and even moments of the partial transpose was proved in Ref.~\cite{Dong2021}. Concerning the multi-entropies, we will use the following lemma.

\begin{lem}
    For $n>1$ and $D>2$, the graphs $\ME_n^D$ corresponding to the multi-entropies are incompatible.
\end{lem}

\begin{proof}
For $\ME_n^D$ to be compatible, there must exist a way to add color-0 edges to obtain a $(D+1)$-colored graph $\widehat{\ME_n^D}$,  such that the color-0 edges form non-crossing pairings on every face of colors $(i,j)$ for $1\le i<j\le D$. If the pairing is chosen to be non-crossing on the faces of colors $(1,2)$ and $(1,3)$, then, by construction of the multi-entropies, the color-0 edges must be parallel to the color-$1$ edges. This configuration, however, fails to be non-crossing on the faces of colors $(2,3)$.
\end{proof}

\begin{figure}[ht]
    \centering
    \includegraphics[height = .6cm]{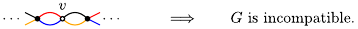}
    \caption{Graphical representation of Lem.~\ref{lem:Cyc_incomp}}
    \label{fig:CorNeck}
\end{figure}

Finally, for $D\geq 4$ the non-melonic $D$-colored cyclic graphs are incompatible due to the following lemma:

\begin{lem} \label{lem:Cyc_incomp}
    Let $D\geq 4$. We consider a $D$-colored graph $G$ and denote by $V(G)$ its set of vertices. If there exists $v\in V(G)$, such that $v$ is on two different faces of size two (see Fig.~\ref{fig:CorNeck}), then the graph is incompatible.
\end{lem}

\begin{proof}
    Assume $G$ is compatible. The edges of color $0$ of $\widehat G \in \MD{}(G)$ must be non-crossing on each face since all Gromov products are vanishing. In particular, it cannot simultaneously be the case for both faces of size two being attached to $v$, since the non-crossing requirement would require pairing both faces. Hence, $G$ is incompatible.
\end{proof}

\paragraph{The tripartite case $D=3$.} In the following table, we consider $\{c_1, c_2, c_3\} = \{1,2,3\}$. We start by studying colored graphs being planar and compatible, \ie $3$-colored melonic graphs by Lem.~\ref{lem:non_melo_and_planar}.
\begin{table}[H]
    \begin{center}
        \begin{tabular}{ |>{\centering\arraybackslash}m{2.6cm}||c|c|c|c|c|c|c|c|c|c| } \hline
            $D=3$            & $\kappa - k$ &  $K_{c_1 c_3}$ &   $K_{c_2c_3}$&  $\frac{1}{2}F_{c_1} - \kappa$ &   $\frac{1}{2}F_{c_3} - \kappa$  & 
            $\Omega_{c_1}$ &   $\Omega_{c_2}$ &   $\Omega_{c_3}$ & $\omega_2 = 2g$ & $\Delta$ \\ \hline \hline
            $2$-vertex graph 
            & $= 0$        & \multicolumn{2}{c|}{$=0$} & \multicolumn{2}{c|}{$= 0$}                     & \multicolumn{3}{c|}{$= 0$}      & \multirow{5}{*}{$=0$}     & \multirow{5}{*}{$=0$} \\ \cline{1-9} 
            cyclic   for $(c_1,c_2)$ vs $c_3$, $k>1$ 
            & $< 0$        &\multicolumn{2}{c|}{$=0$}&  $>0$ &   $=0$   &  \multicolumn{2}{c|}{$= 0$}  & $>0$   &     &  \\ \cline{1-9} 
            $c_1$-melonic, non cyclic
                             & $< 0$        & $>0$ &   $=0$  & \multicolumn{2}{c|}{$> 0$}                         &  $=0$ &   \multicolumn{2}{c|}{$> 0$}  &      &  \\ \cline{1-9} 
            other melonic 
                             & $< 0$        &\multicolumn{2}{c|}{$> 0$}& \multicolumn{2}{c|}{$> 0$}                        & \multicolumn{3}{c|}{$> 0$}    &       &  \\ \hline
        \end{tabular}
    \end{center}
\end{table} 

Now, let us focus on planar non-melonic graphs and particularly on the colored graphs associated with the realignment moments and the Rényi reflected entropies. This family contains incompatible $3$-colored graphs.
\begin{table}[H]
    \begin{center}
        \begin{tabular}{ |>{\centering\arraybackslash}m{2.6cm}||c|c|c|c|c|c|c|c|c|c| } \hline
            $D=3$            & $\kappa - k$ &  $K_{c_1 c_3}$ &   $K_{c_2c_3}$&  $\frac{1}{2}F_{c_1} - \kappa$ &   $\frac{1}{2}F_{c_3} - \kappa$  & 
            $\Omega_{c_1}$ &   $\Omega_{c_2}$ &   $\Omega_{c_3}$ & $\omega_2 = 2g$ & $\Delta$ \\ \hline \hline
            planar non-melonic
                             & \multirow{4}{*}{$<0$}        &\multicolumn{2}{c|}{$>0$}& \multicolumn{2}{c|}{\multirow{4}{*}{$>0$}}                        & \multicolumn{3}{c|}{$>0$}   &  \multirow{4}{*}{$=0$}       & $> 0$ \\ \cline{1-1}\cline{3-4}\cline{7-9}\cline{11-11}
            $\RM_{2n}^{(c_1)}$, $n>1$  
                             &         & $>0$ & $=1$ &  \multicolumn{2}{c|}{}                         & $=1$ & \multicolumn{2}{c|}{$>0$}   &        & $= 1$ \\ \cline{1-1}\cline{3-4}\cline{7-9}\cline{11-11}
            $\RE_{m,n}^{(c_1)}$ $n>1,m\ge 2$
                             &         &\multicolumn{2}{c|}{$>0$}& \multicolumn{2}{c|}{}                         & \multicolumn{3}{c|}{$>0$}    &       & $= 1$ \\ \hline
        \end{tabular}
    \end{center}
\end{table} 

From the introduced colored graph of Sec.~\ref{sub:inv-from-lit}, the graphs being non-planar and compatible are the colored graphs associated with the partial transpose moments and the joint realignment moments. Concerning maximally single-trace invariants, they contain compatible invariants as we will prove later by the use of a binary operation (see Sec.~\ref{s:tree}, Prop.~\ref{prop:MST}). Regarding their combinatorial quantities, we have the following table.
\begin{table}[H]
    \begin{center}
        \begin{tabular}{ |>{\centering\arraybackslash}m{2.6cm}||c|c|c|c|c|c|c|c|c|c| } \hline
            $D=3$            & $\kappa - k$ &  $K_{c_1 c_3}$ &   $K_{c_2c_3}$&  $\frac{1}{2}F_{c_1} - \kappa$ &   $\frac{1}{2}F_{c_3} - \kappa$  & 
            $\Omega_{c_1}$ &   $\Omega_{c_2}$ &   $\Omega_{c_3}$ & $\omega_2 = 2g$ & $\Delta$ \\ \hline \hline
            Maximally single-trace  
                             & \multirow{4}{*}{$<0$}        &\multicolumn{2}{c|}{$= 0$}& \multicolumn{2}{c|}{$= 0$}                          & \multicolumn{3}{c|}{\multirow{4}{*}{$>0$}}   & $> 0$      & $\geq 0$\\ \cline{1-1}\cline{3-6}\cline{10-11}
            $\PT_{2n+1}^{(c_1)}$, $n>0$  
                             &        &\multicolumn{2}{c|}{$= 0$}& \multicolumn{2}{c|}{$= 0$}                          & \multicolumn{3}{c|}{}   & \multirow{2}{*}{$>0$}      &   \multirow{3}{*}{$=0$} \\ \cline{1-1}\cline{3-6}
            $\PT_{2n}^{(c_1)}$, $n>1$  
                             &        &  $=0$ &   $=1$  & $=0$ &   $=\frac{1}{2}$                   & \multicolumn{3}{c|}{}      &         &  \\ \cline{1-1}\cline{3-6}\cline{10-10}
            $\JRM_k^{\vec{i}}$, $k>3$ \tablefootnote{\label{foot:JRM0}We also assume that the color $c_2$ appears at least twice in $\vec i$. This excludes the graph $\PT_3$ (Fig.~\ref{fig:PT_EX}), included above. Up to relabeling of the colors, this is the only excluded case.}
                             &        &  $>0$ &   $\ge 0$ \tablefootnote{\label{foot:JRM}It vanishes if and only if  $c_1$ appears a single time in $\vec i$. The only possibility to have both $K_{c_2c_3}=0$ and $K_{c_1c_2}=0$ is the $k=4$ case with $\vec i=(c_1, c_2, c_3, c_2)$, for which $\frac 1 2 F_{c_2}-\kappa=0$.}  & \multicolumn{2}{c|}{$> 0$ $^{\ref{foot:JRM}}$}                        & \multicolumn{3}{c|}{}   & $= 2$       &  \\ \hline
        \end{tabular}
    \end{center}
\end{table} 

For $D=3$, the colored graphs associated with the Rényi multi-entropies stand out due to the non-vanishing property of all considered combinatorial quantities.
\begin{table}[H]
    \begin{center}
        \begin{tabular}{ |>{\centering\arraybackslash}m{2.6cm}||c|c|c|c|c|c|c|c|c|c| } \hline
            $D=3$            & $\kappa - k$ &  $K_{c_1 c_3}$ &   $K_{c_2c_3}$&  $\frac{1}{2}F_{c_1} - \kappa$ &   $\frac{1}{2}F_{c_3} - \kappa$  & 
            $\Omega_{c_1}$ &   $\Omega_{c_2}$ &   $\Omega_{c_3}$ & $\omega_2 = 2g$ & $\Delta$ \\ \hline \hline
            $\ME_{n}^3$, $n>2$ 
                             & $< 0$        &\multicolumn{2}{c|}{$> 0$}& \multicolumn{2}{c|}{$> 0$}                     & \multicolumn{3}{c|}{$> 0$}      & $> 0$        & $> 0$ \\ \hline
        \end{tabular}
    \end{center}
\end{table} 

The tables above, together with the tables presented at the end of Sec.~\ref{ss:discrimination}, allow one to determine whether the trace-invariants listed in Sec.~\ref{sub:inv-from-lit} discriminate the HT states introduced in Sec.~\ref{ss:discrimination}. This analysis is illustrated by the tables collected in App.~\ref{sec:table-in-appendix}.

We now focus on the families of trace-invariants presented in Sec.~\ref{sec:LUinvariants} that admit a quantity defined by a limit, namely entanglement entropies, the partially transposed entropy, the logarithmic negativity, the reflected entropy, and the multi-entropy.

Throughout, expectation values are evaluated at leading order in the large-$N$ expansion and without considering analytical continuation (the limit as $n$ or $k$ approaches one is taken ``naively''). Moreover, for each family of invariants, we assume that the logarithm and the averaging operation can be interchanged, which relies, with high probability, on the validity of the large-$N$ factorization. The latter is discussed in detail in Sec.~\ref{subsec:averageApprox} and App.~\ref{A:DeltaAB}. 

\begin{itemize}
    \item Entanglement entropies: the \textit{cyclic} graph is considered to be a $\bar{c_3}$-melonic graph.
    \begin{table}[H]
        \centering
        \begin{tabular}{ |>{\centering\arraybackslash}m{2.5cm}||c|c|c|c|c|c|c|c|c| } \hline
            $D=3$ & $\ket{\GHZ}$ & $\ket{\GHZ\vert_{c_1 c_2}}$ & $\ket{\GHZ\vert_{c_1 c_3}}$ & $\ket{\GHZ\vert_{c_2 c_3}}$ & $\ket{\Phi_{c_1}}$ & $\ket{\Phi_{c_2}}$ & $\ket{\Phi_{c_3}}$ & $\ket{\phi_2}$ & $\ket{\varphi}$ \\ \hline \hline
            $\frac{1}{1-k}s_{\rm{cyclic}}$  
                            & 1 & 0 & \multicolumn{2}{c|}{1} & \multicolumn{2}{c|}{$\frac 1 2$} & 1 & 1 & 1 \\ \hline
            $\frac{1}{\ln N} \displaystyle\lim_{k \to 1} \S_{\paa{c_1,c_2}}^{(k)}$
                            & 1 & 0 & \multicolumn{2}{c|}{1} & \multicolumn{2}{c|}{$\frac 1 2$} & 1 & 1 & 1 \\ \hline
            \end{tabular}
    \end{table}
    
    \item Realignment moments: They do not exhibit a quantity defined by a limit; however, scalings are given in the following table.
    \begin{table}[H]
        \centering
        \begin{tabular}{ |>{\centering\arraybackslash}m{2.5cm}||c|c|c|c|c|c|c|c|c| } \hline
            $D=3$ & $\ket{\GHZ}$ & $\ket{\GHZ\vert_{c_1 c_2}}$ & $\ket{\GHZ\vert_{c_1 c_3}}$ & $\ket{\GHZ\vert_{c_2 c_3}}$ & $\ket{\Phi_{c_1}}$ & $\ket{\Phi_{c_2}}$ & $\ket{\Phi_{c_3}}$ & $\ket{\phi_2}$ & $\ket{\varphi}$ \\ \hline \hline
            $s_{\RM_{2n}^{(c_1)}}$
                            & $1 - 2n$ & \multicolumn{2}{c|}{$-n$}  & $2(1-n)$ & $-n$ & \multicolumn{2}{c|}{$1 - \frac{3}{2}n$} & $1-2n$ & $-2n$ \\ \hline
        \end{tabular}
    \end{table}
    
    \item Partially transpose entropy:
    \begin{table}[H]
        \centering
        \begin{tabular}{ |>{\centering\arraybackslash}m{2.5cm}||c|c|c|c|c|c|c|c|c| } \hline
            $D=3$ & $\ket{\GHZ}$ & $\ket{\GHZ\vert_{c_1 c_2}}$ & $\ket{\GHZ\vert_{c_1 c_3}}$ & $\ket{\GHZ\vert_{c_2 c_3}}$ & $\ket{\Phi_{c_1}}$ & $\ket{\Phi_{c_2}}$ & $\ket{\Phi_{c_3}}$ & $\ket{\phi_2}$ & $\ket{\varphi}$ \\ \hline \hline

            $- \frac 1 2 s_{\PT_{2n+1}^{(c_1)}}$  
                             & $n$ & \multicolumn{3}{c|}{$n$} & \multicolumn{3}{c|}{$n$} & $\frac 3 2 n$ & $\frac 3 2 n$ \\ \hline
            $\frac{1}{\ln N} \, \S_{\rm{PTE}}$  
                             & $1$ & \multicolumn{3}{c|}{$1$} & \multicolumn{3}{c|}{$1$} & $\frac 3 2 $ & $\frac 3 2 $ \\ \hline
        \end{tabular}
    \end{table}
    
    \item Logarithmic negativity:
    \begin{table}[H]
        \centering
        \begin{tabular}{ |>{\centering\arraybackslash}m{2.5cm}||c|c|c|c|c|c|c|c|c| } \hline
            $D=3$ & $\ket{\GHZ}$ & $\ket{\GHZ\vert_{c_1 c_2}}$ & $\ket{\GHZ\vert_{c_1 c_3}}$ & $\ket{\GHZ\vert_{c_2 c_3}}$ & $\ket{\Phi_{c_1}}$ & $\ket{\Phi_{c_2}}$ & $\ket{\Phi_{c_3}}$ & $\ket{\phi_2}$ & $\ket{\varphi}$ \\ \hline \hline

            $s_{\PT_{2n}^{(c_1)}}$  
                             & $1 - 2n$ & \multicolumn{2}{c|}{$1 - 2n$} & $2(1-n)$ & $1 - 2n$ & \multicolumn{2}{c|}{$\frac 3 2 - 2n$} & $2 - 3n$ & $2 - 3n$ \\ \hline
            $\frac{1}{\ln N} \, \S_{\rm{LN}}$  
                             & $0$ & \multicolumn{2}{c|}{$0$} & $1$ & $0$ & \multicolumn{2}{c|}{$\frac 1 2$} & $\frac 1 2$ & $\frac 1 2$ \\ \hline
        \end{tabular}
    \end{table}
    
    \item Reflected entropy:
    \begin{table}[H]
        \centering
        \begin{tabular}{ |>{\centering\arraybackslash}m{2.5cm}||c|c|c|c|c|c|c|c|c| } \hline
            $D=3$ & $\ket{\GHZ}$ & $\ket{\GHZ\vert_{c_1 c_2}}$ & $\ket{\GHZ\vert_{c_1 c_3}}$ & $\ket{\GHZ\vert_{c_2 c_3}}$ & $\ket{\Phi_{c_1}}$ & $\ket{\Phi_{c_2}}$ & $\ket{\Phi_{c_3}}$ & $\ket{\phi_2}$ & $\ket{\varphi}$ \\ \hline \hline
            $\frac{1}{1-n}(s_{\RE_{m,n}^{(c_1)}}$ $- n s_{\RE_{m,1}^{(c_1)}})$
                            & $1$ &  \multicolumn{2}{c|}{0} & 2 & 0 & \multicolumn{2}{c|}{1} & 1 & $\frac{n}{n-1}$ \tablefootnote{This expression is valid for $n>1$. However, when computing the reflected entropy and performing the analytic continuation, one must also take into account the contribution from the unbalanced case (\ie when $\dim \H_c = N_c$). As explained in detail in Ref.~\cite{Akers2022}, this requires modifying the expression to $$\frac{1}{1-n}\pa{s_{\RE_{m,n}^{(c_1)}}(\ket{\varphi}) - n s_{\RE_{m,1}^{(c_1)}}(\ket{\varphi})} = \min \paa{2,\frac{n}{n-1}} \,.$$} \\ \hline
            $\frac{1}{\ln N}\,\S_{\RE^{(c_1)}}$  
                             & 1 & \multicolumn{2}{c|}{0}  & 2 & 0 & \multicolumn{2}{c|}{1} & 1 & 2 \\ \hline
            
        \end{tabular}
    \end{table}
    
    \item Multi-entropy: the symbols ?`\,$\cdot$\,? highlight a conjecture of the corresponding value.
    \begin{table}[H]
    \centering
    \begin{tabular}{ |>{\centering\arraybackslash}m{2.5cm}||c|c|c|c|c|c|c|c|c| } \hline
        $D=3$ & $\ket{\GHZ}$ & $\ket{\GHZ\vert_{c_1 c_2}}$ & $\ket{\GHZ\vert_{c_1 c_3}}$ & $\ket{\GHZ\vert_{c_2 c_3}}$ & $\ket{\Phi_{c_1}}$ & $\ket{\Phi_{c_2}}$ & $\ket{\Phi_{c_3}}$ & $\ket{\phi_2}$ & $\ket{\varphi}$ \\ \hline \hline
        
        $\frac{1}{n(1-n)}s_{\ME_{n}^3}$ 
                         & $1 + \frac 1 n$ & \multicolumn{3}{c|}{$1$} & \multicolumn{3}{c|}{$1$} & $\frac 3 2$ & ?`\,$2$\,? \\ \hline
        $\frac{1}{\ln N}\,\S_{\ME^3}$
                        & $2$ & \multicolumn{3}{c|}{$1$} & \multicolumn{3}{c|}{$1$} & $\frac 3 2$ & ?`\,$2$\,? \\ \hline
    \end{tabular}
\end{table}
\end{itemize}

\paragraph{The $D$-partite case $D>3$.}
In the following table, we consider $c_1\in \{1,\ldots, D\}$, and $B_1,B_2, B_3, B,B'$ some subsets of $\{1,\ldots, D\}$ such that $B\subset B_1$ and  $B'\nsubseteq B_1$. The tripartite coarse-grained invariants are considered for blocks $B_1,B_2,B_3$ replacing the colors 1, 2, 3. 

We start with melonic invariants that exhibit vanishing $p$-complete degree, genus, and degree of compatibility. For $B_1$-melonic graphs, we assume that $B_1$ is the largest subset  $\tilde B\subset \{1,\ldots, D\}$ such that the graph is $\tilde B$-melonic.
\begin{table}[H]
    \begin{center}
        \begin{tabular}{ |>{\centering\arraybackslash}m{2.7cm}||c|c|c|c|c|c|c|c|c| } \hline
            $D>3$            &  $K_{\bar{B'}}$ &   $K_{\bar{B}}$ &  $\sum_{c \in B} \pac{F_c - (D-1) \kappa}$  &   $\Omega_{B}$ &   $\Omega_{B'}$ & $\omega_p$ & $\omega_p^{(q)}$ & $g_\tau$ & $\Delta$ \\ \hline \hline
            $2$-vertex graph & \multicolumn{2}{c|}{$0$}        & $= 0$ & \multicolumn{2}{c|}{$=0$}                     & $= 0$      & $= 0$      & $= 0$ & $=0$ \\ \hline
            $B_1$-melonic
            &  $>0$ &   $=0$        &$ \ge 0$ \tablefootnote{The equality occurs when there exists $c_1\in\{1,\ldots, D\}$ such that $B_1 = \bar{c_1}$ and $B = c_1$.} &  $=0$ &   $>0$    &  $=0$  & $= 0$     & $= 0$ & $=0$ \\ \hline
            other melonic 
                             & \multicolumn{2}{c|}{$>0$}        &$> 0$& \multicolumn{2}{c|}{$>0$}                        & $= 0$    & $= 0$       & $= 0$ & $=0$ \\ \hline
        \end{tabular}
    \end{center}
\end{table}     

The cyclic non-melonic are incompatible as explained in Lem.~\ref{lem:Cyc_incomp}. Except for $D=4$, for which the non melonic cyclic graph satisfies $\omega_2^{(3)} = 0$, cyclic graphs also satisfy $\omega_p^{(q)} > 0$.
\begin{table}[H]
    \begin{center}
        \begin{tabular}{ |>{\centering\arraybackslash}m{2.7cm}||c|c|c|c|c|c|c|c|c| } \hline
            $D>3$            &  $K_{\bar{B'}}$ &   $K_{\bar{B}}$ &  $\sum_{c \in B} \pac{F_c - (D-1) \kappa}$  &   $\Omega_{B}$ &   $\Omega_{B'}$ & $\omega_p$ & $\omega_p^{(q)}$ & $g_\tau$ & $\Delta$ \\ \hline \hline
            other cyclic \hspace{1cm} $B_1$ vs $\bar{B_1}$ 
                             & \multicolumn{2}{c|}{$\ge 0$ \tablefootnote{With $K_B = 0$ whenever $B \cap B_1 \neq \emptyset$ and $B \cap \bar{B_1} \neq \emptyset$.}}       &$> 0$& \multicolumn{2}{c|}{$> 0$}                        & $> 0$    & $> 0$       & $\ge 0$ & $>0$ \\ \hline 
        \end{tabular}
    \end{center}
\end{table}     

Regarding the coarse-grained version of the realignment moments and the partial transpose, we have:
\begin{table}[H]
    \begin{center}
        \begin{tabular}{ |>{\centering\arraybackslash}m{2.7cm}||c|c|c|c|c|c|c|c|c| } \hline
            $D>3$            &  $K_{\bar{B'}}$ &   $K_{\bar{B}}$ &  $\sum_{c \in B} \pac{F_c - (D-1) \kappa}$  &   $\Omega_{B}$ &   $\Omega_{B'}$ & $\omega_p$ & $\omega_p^{(q)}$ & $g_\tau$ & $\Delta$ \\ \hline \hline
            
            $\PT_{2n+1}^{(B_1)}$ 
                             &  $\ge 0$ \tablefootnote{With strict inequality whenever $B' = B_i \cup B_j$ for any distinct couple $i,j$ in $\paa{0,1,2}$.} &   $=0$       &$\ge 0$ \tablefootnote{Vanishing if and only if $B = B_i$ and $\abs{B_i} = 1$ for any $i \in \paa{1,2,3}$.}& \multicolumn{2}{c|}{$> 0$}                        & $> 0$   & $> 0$        & $> 0$ & $>0$ \\ \hline
            $\PT_{2n}^{(B_1)}$ 
                             & $\ge 0$ \tablefootnote{With equality whenever $B' = B_1$ or $B' = B_i \cup B_j$ for any distinct couple $i,j$ in $\paa{1,2,3}$.} &   $=0$       &$\ge 0$ \tablefootnote{Equality occurs when $B = B_1$ and $\abs{B_1} = 1$.}& \multicolumn{2}{c|}{$> 0$}                        & $> 0$   & $> 0$        & $> 0$ & $>0$ \\ \hline
            $\RM_{2n}^{(B_1)}$ 
                             & \multicolumn{2}{c|}{$\ge 0$ \tablefootnote{$K_B = 0$ if and only if $B \cap B_i \neq \emptyset$ for all $i \in \paa{1,2,3}$.} }       &$> 0$& \multicolumn{2}{c|}{$> 0$}                         & $> 0$   & $> 0$       &$\ge 0$ & $\ge 1$\tablefootnote{Let $n>1$, $\Delta(\RM_{2n}^{(B_1)}) = 1$ if and only if $\abs{B_1} = D-2$ (see in the middle of Fig.~\ref{fig:RMgen}).} \\ \hline
        \end{tabular}
    \end{center}
\end{table}   

Maximally single-trace invariants are particular in the sense that some combinatorial quantities are always vanishing.
\begin{table}[H]
    \begin{center}
        \begin{tabular}{ |>{\centering\arraybackslash}m{2.7cm}||c|c|c|c|c|c|c|c|c| } \hline
            $D>3$            &  $K_{\bar{B'}}$ &   $K_{\bar{B}}$ &  $\sum_{c \in B} \pac{F_c - (D-1) \kappa}$  &   $\Omega_{B}$ &   $\Omega_{B'}$ & $\omega_p$ & $\omega_p^{(q)}$ & $g_\tau$ & $\Delta$ \\ \hline \hline
            Maximally single-trace
                             & \multicolumn{2}{c|}{$=0$}        & $=0$  & \multicolumn{2}{c|}{$> 0$}                        & $> 0$   & $> 0$       & $> 0$ & $\ge 0$\\ \hline
        \end{tabular}
    \end{center}
\end{table}   

For the joint realignment moments $\JRM_k^{\vec{i}}$, where we suppose that the cyclic sequence of colors $\vec i$ is such that all $D$ colors appear at least once, we can fill the following table.
\begin{table}[H]
    \begin{center}
        \begin{tabular}{ |>{\centering\arraybackslash}m{2.7cm}||c|c|c|c|c|c|c|c|c| } \hline
            $D>3$            &  $K_{\bar{B'}}$ &   $K_{\bar{B}}$ &  $\sum_{c \in B} \pac{F_c - (D-1) \kappa}$  &   $\Omega_{B}$ &   $\Omega_{B'}$ & $\omega_p$ & $\omega_p^{(q)}$ & $g_\tau$ & $\Delta$ \\ \hline \hline
            
            $\JRM_k^{\vec{i}}$, $k \ge D$ 
                             & \multicolumn{2}{c|}{$\ge 0$ \tablefootnote{$K_B = 0$ if and only if $c$ appears once in $\vec i$ and $B = \bar{c}$.}}        & $>0$  & \multicolumn{2}{c|}{$> 0$}                        & $> 0$   & $= 0$       & $\ge 1$ & $=0$\\ \hline
        \end{tabular}
    \end{center}
\end{table}   

Instead of only the Rényi multi-entropies in $D=3$, both the Rényi reflected multi-entropies and the Rényi multientropies stand out for their non-vanishing combinatorial quantities.
\begin{table}[H]
    \begin{center}
        \begin{tabular}{ |>{\centering\arraybackslash}m{2.7cm}||c|c|c|c|c|c|c|c|c| } \hline
            $D>3$            &  $K_{\bar{B'}}$ &   $K_{\bar{B}}$ &  $\sum_{c \in B} \pac{F_c - (D-1) \kappa}$  &   $\Omega_{B}$ &   $\Omega_{B'}$ & $\omega_p$ & $\omega_p^{(q)}$ & $g_\tau$ & $\Delta$ \\ \hline \hline
            
            $\ME_{n}^D$, $n>2$ 
                             & \multicolumn{2}{c|}{$> 0$}        &$> 0$& \multicolumn{2}{c|}{$> 0$}                      & $> 0$      & $> 0$        & $> 0$ & $>0$ \\ \hline
            $\RME_{m,n}^{(c_1)}$ $n>1,m\ge 2$
                             & \multicolumn{2}{c|}{$> 0$}        &$> 0$& \multicolumn{2}{c|}{$> 0$}                          & $> 0$    & $> 0$      & $> 0$ & $> 0$ \\ \hline
        \end{tabular}
    \end{center}
\end{table}    

Regarding the families that exhibit a quantity defined as a limit, we will again compute both the scalings and the limit without taking care of the analytical continuation.

\begin{itemize}
    \item Entanglement entropies: let us consider a cyclic graph associated with the bipartition $\H_{B_1} \ot \H_{\bar{B_1}}$. Furthermore, we assume that $B$ is either a subset of $B_1$ or $\bar{B_1}$, that $B' \cap B_1 \neq \emptyset$ and $B' \cap \bar{B_1} \neq \emptyset$. Let $c\in B_1$, $c' \in \bar{B_1}$ and $\alpha_{p,q,r}$ to be defined by 
    \begin{equation}
        \alpha_{p,q,r} \eqdef \binom{q + r - 1}{p} - \binom{q}{p} - \binom{r}{p} \,.
    \end{equation}
    \begin{table}[H]
        \centering
        \begin{tabular}{ |>{\centering\arraybackslash}m{2cm}||c|c|c|c|c|c|c|c| } \hline
            $D>3$ & $\ket{\GHZ}$ & $\ket{\GHZ\vert_{B}}$ & $\ket{\GHZ\vert_{B'}}$ & $\ket{\Phi_{c}}$ &  $\ket{\Phi_{c'}}$ & $\ket{\phi_p}$ & $\ket{\varphi}$  \\ \hline \hline
            $\frac{1}{1-k}s_{\rm{cyclic}}$
                            & 1 & 0 & 1 & $\frac{D - \abs{B_1}}{D-1}$ & $\frac{D - \abs{\bar{B_1}}}{D-1}$ & $1 - \frac{\alpha_{p,\abs{B_1},\abs{\bar{B_1}}}}{\cal{I}_p}$ & $1 - \frac{\alpha_{2,\abs{B_1},\abs{\bar{B_1}}} - \abs{B_1}(\abs{B_1}-1)}{D-1}$ \\ \hline
            $\frac{1}{\ln N} \displaystyle\lim_{k \to 1} \S_{B_1}^{(k)}$
                            & 1 & 0 & 1 & $\frac{D - \abs{B_1}}{D-1}$ & $\frac{D - \abs{\bar{B_1}}}{D-1}$ & $1 - \frac{\alpha_{p,\abs{B_1},\abs{\bar{B_1}}}}{\cal{I}_p}$ & $1 - \frac{\alpha_{2,\abs{B_1},\abs{\bar{B_1}}} - \abs{B_1}(\abs{B_1}-1)}{D-1}$ \\ \hline
        \end{tabular}
    \end{table}
    
    \item Partially transpose entropy and logarithmic negativity: all scalings can be computed using App.~\ref{sec:table-in-appendix}, and their explicit expressions will not be displayed in the following table. Instead, we restrict our attention to the relevant quantities, such as the partially transposed entropy $\S_\PTE$ and the logarithmic negativity $\S_\LN$. 

    To this end, we consider a tripartition of the Hilbert space $\H$ into the blocks $B_1$, $B_2$, and $B_3$. We assume that $B \subset B_i$ for some $i\in\paa{1,2,3}$, and that the intersections satisfy $B' \cap B_1 \neq \emptyset$, $B' \cap B_2 \neq \emptyset$ and $B'' \cap B_2 \neq \emptyset$, $B'' \cap B_3 \neq \emptyset$. Furthermore, we denote a color in each block by $c_i \in B_i$ for $i\in \paa{1,2,3}$ and set $p_i = \abs{B_i}$. 
    \begin{table}[H]
        \centering
        \begin{tabular}{ |>{\centering\arraybackslash}m{2.5cm}||c|c|c|c|c|c|c|c|c| } \hline
            $D>3$ & $\ket{\GHZ}$ & $\ket{\GHZ\vert_{B}}$ & $\ket{\GHZ\vert_{B'}}$ & $\ket{\GHZ\vert_{B''}}$ & $\ket{\Phi_{c_1}}$ & $\ket{\Phi_{c_2}}$ & $\ket{\Phi_{c_3}}$ & $\ket{\phi_p}$ & $\ket{\varphi}$  \\ \hline \hline
            $\frac{1}{\ln N} \, \S_{\rm{PTE}}$  
                             & $1$ & $0$ & \multicolumn{2}{c|}{$1$} & $\frac{D-p_1}{D-1}$ & $\frac{D-p_2}{D-1}$ & $\frac{D-p_3}{D-1}$ & \eqref{eq:PTE_phi_p} & \eqref{eq:PTE_varphi}  \\ \hline\hline
            $\frac{1}{\ln N} \, \S_{\rm{LN}}$  
                             & $0$ & \multicolumn{2}{c|}{$0$} & $1$ & $0$ & $\frac{p_3}{D-1}$ & $\frac{p_2}{D-1}$ & $\frac{\alpha_{p,p_2,p_3}}{\cal{I}_p}$ & \eqref{eq:LN_varphi}\\ \hline
        \end{tabular}
    \end{table}
    The computation of the partially transposed entropy for a $p$-complete state yields
    \begin{equation} \label{eq:PTE_phi_p}
        \frac{1}{\ln N} \, \S_{\rm{PTE}}(\ket{\phi_p}) = \frac{D}{p} - \frac{1}{2 \cal{I}_p} \pac{\binom{p_1 + p_2}{p} + \binom{p_1 + p_3}{p} + \binom{p_2}{p} + \binom{p_3}{p}} \,,
    \end{equation}
    and for a Haar-random state, the leading order reads
    \begin{equation} \label{eq:PTE_varphi}
        \frac{1}{\ln N} \pac{\mean{\S_\PTE(\ket{\varphi})} - \S_{\rm{PTE}}(\ket{\phi_2})} \underset{N \to \infty}{\sim} \frac{1}{D-1} \left\{
        \begin{array}{ll} 
            2 \pa{\binom {p_i}{2} + \binom {p_j}{2}  + \frac{p_i p_j}{2}} \textrm{ if } p_k > \frac D 2\, \textrm{ for }\{i,j,k\}=\{1,2,3\}, \\
            \sum_{i=1}^3\binom{p_i}{2} \textrm{ otherwise}. \,.
        \end{array}\right.
    \end{equation}

    Regarding the leading order of the average of the logarithmic negativity, a computation based on App.~\ref{sec:table-in-appendix} gives
    \begin{equation} \label{eq:LN_varphi}
        \frac{1}{\ln N} \pac{\mean{\S_\LN(\ket{\varphi})} - \S_{\rm{LN}}(\ket{\phi_2})} \underset{N \to \infty}{\sim} - \frac{2}{D-1} \left\{
        \begin{array}{lll} 
            - \pa{\binom {p_1}{2} + \binom {p_2}{2}  + \frac{p_1 p_2}{2}} + \binom{p_1}{2}\textrm{ if } \abs{B_3} > \frac D 2\,  \textrm{ or }2\leftrightarrow 3, \\
            \frac{p_2p_3}{2} \textrm{ if } \abs{B_1} > \frac D 2\,, \\
            - \frac{1}{2} \sum_{i=1}^3\binom{p_i}{2} + \binom{p_1}{2}\textrm{ otherwise}.
            \end{array}
        \right.
    \end{equation}
    
    \item Multi-entropy: owing to their symmetric construction, the multi-entropies exhibit the following scaling behavior.
    \begin{table}[H]
        \centering
        \begin{tabular}{ |>{\centering\arraybackslash}m{2.5cm}||c|c|c|c|c|c| } \hline
            $D>3$ & $\ket{\GHZ}$ & $\ket{\GHZ\vert_{B}}$ & $\ket{\Phi_{c}}$ or $\ket{\Phi_B}$ & $\ket{\Phi_{\tau}}$  & $\ket{\phi_p}$ & $\ket{\varphi}$  \\ \hline \hline
            $\frac{1}{(1-n)n^{D-2}}\,s_{\ME_{n}^D}$ 
                             & $\sum_{a = 0}^{D-2} n^{-a}$ & $\sum_{a = 0}^{\abs{B}-2} n^{-a}$ & 1 & $\frac D 2$ & $\frac{D}{p} \sum_{a = 0}^{p-2} n^{-a}$ & ?`\,$D-1$\,? \tablefootnote{Based on the study of $\ME_2^3$ in Refs.~\cite{Lionni2018,Penington2023} and supported by numerical analysis, we conjecture $$\Delta(\ME_n^D) = (n-1)n^{D-2}\frac{(D-1)(D-2)}{4} \,.$$} \\ \hline
            $\frac{1}{\ln N}\,\S_{\ME^D}$
                            & $D-1$ & $\abs{B}-1$ & $1$ & $\frac D 2$ & $\frac D p (p-1)$ & ?`\,$D-1$\,? \\ \hline
        \end{tabular}
    \end{table}
    
    \item Reflected multi-entropy: we consider the colored graphs associated with the reflected multi-entropies in the case where the subsystem $c_1$ is traced out. To discuss the scaling of the HT states, we assume that $c_1 \in B$ and $c_1 \notin B'$. Futhermore, since $\S_{\RME^{(c_1)}}(\Phi_{c_1}) = 0$, we restrict our attention to colors $c \neq c_1$. In terms of scalings, the renormalized expression of the Rényi reflected multi-entropies is given by 
    \begin{equation}
        \frac{1}{\ln N}\, \S_{\RME_{m,n}^{(c_1)}} = \frac{1}{(1-n)n^{D-3}}\pa{s_{\RME_{m,n}^{(c_1)}} - n^{D-2} s_{\RME_{m,1}^{(c_1)}}}
    \end{equation}

    \begin{table}[H]
        \centering
        \begin{tabular}{ |>{\centering\arraybackslash}m{2.1cm}||c|c|c|c|c|c|c| } \hline
            $D>3$ & $\ket{\GHZ}$ & $\ket{\GHZ\vert_{B}}$ & $\ket{\GHZ\vert_{B'}}$ & $\ket{\Phi_{c}}$ & $\ket{\Phi_{\tau}}$  & $\ket{\phi_p}$ & $\ket{\varphi}$  \\ \hline \hline
            $\frac{1}{\ln N}\, \S_{\RME_{m,n}^{(c_1)}}$
                            & $\sum_{a = 0}^{D-3} n^{-a}$ & $\sum_{a = 0}^{\abs{B}-3} n^{-a}$ & $2\sum_{a = 0}^{\abs{B}-2} n^{-a}$ & $2\frac{D-2}{D-1}$ & $D-2$ & \eqref{eq:RME_phi_p} & ?`\,$(D-2)\frac{n}{n-1}$\,? \tablefootnote{The expression was computed in analogy with the case $D=3$ without proof. Moreover, by analogy with the procedure of Ref.~\cite{Akers2022} in $D=3$ -- although not proven for $D>3$ -- the extension to an unbalanced Hilbert space is expected to require a modification of the expression to perform, at least naively, its analytic continuation in the balanced case. Namely, we expect $$\frac{1}{(1-n)n^{D-3}}\pa{s_{\RME_{m,n}^{(c_1)}}(\ket{\varphi}) - n^{D-2} s_{\RME_{m,1}^{(c_1)}}(\ket{\varphi})} = (D-2) \, \min \paa{2,\frac{n}{n-1}} \,.$$} \\ \hline
            $\frac{1}{\ln N}\,\S_{\RME^{(c_1)}}$  
                        & $D-2$ & $\abs{B}-2$ & $2(\abs{B}-1)$ & $2 \frac{D-2}{D-1}$ & $D-2$ & $\frac{2}{p}(p-1)D - p$ &  ?`\,$2(D-2)$\,?  \\ \hline
        \end{tabular}
    \end{table}
    
    \begin{equation} \label{eq:RME_phi_p}
        \frac{1}{\ln N} \, \S_{\RME_{m,n}^{(c_1)}}(\ket{\phi_p}) = \sum_{a = 0}^{p-3} n^{-a} + \frac{2}{p} (D-p)\sum_{a = 0}^{p-2} n^{-a} \,.
    \end{equation}
\end{itemize}

As we have seen in this section, the various combinatorial quantities associated to a $D$-colored graph $G$ determine which features of the multipartite entanglement structure of HT states the invariant $\tr_G$ is able to reveal. In other words, the combinatorial quantities shed light on the structural properties of colored graphs (and trace-invariants) that enable the separation of $\LU$-inequivalent HT states. In particular, they demonstrate that Rényi multi-entropies can distinguish the \textit{global structure} of HT states, but not subsystems. Indeed, while $\tr_{\ME_n^D}(\ket{\GHZ_B}) \neq \tr_{\ME_n^D}(\ket{\GHZ_{B'}})$ for $\abs{B} \neq \abs{B'}$, we have $\tr_{\ME_n^D}(\ket{\GHZ_B}) = \tr_{\ME_n^D}(\ket{\GHZ_{B'}})$ whenever $\abs{B} = \abs{B'}$. Furthermore, a naive analytic continuation seems to suggest that, at leading order in the large-$N$ limit, multi-entropies do not distinguish a Haar-random state from a GHZ state.\footnote{The same goes for the von Neumann entropy $\S$ in the bipartite context since $\mean{\S(\ket{\varphi})} - \S(\ket{\Bell}) \underset{N \to \infty}{=} - 1/2 + o(1)$.} Although multi-entropies are prevalent in the literature (see \eg Refs.~\cite{Penington2023,Gadde2023,Harper2024,Iizuka:2025ioc,Iizuka2025BH,Iizuka:2025caq,Gadde2024,Iizuka:2025caq}), it would be interesting to perform a precise analytic continuation to confirm this statement and, if necessary, compute the next-to-leading order correction in the large dimension limit. Regarding Rényi reflected multi-entropies, the same conclusion as for Rényi multi-entropies holds, except that the former can make a partial distinction about the subsystem. In the next section, we will focus on $\LO$ and $\LOCC$ transformations. We will see how the nonnegative property of combinatorial quantities clarifies the potential relations between a subset of HT states considered in this section. Subsequently, the use of trace-invariants will yield precise statements about the $\LO$ and $\LOCC$ transformations between HT states.

\section{Determining \texorpdfstring{$\LO$}{LO} and \texorpdfstring{$\LOCC$}{LOCC} relations between hypergraph-tensor  states by means of trace-invariants} \label{s:LO_LOCC_invTr}

In general, addressing the existence of $\LO$ or $\LOCC$ maps between pairs of quantum states is a difficult problem. We will first show that combinatorial quantities provide a useful tool to rule out the possibility of certain $\LO$ maps between the subset of HT states considered Sec.~\ref{sec:LU-and-ref-states}. More generally, trace-invariants can be employed to obtain concrete results in the classification of $\LO$ transformations between HT states. Finally, this section concludes with a preliminary investigation of $\LOCC$ maps between HT states, by recalling and applying known $\LOCC$ protocols to these states.

\subsection{First combinatorial constraints on \texorpdfstring{$\LO$}{LO} relations between hypergraph-tensor states} \label{subsec:combLO}

Let us start by gathering some partial information about $\LO$ preorder relations among HT states; a complete characterization will be provided in the next subsection.

Introducing combinatorial quantities that characterize the subset of HT states investigated in Sec.~\ref{sec:LU-and-ref-states} enables us to use their known properties to readily exclude the existence of $\LO$ transformations between most of those states, thereby highlighting their practical relevance.  
\begin{prop} \label{prop:LO_by_comb}
    Let $D\geq 2$, $B \subset \{ 1, \ldots , D\}$ with $2 \leq |B| \leq D-1$, and $\tau \in S_D$ a cyclic permutation. Except for the following potentially non-excluded $\LO$ transformations when $D \geq 3$,
    \begin{equation} \label{eq:UnknownLOcomb}
        \ket{\phi_p} \underset{\text{?`}\LO\text{?}}{\longrightarrow} \ket{\GHZ\vert_B} \qquad \rm{and} \qquad \ket{\Phi_\tau} \underset{\text{?`}\LO\text{?}}{\longrightarrow} \ket{\GHZ\vert_B} \,,
    \end{equation}
    none of the HT states $\ket{\GHZ}$, $\ket{\GHZ\vert_B}$, $\ket{\Phi_B}$, $\ket{\Phi_\tau}$, and $\ket{\phi_p}$ are related by $\LO$.
\end{prop}

\begin{proof}
    The proof relies on Cor.~\ref{cor:LO_monotones} and Prop.~\ref{prop:states_not_LO}. Indeed, the non-negativity of the combinatorial quantities introduced in Sec.~\ref{ss:discrimination} allows one to exclude $\LO$ transformations. For instance, we recall Eq.~\eqref{eq:def-KB}:
    \begin{equation}
        s_G(\ket{\GHZ\vert_B}) - s_G(\ket{\GHZ}) = K_B(G) \geq 0\,,
    \end{equation}
    which implies no $\LO$ transformation from (1) $\ket{\GHZ\vert_B}$ to $\ket{\GHZ}$. One can exclude the $\LO$ transformations from (2) $\ket{\GHZ}$ or $\ket{\GHZ\vert_B}$ to $\ket{\Phi_B}$, $\ket{\Phi_\tau}$ or $\ket{\phi_p}$ using the same argument.
    
    Moreover, the existence for any $D$ of a genuinely $D$-partite maximally single trace colored graph (see Sec.~\ref{sec:Trace_Lit_Ref_States}), which satisfies $K_B = 0$ for all subsets $B$, rules out the existence of a $\LO$ transformation from: (3) $\ket{\GHZ\vert_B}$ or $\ket{\GHZ}$ to $\ket{\GHZ\vert_{B'}}$; and (4) from $\ket{\Phi_B}$ to either $\ket{\GHZ}$, $\ket{\GHZ\vert_{B'}}$  or $\ket{\Phi_{B'}}$. From genuinely $D$-partite melonic graphs, we further deduce that: (5) $\ket{\Phi_\tau}$ and $\ket{\phi_p}$ cannot be transformed to $\ket{\GHZ}$ by $\LO$, as well as (6) $\ket{\Phi_\tau}$ cannot me mapped to $\ket{\Phi_{\tau'}}$ by $\LO$. For $2 \leq p \leq q \leq D$, one finds $s_G(\ket{\phi_p}) - s_G(\ket{\phi_q}) \, \propto \, \omega_p^{(q)}(G)$, which vanishes for the genuinely $D$-partite invariant $\JRM^{\vec{i}}_k$ for a cyclic sequence $(1,2,\dots,D)$; this implies that: (7) $\ket{\phi_q} \nottoLO \ket{\phi_p}$ and $\ket{\phi_p} \nottoLO \ket{\phi_q}$.
    
    Evaluating the scaling differences for the transformations $\ket{\Phi_B} \stackrel[]{}{\leftrightarrows} \ket{\Phi_\tau}$ and $\ket{\Phi_B} \stackrel[]{}{\leftrightarrows} \ket{\phi_p}$ on melonic and maximally single trace graphs yields both positive and negative values, thereby excluding any $\LO$ relation among those states (8). Finally, consider $\ket{\Phi_\tau} \stackrel[]{}{\leftrightarrows} \ket{\phi_p}$ for some cyclic permutation $\tau\in S_D$. From Eq.~\eqref{eq:zeta_pq},
    \begin{equation}
        \frac{\omega_{D-1}(G)}{D-1} - g_\tau(G) \leq s_G(\ket{\Phi_\tau}) - s_G(\ket{\phi_p}) \leq \frac{\omega_2(G)}{D-1} - g_\tau(G) \,.
    \end{equation}
    The unique connected cyclic graph $G$ with $k(G)=2$ corresponding to the bipartition $\{A, \bar A \}$ with $A=\paa{1,\tau(1)}$ (see the example on the left of Fig.~\ref{fig:jackNeck} for $\tau = \pa{1\,2 \dots D}$) has $\omega_{D-1}(G) >0$ and $g_\tau(G) = 0$, and therefore $s_G(\ket{\Phi_\tau}) > s_G(\ket{\phi_p})$. On the other hand, the unique connected cyclic graph $H$ with $k(H)=2$ corresponding to the bipartition $\{B, \bar B \}$ with $B = \paa{1,\tau^2(1)}$
     (right of Fig.~\ref{fig:jackNeck}) has $g_\tau(H) = 1$ and $\omega_2(H) = D-3$, so that 
    \begin{equation} \label{eq:neck_neg}
        s_H(\ket{\Phi_\tau}) - s_H(\ket{\phi_p}) \le \frac{\omega_2(H)}{D-1} - g_\tau(H) = - \frac{2}{D-1} < 0 \,.
    \end{equation}
    This rules out any $\LO$ relation between $\ket{\Phi_\tau}$ and $\ket{\phi_p}$ (9).
    
    The conditions (1)--(9) exclude any $\LO$ relation between the said HT states, except for the two tentative relations of Eq.~\eqref{eq:UnknownLOcomb}. 
    
    \begin{figure}[ht]
		\centering
		\includegraphics[height = 3cm]{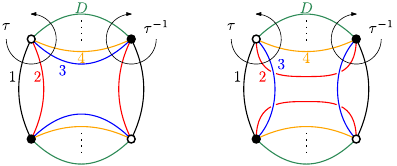}
		\caption{Fixing $\tau = \pa{1\,2 \dots D}$, the cyclic graph $G$ on the left is such that the jacket $G_\tau$ is planar, and $H$ on the right is such that $H_\tau$ has genus one.}
		\label{fig:jackNeck}
    \end{figure}
\end{proof}

As a remark, given two quantum states $\ket{\psi}$ and $\ket{\phi}$, Prop.~\ref{prop:LO_by_comb} shows that
\begin{equation} \label{eq:non_implication1_LO}
    \forall G \in \cG_D^{\conn} \,, \quad \abs{\tr_G(\ket{\psi})} \leq \abs{\tr_G(\ket{\phi})} \qquad \notimplies \qquad \ket{\psi} \toLO \ket{\phi} \,.
\end{equation}
For instance, $\ket{\phi_p} \nottoLO \ket{\GHZ}$ even though $\abs{\tr_G(\ket{\phi_p})} \leq \abs{\tr_G(\ket{\GHZ})}$ for any $G \in \cG_D^{\conn}$. Consequently, albeit effective in excluding certain $\LO$ relations, order relations between combinatorial quantities or, more generally, between trace-invariants, cannot characterize the $\LO$ preorder in full: they only provide necessary conditions. Said differently, the infinite family of $\LO$-monotones $\{ R_G\}_{G \in \cG_D}$ introduced in Ex.~\ref{ex:renyi_higher-D} (which can be interpreted as multipartite generalizations of R\'{e}nyi entanglement entropies) does not characterize the $\LO$-preorder. 

To go beyond Eq.~\eqref{eq:non_implication1_LO}, let us consider the potential $\LO$ map $\ket{\phi_p} \underset{\text{?`}\LO\text{?}}{\longrightarrow} \ket{\GHZ_B}$ introduced in Prop.~\ref{prop:LO_by_comb}. Since all genuinely $D$-partite melonic graphs satisfy $K_B>0$, it follows that 
\begin{equation}
    \forall G\in\cG_D^{\conn,\neq}\,, \quad \abs{\tr_G(\ket{\phi_p})} < \abs{\tr_G(\ket{\GHZ_B})}\,.
\end{equation}
Thus, even when the inequalities associated with trace-invariants are strictly satisfied for all genuinely $D$-partite colored graphs, these conditions are, \lat{a~priori}, merely necessary: they neither guarantee the existence of an $\LO$ transformation nor allow one to rule it out. Nevertheless, by exploiting the properties of trace-invariants in relation to the characterization of the $\LO$ preorder given in Prop.~\ref{prop:charac_LO}, one can recover a complete classification of $\LO$ relations amongst HT states, which is the purpose of the next subsection.

\subsection{Complete characterization of the \texorpdfstring{$\LO$}{LO} preorder on hypergraph-tensor states}\label{sec:complete_LO_ref_states}

To address the characterization problem, we now turn our attention to HT states in general (as introduced in Sec.~\ref{sec:ref_states}), which we will parametrize by their weight functions (as introduced in Eq.~\eqref{eq:def_alpha_states}). In doing so, we will show that a judicious use of trace-invariants allows for a complete characterization of all $\LO$ transformations among HT states, for any number of parties $D$. This will, in particular, enable us to answer the questions raised in the previous subsection, namely Eq.~\eqref{eq:UnknownLOcomb}. 

To begin with, we introduce an order relation on the set of weight functions labeling HT states.
\begin{defi}
    Let $D\ge 2$, $\alpha: \{ B \subset \{ 1, \dots, D\} \,|\, |B| \geq 2\} \to \mathbb{N}^*$ and $\beta : \{ B \subset \{ 1, \dots , D\} \,|\, |B| \geq 2\} \to \mathbb{N}^*$. We say that \emph{$\alpha$ divides $\beta$} -- noted $\alpha \divides \beta$ -- whenever: for any $B \subset\{1,\ldots , D\}$ with $|B|\geq 2$,
    \begin{equation}
        \alpha(B) \divides \beta(B)\,.
    \end{equation}
\end{defi}
\noindent This defines a partial order on the set of weight functions. Our general theorem states that this partial order is, in fact, isomorphic to the $\LO$ partial order on the set of $\LU$-classes of HT states.
\begin{theo} \label{th:LOD}
     Let $D\ge 2$, $\alpha: \{ B \subset \{ 1, \dots, D\} \,|\, |B| \geq 2\} \to \mathbb{N}^*$ and $\beta : \{ B \subset \{ 1, \dots , D\} \,|\, |B| \geq 2\} \to \mathbb{N}^*$. Then:
     \begin{equation}
         \ket{\psi_\beta} \toLO \ket{\psi_\alpha} \qquad \Longleftrightarrow \qquad \alpha \divides \beta \,.
     \end{equation}
\end{theo}

The proof of this result relies on the following property of finite-dimensional Hermitian matrices.

\begin{lem}\label{lem:Resolvent_pap}
    Let $M$ be a finite-dimensional Hermitian matrix and $a \in \mathbb{R}_+^*$. If
    \begin{equation}\label{eq:constraint_moments}
        \forall m \in \mathbb{N}^*\,, \qquad \tr\left( M^m\right) = a^{1-m}\,,
    \end{equation}
    then $a \in \mathbb{N}^*$.
\end{lem}

\begin{proof}
    For any $z \in \bb{C}$ such that $\displaystyle\abs{z} > \max_{\lambda \in \spec(M)} \abs{\lambda}$, we can define the resolvent matrix $R(z) \eqdef \pa{z\id - M}^{-1}$, which satisfies the identities\footnote{The first equality is a consequence of the spectral theorem, which allows us to write $\displaystyle MR(z)= \sum_{\lambda \in \spec(M)} \frac{\lambda}{z - \lambda} P_\lambda$, where for any $\lambda \in \spec(M)$, $P_\lambda$ is the orthogonal projector on the eigenspace with eigenvalue $\lambda$ (which obeys $\tr(P_\lambda)=n_\lambda$). To obtain the second equality, we note that $\| M/z\|<1$ and express $\pa{\id - M/z}^{-1}$ as the convergent power series $\displaystyle\pa{\id - M/z}^{-1}= \sum_{m \in \mathbb{N}} \left( M/ z\right)^m$.}
    \begin{equation}
         \sum_{\lambda \in \spec(M)\setminus \{ 0\}} \frac{\lambda n_\lambda}{z - \lambda} = \tr(M R(z))  = \sum_{m = 1}^{+\infty} \frac{\tr(M^m)}{z^m} \,,
    \end{equation}
    where, for any $\lambda \in \spec(M)$, $n_\lambda \in \bb{N}^*$ denotes the multiplicity of $\lambda$. 
    Now, assume the existence of $a \in \mathbb{R}^*_+$ such that Eq.~\eqref{eq:constraint_moments} holds, the previous identities imply that: for any $z \in \bb{C}$ such that $\displaystyle\abs{z} > \max_{\lambda \in \spec(M)} \abs{\lambda}$,
    \begin{equation}
        \sum_{\lambda \in \spec(M)\setminus \{ 0\}} \frac{\lambda n_\lambda}{z - \lambda}= \sum_{m=1}^{+\infty}  \frac{a^{1-m}}{z^m} = \frac{1}{z - a^{-1}} \,.
    \end{equation}
As a result, we must have $\spec(M) \setminus \{ 0\} = \{a^{-1}\}$ and $n_{a^{-1}} = a$. In particular, $a \in \mathbb{N}^*$.
\end{proof}

Having established the necessary preliminaries, we now prove Thm.~\ref{th:LOD}.

\begin{proof} 
    We start by proving the converse implication. Assume that $\alpha \divides \beta$, \ie $\alpha(B)$ divides $\beta(B)$ for any $B \subset\{1, \ldots, D\}$ with $\abs{B}\geq 2$. We can then define the map $\beta/\alpha:  \{ B \subset \{ 1, \ldots , D\} \,|\, |B| \geq 2\} \to \mathbb{N}^*, \, B \mapsto \frac{\beta(B)}{\alpha(B)}$ and denote by $\ket{\psi_{\beta/\alpha}}$ the associated HT state. According to Eq.~\eqref{eq:trace-inv_alpha-states}, for any colored graph $G  \in \cG_D$ we have
    \begin{equation}
        \tr_G\left(\ket{\psi_\beta}\right) = \tr_G\left(\ket{\psi_\alpha}\right)\tr_G\left(\ket{\psi_{\beta/\alpha}}\right)\,.
    \end{equation}
    By Prop.~\ref{prop:charac_LO}, this factorization property implies the relation $\ket{\psi_\beta}\toLO \ket{\psi_\alpha}$.
    
    Now, let us focus on the direct implication, and suppose that $\ket{\psi_\beta} \toLO \ket{\psi_\alpha}$. By Prop.~\ref{prop:charac_LO}, there exists a $D$-partite pure state $\ket{\eta}$ such that: 
    \begin{equation} \label{eq:LOfactorization}
        \forall G \in \cG_D \,, \qquad \tr_G(\ket{\psi_\beta}) = \tr_G(\ket{\psi_\alpha}) \tr_G(\ket{\eta})\,.
    \end{equation}
    Let $B \subset \{ 1,\ldots, D\}$ with $|B|\geq 2$. Ideally, our goal would be to identify a family of colored graphs $\{ G_m \}_{m \in \mathbb{N}^*}$ associated with moments of a finite-dimensional Hermitian matrix $M$, such that: $\tr_{G_m}(\ket{\eta}) = \tr(M^m) = a^{1-m}$, where $a =\beta(B)/\alpha(B)$. Applying Lem.~\ref{lem:Resolvent_pap} would then imply the desired result, namely: $\alpha(B) \divides \beta(B)$. We will not be able to find a suitable family $\{ G_m \}_{m \in \mathbb{N}^*}$ for arbitrary choices of $B$, which will make the proof slightly more involved, but that is the gist of the idea. We distinguish some cases.
    
    If $D=2$, then $B= \{ 1, 2\}$ and the moments of the Hermitian matrix $\rho_1 \eqdef \tr_{\H_2}(\ket{\eta} \bra{\eta})$ are described by the colored graph $\{C_m\}_{m \in \mathbb{N}^*}$. Therefore, using Eq.~\eqref{eq:LOfactorization}, we have
    \begin{equation}
        \forall m \in \mathbb{N}^*\,, \qquad \tr(\rho_1^m) = \tr_{C_m}(\ket{\eta}) = \pa{\frac{\beta(B)}{\alpha(B)}}^{1-m} \,,
    \end{equation}
    which, by application of Lem.~\ref{lem:Resolvent_pap}, implies that $\alpha(B) \divides \beta(B)$ (and therefore $\alpha \divides \beta$). From now on, let us assume that $D \geq 3$.

    If $B= \{ 1, \ldots , D\}$, consider the Hermitian matrix $\tilde{X}_{\paa{1,\dots,D}}$ defined by the contraction of tensors $\eta$ and $\bar\eta$ represented in the upper-left corner of Fig.~\ref{fig:LOproof},\footnote{One observes that $\tr(\tilde{X}_{\paa{1,\dots,D}})$ is the colored graph $\JRM_{2(D-1)}^{\vec{i}}$ with the cyclic sequence of colors being $\vec i = (1,\,2,\dots,\,D-1,\,D,\,D-1,\dots,\,2)$. The symmetry of the sequence $\vec i$ ensure the Hermitian property of $\tilde{X}_{\paa{1,\dots,D}}$.} and define the normalized (Hermitian) matrix 
    \begin{equation}
        X \eqdef \frac{\tilde{X}_{\paa{1,\dots,D}}}{\tr\pa{\tilde{X}_{\paa{1,\dots,D}}}} \,.
    \end{equation}
    For any $m \in \mathbb{N}^*$, let $G_{1,m}$ (resp.~$\tilde{G}_{1,m}$) denote the colored graph representing $\tr(X^m)$ (resp.~$\tr(\tilde{X}_{\paa{1,\dots,D}}^m)$): hence, we have $\tr(X^m)= \tr_{G_{1,m}}(\ket\eta)$ and $\tr(\tilde{X}^m_{\paa{1,\dots,D}})= \tr_{\tilde{G}_{1,m}}(\ket\eta)$. A direct inspection of the combinatorial structure of $\tilde{G}_{1,m}$ restricted to a subset $C\subset\{1,\dots,D\}$ with $\abs{C}>1$, yields
    \begin{equation} \label{eq:kappa-tilde-G1}
        \kappa(\tilde{G}_{1,m}\vert_C) = 
        \begin{cases}
            1 \, &\rm{ if } C = \paa{1,\dots,D} \,,\\
            \mu_C \, m &\rm{ otherwise,}
        \end{cases}
    \end{equation}
    where $\mu_C \in \mathbb{N}^*$ is a constant (that depends on the subset $C$). After normalization, any contribution proportional to $m$ in Eq.~\eqref{eq:LOfactorization} cancels, leading to
    \begin{equation} \label{eq:trX^k}
        \tr(X^m) = \tr_{G_{1,m}}(\ket{\eta}) = \frac{\tr_{\tilde{G}_{1,m}}(\ket{\eta})}{\left(\tr_{\tilde{G}_{1,1}}(\ket{\eta})\right)^m} =  \pa{\frac{\beta(\paa{1,\dots,D})}{\alpha(\paa{1,\dots,D})}}^{1-m} \,.
    \end{equation}
    By the argument of Lem.~\ref{lem:Resolvent_pap}, we conclude that $\alpha(\paa{1,\dots,D}) \divides \beta(\paa{1,\dots,D})$.
        
    \begin{figure}[!ht]
        \centering
        \includegraphics[width = \textwidth]{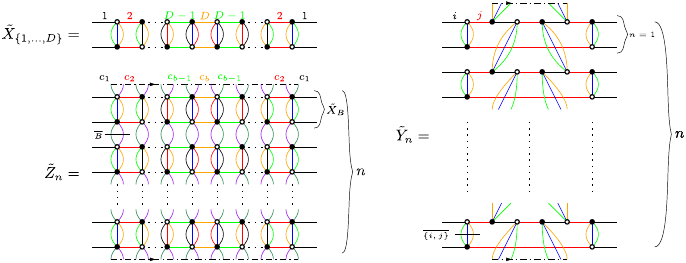}
        \caption{Matrices considered in the proof of Thm.~\ref{th:LOD}. In this figure, each white (resp.~black) vertex represents a tensor $\eta$ (resp.~$\bar\eta$). On the upper left, the matrix $\tilde{X}_{\paa{1,\dots,D}}$, where its relation with the joint realignment moment is easier to see. The latter is particularly drawn for $D=5$. On the right, the matrix $\tilde{Y}_n$ where the upper edges are connected to the bottom ones as exemplified by the dash-dotted lines. Again, here we show the case $D=5$. On the bottom left, the matrix $\tilde{Z}_n$ is presented where, for $B = \paa{c_1,\dots,c_b}$, we stressed the used of $n$ matrices $\tilde{X}_B$ connected by the colors of $\bar{B}$. Again, the upper edges are connected to the bottom ones as exemplified by the dash-dotted lines. On the figure, the matrix $\tilde{Z}_n$ is given for $D=7$.}
        \label{fig:LOproof}
    \end{figure}
    
    If $B = \paa{i,j}$ with $i,j \in \{ 1, \ldots , D\}$ distinct, we consider the one-parameter family of Hermitian unit-trace matrices $\{Y_n\}_{n \in \mathbb{N}^*} = \{\tilde{Y}_n / \tr(\tilde{Y}_n)\}_{n \in \mathbb{N}^*}$ depicted on the right-hand side of Fig.~\ref{fig:LOproof}. For any $n, m \in \mathbb{N}^*$, let $G_{2, n, m}$ (resp.~$\tilde{G}_{2,n,m}$) denote the colored graph associated with $\tr(Y_n^m)$ (resp.~$\tr(\tilde{Y}_n^m)$). A direct combinatorial analysis shows that
    \begin{equation}\label{eq:comb_G2nm}
        \kappa(\tilde{G}_{2,n,m}\vert_C) = 
        \begin{cases}
            2n  &\rm{ if } C = \paa{i,j} \,,\\
            1 &\rm{ if } C \supsetneq \paa{i,j}\,, \\
            \mu_{C,n} m &\rm{ otherwise}\,,
        \end{cases}
    \end{equation}
where $\mu_{C,n}\in \mathbb{N}^*$ is independent of $m$. Normalization again removes any term proportional to $m$ in Eq.~\eqref{eq:comb_G2nm}, yielding
    \begin{equation}
        \tr(Y_n^m) = \tr_{G_{2,n,m}}(\ket{\eta})  = \frac{\tr_{\tilde{G}_{2,n,m}}(\ket{\eta})}{\left(\tr_{\tilde{G}_{2,n,1}}(\ket{\eta})\right)^m} = \pac{ \pa{ \frac{ \beta(\paa{i,j}) }{ \alpha(\paa{i,j}) }}^{2n} \prod_{C \supsetneq \paa{i,j}} \frac{\beta(C)}{\alpha(C)} }^{1-m} \,.
    \end{equation}
    As a consequence of Lem.~\ref{lem:Resolvent_pap}, the quantity in square brackets must always be an integer, which translates into the arithmetic conditions 
    \begin{equation}\label{eq:divides_ij}
         \forall n \in \mathbb{N}^*\,, \qquad \left( \alpha(\paa{i,j})^{2n} \prod_{C \supsetneq \paa{i,j}} \alpha(C) \right) \divides \left( \beta(\paa{i,j})^{2n} \prod_{C \supsetneq \paa{i,j}} \beta(C)\right)\,.
    \end{equation}
 Finally, taking the limit $n \to +\infty$, this implies that $\alpha(\paa{i,j}) \divides \beta(\paa{i,j})$.\footnote{Indeed, for any prime integer $p$ and any $n \in \mathbb{N}^*$, we have from Eq.~\eqref{eq:divides_ij} that
    \begin{equation}
         2n\pac{v_p(\beta(\paa{i,j})) - v_p(\alpha(\paa{i,j}))} + \sum_{C \supsetneq \paa{i,j}} \pac{v_p(\beta(C)) - v_p(\alpha(C))} \geq 0 \,,
    \end{equation}
    where $v_p(\cdot)$ denotes the $p$-adic valuation. 
    Taking the limit $n \to +\infty$ in the previous equation leads to $v_p(\beta(\paa{i,j})) \geq v_p(\alpha(\paa{i,j}))$. But since this must hold for any prime $p$, one must have $\alpha(\paa{i,j}) \divides \beta(\paa{i,j})$.} 
    
    Finally, if $B= \{ c_1 , c_2, \ldots , c_b\}$ is such that $2<b=\abs{B}<D$, we consider the one-parameter family of Hermitian unit-trace matrices $\paa{Z_n}_{n \ge 1}= \{\tilde{Z}_n / \tr(\tilde{Z}_n)\}_{n \in \mathbb{N}^*}$ represented in the bottom-left corner of Fig.~\ref{fig:LOproof}. Given the structure of each $\tilde{Z}_n$, we can view it as the trace of some matrix $\tilde{X}_B$, where $\tilde{X}_B$ is defined similarly as the previously discussed $\tilde{X}_{\{1, \ldots , D\}}$ (see Fig.~\ref{fig:LOproof}), which we can express as:   
    \begin{equation}
        \tilde{Z}_n \eqdef \tr_{\bar{B}}(\tilde{X}_B^n) \,, \qquad \rm{and} \qquad Z_n \eqdef \frac{\tilde{Z}_n}{\tr \pa{\tilde{Z}_n}} \,.
    \end{equation}
    For any $n,m \in \mathbb{N}^*$, let $G_{3,n,m}$ (resp.~$\tilde{G}_{3,n,m}$)\footnote{$\tilde{G}_{3,n,m}$ can be understood as a lattice extension of the joint realignment moment, namely the colored graph $L_{m,n}^{\vec i}$ with $\vec i = (c_1,\dots,c_{\abs{B}-1},c_{\abs{B}},c_{\abs{B}-1},\dots,c_2)$.} denote the colored graph associated with $\tr(Z_n^m)$ (resp.~$\tr(\tilde{Z}_n^m)$). The graph $\tilde{G}_{3,n,m}$ satisfy
    \begin{equation}
        \kappa(\tilde{G}_{3,n,m}\vert_C) = 
        \begin{cases}
            n \,& \rm{ if } C = B \,,\\
            1 \, & \rm{ if } C \supsetneq B\,, \\
            \mu_{C,n} m \,& \rm{ otherwise,}
        \end{cases}
    \end{equation}
where $\mu_{C,n}\in \mathbb{N}^*$ is again independent from $m$.    
As before, normalization eliminates all terms proportional to $m$, and Lem.~\ref{lem:Resolvent_pap} yields 
    \begin{equation}
        \tr(Z_n^m) = \tr_{G_{3,n,m}}(\ket{\eta}) = \frac{\tr_{\tilde{G}_{3,n,m}}(\ket{\eta})}{\left(\tr_{\tilde{G}_{3,n,1}}(\ket{\eta})\right)^m} = \pac{ \pa{ \frac{ \beta(B) }{ \alpha(B)} }^{n} \prod_{C \supsetneq B} \frac{\beta(C)}{\alpha(C)} }^{1-m} \,.
    \end{equation}
By Lem.~\ref{lem:Resolvent_pap}, the quantity appearing in square brackets must be an integer for any $n\in \mathbb{N}^*$, and taking the limit $n \to + \infty$, this implies the condition $\alpha(B) \divides \beta(B)$.
    
    We have shown that $\ket{\psi_\beta} \toLO \ket{\psi_\alpha}$
    implies $\alpha(B)\divides\beta(B)$ for any $B \subset \{1, \ldots , D\}$ with $\abs{B}\geq2$, completing the proof.
\end{proof}

\begin{figure}[ht]
	\centering
	\includegraphics[height = 3cm]{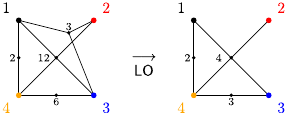}
	\caption{Example of a $\LO$ transformation between HT states.}
	\label{fig:LO}
\end{figure}

As a concrete example, Fig.~\ref{fig:LO} illustrates Thm.~\ref{th:LOD}.

As a first consequence, Thm.~\ref{th:LOD} allows us to go beyond what was inferred from the combinatorial quantities in Sec.~\ref{subsec:combLO}. Indeed, we first observe that if $\abs{B} \neq p$, there can be no $\LO$ map between $\ket{\phi_p}$ and $\ket{\GHZ\vert_B}$, as follows directly from Thm.~\ref{th:LOD}. Similarly, if $\abs{B} \neq 2$, there can be no $\LO$ map from $\ket{\Phi_\tau}$ to $\ket{\GHZ\vert_B}$. 

Moreover, whenever $\abs{B} = p$ (resp.~$\abs{B} = 2$), the conditions of Thm.~\ref{th:LOD} require that $N$ divides $N^{1/\cal{I}_p}$ (resp.~$N$ divides $N^{1/2}$), which is impossible except for the trivial case $N = 1$. Thus, we can conclude the first corollary of Thm.~\ref{th:LOD}:

\begin{cor}
    There are no $\LO$ maps between the HT states $\ket{\GHZ}$, $\ket{\GHZ\vert_B}$, $\ket{\Phi_B}$, $\ket{\Phi_\tau}$, and $\ket{\phi_p}$.
\end{cor}

By including lower-dimensional GHZ states, we find that for any subsets $\{c_1,\dots,c_p\} \subset \{1,\dots,D\}$ and $\{i,j\} \subset \{1,\dots,D\}$, the following $\LO$ maps exist:
\begin{equation}
    \ket{\phi_p} \toLO \lvert\GHZ^{\frac{1}{\cal{I}_p}}\vert_{\paa{c_1,\dots,c_p}}\rangle \qquad \rm{ and } \qquad \ket{\Phi_\tau} \toLO \lvert\GHZ^{\frac{1}{2}}\vert_{\paa{i,j}}\rangle \,.
\end{equation}

The now-excluded $\LO$ map from $\ket{\phi_p}$ to $\ket{\GHZ\vert_B}$ shows that trace-invariants by themselves are not sufficient to characterize $\LO$ order relations, even when genuinely $D$-partite trace-invariants satisfy the inequality of Cor.~\ref{cor:LO_monotones} strictly. This observation leads to the following corollary:
\begin{cor}
    For any $D$-partite pure states $\ket{\psi}$ and $\ket{\phi}$, we have 
    \begin{equation}
        \begin{cases}
            \forall G \in \cG_D^{\conn}\,, \quad \abs{\tr_G(\ket{\psi})} \leq \abs{\tr_G(\ket{\phi})} \\
            \forall G \in \cG_D^{\conn, \neq}\,, \quad \abs{\tr_G(\ket{\psi})} < \abs{\tr_G(\ket{\phi})}
        \end{cases}  \qquad \notimplies \qquad \ket{\psi} \toLO \ket{\phi}\,.
    \end{equation}
\end{cor}

Finally, we note that we can recover the main statement of Thm.~\ref{th:RefState_Equiv} (that weight functions label the $\LU$-orbits of HT states) as a consequence of Thm.~\ref{th:LOD} and Cor.~\ref{cor:LO_classes}. Indeed, given two weight functions $\alpha$ and $\beta$, one has $\ket{\psi_\alpha}\simLU \ket{\psi_\beta}$ if and only if $\ket{\psi_\alpha} \toLO \ket{\psi_\beta}$ and $\ket{\psi_\beta} \toLO \ket{\psi_\alpha}$ (by Cor.~\ref{cor:LO_classes}), and those two conditions are themselves equivalent to $\alpha = \beta$ (by Thm.~\ref{th:LOD}). This provides an alternative proof of this classification result, which is independent from the one given in Sec.~\ref{sub:LU-inequivalent-reference-states}.

\subsection{Towards a characterization of the \texorpdfstring{$\LOCC$}{LOCC} preorder on hypergraph-tensor states} \label{subsubsec:FlowLOCC}

We start by observing that an order relation between two weight functions implies a $\LOCC$ relation between the two $\LU$-classes of HT states they represent. 
\begin{prop}\label{prop:move_LOCC-0}
    Let $D\geq 2$. For any weight functions $\alpha: \{ B \subset \{ 1, \ldots , D\} \,|\, |B| \geq 2\} \to \mathbb{N}^*$ and $\beta: \{ B \subset \{ 1, \ldots , D\} \,|\, |B| \geq 2\} \to \mathbb{N}^*$, one has
    \begin{equation}
\alpha \leq \beta \quad \Rightarrow \quad         \ket{\psi_\beta} \toLOCC \ket{\psi_\alpha}\,.
    \end{equation}
\end{prop}
\begin{proof}
    If $\alpha = \beta$, the statement is trivial; let us therefore assume the existence of $B \subset\{1, \ldots , D\}$ with $p \eqdef |B|\geq 2$ such that $N \eqdef \beta(B)> \alpha(B) \eqdef N'$.
    
    To begin with, let us establish that $\ket{\GHZ}_{B , N} \toLOCC \ket{\GHZ}_{B , N-1}$ (which is illustrated in Fig.~\ref{fig:LOCC_1}).
    
    \begin{figure}[ht]
		\centering
		\includegraphics[height = 2.5cm]{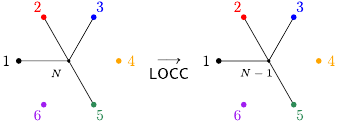}
		\caption{Graphical representation of the elementary step introduced in the proof of Prop.~\ref{prop:move_LOCC-0} for $D = 6$ and $B = \paa{1,2,3,5}$.}
		\label{fig:LOCC_1}
    \end{figure}
    
    Up to an $\LU$ transformation, we can choose to represent those two $\LU$-classes of states on the $|B|$-partite Hilbert space $\H_B = \bigotimes_{c \in B} \H_c$ as:
    \begin{equation}
        \ket{\GHZ}_{B , N} = \frac{1}{\sqrt{N}} \sum_{i=1}^N \ket{i}^{\otimes p}\,, \qquad \ket{\GHZ}_{B , N-1} = \frac{1}{\sqrt{N-1}} \sum_{i=1}^{N-1} \ket{i}^{\otimes p}\,,
    \end{equation}
where the local Hilbert spaces $\{\H_c\}$ are taken to be identical, of dimension $N$, and spanned by the orthonormal basis $(\ket{i})_{1 \leq i \leq N}$.  Now, let us fix $c \in B$ and introduce the following operators on $\H_{c}$: for any $1 \leq j \leq N$,
\begin{equation}
    M_j \eqdef \frac{1}{\sqrt{N - 1}}\left( \id_{\H_{c}}- \ket{j}\bra{j}\right)\,.
\end{equation}
$\{M_j\}_{1 \leq j \leq N}$ defines a generalized measurement on $\H_c$, since
\begin{equation}
    \sum_{j=1}^N M_j^\dagger M_j = \id_{\H_{c}}\,.
\end{equation}
A local observer in subsystem $c$ (say, Alice) may perform this measurement on $\ket{\GHZ}_{B , N}$: for any $j\in \{1,\ldots , N\}$, if outcome $j$ has been obtained, the state is updated to\footnote{On the left-hand side, tensor factors of identity maps acting on subsystems indexed $B \setminus \{ c\}$ are kept implicit to avoid cluttering the equation.} 
\begin{equation}
    \frac{M_j \ket{\GHZ}_{B , N}}{\sqrt{\bra{\GHZ}_{B , N}M_j^\dagger M_j\ket{\GHZ}_{B , N}}} = \frac{1}{\sqrt{N-1}}\sum_{i\in\{1, \ldots , N\}\setminus \{j\}} \ket{i}^{\otimes p} \,, 
\end{equation}
which is $\LU$-equivalent to $\ket{\GHZ}_{B , N-1}$. As a result, if Alice communicates her measurement outcome to local observers of the $p-1$ subsystems labeled by $B \setminus \{c\}$, each such observer can perform a  suitable (outcome-dependent) unitary transformation on his subsystem to guarantee that the state $\ket{\GHZ}_{B , N-1}$ is obtained in all cases. This protocol is clearly in $\LOCC$, therefore $\ket{\GHZ}_{B , N} \toLOCC \ket{\GHZ}_{B , N-1}$.

Iterating the previous protocol immediately yields
\begin{equation}
    \ket{\GHZ}_{B , N} \toLOCC \ket{\GHZ}_{B , N-1} \toLOCC \cdots \toLOCC \ket{\GHZ}_{B , N'}\,,
\end{equation}
and therefore $\ket{\GHZ}_{B , \beta(B)} = \ket{\GHZ}_{B , N} \toLOCC \ket{\GHZ}_{B , N'}=\ket{\GHZ}_{B , \alpha(B)}$ by transitivity.

Finally, we can apply the previous arguments successively to any $B \subset\{1, \ldots , D\}$ with $|B|\geq 2$ such that $\beta(B)>\alpha(B)$, leading (again by transitivity) to the desired result: $\ket{\psi_\beta}\toLOCC \ket{\psi_\alpha}$. 
\end{proof}

\begin{rem}
    When $D=2$, the $\LOCC$ preorder is characterized by majorization conditions on entanglement spectra (see Ref.~\cite{Nielsen:1999zza}). In our setup, this result implies the equivalence relation: 
    \begin{equation}
    \forall \alpha, \beta \in \mathbb{N}^*  \,, \qquad   \alpha \leq \beta \quad \Longleftrightarrow \quad  \ket{\psi_\beta} \toLOCC \ket{\psi_\alpha}\,,
    \end{equation}    
where $\ket{\psi_\beta} \simLU \ket{\Bell}_{\beta}$ and $\ket{\psi_\alpha} \simLU \ket{\Bell}_{\alpha}$. On the other hand, as the next results will show, the implication of Prop.~\ref{prop:move_LOCC-0} cannot be upgraded to an equivalence when $D \geq 3$.
\end{rem}

\medskip

In the remainder of this subsection, we illustrate how familiar $\LOCC$ protocols act on HT states, leaving the full characterization of the $\LOCC$ preorder on HT states as an open question for future work.

\paragraph{Local measurement reduction.} Let $D \geq 2$. Given two subsets $C \subsetneq B \subset \paa{1,\dots,D}$ with $|C| \geq 2$, Thm.~\ref{th:LOD} implies that, for any $N\geq 2$, we must have 
\begin{equation}
    \ket{\GHZ}_{B, N} \nottoLO  \ket{\GHZ}_{C, N}\,.
\end{equation}
However, it is known (see \eg Ref.~\cite{Bennett2000}) that such states are related via $\LOCC$ transformations, \ie 
\begin{equation}
    \ket{\GHZ}_{B, N} \toLOCC  \ket{\GHZ}_{C, N} \,.
\end{equation}
To determine what this type of relation implies for our full set of HT states, we need to introduce further structure and definitions. Let us consider the poset for the inclusion $(\pos_D, \subset ) \eqdef \left( \{ B \subset \{ 1, \ldots , D\} \, | \, |B|\geq 2\} , \subset \right)$, and denote by $H_D$ its Hasse diagram. Hence, $H_D$ is a directed graph with $|\pos_D| = 2^D - (D+1)$ vertices (labeled by the elements of $\pos_D$), which has a directed edge from $B$ to $C$ if and only if: $C \subset B$ and $|C|= |B|-1$. Let us denote by $\cE(H_D)$ the set of edges of $H_D$, and use the notation $(B,C) \in \cE(H_D)$ whenever there is an edge from $B$ to $C$. We can then define a \emph{flow} $\gamma$ on $H_D$ to be an assignment of integer weight $\gamma(B,C)\in \mathbb{N}^*$ to any edge $(B,C) \in \cE(H_D)$. Given a flow $\gamma$, we further introduce the notion of local ingoing (resp.~outgoing) flow to (resp.~from) a vertex:
\begin{equation}
  \forall B \in \pos_D\,, \qquad \gamma_{\iin}(B) \eqdef \prod_{(C,B) \in \cE(H_D)} \gamma(C,B) \quad \mathrm{and} \quad 
        \gamma_{\oout}(B) \eqdef \prod_{(B,C) \in \cE(H_D)} \gamma(B,C)
\end{equation}
Note, in particular, that:
\begin{equation}\label{eq:global_constr_flow}
    \prod_{B\in \pos_D} \gamma_{\iin}(B) = \prod_{(B,C) \in \cE(H_D)} \gamma(B,C) = \prod_{B\in \pos_D} \gamma_{\oout}(B)\,,
\end{equation}
as a consequence of the fact that any edge has exactly one ingoing vertex and one outgoing vertex. 

With this definition at hand, we can introduce a new partial order relation on the set of weight functions, which we denote 
\begin{equation}
    W_D \eqdef \{ \alpha : \pos_D \to \mathbb{N}^* \}\,.
\end{equation}
\begin{defi}
    For any weight functions $\alpha , \beta \in W_D$, we will say that $\alpha \preceq \beta$ whenever there exists a flow $\gamma$ on $H_D$ such that
    \begin{equation}\label{eq:def_order_flow}
        \forall B \in \pos_D\,, \qquad \alpha(B) \gamma_{\oout} (B) = \beta(B) \gamma_{\iin} (B)\,. 
    \end{equation}
    $\preceq$ defines a \emph{partial order}\footnote{$\preceq$ is clearly reflexive and transitive. To check that it is antisymmetric, assume that $\alpha \preceq \beta$ and $\beta \preceq \alpha$, and let $\gamma^{(1)}$ and $\gamma^{(2)}$ denote the two flows on $H_D$ underlying these relations; one then finds: $\forall B \in \pos_D$, $\gamma_{\iin}^{(1)}(B)\gamma_{\iin}^{(2)}(B) = \gamma_{\oout}^{(1)} (B)\gamma_{\oout}^{(2)} (B)$. Focusing first on $B = \paa{1,\dots,D}$, which has no ingoing edge (\ie $\gamma_{\iin}^{(1)}(B) = \gamma_{\iin}^{(2)}(B) = 1$) leads to $\gamma_{\oout}^{(1)}(B) = \gamma_{\oout}^{(2)}(B) = 1$. Equivalently, for both flows, any edge from $B$ to $C$ with $\abs{C} = D-1$ has a unit weight. A downward induction on the size of $B$ leads to $\gamma^{(1)}(B) = \gamma^{(2)}(B) = 1$ for any $B \in \pos_D$, \ie $\alpha = \beta$.} on $W_D$. 
\end{defi}

\begin{rem}
    Note, in particular, that 
    \begin{equation}\label{eq:flow_cond-1}
        \alpha \preceq \beta \quad \Rightarrow \quad \prod_{B \in \pos_D} \alpha(B) = \prod_{B \in \pos_D} \beta(B)\,,
    \end{equation}
as a consequence of \eqref{eq:global_constr_flow}.    
More generally, for any $B\in \pos_D$, the elements of the upper set $\uparrow\{B\} \eqdef \{ C \in \pos_D \, | \, C \supset B\}$ only have outgoing arrows to elements that are not in $\uparrow\{B\}$; as a result, we have:
\begin{equation}\label{eq:flow_cond-2}
        \alpha \preceq \beta \quad \Rightarrow \quad  \forall B \in \pos_D\,, \; \prod_{C \supset B} \alpha(C)  \divides  \prod_{C \supset B} \beta(C) \,.
\end{equation}
\end{rem}

We then have the following result.
\begin{prop} \label{prop:LOCC_second_order-rel}
    Let $D\geq 2$ and $\alpha, \beta \in W_D$ two weight functions. We then have:
        \begin{equation}\label{eq:LOCC_second_order-rel}
\alpha \preceq \beta \quad \Rightarrow \quad         \ket{\psi_\beta} \toLOCC \ket{\psi_\alpha}\,.
    \end{equation}
\end{prop}
\begin{proof}
Let us assume that $\alpha \preceq \beta$, and let $\gamma$ be a flow such that Eq.~\eqref{eq:def_order_flow} holds. We will prove that $\ket{\psi_\beta} \toLOCC \ket{\psi_\alpha}$ by induction on
   \begin{equation}
       p\eqdef \#\{(B,C) \in \cE(H_D) \,| \, \gamma (B,C ) \neq 1\}\,. 
   \end{equation}
    
    If $p = 0$, $\alpha = \beta$ and one trivially has $\ket{\psi_\beta} \toLOCC \ket{\psi_\alpha}$.
    
    Suppose that $p \geq 1$ and that the implication \eqref{eq:LOCC_second_order-rel} holds for any flow $\tilde{\gamma}$ such that $\#\{(B,C) \in \cE(H_D) \,| \, \tilde{\gamma} (B,C ) \neq 1\} = p-1$. Given that $p \geq 1$, we can find $B, C \subset\{1, \ldots ,D \}$ with $|C|= |B|-1 \geq 2$ such that: $N \eqdef \gamma(B,C)\geq 2$. Let us start by proving that:  
    \begin{equation}\label{eq:LOCC_B_C}
        \ket{\GHZ}_{B,N} \toLOCC \ket{\GHZ}_{C,N} \,,
    \end{equation} 
    which is illustrated in Fig.~\ref{fig:LOCC_2}.
    \begin{figure}[ht]
		\centering
		\includegraphics[height = 2.5cm]{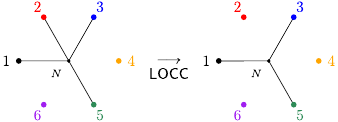}
		\caption{Graphical representation of the elementary step introduced in the proof of Prop.~\ref{prop:LOCC_second_order-rel} for $D = 6$, $B = \paa{1,2,3,5}$ and $C = \paa{1,3,5}$.}
		\label{fig:LOCC_2}
    \end{figure}
    
    To this effect, let $c \in B \setminus C$ denote the unique subsystem that is in $B$ but not in $C$; we can introduce the Fourier basis $\{+_k\}_{k\in\paa{1,\dots,N}}$ in $\H_c$, defined by
    \begin{equation}
\forall k\in\paa{1,\dots,N}\,, \qquad         \ket{+_k} \eqdef \frac{1}{\sqrt{N}} \sum_{j=1}^N \theta_N^{jk} \ket{j} \,, \qquad \rm{ with } \qquad \theta_{N} \eqdef \e{2i\pi/N}\,.
    \end{equation}
   One can express $\ket{\GHZ}_{B,N}$ in terms of this new basis as
    \begin{equation}
        \ket{\GHZ}_{B,N} = \frac{1}{\sqrt{N}} \sum_{j=1}^N \ket{j}^{\otimes |B|} =  \frac{1}{\sqrt{N}} \sum_{k=1}^N \pac{\frac{1}{\sqrt{N}} \sum_{j=1}^N \theta_N^{-jk} \ket{j}^{\ot (\abs{B}-1)}} \ot \ket{+_k} \,.
    \end{equation}
Now, a local observer in subsystem $c$ (say Alice) may perform the projective measurement $\{P_k \eqdef \ket{+_k}\bra{+_k} \}_{k \in \{1, \ldots , N\}}$ on $\ket{\GHZ}_{B,N}$. If outcome $k_0 \in \paa{1,\dots,N}$ is recorded by Alice, the post-measurement state is
    \begin{equation}
         \frac{1}{\sqrt{N}} \sum_{j=1}^N \theta_N^{-jk_0} \ket{j}^{\ot (\abs{B}-1)} \ot \ket{+_{k_0}} \,.
    \end{equation}
    Let us fix a second subsystem $c' \in C = B \backslash \paa{c}$. If made aware of the outcome $k_0$ through classical communication from Alice, a local observer in subsystem $c'$  (say Bob) may perform the unitary transformation $U_{c'}^{(k_0)}$ on $\H_{c'}$ defined as
    \begin{equation}
        U_{c'}^{(k_0)} \eqdef \sum_{l=1}^N \theta_N^{lk_0} \ket{l} \bra{l} \,.
    \end{equation}
    This has the effect of correcting the phase appearing after Alice's measurement, yielding $\ket{\GHZ}_{C,N}$ as the end-state of this $\LOCC$ protocol. This establishes Eq.~\eqref{eq:LOCC_B_C}, which in turn implies that
    \begin{equation}\label{eq:LOCC_beta_beta-tilde}
        \ket{\psi_\beta} \toLOCC \vert\psi_{\tilde{\beta}}\rangle\,,
    \end{equation}
    where $\tilde{\beta}$ is a new weight function defined by
    \begin{equation}
        \begin{cases}
            \tilde{\beta}(B) \eqdef \beta(B)/N \\
            \tilde{\beta}(C) \eqdef \beta(C) N \\
         \tilde{\beta}(B') \eqdef \beta(B')\,, \quad \forall B' \in \pos_D \setminus \{B,C\}  
        \end{cases}
    \end{equation}
Let us finally introduce a new flow $\tilde{\gamma}$, defined as
\begin{equation}
        \begin{cases}
            \tilde{\gamma}(B,C) \eqdef \gamma(B,C)/N= 1 \\
         \tilde{\gamma}(B', C') \eqdef \gamma(B', C')\,, \quad \forall (B',C') \in \cE(H_D) \setminus \{(B,C)\} 
        \end{cases}
    \end{equation}
One then has
\begin{equation}
        \forall B' \in \pos_D\,, \qquad \alpha(B') \tilde{\gamma}_{\oout} (B') = \tilde{\beta}(B) \tilde{\gamma}_{\iin} (B') 
    \end{equation}
and
\begin{equation}
    \#\{(B',C') \in \cE(H_D) \,| \, \tilde{\gamma} (B',C' ) \neq 1\} = p-1\,.
\end{equation}
We can thus apply the induction hypothesis to infer that
\begin{equation}
    \ket{\psi_{\tilde{\beta}}} \toLOCC \ket{\psi_\alpha}\,.
\end{equation}
Together with Eq.~\eqref{eq:LOCC_beta_beta-tilde}, this yields the looked for relation $\ket{\psi_{\beta}} \toLOCC \ket{\psi_\alpha}$.
\end{proof}

\begin{ex} \label{ex:LOCC}
    With $D =3$, consider the weight functions $\alpha , \beta \in W_3$ defined as follows:
    \begin{align}
        \beta(\{ 1,2,3\}) = 6\,, \quad \beta(\{ 1,2\}) = 2\,, \quad \beta(\{ 1,3\}) = \beta(\{ 2,3\}) = 1\,,\\
        \alpha(\{ 1,2,3\}) = 2\,, \quad \alpha(\{ 1,2\}) = 6\,, \quad \alpha(\{ 1,3\}) = \alpha(\{ 2,3\}) = 1\,.
    \end{align}
    We have the relation $\alpha \preceq \beta$, which can be established by introducing the following flow $\gamma$ on $H_3$:
    \begin{equation}
        \gamma(\{1,2,3\}, \{1,2\}) = 3\,, \quad \gamma(\{1,2,3\}, \{1,3\}) = \gamma(\{1,2,3\}, \{2,3\}) = 1\,.  
    \end{equation}
    As a result, we may conclude that $\ket{\psi_\beta} \toLOCC \ket{\psi_\alpha}$. Moreover, $\alpha \leq \beta$ does not hold (and neither does $\beta \leq \alpha$). This simple example, therefore, confirms that the reciprocal of Prop.~\ref{prop:move_LOCC-0} does not hold when $D\geq 3$. See Fig.~\ref{fig:Hasse} for a graphical illustration of this example. 
    \begin{figure}[ht]
		\centering
		\includegraphics[height = 3cm]{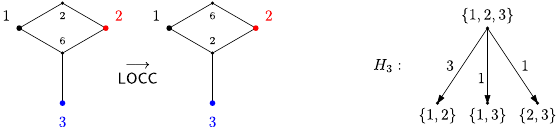}
		\caption{Left: The $\LOCC$ transformation described in Ex.~\ref{ex:LOCC} and illustrating Prop.~\ref{prop:LOCC_second_order-rel}. Right: The associated flow on the Hasse diagram $H_3$.}
		\label{fig:Hasse}
    \end{figure}
\end{ex}

A notable feature of the protocol of Prop.~\ref{prop:LOCC_second_order-rel} is that it preserves the monotonicity of all trace-invariants. Indeed, if $\alpha \preceq \beta$ and $\gamma$ is a flow realizing this relation, then: for any $G \in \cG_D$,
\begin{equation}
    \frac{\abs{\tr_G(\ket{\psi_\alpha})}}{\abs{\tr_G(\ket{\psi_\beta})}} = \prod_{B \in \pos_D} \left( \frac{\gamma_{\iin}(B)}{\gamma_{\oout}(B)} \right)^{k(G)- \kappa(G\vert_B)} = \prod_{(C,B) \in \cE(H_D)} \gamma(C,B)^{\kappa(G\vert_B) - \kappa(G\vert_C)} \geq 1\,,
\end{equation}
since for any $(C,B)\in \cE(H_D)$, $\gamma(C,B) \geq 1$ and $\kappa(G\vert_B) \geq \kappa(G\vert_C)$ (given that $C\subset B$). Hence, the modulus of any trace-invariant is non-decreasing under this particular type of $\LOCC$ transformation, as it is for an arbitrary $\LO$ transformation (see Cor.~\ref{cor:LO_monotones}). Equivalently, the (generalized) R\'enyi entropies introduced as $\LO$ monotones in Exs.~\ref{ex:Rényi} and \ref{ex:renyi_higher-D} are also non-decreasing under this special class of $\LOCC$ transformations.

\paragraph{Quantum teleportation.} The quantum teleportation protocol provides a mechanism by which a collection of GHZ states shared between a reference subsystem and each of the remaining subsystems can be merged into a single multipartite GHZ state using only $\LOCC$ operations. This mechanism extends naturally to the class of HT states considered in this paper.

\begin{prop} \label{prop:LOCC_QT}
    Let $D\geq 2$ and $B_1, B_2 \in \pos_D$ such that $B_1 \cap B_2 \neq \emptyset$. For any weight functions $\alpha, \beta \in W_D$ and integer $N \in \bb{N}^*$ obeying the conditions
    \begin{equation}\label{eq:LOCC-3}
        \beta(C) = 
        \begin{cases}
            N \alpha(C)\; \rm{ if }\, C \in  \{ B_1 , B_2 \}  \\
            \alpha(C)/N \; \rm{ if }\, C = B_1 \cup B_2 \;,\\
            \alpha(C) \; \rm{ otherwise} 
        \end{cases}
    \end{equation}
    one has: $\ket{\psi_\beta} \toLOCC \ket{\psi_\alpha}$.
\end{prop}

\begin{proof}
    Let us fix $B_1, B_2 \in \pos_D$, $\alpha , \beta \in W_D$ and $N \in \mathbb{N}^*$ obeying the conditions \eqref{eq:LOCC-3}, and define $B \eqdef B_1 \cup B_2$. By assumption, one can write 
    \begin{align}
        \ket{\psi_\beta} &= \ket{\GHZ}_{B_1,N\alpha(B_1)} \ot \ket{\GHZ}_{B_2,N\alpha(B_2)} \bigotimes_{C \in \pos_D \setminus \{B_1, B_2\}} \ket{\GHZ}_{C,\beta(C)} \\
        &\simLU \ket{\GHZ}_{B_1,N} \ot \ket{\GHZ}_{B_2,N} \bigotimes_{C \in \pos_D \setminus \{B_1, B_2\}} \ket{\GHZ}_{C,\beta(C)} \ot \ket{\GHZ}_{B_1,\alpha(B_1)} \ot \ket{\GHZ}_{B_2,\alpha(B_2)}\,.
    \end{align}
    We then start by establishing the following $\LOCC$ relation 
    \begin{equation}
        \ket{\psi_\rm{ini}} \eqdef \ket{\GHZ}_{B_1,N} \ot \ket{\GHZ}_{B_2,N} \toLOCC \ket{\GHZ}_{B,N} \,,
    \end{equation}
    as illustrated in Fig.~\ref{fig:QTLOCC}. The proof of Prop.~\ref{prop:LOCC_QT} follows straightforwardly from this elementary step.
    
    \begin{figure}[ht]
		\centering
		\includegraphics[height = 2.5cm]{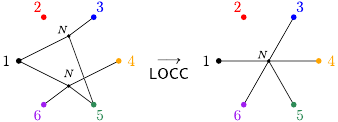}
		\caption{Graphical representation of the elementary step introduced in the proof of Prop.~\ref{prop:LOCC_QT} for $D=6$, $B_1 = \paa{1,3,5}$, $B_2 = \paa{1,4,5,6}$ and then $B = \paa{1,3,4,5,6}$.}
		\label{fig:QTLOCC}
    \end{figure}
    
    For that matter, let us fix a subsystem $c \in B_1 \cap B_2$ and suppose that a local observer in $c$ (say Alice) prepares a separable state $\ket{\psi}_{\H_{c,\rm{sep}}}$ in an ancilla Hilbert space $\H_{c,\rm{sep}}$. We have the $\LU$-equivalence relation
    \begin{equation}
        \ket{\psi_\rm{ini}} \simLU \ket{\GHZ}_{B_1,N} \ot \ket{\GHZ}_{B_2,N} \ot \ket{\psi}_{\H_{c,\rm{sep}}} \in  \H_{c,1} \ot \H_{c,2} \ot \H_{c,\rm{sep}} \,,
    \end{equation}
    where, $\H_c = \H_{c,1} \ot \H_{c,2} \ot \H_{c,\rm{sep}}$ is the fine-grained Hilbert space associated with subsystem $c$, and $\H_{c,1}, \H_{c,2}$ are both taken to be $N$-dimensional. To implement the quantum teleportation protocol, Alice must prepare a $\GHZ$ state in her local laboratory: namely, one assumes the ancilla Hilbert space to have the tensor structure $\H_{c,\rm{sep}} = \H_{c,\rm{sep}_0} \ot \H_{c,\rm{sep}_1} \ot \H_{c,\rm{sep}_2}$, where each tensor factor has dimension $N$, and
    \begin{equation} \label{eq:ini_QT}
        \ket{\psi}_{\H_{c,\rm{sep}}} \eqdef \pac{\frac{1}{\sqrt{N}} \sum_{j=1}^N \ket{j}_{\H_{c,\rm{sep}_0}} \ot \ket{j}_{\H_{c,\rm{sep}_1}} \ot \ket{j}_{\H_{c,\rm{sep}_2}}} \,.
    \end{equation}
    
    We now introduce the generalized Bell basis states defined, for any $(m,n) \in \paa{1,\dots,N}^2$, as
    \begin{equation}
        \ket{\Phi_{m,n}} \eqdef \frac{1}{\sqrt{N}} \sum_{k = 1}^{N} \theta_{N}^{kn} \ket{k} \ot \ket{k+m} \,,
    \end{equation}
    where, here and in the rest of the proof, we identify $N+1$ with $1$ such that whenever $N < k+m \leq 2N$ we identify $k+m$ with $k+m - N$. Introducing the (unitary) Pauli operator $X$ defined by $X\ket{i} = \ket{i+1}$ for any $i\in\paa{1,\dots,N}$, one can express the initial quantum state in terms of the Bell basis states as
    \begin{align}
        \ket{\psi_\rm{ini}} &\simLU \ket{\GHZ}_{B_1,N} \ot \ket{\GHZ}_{B_2,N} \ot \pac{\frac{1}{\sqrt{N}} \sum_{j=1}^N \ket{j}_{\H_{c,\rm{sep}_0}} \ot \ket{j}_{\H_{c,\rm{sep}_1}} \ot \ket{j}_{\H_{c,\rm{sep}_2}}} \\
        &= \frac{1}{\sqrt{N}^3} \sum_{j,a,b = 1}^N \ket{a}^{\ot (\abs{B_1} - 1)} \ot \ket{b}^{\ot (\abs{B_2} - 1)} \ot \pa{\ket{a}_{\H_{c,1}} \ot \ket{j}_{\H_{c,\rm{sep}_1}}} \ot \pa{\ket{b}_{\H_{c,2}} \ot \ket{j}_{\H_{c,\rm{sep}_2}}} \ot \ket{j}_{\H_{c,\rm{sep}_0}} \,, \\
        &= \frac{1}{\sqrt{N}^5} \sum_{\substack{j = 1 \\ m_1,n_1 = 1 \\ m_2, n_2 = 1}}^N  \theta_N^{(m_1-j)n_1 + (m_2-j)n_2} \bigotimes_{v = 1}^2 \pac{(X^{-m_v}\ket{j})^{\ot (\abs{B_v} - 1)} \ot \ket{\Phi_{m_v,n_v}}_{\H_{c,v}\ot\H_{c,\rm{sep}_v}} } \ot \ket{j}_{\H_{c,\rm{sep}_0}} \,.
    \end{align}
    Alice may then perform a local projective measurement on the tensor product space  $(\H_{c,1}\ot\H_{c,\rm{sep}_1})\ot(\H_{c,2}\ot\H_{c,\rm{sep}_2})$, defined by the $N^4$ mutually orthogonal projectors: 
    \begin{equation}
    \forall m_1, n_1, m_2, n_2 \in \mathbb{N}^*\,, \qquad     P_{m_1,n_1,m_2,n_2} \eqdef \ket{\Phi_{m_1,n_1}}\bra{\Phi_{m_1,n_1}} \ot \ket{\Phi_{m_2,n_2}}\bra{\Phi_{m_2,n_2}} \,.
    \end{equation}
    Assuming the outcome to be $(m_1,n_1,m_2,n_2) \in \paa{1,\dots,N}^4$, the post-measurement state is
    \begin{equation}
        \frac{1}{\sqrt{N}}\sum_{j=1}^N \theta_N^{(m_1-j)n_1 + (m_2-j)n_2} \bigotimes_{v=1}^2\pac{(X^{-m_v} \ket{j})^{\ot \abs{B_v}-1}  \ot\ket{\Phi_{m_v,n_v}}_{\H_{c,v}\ot\H_{c,\rm{sep}_v}} } \ot \ket{j}_{\H_{c,\rm{sep}_0}} \,.
    \end{equation}
    Once local observers in every subsystem of $B_1 \backslash \paa{c}$ (resp.~$B_2\backslash\paa{c}$) are made aware of the outcome by classical communication, they can each apply the unitary correction $X^{m_1}$ (resp.~$X^{m_2}$) in their local laboratory. Finally, Alice may perform an $\LU$ operation to correct the phase (see the $\LU$ correction applied in the proof of Prop.~\ref{prop:LOCC_second_order-rel}). At the end of this $\LOCC$ protocol, the resulting quantum state is: \begin{equation}\ket{\GHZ}_{B,N} \otimes \ket{\Phi_{m_1,n_1}}_{\H_{c,1}\ot\H_{c,\rm{sep}_1}} \otimes \ket{\Phi_{m_2,n_2}}_{\H_{c,2}\ot\H_{c,\rm{sep}_2}}  \simLU \ket{\GHZ}_{B,N}\,.\end{equation}
This concludes the proof.    
\end{proof}

\begin{figure}[ht]
	\centering
	\includegraphics[width = \textwidth]{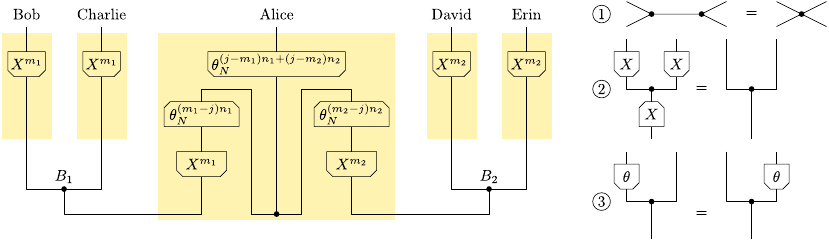}
	\caption{Left: Penrose diagram illustrating the quantum teleportation protocol for five subsystems symbolized by the local observers Alice, Bob, Charlie, David, and Erin. The gold regions represent their respective laboratories into which one first identifies the dot in the bottom center of Alice's laboratory as the prepared GHZ state in $\H_{c,\rm{sep}}$. Moreover, the measurement in Alice's laboratory is illustrated by the $\LU$ transformations (namely, $X^{m_1}$, $X^{m_2}$, and the phase correction) it generates on $B_1$ and $B_2$. Finally, the $\LU$-corrections applied by the local observers after the measurement (either $X^{m_1}$, $X^{m_2}$, or the phase correction for Alice) are depicted at the top of each laboratory. Since the $\LU$-corrections for Bob, Charlie, David, and Erin depend on Alice's measurement, classical communication is needed, proving diagrammatically that the protocol is indeed $\LOCC$. Right: Conventions used in the picture that allow one to recover the GHZ state on $B_1 \cup B_2$, \ie on the system shared by Alice, Bob, Charlie, David, and Erin.}
	\label{fig:QT_Penrose}
\end{figure}

In its diagrammatic form, \ie using the Penrose graphical notation introduced in Ref.~\cite{Coecke2009}, the protocol of Prop.~\ref{prop:LOCC_QT} can be depicted as shown in Fig.~\ref{fig:QT_Penrose}.

\

Albeit more subtle than for Prop.~\ref{prop:LOCC_second_order-rel}, one may again verify that the $\LO$ monotonicity property of trace-invariants (Cor.~\ref{cor:LO_monotones}) extends to the particular $\LOCC$ quantum teleportation protocol on HT states invoked in Prop.~\ref{prop:LOCC_QT}. As a matter of fact, assuming that the conditions of Prop.~\ref{prop:LOCC_QT} are satisfied, \ie $\ket{\psi_\beta} \toLOCC \ket{\psi_\alpha}$ by quantum teleportation, one finds that for any colored graph $G \in \cG_D$ 
\begin{equation}
    \frac{\abs{\tr_G(\ket{\psi_\alpha})}}{\abs{\tr_G(\ket{\psi_\beta})}} = N^{\kappa(G\vert_B) + k(G) - \kappa(G\vert_{B_1}) - \kappa(G\vert_{B_2})} \geq 1 \,.
\end{equation}
This is due to the following ``triangle inequality'' that holds for any colored graph $G$:
\begin{equation} \label{eq:ToProofQT_TI}
    \kappa(G\vert_B) \geq \kappa(G\vert_{B_1}) + \kappa(G\vert_{B_2}) - k(G) \,.
\end{equation}
To see this, one can construct an abstract bipartite graph $H$ whose white vertices (resp.~black vertices) represent the connected components of $G\vert_{B_1}$ (resp.~$G\vert_{B_2}$). In particular, the number of vertices of $H$ is $V(H) = \kappa(G\vert_{B_1}) + \kappa(G\vert_{B_2})$. To define the edges of $H$, we note that any white vertex $v$ of $G$ belongs to a unique connected component $C_1$ of $G\vert_{B_1}$ and a unique connected component $C_2$ of $G\vert_{B_2}$: we represent this situation by an edge connecting $C_1$ and $C_2$ in $H$. As a result of this construction, $H$ has $E(H)= k(G)$ edges, and it is also clear that it is has $\kappa(H)= \kappa(G\vert_{B})$ connected components. Henceforth, the quantity
\begin{equation}
    L(H) \eqdef E(H) - V(H) + \kappa(H) = \kappa(G\vert_B) - \left( \kappa(G\vert_{B_1}) + \kappa(G\vert_{B_2}) \right) - k(G)
\end{equation}
computes the number of loops of the graph $H$, and is therefore non-negative. This yields Eq.~\eqref{eq:ToProofQT_TI}.

\

\begin{figure}[ht]
	\centering
	\includegraphics[height = 3.7cm]{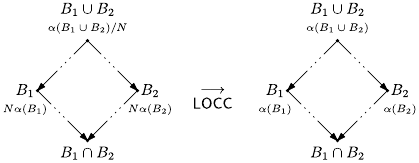}
	\caption{Illustration of the effect of the quantum teleportation protocol at the level of the Hasse diagram. The dotted lines represent omitted parts of the Hasse diagram.}
	\label{fig:losange}
\end{figure}

Finally, we note that the content of Prop.~\ref{prop:LOCC_QT} can be conveniently visualized in the Hasse diagram $H_D$. As highlighted by Fig.~\ref{fig:losange}, the quantum teleportation protocol authorizes to move weights upstream in the Hasse diagram with $B_1$, $B_2$, $B_1 \cup B_2$ and  $B_1 \cap B_2$ forming a diamond. 
This is to be contrasted to the protocol of Prop.~\ref{prop:LOCC_QT}, which only allows us to move weights downstream of the Hasse diagram. At the moment, it is not clear to us how to efficiently describe the net effect of iterated quantum teleportation protocols at the level of the Hasse diagram (as we did in Prop.~\ref{prop:LOCC_QT} for iterated applications of local measurement reductions). We leave this problem, as well as the broader question of how to fully characterize the $\LOCC$ preorder on HT states, open for future work. 

\section{Binary operation and tree construction of trace-invariants}\label{s:tree}

\subsection{Binary operations} \label{ss:operation}

This section aims to give a concrete way to compute and track the combinatorial quantities appearing 
in the distinction conditions imposed on trace-invariants. 

Indeed, computing the degree of compatibility and the other combinatorial quantities associated with a colored graph might be a challenging task. However, in the presence of specific binary operations denoted by
\begin{align}
	\cal{O} \colon \cG_D \times \cG_D & \longrightarrow \cG_D \\
	(G_1,G_2) & \longmapsto G_1 \cal{O} G_2 \nonumber 
\end{align}
the values of, \eg $\omega_p(G_1), \Delta(G_1), \mu_{G_1}$ and $ \omega_p(G_2), \Delta(G_2), \mu_{G_2}$ can be used directly to infer those of the combined object $\omega_p(G_1 \, \cal{O} \,G_2) , \Delta(G_1 \, \cal{O} \, G_2), \mu_{G_1 \, \cal{O} \, G_2}$. This observation motivates the search for binary operations allowing for tractable expressions of those quantities. Among the most prominent examples of such operations are the \textit{union}, the \textit{flip}, and the \textit{vertex contraction}, which we now introduce.

\begin{defi} \label{def:operations}
	Let $G_1,G_2 \in \cG_D$ be two $D$-colored graphs having respectively $k(G_1)$ and $k(G_2)$ pairs of vertices. 
	\begin{itemize}
		\item \textbf{Union.} The disjoint union of $G_1$ and $G_2$ -- denoted $G_1 \sqcup G_2$ -- is the $D$-colored graph whose connected components are exactly those of $G_1$ and those of $G_2$.\footnote{This definition is in accordance with the multiplicativity of the map $H \ni \cG_D \to \tr_H$ under disjoint unions, which was already commented on in Eq.~\eqref{eq:prodUnion}.}
		
		\item \textbf{Flip.} Let $c \in \{1, \ldots , D\}$ and $e_{G_1}$ (resp. $e_{G_2}$) be an edge of color $c$ in $G_1$ (resp. $G_2$). Performing a flip on the pair of edges $(e_{G_1} , e_{G_2})$ amounts to: a) cutting those two edges open in the graph $G_1 \sqcup G_2$; and b) reconnecting the resulting half-edges in the only other way allowed by 
		$\LU$-invariance (\ie in a way that preserves the bipartite structure of the graph). The resulting $D$-colored graph is denoted $G_1 \flip_{(e_{G_1},e_{G_2})}G_2$, and has $k(G_1) + k(G_2)$ vertices. When the specific choice of pair of edges $(e_{G_1},e_{G_2})$ on which the flip is being performed is not relevant to the discussion, we will simply write $G_1 \flip G_2$, or $G_1 \flip_c G_2$ in case we want to keep track of the color of the edge being flipped. The operation is represented graphically in Fig.~\ref{fig:flip}.
    	\begin{figure}[ht]
        		\centering
        		\includegraphics[height =2cm]{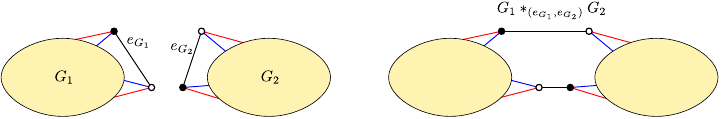}
        		\caption{Left: the graphs $G_1$ and $G_2$ with the edges $e_{G_1}$ and $e_{G_2}$ highlighted. Right: the result of the operation $G_1 \flip_{(e_{G_1}, e_{G_2})} G_2$.}
        		\label{fig:flip}
    	\end{figure}
	
	\item \textbf{Vertex contraction.} Let $v_{G_1}$ be a white (resp.~black) vertex of $G_1$ and $v_{G_2}$ a black (resp.~white) vertex of $G_2$. Contracting $G_1$ and $G_2$ along the pair $(v_{G_1}, v_{G_2})$ consists in: a) removing the vertex $v_{G_1}$ from $G_1$ and the vertex $v_{G_2}$ from $G_2$, and b) reconnecting the resulting half-edges in the only way that preserves the $D$-colored nature of the graph. This procedure yields a $D$-colored graph with $k(G_1) + k(G_2) - 1$ vertices, denoted as $G_1 \op_{(v_{G_1}, v_{G_2})} G_2$. When the specific choice of pairs of vertices $(v_{G_1}, v_{G_2})$ is not essential, we will simply write $G_1 \op G_2$ to refer to the contraction operation in a general sense. This operation is illustrated in Fig.~\ref{fig:op}.
	\begin{figure}[ht]
    		\centering
    		\includegraphics[height = 1.65cm]{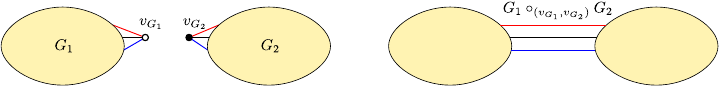}
    		\caption{Left: the graphs $G_1$ and $G_2$ with the vertices $v_{G_1}$ and $v_{G_2}$ highlighted. Right: the result of the operation $G_1 \op_{(v_{G_1},v_{G_2})} G_2$.}
    		\label{fig:op}
	\end{figure}
	\end{itemize} 
\end{defi}

Given a set of operations $\bb{O}$ and explicit transformation rules for $\Delta$ and $\mu$, a natural strategy for constructing families of invariants is to consider tree-like compositions built recursively from a fixed set of generators. More precisely, let $\bb{A}(\bb{B}, \bb{O})$ denote the set of all graphs that can be generated by applying a collection of $s$ binary operations $\bb{O} = \paa{\cal{O}_1, \dots, \cal{O}_s}$ to elements of a finite set of $r$ invariants $\bb{B} = \paa{B_1, \dots, B_r}$. 

Any element $G \in \bb{A}(\bb{B}, \bb{O})$ can be obtained recursively given $B \in \bb{B}$, $\cal{O} \in \bb{O}$ and $\tilde{G} \in \bb{A}(\bb{B},\bb{O})$ and the relation
\begin{equation}
    G = \tilde{G} \, \cal{O} \, B \,.
\end{equation}

\subsection{Combinatorial quantities under binary operations} \label{subsec:Evolution}

Let us review the main combinatorial quantities and their behavior under a binary operation. For the rest of this section, we suppose $G_1$ and $G_2$ to be $D$-colored graphs with, respectively, $k(G_1)$ and $k(G_2)$ pairs of vertices. We study the binary operations presented in the last section, \ie the union, the flip and the vertex contraction. 

\paragraph{Connected components of the subgraphs.} Let us consider a subset $B \in \paa{1,\dots,D}$. The connected components of the graph $G_1 \sqcup G_2\vert_B$ satisfy the trivial relation
\begin{equation}
	\kappa(G_1 \sqcup G_2\vert_{B}) = \kappa(G_1\vert_{B}) + \kappa(G_2 \vert_{B}) \qquad \rm{or} \qquad K(G_1\sqcup G_2\vert_B) = K(G_1\vert_B) + K(G_2\vert_B) \,.
\end{equation}
Regarding the flip, the result depends on the color of the flip. Indeed, one can compute 
\begin{equation} \label{eq:relKappa_flip}
	\kappa(G_1 \flip_c G_2\vert_{B}) = \left\{
        \begin{array}{lr}
            \kappa(G_1\vert_{B}) + \kappa(G_2 \vert_{B}) \,, \rm{ if } c \notin B \,, \\
            \kappa(G_1\vert_{B}) + \kappa(G_2 \vert_{B}) - 1 \,, \rm{ if } c \in B \,.
        \end{array}
        \right.
\end{equation}
Equivalently, one has 
\begin{equation}
    K_B(G_1 \flip_c G_2) = \left\{
        \begin{array}{lr}
            K_B(G_1) + K_B(G_2) + 1 \,, \rm{ if } c \notin B \,, \\
            K_B(G_1) + K_B(G_2) \,, \rm{ if } c \in B \,.
        \end{array}
        \right.
\end{equation}
Concerning the vertex contraction, the following relation applies
\begin{equation} \label{eq:relKappa_op}
	\kappa(G_1 \op G_2\vert_{B}) = \kappa(G_1\vert_{B}) + \kappa(G_2 \vert_{B}) - 1  \qquad \rm{or} \qquad K_B(G_1 \op G_2) = K_B(G_1) + K_B(G_2 )\,.
\end{equation}

\paragraph{Genus of the jackets.} Knowing the transformation on the connected components of the subgraphs and then of the faces leads to the following lemma.

\begin{lem} \label{lem:addGen}
	For two $D$-colored graphs $G_1$ and $G_2$, and a cyclic permutation $\tau \in S_D$, the genus of a jacket evolves according to 
	\begin{equation} \label{eq:TreeGenus}
		g_\tau(G_1\sqcup G_2) = g_\tau(G_1 \flip G_2) = g_\tau(G_1 \op G_2) = g_\tau(G_1) + g_\tau(G_2) \,.
	\end{equation}
\end{lem}

\begin{proof}
	We start with the union operation. The computation of the genus is given by
	\begin{equation}
		g_\tau(G_1\sqcup G_2) = \kappa(G_1 \sqcup G_2) + \frac{D-2}{2} k(G_1 \sqcup G_2) - \frac{1}{2} \sum_{c=1}^D F_{c \tau(c)}(G_1 \sqcup G_2) \,.
	\end{equation}
	However, since the number of connected components, the number of vertices, or faces are all additive under the union operation, Eq.~\eqref{eq:TreeGenus} is proven for the union.
	
	For the flip, let us consider a flip with the color $c$. Then, since $\kappa(G_1 \flip_c G_2) = \kappa(G_1) + \kappa(G_2) -1$, $k(G_1 \flip_c G_2) = k(G_1) + k(G_2)$ and 
	\begin{equation}
		\sum_{c=1}^D F_{c \tau(c)}(G_1 \flip_c G_2) = \sum_{c=1}^D F_{c \tau(c)} (G_1) + \sum_{c=1}^D F_{c \tau(c)} (G_2) - 2 \,,
	\end{equation}
	it yields the additivity of the genus.
	
	Finally, for the vertex contraction, having 
	\begin{equation}
		\kappa(G_1 \op G_2) + \frac{D-2}{2}k(G_1 \op G_2) = \kappa(G_1) + \kappa(G_2) + \frac{D-2}{2}\pa{k(G_1) + k(G_2)} - \frac{D}{2}  \,,
	\end{equation}
	along with $\sum_{c=1}^D F_{c \tau(c)}(G_1 \op G_2) = \sum_{c=1}^D F_{c \tau(c)} (G_1) + \sum_{c=1}^D F_{c \tau(c)} (G_2) - D$ proves the additivity of the genus under the vertex contraction operation. 
\end{proof}

\paragraph{$p$-complete degrees.} The Gurau degree being a sum of genera of jackets (see Eq.~\eqref{eq:GurauDegGenus}), it is additive under all the considered operations. Indeed, using Lem.~\ref{lem:addGen}, we have
\begin{equation}
	\omega_2(G_1\sqcup G_2) = \omega_2(G_1 \flip G_2) = \omega_2(G_1 \op G_2) = \omega_2(G_1) + \omega_2(G_2) \,.
\end{equation}
What is more, since the quantity $\omega_2^{(p)}$ is defined as a sum over Gurau degrees of subgraphs, the quantity $\omega_2^{(p)}$ satisfies
\begin{equation}
	\omega_2^{(p)}(A\sqcup B) = \omega_2^{(p)}(A \flip B) = \omega_2^{(p)}(A \op B) = \omega_2^{(p)}(A) + \omega_2^{(p)}(B) \,.
\end{equation}

By the use of Eq.~\eqref{eq:zeta_pq}, one can relate $\omega_p$ to $\omega_2$ and $\omega_2^{(p)}$ by the formula
\begin{equation}
    \omega_p(G) = \frac{\cal{I}_p}{D-1} \pa{\omega_2(G) - \frac{1}{\binom{D-2}{p-2}} \omega_2^{(p)}(G)} \,.
\end{equation}
However, since the right-hand side is additive with respect to all binary operations under consideration, the additivity of $\omega_p$ follows.

Finally, $\omega_p^{(q)}$ being defined as a sum over $p$-complete degrees of subgraphs, one is now able to show that the combinatorial quantities $\omega_p^{(q)}$ for $p\leq q$ are also additive.

\paragraph{$c$-degree.} Recall that, in order to distinguish the $c$-star state $\ket{\Phi_c}$ from $\ket{\GHZ}$ or from $\lvert \GHZ^{\frac{1}{D-1}}\rangle$, one must respectively evaluate the quantities
\begin{equation} \label{eq:c-star_state_qtt}
    \frac{1}{D-1}F_c(G) - \kappa(G) \qquad \rm{and} \qquad \Omega_c(G) = \kappa(G) + (D-2) k(G) - F_c(G) \,,
\end{equation}
as defined in Sec.~\ref{sec:LU-and-ref-states}. Since $F_c$, $\kappa$, and $k$ are all additive under the disjoint union of graphs, both quantities in Eq.~\eqref{eq:c-star_state_qtt} are themselves additive. An analogous statement holds for the vertex contraction operation, as the extra terms in Eq.~\eqref{eq:relKappa_op} cancel each other.

Concerning the flip operation, a distinction must be made.
Suppose that a flip is performed along an edge of color $i$, \ie $G_1 \flip_i G_2$. Using Eq.~\eqref{eq:relKappa_flip}, one finds, on the one hand,
\begin{equation}
    \frac{1}{D-1} F_c(G_1 \flip_i G_2) - \kappa(G_1\flip_i G_2) = \left\{
        \begin{array}{lr}
            \frac{1}{D-1} F_c(G_1) - \kappa(G_1) + \frac{1}{D-1} F_c(G_2) - \kappa(G_2) + \frac{D-2}{D-1} \,, \rm{ if } i \neq c\,, \\
            \frac{1}{D-1} F_c(G_1) - \kappa(G_1) + \frac{1}{D-1} F_c(G_2) - \kappa(G_2) \,, \rm{ if } i=c \,.
        \end{array}
        \right.
\end{equation}
On the other hand, the $c$-degree satisfies
\begin{equation}
    \Omega_c(G_1 \flip_i G_2) = \left\{
        \begin{array}{lr}
            \Omega_c(G_1) + \Omega_c(G_2) \,, \rm{ if } i \neq c\,, \\
            \Omega_c(G_1) + \Omega_c(G_2) + D-2 \,, \rm{ if } i=c \,.
        \end{array}
        \right.
\end{equation}

\paragraph{Degree of compatibility and combinatorial constant.} In general, the degree of compatibility $\Delta(G)$ of a graph $G$ (see Eq.~\eqref{eq:Delta}) or the combinatorial constant $\mu_G(\varphi)$ associated with $G$ and a sequence of pure states $\varphi$ (see Def.~\ref{def:LargeN_LU_eq}), are difficult to compute. However, thanks to the binary operations studied in this paper, we will be able to compute the degree of  compatibility given the degrees of the graphs involved in the binary operation. 

To simplify the statement of the following theorem, we first recall the definition of the set $\MD{}(G)$ introduced in Eq.~\eqref{eq:MD}
\begin{equation}  \label{eq:MD_2}
    \MD{}(G) = \paa{\widehat G \in\cG_{D+1}(G) \;\; \rm{s.t.} \;\; F_{0}(\widehat G) = \max_{\widehat H \in\cG_{D+1}(G)} F_0(\widehat H)}\,.
\end{equation} 
Secondly, let us define the following notions. 
\begin{defi}\label{def:tree-like}
    Let $D \geq 3$, $p \in \bb{N}^*$, $G_1, \ldots , G_p \in \cG_D$ (not necessarily connected), and $G= G_1 \sqcup \cdots \sqcup G_p$.  
    We define the following notions:
    \begin{enumerate}
        \item A \emph{two-cut} in a graph is a pair of edges whose removal increases the number of connected components. 
        \item Let $i\in \{1,\ldots , p\}$ and $\widehat G\in\cG_{D+1}(G)$. $G_i$ satisfies the \emph{maximal two-cut property} in $\widehat G$ if there exists $(D+1)$-colored graphs $\widehat G_i \in \MD{}(G_i)$, $\widehat H \in \cG_{D+1}(G_1 \sqcup \cdots \sqcup G_{i-1} \sqcup G_{i+1} \sqcup \cdots \sqcup G_p)$ and $2\ell \leq 2\min\{k(\widehat G_i),k(\widehat H)\}$ edges of color $0$, denoted in $\widehat G_i$ by  $(e_1,\dots,e_\ell)$ and denoted in $\widehat H$ by $(f_1,\dots,f_\ell)$, such that 
        \begin{equation}
            \widehat G = \widehat G_i \flip_{(e_1,f_1)} \cdots \flip_{(e_\ell,f_\ell)} \widehat H  \,.
        \end{equation}
        \item We say that $\widehat G \in\cG_{D+1}(G)$ is \emph{tree-like on $G$} if $\widehat G$ is connecting the $G_i$'s (without necessarily satisfying $\kappa(\widehat G) = \kappa(G)$) and every $G_i$ satisfies the maximal two-cut property in $\widehat G$.
        \item $G$ is said to have \emph{tree-like dominant graphs} if there exists a graph $\widehat G \in \cG_{D+1}(G)$ minimizing the degree of compatibility while connecting the $G_i$'s (without necessarily connecting the connected components of the $G_i$'s) that is tree-like on $G$.
        \item $G$ is said to have \emph{only tree-like dominant graphs} if any graph $\widehat G \in \cG_{D+1}(G)$ minimizing the degree of compatibility while connecting the $G_i$'s (without necessarily connecting the connected components of the $G_i$'s) is tree-like on $G$.
    \end{enumerate}
\end{defi}

We are now able to state the following statement regarding the evolution of the degree of compatibility under the binary operations.

\begin{theo} \label{th:TreeDegComp}
    Let $\varphi \eqdef (\ket{\varphi_N})_{N \in \mathbb{N}^*}$, where for every $N \in \mathbb{N}^*$, $\ket{\varphi_N}$ denotes the Haar-random state of local dimension $N$. Let $D\geq 2$, $G_1 , G_2 \in \cG_D$, $\cal{O}$ a binary operation and let $v_{G_1},v_{G_2}$ be two vertices of distinct colors of respectively $G_1$ and $G_2$. The degree of compatibility is additive and the combinatorial constant is multiplicative under the binary operation $\cal{O}$
    \begin{equation}
    		\Delta(G_1 \, \cal{O} \, G_2) = \Delta(G_1) + \Delta(G_2) \qquad \rm{and} \qquad \mu_{G_1 \, \cal{O} \, G_2}(\varphi) = \mu_{G_1}(\varphi) \mu_{G_2} (\varphi) \,,
    \end{equation}
    if one of the following assumptions is satisfied
    \begin{enumerate}
        \item \label{it:Delta_haveTree} $\cal{O} \in 
        \paa{\sqcup, \flip}$ and $G_1 \sqcup 
        G_2$ have tree-like dominant graphs (see Ref.~\cite{Factorization2026} and Prop.~\ref{prop:tree_union_flip}).
        \item \label{it:Delta_haveOnlyTree} $\cal{O} = \op$ and $G_1 \sqcup G_2$ have only tree-like dominant graphs (see Prop.~\ref{prop:Tree_only}).
        \item \label{it:DADB_cup} $\cal{O} = \sqcup$ and: $\Delta(G_1) + \Delta(G_2) < \frac{D(D-1)}{2}$ or $\Delta(G_1 \sqcup G_2) < \frac{D(D-1)}{2}$ (see Prop.~\ref{prop:UnionFlipAdd}).
        \item \label{it:DADB_flip} $\cal{O} = \flip$ and: $\Delta(G_1) + \Delta(G_2) < \frac{(D-1)(D-2)}{2}$ or $\Delta(G_1 \flip G_2) < \frac{(D-1)(D-2)}{2}$ (see Prop.~\ref{prop:UnionFlipAdd}).
        \item \label{it:DADB_op} $\cal{O} = \op_{(v_{G_1},v_{G_2})}$ and $\Delta(G_1) = \Delta(G_2) = 0$ with $v_{G_1}$ or $v_{G_2}$ belonging to, at least, one face of size four or less\footnote{The \textit{size} of a face is defined as the number of edges along it.} (see  Prop.~\ref{prop:comp}).
    \end{enumerate}
\end{theo}

\begin{proof}
    See App.~\ref{A:DeltaAB}.
\end{proof}

Determining whether a given graph has, exactly or not, tree-like dominant contributions is a difficult task. Nevertheless, the literature provides several families of graphs for which tree dominance is well understood, thus providing examples for which point~\ref{it:Delta_haveTree} and point~\ref{it:Delta_haveOnlyTree} of Thm.~\ref{th:TreeDegComp} can be applied. \\
Indeed, in the tripartite case, it was shown in Ref.~\cite{bonzom2018maximizingnumberedgesthreedimensional} that $3$-colored \textit{planar} graphs have only tree-like dominant graphs. From the trace-invariants introduced in Sec.~\ref{sec:Trace_Lit_Ref_States}, let us recall that the colored graphs associated with the \textit{realignment moments} (and particularly $\RM_4 = \ME_2^3 = \RE_{2,2}$), the one associated with \textit{Rényi reflected entropies}, and the $3$-colored \textit{melonic} graphs are all planar graphs. In Ref.~\cite{Lionni2018}, it was proven that the colored graphs $\JRM_k^{\vec i}$ associated with the \textit{joint realignment moments} have tree-like dominant graphs for $3 \leq k \leq 4$ and only tree-like dominant graphs for $k > 4$.  \\
In the multipartite setting, it was shown in Ref.~\cite{Gurau2011_1,Gurau2011_2} that the union of \textit{melonic graphs} has only tree-like dominant contributions. Moreover, from Ref.~\cite{Lionni2018}, it was shown that union of the tripartite block version of the joint realignment moment $\JRM_k^{\vec i}$ with $\vec i$ a cyclic sequence of colors, or of the realignment moments $\RM_{2n}^{(B)}$ with $\abs{B} = D-2$, have also only tree-like dominant graphs. Regarding the disjoint union of cyclic graphs, the behaviour of the dominant contributions depends on the size of the bipartitions associated with the graphs appearing in the union. In particular, if there are at least two cyclic graphs associated with a \textit{balanced} bipartition (\ie a bipartition of the form $\H=\H_B\ot\H_{\bar B}$ with $\abs{B}=D/2$ and $D$ even), then the disjoint union does not have exclusively tree-like dominant graphs. Otherwise, the dominant graphs are tree-like. This result arises because cyclic graphs associated with a balanced bipartition behave like square matrix models, whereas cyclic graphs associated with an unbalanced bipartition behave like rectangular matrix models (see Ref.~\cite{Bonzom:2013lda,bonzom2014tensormodelsviewpointmatrix}).
\\
Before proceeding, we highlight an important result concerning tree-like dominant families of colored graphs. If $G_1$ has (only) tree-like dominant graphs and $G_2$ is an arbitrary $D$-colored graph, then the disjoint union $G_1 \sqcup G_2$ also has (only) tree-like dominant graphs. As an illustration, consider $G_1 = \RM_5^{(1,2)} \sqcup \JRM_6^{(1,2,3,4,2,3)}$ for which $\Delta(G_1) = 1$, and let $G_2 \in \cG_4$ be an arbitrary $4$-colored graph. Then, the colored graph $G_1 \sqcup G_2$ has only tree-like dominant graphs leading to $\Delta(G_1 \sqcup G_2) = 1 + \Delta(G_2)$.\footnote{We emphazis that if $G_2$ is disconnected with $H_1,\dots,H_p$ its connected components, then, there is \apriori no reason for $\Delta(G_2)$ to be equal to $\sum_{i=1}^p \Delta(H_i)$.}

\ 

We observe that points~\ref{it:DADB_cup} and~\ref{it:DADB_flip} in Thm.~\ref{th:TreeDegComp} apply in particular to compatible invariants. Among the invariants of Sec.~\ref{sec:Trace_Lit_Ref_States}, it is known that melonic invariants, moments of the partial transpose and joint realignment moments are all compatible. Moreover, we recall that colored graphs associated with realignment moments and Rényi reflected entropies have a degree of compatibility equal to one.

\ 

\begin{figure}[ht]
	\centering
	\includegraphics[height = 6cm]{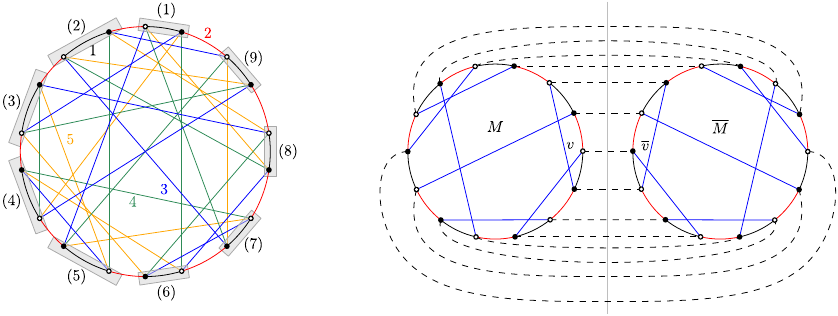}
	\caption{Left: an example of incompatible, maximally single-trace $5$-colored graph. Using the vertex labeling in the figure, we have: $\sigma_1 = \id$, $\sigma_2 = (1\, 2\, 3\, 4\, 5\, 6\, 7\, 8\, 9)$, $\sigma_3 = (1\, 5\, 4\, 9\, 2 \,7\, 6\, 8\, 3)$, $\sigma_4 = (1\, 7\, 4\, 3 \,9 \,5\, 2 \,8\, 6)$, and $\sigma_5 = (1 \,8 \,2 \,9\, 7\, 5\, 3\, 6\, 4)$. Numerically, we found $\Delta = 4$ and $\mu = 4$. Right: for $M$ maximally single-trace, we represented the colored graph $\widehat G$ constructed in the proof of Prop.~\ref{prop:MST}, highlighted the mirror construction with the vertical gray line, and showed a possible pair $(v,\bar{v})$.}
	\label{fig:MST_incomp}
\end{figure}

We were able to check via a numerical analysis that all $3$-colored maximally single-trace invariants with $k \leq 9$ and all $4$-colored maximally single-trace invariants with $k \leq 7$ are \textit{compatible}. Let $M$ be a maximally single-trace graph described by the $D$-tuple of permutations $\vec \sigma$. If $M$ is compatible, then any permutation $\nu$ associated with color $0$ which minimizes the degree must satisfy
\begin{equation}
    \forall i<j\,, \quad d(\sigma_i,\nu) + d(\nu,\sigma_j) = d(\sigma_i,\sigma_j) \quad \Longleftrightarrow \quad \forall i\,, \quad  d(\sigma_i,\nu) = \frac{k-1}{2} \,.
\end{equation}
Hence, for all pairs $(\sigma_i,\sigma_j)$, the permutation $\nu$ is forced to be non-crossing (see Sec.~\ref{subsec:GraphStructure}, paragraph ``The $c$-degree $\Omega_c$.''). These conditions seem satisfiable for all odd $k$ and low $D$, such as $D=3$ (though this is not proven). However, the more we add colors, the more constraints the permutation $\nu$ will have to satisfy. Therefore, as the following example shows (see left-hand side of Fig.~\ref{fig:MST_incomp}), incompatible maximally single-trace invariants are easier to come by as $D$ grows. 

However, we emphasize that for any $D$ one can always construct a \textit{compatible} maximally single-trace invariant, as proven by the following proposition.

\begin{prop} \label{prop:MST}
    Let $D \geq 3$, $M \in \cG_D^{\conn}$ be a maximally single trace-invariant and let us denote by $v$ (resp.~$\bar{v}$) a vertex of $M$ (resp.~the vertex image in a mirror of $v$ in $\bar{M}$).\footnote{For $G \in \cG_D$, recall that the notation of $\bar{G}$ was introduced in Rem.~\ref{rem:barG_1}.} The colored graph $M \op_{(v,\bar{v})} \bar{M}$ is maximally single-trace and compatible. 
\end{prop}

\begin{proof}
    Let $G = M \sqcup \bar{M}$ and $H = M \op_{(v,\bar{v})} \bar{M}$. Since both $M$ and $\bar{M}$ are maximally single-trace, Eq.~\eqref{eq:relKappa_op} yields that: for all $i<j$, $F_{ij}(H) = 1$. So, $H$ is maximally single-trace.
    
    Construct $\widehat G \in \cG_{D+1}^{\conn}(G)$ a $(D+1)$-colored graph for which all edges of color $0$ pair a vertex in $M$ with its image in a mirror in $\bar{M}$ (see the right-hand side of Fig.~\ref{fig:MST_incomp}). The colored graph $\widehat G$ satisfies $F_0(\widehat G) = Dk(M)$, \ie by the use of Eq.~\eqref{eq:Delta_0}
    \begin{equation}
        \Delta_0(\widehat G) = 2 \cdot \frac{D(D-1)}{4} k(M) + 2 \cdot \frac{1}{2}\frac{D(D-1)}{2} - \frac{D-1}{2} Dk(M) = \frac{D(D-1)}{2} \,.
    \end{equation}
    Consider an edge of color $0$ in $\widehat G$ that is pairing $v$ to $\bar{v}$ (by construction, $v$ and $\bar{v}$ are image of each other). The vextex contraction operation, $M \op_{(v,\bar{v})} \bar{M} = H$, has the following effects: $F_0(\widehat H) = F_0(\widehat G)$, $F(H) = F(G) - D(D-1)/2$ and $k(H) = k(G) - 1$. Thus, we have
    \begin{equation}
        \Delta_0(\widehat H) = \Delta_0(\widehat G) - \frac{D(D-1)}{2} = 0 \,,
    \end{equation}
    proving the compatibility of $H$ independently of the compatibility of $M$. 
\end{proof}

\subsection{Applications}\label{sec:applications-binary}

\subsubsection{Large $N$ approximation of $\LO$-monotones in the Haar-random state} \label{subsec:averageApprox}

In this subsection, we fix $D\geq 2$ and let $\varphi \eqdef (\ket{\varphi_N})_{N \in \mathbb{N}^*}$ denote the following sequence of Haar-distributed random states: for any $N \in \mathbb{N}^*$, $\ket{\varphi_N}$ is Haar-distributed on a $D$-partite state space of local dimension $N$. 

\medskip 

In Sec.~\ref{ss:ent-monotones-trace-inv}, we introduced the family of $\LO$-monotones $\{R_G\}_{G \in \cG_D}$, which can be understood as a multipartite generalization of the family of bipartite entanglement R\'{e}nyi entropies (see Exs.~\ref{ex:Rényi} and \ref{ex:renyi_higher-D}). Given $G \in \cG_D$, we will be interested in the following related questions:
\begin{enumerate}
    \item What is the \emph{typical value} of $R_G(\ket{\varphi_N})$ in the asymptotic regime of large local dimension $N \to \infty$?
    \item Under which condition on $G$ the asymptotic equivalence relation 
    \begin{equation}\label{eq:as_equiv_R_G}
    \langle R_G (\ket{\varphi_N})\rangle \underset{N\to \infty}{\sim} - \ln\left( \langle \tr_G (\ket{\varphi_N})\rangle\right)
\end{equation}
may hold?
\end{enumerate}
In the bipartite setting, it is well known that the typical value of the entanglement R\'{e}nyi-$k$ entropy (with $k\geq 2$ an integer),  
\begin{equation}
    S_\rm{R}^{(k)} \left( \ket{\varphi_N}\right) = \frac{1}{k-1} R_{C_k} \left( \ket{\varphi_N} \right)\,,
\end{equation}
is $- \frac{1}{k-1}\ln \left( \mean{\tr_{C_k} (\ket{\varphi_N})} \right) = \ln(N)- \frac{1}{k-1} \ln \left(\rm{Cat}_k \right)+O\left(1/N\right)$ (see Ref.~\cite{Page:1993df}). Moreover, Eq.~\eqref{eq:as_equiv_R_G} does hold with $G= C_k$. By contrast, we will see that answering the previous two questions is more difficult when $D \geq 3$: arguments based on concentration phenomena that allow us to answer the first question does not apply to any graph $G$; furthermore, the validity of Eq.~\eqref{eq:as_equiv_R_G} is not directly implied by a good control over the first question. 

\medskip

While our main focus will be on the sequence $\varphi$ of Haar-distributed states introduced at the beginning of this section, we will prove general results that may be applied to other sequences of random states. In particular, random tensor networks \cite{Hayden2016} (see also \eg Refs.~\cite{KudlerFlam2022, Cheng:2022ori, Penington2023}), which are defined in terms of collections of independently and identically distributed Haar (or Gaussian) random tensors, fall in the category of models that are computable enough for the key large-$N$ factorization assumption made in Eq.~\eqref{eq:factorization_criterion} to be rigorously established (for simple enough families of graphs). Furthermore, the Haar-distributed random sequence $\varphi$ can be understood as the simplest possible random tensor network, comprising one bulk node and $D$ boundary nodes. Factorization results have already been investigated in some detail for specific invariants in this context (\eg cyclic graphs for general random tensor networks in Ref.~\cite{Hayden2016}, or multi-entropy invariants for Haar-distributed states in Ref.~\cite{Iizuka2025BH}), but it would be valuable to explore this question more systematically in the future. The next definition introduces the main restriction we will impose on the choices of distribution and trace-invariant. 
\begin{defi}
Let $\psi= (\ket{\psi_N})_{N \in \mathbb{N}^*}$ denote a sequence of random $D$-partite states and $G \in \cG_D$. We will say that $(G, \psi)$ obeys the \emph{large-$N$ factorization criterion} whenever:\footnote{Recall that the notation $\bar G$ was introduced in Rem.~\ref{rem:barG_1}.}
    \begin{equation}\label{eq:factorization_criterion}
    \mean{\tr_G(\ket{\psi_N}) \tr_{\bar G}(\ket{\psi_N})} \underset{N \to \infty}{ =} \mean{\tr_G(\ket{\psi_N})} \mean{\tr_{\bar G}(\ket{\psi_N})} \left( 1+ O\left(1/N\right)\right)\,.
\end{equation}
\end{defi}
\begin{rem}\label{rem:barG_2} One trivially has that: $(G, \psi)$ obeys the large-$N$ factorization criterion if and only if $(\bar G , \psi )$ does.
\end{rem}
Assuming a large-$N$ Ansatz, the next elementary lemma expresses the large-$N$ factorization criterion in terms of the coefficients appearing in the large-$N$ Ansatz.
\begin{lem}
Let $\psi= (\ket{\psi_N})_{N \in \mathbb{N}^*}$ a sequence of random $D$-partite states and $G \in \cG_D$, such that $(G,\psi)$ obeys the large-$N$ Ansatz (see Eq.~\eqref{eq:Ansatz_large-N}). Then, $(G, \psi)$ obeys the large-$N$ factorization criterion \eqref{eq:factorization_criterion} if and only if  $(G\sqcup \bar G , \psi)$ obeys the large-$N$ Ansatz  
\begin{equation}
\mean{\tr_{G \sqcup \bar G}\left( \ket{\psi_N}\right)} \underset{N \to \infty}{=} \mu_{G\sqcup \bar G} (\psi) N^{s_{G\sqcup \bar G} (\psi)}    \left( 1+ O\left(1/N\right)\right)
\end{equation}
with
\begin{equation}   
\label{eq:largeNfactoproof}
\mu_{G \sqcup \bar G}(\psi) = \mu_G(\psi) \mu_{\bar{G}}(\psi)=  \abs{\mu_G(\psi)}^2 \,, \qquad   s_{G\sqcup \bar G}(\psi) = s_G(\psi) + s_{\bar G}(\psi) = 2 s_G(\psi) \,.
\end{equation}
\end{lem}
\begin{proof}
This is a direct consequence of the definitions (together with Rems.~\ref{rem:barG_1} and \ref{rem:barG_2}).
\end{proof}
Specializing to the sequence of Haar-distributed states $\varphi$, the large-$N$ Ansatz holds for any graph $G \in \cG_D$. As a result, the large-$N$ factorization criterion for $(G, \varphi)$ can be directly expressed in terms of the coefficients $\{s_H (\varphi),  \mu_H (\varphi)\}_{H \in \{G, G\sqcup \bar G\}}$.
\begin{lem}
    For any $G \in \cG_D$, we have:
    \begin{enumerate}
        \item $(G , \varphi)$ obeys the large-$N$ Ansatz;
        \item $(G , \varphi)$ obeys the large-$N$ factorization criterion if and only if:
        \begin{equation}   
\label{eq:largeNfacto_Haar}
\mu_{G \sqcup \bar G}(\varphi) = \mu_G(\varphi) \mu_{\bar{G}}(\varphi)=  \mu_G(\varphi)^2 \,, \qquad  \rm{ and } \qquad s_{G\sqcup \bar G}(\varphi) = s_G(\varphi) + s_{\bar G}(\varphi) = 2 s_G(\varphi)  \,.
\end{equation}
    \end{enumerate}
\end{lem}
\begin{proof}
    Let $G\in \cG_D$, $k\eqdef  k(G)$ and $N \in \mathbb{N}^*$. According to Wick's theorem, Eq.~\eqref{eq:meanLUWeing} and Eq.~\eqref{eq:meanLUWeing2}, we have:
    \begin{equation}
        \mean{ \tr_G \left( \ket{\varphi_N}\right)} = f_{k,D,N} \mu_G(\varphi) N^{s_G (\varphi)} + f_{k,D,N} \sum_{\substack{\widehat{G}\in \cG_{D+1}(G) \\
        F_0(\widehat{G})\leq s_G(\varphi)+Dk - 1}} \Xi(\widehat G) N^{F_0 (\widehat{G}) - Dk} = \mu_G (\varphi) N^{s_G(\varphi)}\left( 1 + O(1/N)\right)\,.
    \end{equation}
Furthermore, $\mu_G(\varphi)$ is a positive and nonzero integer counting the number of Wick contractions contributing to the leading order, hence the conditions from Eq.~\eqref{eq:largeNfactoproof} specialize to those of Eq.~\eqref{eq:largeNfacto_Haar}.
\end{proof}
\begin{rem} \label{rem:MST_incomp}
    In the case of Haar-distributed states, the large-$N$ factorization criterion does not hold for all $D$-colored graphs. The companion paper Ref.~\cite{Factorization2026} will provide a detailed example for $D=6$. 
\end{rem}

\paragraph{Typical value of $R_G$ in the Haar-random state.} Let us start out by quoting the following concentration phenomenon result.
\begin{prop}\label{prop:concentration} 
    Let $\psi= (\ket{\psi_N})_{N\in \mathbb{N}^*}$ be a sequence of random $D$-partite states and $G \in \cG_D$. If $(G,\psi)$ obeys both the large-$N$ Ansatz \eqref{eq:Ansatz_large-N} and the large-$N$ factorization criterion \eqref{eq:factorization_criterion} (or equivalently \eqref{eq:largeNfactoproof}), then: for any $\varepsilon > 0$, there exists a constant $N_c > 0$ such that 
    \begin{equation}\label{eq:concentration}
      \forall N \in \mathbb{N}^*\,,\qquad  \Prob{ \abs{\frac{\abs{\tr_G(\ket{\psi_N})}}{\abs{\mu_G(\psi)}N^{s_G(\psi)}} - 1}  < \varepsilon}  \geq 1 - \frac{N_c}{N} \,.
    \end{equation} 
In particular, if the invariant $\tr_G$ does not admit zeroes (so that $R_G$ is always real-valued), the typical value of $R_G(\ket{\psi_N})$ in the large-$N$ regime is:
\begin{equation}
    \abs{s_G (\psi)} \ln(N)- \ln\abs{\mu_G (\psi)}\,.
\end{equation}
\end{prop}
\begin{proof}
The proof takes its inspiration from Ref.~\cite[33]{Hayden2016}, in that its key ingredient is Markov's inequality, applied to the random variable $\abs{\frac{\abs{\tr_G(\ket{\psi_N})}}{\abs{\mu_G(\psi)}N^{s_G(\psi)}} - 1}$. The interested reader is referred to the companion paper \cite{Factorization2026}. 
\end{proof}
Specializing to the Haar-distributed state, we obtain the following immediate Corollary.
\begin{cor}\label{cor:typical-value_Haar}
    Let $G \in \cG_D$ be such that $(G,\varphi)$ obeys the large-$N$ factorization criterion (Eq.~\eqref{eq:largeNfacto_Haar}). Then, the typical value of $R_G(\ket{\varphi_N})$ in the large-$N$ regime is 
    \begin{equation}
        \abs{s_G (\varphi)} \ln(N)- \ln\abs{\mu_G (\varphi)}\,.
    \end{equation}
\end{cor}
In the bipartite case, Eq.~\eqref{eq:largeNfacto_Haar} holds for \emph{any} colored graph $G \in \cG_2$. However, as recently proven in Ref.~\cite{Gurau2025}, in the multipartite context there exist trace-invariants for which Eq.~\eqref{eq:largeNfacto_Haar} does not hold.\footnote{In fact, Ref.~\cite{Gurau2025} relies on a probabilistic proof, that establishes a much stronger result: in informal terms, a large graph $G$ chosen uniformly at random will fail to obey the conditions of Eq.~\eqref{eq:largeNfactoproof} with probabilty close to $1$ (we refer the interested reader to Ref.~\cite{Gurau2025} for a precise statement).} This has two interesting consequences. First, there does not exist any sequence of deterministic quantum states $\psi=(\ket{\psi_N})_{N \in \mathbb{N}^*}$ that is $\LU$-equivalent in scaling (resp.~asymptotically $\LU$-equivalent) to the sequence of Haar-random states $\varphi= (\ket{\varphi_N})_{N \in \mathbb{N^*}}$, as was already pointed out in Sec.~\ref{sec:examples_asymptotic_rel}. 
Second, using concentration of measure phenomena to approximate averages of non-linear functions of trace-invariants is a viable strategy for those invariants for which Eq.~\eqref{eq:largeNfacto_Haar} is known to hold. In the companion paper Ref.~\cite{Factorization2026}, we provide a number of sufficient conditions for such large-$N$ factorization criterion to hold. 

\begin{cor}\label{cor:Factorization}
    Let $D \geq 2$ and $G \in \cG_D$. The large-$N$ factorization criterion is satisfied if either $G \sqcup \bar G$ have tree-like dominant graphs or $\Delta(G) < \frac{D(D-1)}{4}$.
\end{cor}

\begin{proof}
    It is a direct application of Thm.~\ref{th:TreeDegComp} regarding the disjoint union between $G$ and $\bar G$.
\end{proof}

\begin{ex}
    In particular, by Cor.~\ref{cor:Factorization}, the large-$N$ factorization criterion is satisfied by a subset of the trace-invariants discussed in this paper, namely: the cyclic graphs (see Sec.~\ref{sss:necklace}); the melonic graphs (see Sec.~\ref{sss:Melo}); the planar graphs for $D=3$ (see Sec.~\ref{sss:planar}); the compatible maximally single-trace invariants (see Sec.~\ref{sss:MST}); the moments of the partial tranpose (see Sec.~\ref{subsubsec:PT}); the realignment moments and their tripartite generalization $\RM_{2n}^{(B)}$ for $n \in \bb{N}^*$ and $\abs{B} = D-2$ (see Sec.~\ref{sss:RM_JRM}); the joint realignment moments and their tripartite generalization $\JRM_{k}^{\vec i}$ for any $k \in \bb{N}^*$ and any sequence of colors $\vec i$ (see Sec.~\ref{sss:RM_JRM}); the reflected entropies (see Sec.~\ref{ss:RE}).
\end{ex}

For any graph $G$ covered by Cor.~\ref{cor:Factorization}, the concentration phenomenon of Prop.~\ref{prop:concentration} (or Cor.~\ref{cor:typical-value_Haar}) allows establishing that the \emph{typical value} of $R_G \left( \ket{\varphi_N} \right)$ is 
\begin{equation}
    - \ln\left( \langle \tr_G (\ket{\varphi_N})\rangle\right)\underset{N\to \infty}{=} \abs{ s_G(\varphi)} \ln\left(N\right) - \ln\left(\mu_G(\varphi)\right)+o(1)
\end{equation}
in the asymptotic regime of large local dimension. However, since we cannot \emph{a priori} exclude the existence of low probability deviations with arbitrarily large amplitudes, this result is not quite sufficient to establish an asymptotic equivalence relation of the form \eqref{eq:as_equiv_R_G}. In the next paragraph, we impose further restrictions on $G$ that are sufficient to establish this stronger result.

\paragraph{Asymptotic expression for the expectation value of  $R_G$ in the Haar-random state.} In order to control potentially large deviations from the large-$N$ typical value of $R_G$, we assume the existence of a uniform lower bound on $\abs{\tr_G (\cdot)}$ that is polynomial in $N$.
\begin{defi}
    Let $G \in \cG_D$. For any $N \in \mathbb{N}^*$, we introduce:
    \begin{equation}
        m_G(N)\eqdef \min\Bigl\{ \abs{\tr_G \left( \ket{\psi}\right)} \, \big\vert \, \ket{\psi} \in S\left((\mathbb{C}^N)^{\otimes D} \right)\Bigr\}\,.
    \end{equation}
We will say that $\tr_G$ (or $G$) follows a \emph{power law} if there exists $K > 0$ and $r\geq 0$ such that:
\begin{equation}\label{eq:power_law}
    \forall N \in \mathbb{N}^*\,, \qquad m_G(N) \geq K N^{-r}\,.
\end{equation}
\end{defi}

\begin{figure}[ht]
    	\centering
    	\includegraphics[width = \textwidth]{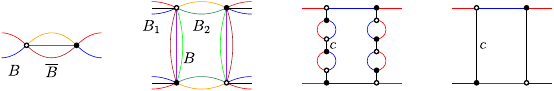}
    	\caption{From left to right: the matrices $P$, $Q$, $R$ (for $m=6$) and $S$ considered in the proof of Prop.~\ref{prop:poly_bound}.}
    	\label{fig:pol_bound}
    \end{figure}

\begin{prop}\label{prop:poly_bound} 
    Let $G \in \cG_D$. If any connected component of $G$ belongs to\footnote{The second (resp.~third) family of graphs appearing in this union is nonempty only if $D\geq 3$ (resp.~$D=3$).} 
    \begin{align} \label{eq:set_poly_bound}
        \{ H \in \cG_D^{\conn}\, \big\vert \, H \;\textrm{cyclic}\}\cup \{\RM_{2n}^{(B)}\, \big\vert \, B \subset \{1, \ldots, D \},\, B\neq \emptyset ,\, n \in \mathbb{N}^*\} &\cup \{\RE_{m,n}^{(c)}\, \big\vert \, c \in \{1, 2, 3 \},\, n \in \mathbb{N}^*,\, m \in 2\mathbb{N}^*\} \nonumber \\
        &\cup \{\PT_{2n}^{(c)}\, \big\vert \,  c \in \{1, 2, 3 \},\, n \in \mathbb{N}^*\}\,,
    \end{align}
    then $\tr_G$ is real, positive, and follows a power law. 
\end{prop}
\begin{proof}
    The real, positive and power law properties of trace-invariants are clearly stable under dijoint unions, hence we can assume without loss of generality that $G$ is connected. Let $\ket{\psi}$ denote a $D$-partite pure state.
    
    Let us first suppose that $G$ is (connected) cyclic and let $k \eqdef k(G)$. We can then find a bipartition $\{B , \bar{B}\}$ of $\{1, \ldots , D\}$ such that: 
    \begin{equation}
        \tr_G (\ket{\psi}) = \tr\left( P^k \right)\,,
    \end{equation}
    where $P$ is a positive, Hermitian matrix on $(\mathbb{C}^N)^{\otimes |B|}$ (see left of Fig.~\ref{fig:pol_bound}) obeying
    \begin{equation}
        \tr(P) = \langle\psi \vert \psi \rangle=1\,. 
    \end{equation}
    It is thus clear that $\tr_G(\ket{\psi})$ is both real and positive. Next, let us prove a polynomial lower-bound of the form $\tr_G (\ket{\psi}) \leq K N^{-r}$, with $K >0$ and $r\geq 0$. If $k=1$, such a bound holds with $K=1$ and $r=0$. Suppose now that $k\geq 2$. By H\"{o}lder's inequality, we have
    \begin{equation}
         1=\tr(P)=\tr(P \cdot \id) \leq \tr(P^k)^{1/k} \tr(\id^{\frac{k}{1-k}})^{1-\frac{1}{k}}=\tr(P^k)^{1/k} N^\frac{(k-1)|B|}{k}\,,
    \end{equation}
    which leads to
    \begin{equation}
        \tr_G\left( \ket{\psi}\right) = \tr(P^k) \geq N^{(1-k)|B|}\,.
    \end{equation}
    Hence, the looked-for bound holds with $K=1$ and $r=(k-1)|B|$. We conclude that $\tr_G$ follows a power law.
    
    Next, let us assume that $G=\RM_{2n}^{(\bar{B_1,B_2})}$ with $\{B,B_1,B_2\}$ a partition of $\paa{1,\dots,D}$, $B\neq \emptyset$ and $n \in \mathbb{N}^*$. We then have 
    \begin{equation}
        \tr_G \left( \ket{\psi}\right) = \tr\left( Q^k\right)\,,
    \end{equation}
    where $Q$ is a positive Hermitian matrix on $(\mathbb{C}^N)^{\otimes 2 \abs{ B_1 }}$ (see the middle-left panel of Fig.~\ref{fig:pol_bound}) such that $\tr\left( Q\right)=\tr_{\RM_{2}^{(\bar{B_1,B_2})}}(\ket{\psi})$. This readily establishes positivity of $\tr_G$. Moreover, given that $\RM_{2}^{(\bar{B_1,B_2})}$ is a connected cyclic graph, the previous paragraph ensures that $\tr(Q)$ can be bounded from below by $\tilde{K} N^{-\tilde{r}}$, where $\tilde{K}>0$ and $\tilde{r}\geq 0$ are independent from the choice of state $\ket{\psi}$. This proves the required bound for $n=1$. If $n \geq 2$, we can apply H\"{o}lder's inequality to obtain:
    \begin{equation}
        \tilde{K} N^{-\tilde{r}}\leq \tr(Q) \leq \tr(Q^n)^{1/n} \tr(\id^{\frac{n}{1-n}})^{1-\frac{1}{n}}=\tr(Q^n)^{1/n} N^{\frac{2(n-1)|B_1|}{n}}\,,
    \end{equation}
    and therefore
    \begin{equation}
        \tr_G\left( \ket{\psi}\right)= \tr(Q^n) \geq \tilde{K}N^{- \tilde{r}+2(1-n)|B_1|}\,.
    \end{equation}
    Hence $\tr_G$ is follows a power law. 

    Next, let us assume that $D=3$ and consider the case $G = \RE_{m,n}^{(c)}$, with $c \in \{1, 2,3 \}$, $n \in \mathbb{N}^*$, and $m \in 2\mathbb{N}^*$. In this case, we have 
    \begin{equation}
        \tr_G \left( \ket{\psi}\right) = \tr\left( R^n \right)\,,
    \end{equation}
    where $R$ is a Hermitian and positive operator on $\left( \mathbb{C}^N\right)^{\otimes 2}$ (see the middle-right panel of Fig.~\ref{fig:pol_bound}), and
    \begin{equation}
        \tr(R) = \tr_{C_m}\left( \ket{\psi}\right)\,,
    \end{equation}
    where $C_m$ denotes the unique connected cyclic graph relative to the bipartition $\{1,2,3\}= \{c\}\cup (\{1,2,3\}\setminus\{c\})$ with $k(C_m)=m$. Proceeding as in the previous paragraph, we conclude that $\tr_G$ is real, positive, and follows a power law.

    Finally, let us assume that $G= \PT_{2n}^{(c)}$, with $c \in \paa{1,2,3}$ and $n \in \mathbb{N}^*$. We then have $\tr_{\PT_{2n}^{(c)}}(\ket{\psi}) = \tr(S^{n})$ with $S$ Hermitian and positive (see the rightmost graph of Fig.~\ref{fig:pol_bound}). Furthermore, $\tr_{\PT_{2}}(\ket{\psi}) = \tr(S)$ is a cyclic graph, so it follows a power law. Applying Holder's inequality as before allows concluding that $G$ is real, positive, and follows a power law. 
\end{proof}

\begin{rem}
In particular, the previous proposition implies that the $3$-partite graph $\RM_4 = \RE_{2,2} = \ME_{2}^3$ represented in Fig.~\ref{fig:ME_RE_RM} follows a power law. This was already established by means of a Cauchy-Schwarz inequality in Ref.~\cite{Penington2023}. 
\end{rem}

\begin{prop}\label{prop:asymptotic_R_G}   
Let $\psi=(\ket{\psi_N})_{n \in \mathbb{N}^*}$ and $G \in \cG_D$ such that: 1) $(G,\psi)$ obeys both the large-$N$ Ansatz \eqref{eq:Ansatz_large-N} and the large-$N$ factorization criterion \eqref{eq:factorization_criterion} (or equivalently \eqref{eq:largeNfactoproof}); and 2) $\tr_G$ follows a power law. Then:
   \begin{equation}
       \langle R_G (\ket{\psi_N})\rangle \underset{N\to \infty}{=} - \ln \abs{\langle \tr_G (\ket{\psi_N})\rangle}+o(1) \underset{N\to \infty}{=} \abs{s_G(\psi)} \ln\left(N\right) - \ln\abs{\mu_G(\psi)}+o(1)\,.
\end{equation}
\end{prop}
\begin{proof}
The proof combines the concentration result of Prop.~\ref{prop:concentration} with the uniform bound of Eq.~\eqref{eq:power_law} to show that $R_G (\ket{\psi_N})- \abs{s_G(\psi)} \ln\left(N\right) + \ln\abs{\mu_G(\psi)}$ converges to $0$ as $N$ goes to $+\infty$. The interested reader is referred to our companion article \cite{Factorization2026}. 
\end{proof}
\begin{rem}
    All the examples of invariants obeying a power law we know of are also real and positive (see Prop.~\ref{prop:poly_bound}). But since positivity is not necessary for Prop.~\ref{prop:asymptotic_R_G} to hold, it would be interesting to investigate whether or not there exists invariants that are not positive but still follow a power law.
\end{rem}

\begin{cor}  \label{cor:Lit_inv_and_exchange}
    Let $\varphi \eqdef (\ket{\varphi_N})_{N \in \mathbb{N}^*}$, where for every $N \in \mathbb{N}^*$, $\ket{\varphi_N}$ denotes the Haar-random state of local dimension $N$. If $G$ is a $D$-colored graph whose connected components belong to
    \begin{align*}
        \{ H \in \cG_D^{\conn}\, \big\vert \, H \;\textrm{cyclic}\}\cup \{\RM_{2n}^{(B)}\, \big\vert \, B \subset \{1, \ldots, D \},\, \abs{B} = D-2 ,\, n \in \mathbb{N}^*\} &\cup \{\RE_{m,n}^{(c)}\, \big\vert \, c \in \{1, 2, 3 \},\, n \in \mathbb{N}^*,\, m \in 2\mathbb{N}^*\} \\
        \nonumber
        &\cup  \{\PT_{2n}^{(c)}\, \big\vert \,  c \in \{1, 2, 3 \},\, n \in \mathbb{N}^*\} \,,
    \end{align*}
    then, the following asymptotic relation holds:
    \begin{equation}
        \langle R_G (\ket{\varphi_N})\rangle \underset{N\to \infty}{=} \abs{s_G(\varphi)} \ln\left(N\right) - \ln\left(\mu_G(\varphi)\right)+o(1)\,.
    \end{equation}
\end{cor} 

\begin{proof}
    This follows straightforwardly from the concentration phenomenon of Prop.~\ref{prop:concentration}, Cor.~\ref{cor:Factorization}, Prop.~\ref{prop:poly_bound} and Prop.~\ref{prop:asymptotic_R_G}.
\end{proof}

\begin{ex} 
    Concretely, let us consider the $3$-colored graph $G$ constructed by the disjoint unions of $p$ copies of $\RM_4$ and $q$ copies of $\PT_{2n}^{(3)}$, with $n\in\bb{N}^*$. By the use of Sec.~\ref{subsec:Evolution} and especially Thm.~\ref{th:TreeDegComp}, one computes $\abs{s_G(\varphi)} = 4p + (2 - 3n)q$ and $\mu_G(\varphi) = 3^p \cdot \mathrm{Cat}_{n}^q$, thus leading, thanks to Cor.~\ref{cor:Lit_inv_and_exchange}, to the asymptotic relation:
    \begin{equation}
        \mean{R_G(\ket{\varphi_N})} \underset{N \to \infty}{=} \pac{4p + (2 - 3n)q} \ln N - p \ln 3 - q \ln \mathrm{Cat}_{n} + o(1) \,. 
    \end{equation}
    Being incompatible, the example presented here is able to distinguish a Haar-random state from the HT states.
\end{ex}

\subsubsection{$\LU$-inequivalent hypergraph-tensor states}\label{sec:appli_LU_refs}

In the search for characterization of $\LU$-equivalence classes of HT states, we relied in Thm.~\ref{th:RefState_Equiv} on the colored graphs associated with the multi-entropy and reflected multi-entropy families. Our goal here is to clarify if and how the tree construction of trace-invariants can be used for the same purpose. 

The guiding idea in Thm.~\ref{th:RefState_Equiv} was to characterize the $\LU$-equivalence between $\ket{\psi_\alpha}$ and $\ket{\psi_{\beta}}$ by analyzing the sum
\begin{equation}
    \Sigma(G) \eqdef \ln(\abs{\tr_G(\ket{\psi_\alpha})}) - \ln(\abs{\tr_G(\ket{\psi_\beta})}) = \sum_{\substack{C \subset \paa{1,\dots,D} \\ \abs{C} > 1}} \pa{k(G) - \kappa(G\vert_C)} \ln \pa{\frac{\beta(C)}{\alpha(C)}} \,,
\end{equation}
and requiring $\Sigma(G) = 0$, $\forall G \in \cG_D^{\conn}$.

For graphs obtained by union or vertex contraction, $\Sigma(G)$ contains no information beyond that already present in the generators. Indeed, for any $G \in \bb{A}(\bb{B},\sqcup)$ or $G \in \bb{A}(\bb{B},\op)$, one simply has
\begin{equation}
    \Sigma(G) = \sum_{B \in \bb{B}} n_B(G) \Sigma(B) \,,
\end{equation}
where $n_B(G)$ counts the multiplicity of the generator $B \in \bb{B}$ appearing in the construction of $G$. Thus, these operations merely produce linear combinations of the constraints encoded in the elementary building blocks.

For a tree-constructed family of trace-invariants $G \in \bb{A}(\bb{B},\paa{\flip_c}_{1 \leq c \leq D})$, one has 
\begin{equation}
    \Sigma(G) = \sum_{B \in \bb{B}} n_B(G) \Sigma(B) + \sum_{c = 1}^D \pa{n_{\flip_c}(G) \sum_{\substack{C \subset \paa{1,\dots,D} \\ \abs{C} > 1 \rm{ and } c \in C}} \ln \pa{\frac{\beta(C)}{\alpha(C)}}} \,,
\end{equation}
where $n_{\cal{O}}(G)$ counts the multiplicity of the operation $\cal{O} \in \bb{O}$ appearing in the construction of $G \in \bb{A}(\bb{B}, \bb{O})$. For that matter, $\Sigma(G)$ contains a linear combination of the generator contributions plus $D$ independent terms, one for each color.

\begin{rem}
    These $D$ equations coincide with those produced by the melonic family: the tree construction with flips automatically encodes the melonic constraints, so that any non-melonic generator is sufficient to generate them. Consequently, flips allow one to reduce the number of independent trace-invariants that must be specified, although the total number of independent equations required remains $2^D - D - 1$.
\end{rem}

In the proof of Thm.~\ref{th:RefState_Equiv}, a naive choice involves, for $D=3$, five invariants $\ME_2^3$, $\ME_3^3$, and the three $\RE_{2,1}^{(c)}$ yielding one redundant equation. More economical choices are possible, for instance by discarding one reflected Rényi invariant, or by using $\PT_3$ together with the three melonic invariants at $k=2$, both leading to four equations for four unknowns. The flip-based tree construction shows that even this can be simplified further: a single non-melonic invariant, such as $\PT_3$, is sufficient. Indeed, flips automatically generate the graphs
\begin{equation}
    \PT_3 \,, \qquad \PT_3 \flip_1 \PT_3 \,, \qquad \PT_3 \flip_2 \PT_3 \,, \qquad \rm{and} \qquad \PT_3 \flip_3 \PT_3 \,,
\end{equation}
which yield four independent equations, including the three melonic constraints. Thus, while flips do not reduce the total number of required equations, they minimize the number of independent trace-invariants that must be chosen.

\subsubsection{Distinguishing power of tree-based trace-invariants} \label{subsubsec:DistinctionTree}

Let $\varphi \eqdef (\ket{\varphi_N})_{N \in \mathbb{N}^*}$, where for every $N \in \mathbb{N}^*$, $\ket{\varphi_N}$ denotes the Haar-random state of local dimension $N$.

\paragraph{Tripartite case.} In Fig.~\ref{fig:ExDistinction}, we previously exhibited a colored graph obtained via numerical exploration that is capable of distinguishing the states $\ket{\GHZ}$, $\ket{\GHZ\vert_B}$, $\ket{\phi_p}$, $\ket{\Phi_\tau}$ and $\ket{\varphi_N}$. Let us discuss how the tree-based construction provides a systematic way to generate such trace-invariants, and in fact produce entire families of graphs satisfying the required distinction properties in the tripartite setting.

\begin{figure}[ht]
	\centering
	\includegraphics[height = 3.5cm]{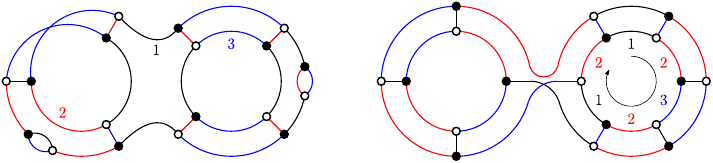}
	\caption{Left: an example of a $3$-colored graph constructed using flips as binary operations. One recognizes $\PT_3$ on the left and $\RM_4$ on the right. Right: another example of a $3$-colored graph constructed using the vertex contraction operation. One recognizes $\RM_4$ on the left and $\JRM_6^{(1,2,3,2,1,2)}$ on the right.}
	\label{fig:treeD3}
\end{figure}

As a representative example, consider a construction based on flips involving the partial transpose moment $\PT_3$, the realignment moment $\RM_4^{(2)}$, and melons. The role of these ingredients is complementary: $\RM_4^{(2)}$, being planar and incompatible, increases the degree of incompatibility; $\PT_3$, non-planar but compatible, contributes to the Gurau degree; and melons act as local corrections enhancing subsystem distinguishability. A simple tree construction is
\begin{equation} \label{eq:extreeD3}
    G = \pac{\PT_3} \flip_1 \RM_4 \flip_1 m \flip_2 m \,.
\end{equation}
where $\PT_3$ must be inserted first due to the absence of general results on its addition (Thm.~\ref{th:TreeDegComp}). The resulting connected graph satisfies
\begin{equation}
    k(G) = 9\,,  \quad F_{12}(G) = 2\,, \quad F_{13}(G) = 3 \,, \quad F_{23}(G) = 4\,, \quad g(G) = 1 \,, \quad \Delta(G) = 1\,, \quad \mu_G(\varphi) = 12\,.
\end{equation}
and is shown on the left of Fig.~\ref{fig:treeD3}. Unlike the earlier numerical example, these properties follow directly from the general evolution rules of Sec.~\ref{subsec:Evolution} and Thm.~\ref{th:TreeDegComp}.

This construction is far from unique: since flips may be applied to any edge of a given color, a single tree already generates multiple inequivalent graphs.

\ 

As a second illustration, one may consider a vertex contraction between the incompatible graph $\RM_4$, and the compatible non-planar graph $\JRM_6^{(1,2,3,2,1,2)}$, yielding
\begin{equation}
    G = \JRM_6^{(1,2,3,2,1,2)} \op \RM_4 \,.
\end{equation}
Since $\JRM_6^{(1,2,3,2,1,2)}$ already has pairwise-disjoint faces, this contraction produces a graph that distinguishes all GHZ states supported on subsystems from the global GHZ state, and separates both the 2-complete and Haar-random states from the remaining ones. The construction is shown on the right of Fig.~\ref{fig:treeD3}, and the resulting graph satisfies:
\begin{equation}
    k(G) = 9\,, \quad F_{12}(G) = 2\,, \quad F_{13}(G) = 4 \,, \quad F_{23}(G) = 3\,, \quad g(G) = 1 \,, \quad \Delta(G) = 1\,, \quad \mu_G(\varphi) = 3\,.
\end{equation}

\paragraph{$\mathbf{D=4}$ case.} A direct numerical search may be difficult to carry out, as discussed at the end of Sec.~\ref{ss:discrimination}. Nevertheless, the tree-construction method allows one to efficiently generate trace-invariants satisfying all the required distinction conditions to distinguish, \eg $\ket{\GHZ}$, $\ket{\GHZ\vert_B}$, $\ket{\phi_p}$, $\ket{\Phi_\tau}$ and $\ket{\varphi}$. We present here an explicit example in the case $D=4$, keeping in mind that the construction applies more generally.

We take as a starting point the $4$-complete colored graph $\K_{4,4}$, whose combinatorial features make it a convenient reference for the construction.\footnote{Let us briefly review some combinatorial properties about $\K_{4,4}$. The latter is connected and described (up to relabelling) by the permutations $\sigma_1 = \id$, $\sigma_2 = (1\,2\,3\,4)$, $\sigma_3 = (1\,3)(2\,4)$ and $\sigma_4 = (1\,4\,3\,2)$. It has genera of jackets equal to $g_{(1\,2\,3\,4)}(G) = 3$, $g_{(1\,2\,4\,3)}(G) = 2$ and $g_{(1\,3\,2\,4)}(G) = 2$. Moreover, one has $\omega_2(\K_{4,4}) = 7$, $\omega_3(\K_{4,4}) = 3$ and then $\omega_2^{(3)}(G) = \omega_2(G) - \omega_3(G) = 4$. Finally, its degree of compatibility is $\Delta(\K_{4,4}) = 1$, and its degeneracy is $\mu_{\K_{4,4}}(\varphi) = 4$ with $\MD{}(\K_{4,4}) = \paa{\nu_1,\nu_2,\nu_3,\nu_4}$ where $\nu_1 = (1)(2\,4)(3)$, $\nu_2 = (1\,2)(3\,4)$, $\nu_3 = (1\,3)(2)(4)$ and $\nu_4 = (1\,4)(2\,3)$.} 

By Thm.~\ref{th:TreeDegComp}, assumption~\ref{it:DADB_flip}, 
$\K_{4,4}$ can be flipped with a $4$-vertex cyclic graph associated with the bipartition $\H_{\paa{1,3}}\ot\H_{\paa{2,4}}$. We denote this graph by $\rm{N}^{(1,3)}=\rm{N}^{(2,4)}$, where edge $1$ is parallel to edge $3$ (and similarly $2$ to $4$). Additional flips, using assumption~\ref{it:Delta_haveTree} with $\RM_4^{(1,3)}$, $\RM_4^{(1,4)}$, and $\JRM_4^{(1,2,4,2)}$, allow one to break the jacket genus degeneracies and enforce the desired constraints on the degrees. Finally, melonic insertions, as prescribed by assumption~\ref{it:Delta_haveTree}, ensure that the resulting trace-invariant distinguishes subsystems.

\begin{figure}[ht]
	\centering
	\includegraphics[width = \textwidth]{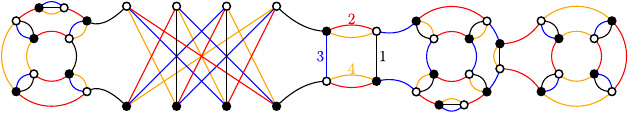}
	\caption{An example of a $4$-colored graph that satisfies the distinction conditions. From left to right, one recognize successively $\JRM_4^{(1,2,4,2)}$, $\K_{4,4}$, $\rm{N}^{(1,3)}$, $\RM_4^{(1,4)}$ and $\RM_4^{(1,3)}$.}
	\label{fig:treeD4}
\end{figure}

Altogether, this leads to the colored graph 
\begin{equation}
    G = \pac{\K_{4,4} \flip_1 \rm{N}^{(1,3)}} \flip_2 \RM_4^{(1,3)} \flip_3 \RM_4^{(1,4)} \flip_1 \JRM_4^{(1,2,4,2)} \flip_2 m \flip_2 m \flip_3 m\,.
\end{equation}
Up to the additional melons, the operation inside the parentheses must be performed first, in accordance with assumption~\ref{it:DADB_flip} of Thm.~\ref{th:TreeDegComp}. An explicit realization of the graph $G$ is shown in Fig.~\ref{fig:treeD4}. It is connected and, using the results of Sec.~\ref{subsec:Evolution}, one verifies that it satisfies the following properties:
\begin{equation}
    \begin{array}{lll}
        k(G) = 21\,, \qquad & \qquad F_{12}(G) = 5\,, \qquad & \qquad \kappa(G\vert_{123}) = 2\,, \\
        g_{(1\,2\,3\,4)}(G) = 6\,, \qquad & \qquad F_{13}(G) = 12\,, \qquad & \qquad \kappa(G\vert_{124}) = 4\,, \\
        g_{(1\,2\,4\,3)}(G) = 4\,, \qquad & \qquad F_{14}(G) = 11\,, \qquad & \qquad \kappa(G\vert_{134}) = 7\,, \\
        g_{(1\,3\,2\,4)}(G) = 3\,, \qquad & \qquad F_{23}(G) = 6\,, \qquad & \qquad \kappa(G\vert_{234}) = 3\,, \\
        \omega_2(G) = 13\,, \qquad & \qquad F_{24}(G) = 9\,, \qquad & \qquad \Delta(G) = 4\,, \\
        \omega_3(G) = 8\,, \qquad & \qquad F_{34}(G) = 10\,, \qquad & \qquad \mu_G(\varphi) = 8\,.
    \end{array}
\end{equation}

\subsubsection{Preserving or breaking distinction} \label{subsubsec:symmViaTree}

\paragraph{Preserving distinction.} Let $D\geq 2$, $\alpha: \{ B \subset \{ 1, \ldots , D\} \,|\, |B| \geq 2\} \to \mathbb{N}^*$ and $\beta: \{ B \subset \{ 1, \ldots , D\} \,|\, |B| \geq 2\} \to \mathbb{N}^*$ be two weight functions. We let $\ket{\psi_\alpha}$ and $\ket{\psi_\beta}$ be two HT states that are not $\LU$-equivalent and assume that we have found $G$ a $D$-colored graph such that
\begin{equation}
    \tr_G(\ket{\psi_\alpha}) \neq \tr_G(\ket{\psi_\beta}) \,,
\end{equation}
\ie $G$ is able to distinguish $\ket{\psi_\alpha}$ from $\ket{\psi_\beta}$ (the existence of such $G$ is guaranteed by Prop.~\ref{prop:charac_LU-1}). So, starting from a graph $G$ satisfying the distinction conditions described in Sec.~\ref{ss:discrimination} and \ref{subsubsec:DistinctionTree}, one may generate new trace-invariants while preserving these conditions by a suitable use of binary operations.

In practice, tracking all combinatorial quantities can be complicated. A convenient simplification relies on the use of the vertex contraction operation. Concretely, if $G$ satisfies the distinction conditions, so does the graphs
\begin{equation}
    G_n = \underbrace{G \op \cdots \op G}_{G \rm{ appears } n \rm{ times}} \,.
\end{equation}
Said differently, for $n \in \bb{N}^*$, a straightforward computation yields:
\begin{equation}
    \tr_{G_n}(\ket{\psi_\alpha}) = \tr_G^n(\ket{\psi_\alpha}) \neq \tr_G^n(\ket{\psi_\beta}) = \tr_{G_n}(\ket{\psi_\beta}) \,.
\end{equation}
Similarly, any  construction  of the type $G_n = G \sqcup \cdots \sqcup G$, where $G$ appears $n$ times, generates a family of trace-invariants with identical distinction properties.

\paragraph{Breaking distinction.}
In some situations, one may wish to distinguish only the \emph{structure} of HT states rather than their specific subsystems. This motivates the symmetry requirement
\begin{equation}
    \label{eq:symm}
    \kappa(G\vert_B) = f(|B|) \,,
\end{equation}
for some function $f$, which enforces invariance under color permutations.

A general symmetrization procedure is obtained via the union operation. Given a colored graph $G$ described by permutations $(\sigma_1,\dots,\sigma_D)$, consider the generating set
\begin{equation}
    \bb{B} = \paa{G \in \cG_D \, \vert \, (\sigma_{\tau(1)},\dots,\sigma_{\tau(D)}) \in S_{k(G)}^D (G) \quad \forall \tau \in S_D }\,.
\end{equation}
The graph
\begin{equation}
    H = \bigsqcup_{\tilde{G} \in \bb{B}} \tilde{G}
\end{equation}
is symmetric in the sense of Eq.~\eqref{eq:symm}. Indeed, for any subset $B \subset \{1,\dots,D\}$ with $\abs{B}=p$, one finds
\begin{equation}
    \kappa(H\vert_B)
    = p!(D-p)! \sum_{1 \le c_1 \le \cdots \le c_p \le D}
    \kappa(G\vert_{c_1,\dots,c_p}) = p!(D-p)! \kappa^{(p)}(G) \,,
\end{equation}
where $\kappa^{(p)}(G)$ was defined Eq.~\eqref{eq:def-of-kappa-p}. It is clear that $\kappa(H\vert_B)$ depends only on $\abs{B}$.

Once symmetric generators are available, the vertex contraction operation preserves the symmetry condition~\eqref{eq:symm}. Consequently, any composition built from symmetric generators using vertex contractions remains symmetric. For example, with
$\bb{B} = \paa{\RM_4,\PT_3,\ME_5^3}$, the graph
\begin{equation}
    \RM_4 \op \RM_4 \op \PT_3 \op \ME_5^3 \op \RM_4 \op \PT_3
\end{equation}
satisfies Eq.~\eqref{eq:symm}.

By contrast, the flip operation is not symmetric, as it depends explicitly on the chosen color. To preserve symmetry, flips must therefore be applied once on each color. For instance, given two symmetric $3$-colored graphs $G_1$ and $G_2$, the construction
\begin{equation}
    \pac{\pac{G_1 \flip_1 G_2 \flip_2 G_2 \flip_3 G_2}
    \flip_1 G_1 \flip_2 G_1 \flip_3 G_1}
\end{equation}
is symmetric in the sense of Eq.~\eqref{eq:symm}. A simple example is obtained by taking $G_1=\RM_4$ and $G_2=m$.

\newpage

\appendix

\section{Calculus with Haar-random states and Gaussian tensors} \label{A:Wick}

\paragraph{Haar-random state.}
In Sec.~\ref{sub:intro-of-ref-states}, we introduced Haar-random states $\ket{\varphi} = U \ket{\varphi_0}$ where $U$ is a Haar-distributed $N^D$ by $N^D$ unitary matrix and $\ket{\varphi_0}$ is any pure state, and whose components can be shown to be normalized complex Gaussian variables $\varphi_{i_1 \ldots i_D} = T_{i_1 \dots i_D} / \norm{T}$ (see Ref.~\cite{Nechita2007}).

To compute the average of a polynomial in these states, one may use the following analogue of Wick's theorem: 
\begin{equation} 
\label{eq:wick-for-wein}
    \mean{\prod_{s=1}^k  \varphi_{i_1^s \ldots i_D^s} {\bar \varphi}_{j_1^s \ldots j_D^s}   } =  \frac{f_{k,D,N}}{N^{kD}}  \sum_{\sigma\in S_k} \prod_{s=1}^k \prod_{c=1}^D\delta_{i^s_c\,j^{\sigma(s)}_c}  \,,
\end{equation}
where $f_{k,D,N}=\frac{N^{Dk}\pa{N^D - 1}!}{\pa{N^D - 1 + k}!}$.  Eq.~\eqref{eq:wick-for-wein} implies the expression of the average of trace-invariants of Haar-random states given in  Eq.~\eqref{eq:meanLUWeing}: indeed, from the definition of Eq.~\eqref{eq:LUinv}, the products of Kronecker deltas give rise to a factor of $N$ for every cycle of $\sigma_c\sigma^{-1}$.

Indeed, the left-hand side of Eq.~\eqref{eq:wick-for-wein} reads
\begin{equation} 
\label{eq:wick-for-wein-1}
    \mean{\prod_{s=1}^k  \varphi_{i_1^s \ldots i_D^s} {\bar \varphi}_{j_1^s \ldots j_D^s}   } =  \sum_{\{\vec{a^s}, \vec{ b^s}\}_s}  \prod_{s=1}^k  (\varphi_0)_{a_1^s \ldots a_D^s} (\bar{\varphi_0})_{b_1^s \ldots b_D^s} \mean{\prod_{s=1}^k  U_{i_1^s \ldots i_D^s\;;\;a_1^s \ldots a_D^s} {\bar U}_{j_1^s \ldots j_D^s\;;\;b_1^s \ldots b_D^s}    }  \,,
\end{equation}
and the average on the right-hand side can be computed using Weingarten calculus. 
One has, for any $k\in \mathbb{N}^*$ (see Refs.~\cite{Collins:2003ncs, Collins:2006jgn}): 
\begin{align}
    \label{eq:WeinDef}
    \mean{\prod_{s=1}^k  U_{i_1^s \ldots i_D^s\;;\;a_1^s \ldots a_D^s} {\bar U}_{j_1^s \ldots j_D^s\;;\;b_1^s \ldots b_D^s}    } &= \int dU \; \left(\prod_{s=1}^k  U_{i_1^s \ldots i_D^s\;;\;a_1^s \ldots a_D^s} {\bar U}_{j_1^s \ldots j_D^s\;;\;b_1^s \ldots b_D^s} \right)  \nonumber\\
    &=
    \sum_{\sigma,\tau\in S_k}
     \prod_{c=1}^D  \left(  \prod_{s=1}^k \delta_{i_c^s , j_c^{\sigma(s)}}  \right)
      \left( \prod_{s=1}^k \delta_{a_c^s , b_c^{\tau(s)}} \right)  \, W_{N,D} (\sigma\tau^{-1})  \,,
\end{align}
where $dU$ is the normalized Haar measure on the group $U(N^D)$ of $N^D\times N^D$ unitary matrices, and $W_{N,D}$ is its Weingarten function, which satisfies \cite{Collins:2003ncs, Collins:2006jgn}:
\begin{equation}
\label{eq:summation-of-wein}
    \sum_{\nu\in S_k} W_{N,D} (\nu) = \frac{(N^D-1)!}{(N^D+k-1)!}\,. 
\end{equation}
One can therefore rewrite~\eqref{eq:wick-for-wein-1} as:
\begin{equation} 
    \mean{\prod_{s=1}^k  \varphi_{i_1^s \ldots i_D^s} {\bar \varphi}_{j_1^s \ldots j_D^s}   } =  \sum_{\sigma, \tau \in S_k} W_{N,D}(\sigma\tau^{-1})\prod_{s=1}^k \prod_{c=1}^D \delta_{i^s_c\,j^{\sigma(s)}_c} \sum_{\{\vec{a^s}, \vec{ b^s}\}_s}  \prod_{s=1}^k  (\varphi_0)_{a_1^s \ldots a_D^s} (\bar{\varphi_0})_{b_1^s \ldots b_D^s}  \prod_{c=1}^D \delta_{a^s_c\,b^{\tau(s)}_c} \,.
\end{equation}
The rightmost sum is $(\innprod{\varphi_0}{\varphi_0})^k=1$, and instead of summing over $\tau$, one may sum over $\nu=\sigma\tau^{-1}$, therefore: 
\begin{equation} 
    \mean{\prod_{s=1}^k  \varphi_{i_1^s \ldots i_D^s} {\bar \varphi}_{j_1^s \ldots j_D^s}   } =  \sum_{\sigma\in S_k}  \left(\sum_{\nu \in S_k} W_{N,D}(\nu)\right)\prod_{s=1}^k \prod_{c=1}^D \delta_{i^s_c\,j^{\sigma(s)}_c}  \,,
\end{equation}
which proves Eq.~\eqref{eq:wick-for-wein} using Eq.~\eqref{eq:summation-of-wein}.\footnote{We note that, though convenient, the full power of Weingarten's calculus is not necessary to establish the previous result. A more elementary proof can be obtained as follows: 1) the operator $\int dU\,( U \ket{\phi_0}\bra{\phi_0}U^\dagger)^{\otimes k}$ commutes with the diagonal action of the unitary group on $\H^{\otimes k}$, and the same is true on the symmetric subspace of $\H^{\otimes k}$ (\ie the subspace of states that are invariant under permutations of the tensor factors); 2) the symmetric subspace is irreducible in that representation, so by Schur's lemma, one may infer that $\int dU\,( U \ket{\phi_0}\bra{\phi_0}U^\dagger)^{\otimes k}$ is proportional to the orthogonal projector on the symmetric subspace; 3) the trace of the previous equation can be easily integrated, which fixes the multiplicative factor; 4) plugging this result into Eq.~\eqref{eq:wick-for-wein-1} directly yields Eq.~\eqref{eq:wick-for-wein}, bypassing the introduction of the Weingarten function. See \eg Ref.~\cite{Harrow:2013nib} for further detail.}

\paragraph{Gaussian tensors.} Here we detail the relation to the Gaussian state $\ket{X}$, whose components $X_{i_1, \ldots, i_D}$ are centered i.i.d.~Gaussian complex variables with variance $1/N^D$. 
The density of the distribution is given by $\cal{Z}_0^{-1} \e{-N^D \norm{X}^2}$ where $\norm{X}^2 =  \sum_{i_1, \ldots, i_D=1}^N X_{i_1, \ldots, i_D} \bar{X}_{i_1, \ldots, i_D}$ and letting $\d X \eqdef \prod_{i,j,k\dots} \d X_{i  j  k\dots}$ and $\d \bar{X} \eqdef \prod_{i,j,k\dots} \d \bar{X}_{i  j  k\dots}$,
\begin{equation}
    \cal{Z}_0 \eqdef \int \e{-N^D \norm{X}^2} \d X \d \bar{X}\,.
\end{equation}
The expression of Eq.~\eqref{eq:meanLUinv}  for the averages of the trace-invariants is obtained using Wick's theorem, which can be formulated as:
\begin{equation} 
\label{eq:wick}
    \mean{\prod_{s=1}^k  X_{i_1^s \ldots i_D^s} {\bar X}_{j_1^s \ldots j_D^s}   } = \frac{1}{N^{kD}} \sum_{\sigma\in S_k} \prod_{s=1}^k \prod_{c=1}^D \delta_{i^s_c\,j^{\sigma(s)}_c}  \,.
\end{equation}
The normalization implies that  $\mean{\norm{X}}^2 = 1$, but this normalization holds only on average, unlike the Haar-random state $\ket{\varphi}$ for which $\norm{\ket{\varphi}}=1$ with probability $1$, so that $\ket{X}$ cannot be considered a random quantum state. 

The two tensor distributions agree at first order at large $N$ though, in the sense that Eq.~\eqref{eq:wick} and Eq.~\eqref{eq:wick-for-wein} differ only by a factor $f_{k,D,N}$, which for $k,D$ fixed goes to 1 when $N$ goes to infinity, so that the moments characterizing the distributions differ by the same factor:
\begin{equation}
 \mean{\tr_{G}(\ket{\varphi})} =  f_{k(G),D,N}      \mean{\tr_{G}(\ket{X})}\,. 
\end{equation}
In particular, the dominant contributions when $N$ goes to infinity coincide for the two distributions for any $G$: 
\begin{equation}
 \mean{\tr_{G}(\ket{\varphi})} \underset{N\to\infty}{\sim}      \mean{\tr_{G}(\ket{X})}\,, 
\end{equation}
so that in the same way as in Def.~\ref{def:LargeN_LU_eq}, the two $\LU$-invariant random tensors $\varphi$ and $X$ may be said to  be asymptotically $\LU$-equivalent.\footnote{More precisely, we can introduce sequences $\varphi =(\ket{\varphi_N})_{N \in \mathbb{N}^*}$ and $X =(\ket{X_N})_{N \in \mathbb{N}^*}$, where $\ket{\varphi_N}$ (resp.~$X_N$) denotes a Haar-distributed (resp.~Gaussian) random state on a $D$-partite state space with local dimension $N$ (as constructed above). Then the claim is that those two sequences are asymptotically $\LU$-equivalent in the sense of Def.~\ref{def:LargeN_LU_eq}.}

\section{On the \texorpdfstring{$p$}{p}-complete degree} \label{A:GurauDeg}

Let $G \in \cG_D$ be a $D$-colored graph with $k(G)$ white vertices. This section is focused on a generalized version of the Gurau degree called $p$-complete degree and denoted by $\omega_p(G)$,  $2\leq p \leq D-1$, introduced in Sec.~\ref{ss:discrimination}, Eq.~\eqref{eq:GenGurauDeg} as:  
\begin{equation}
    \omega_p(G) =\binom{D-1}{p-1} \kappa(G) + \binom{D-1}{p} k(G) - \kappa^{(p)}(G),\hspace{1cm} \kappa^{(p)}(G) = \sum_{1 \leq c_1 < \cdots < c_p \leq D} \kappa(G\vert_{{c_1,\dots,c_p}}) \,, 
\end{equation}
Due to their trivial definitions, we do not include in this appendix the cases $p=1$ and $p=D$ introduced in Sec.~\ref{sec:LU-and-ref-states}.

From the definition, it is clear that $\omega_p(G) \in \bb{Z}$ for any $D$-colored graph. Melonic and planar graphs have been introduced in Sec.~\ref{sub:inv-from-lit}. Inspired by the proofs of Ref.~\cite{Gurau2011_2}, this appendix aims to prove the following theorem:
\begin{theo} \label{th:GenGurauDegPos}
    For all $D$-colored graph $G$ and $2 \leq p \leq D-1$, the $p$-complete degree is non-negative and vanishes if and only if $G$ is  melonic ($D>3$) or planar  ($D=3$).
\end{theo}

\subsection{Preliminaries on the \texorpdfstring{$p$}{p}-complete degree}

We remind the following expressions of $\omega_p^{(q)}(G)$ for $2\le p < q\le D$  (Eq.~\eqref{eq:zeta_pq} and Eq.~\eqref{eq:def-omega-pq}): 
\begin{align} 
\label{eq:recall-omegq-pq}
    \omega_p^{(q)}(G) &= \binom{q-1}{p-1} \kappa^{(q)}(G) + \binom{q-1}{p} \binom{D}{q} k(G) - \binom{D-p}{q-p} \kappa^{(p)}(G)\\&=  \binom{D-p}{q-p} \omega_p(G) - \binom{q-1}{p-1}\omega_q(G) \,.
    \label{eq:recall-omegq-pq-diff}
\end{align}
In this appendix, we exclude the case $p=q$, since $\omega_p^{(p)} = 0$.

\begin{lem} \label{lem:omega_p,q}
    Let $G$ be a $D$-colored graph. For $2 \leq p < q \leq D$, we have
        \begin{equation}
        \label{eq:toprove-omega-pq-appendix}
        \omega_p^{(q)}(G) = \sum_{c_1< \cdots < c_q} \omega_p(G\vert_{c_1,\dots,c_q}) \,.
    \end{equation}
\end{lem}

\begin{proof}For $1 \leq c_1 < \cdots < c_q \leq D$, one has $k(G\vert_{c_1,\dots,c_q}) = k(G)$ and therefore, by definition of $\omega_p$: 
    \begin{equation}
    \label{eq:in-the-proof-omega-pq-appendix}
        \omega_p(G\vert_{c_1,\dots,c_q}) = \binom{q-1}{p-1} \kappa(G\vert_{c_1,\dots,c_q}) + \binom{q-1}{p} k(G) - \kappa^{(p)}(G\vert_{c_1,\dots,c_q}) \,.
    \end{equation}
Summing Eq.~\eqref{eq:in-the-proof-omega-pq-appendix} over $c_1<\cdots<c_q$ and comparing it to Eq.~\eqref{eq:recall-omegq-pq}, we see that proving Eq.~\eqref{eq:toprove-omega-pq-appendix} boils down to showing
   \begin{equation}
        \sum_{c_1< \cdots < c_q} \kappa^{(p)}(G\vert_{c_1,\dots,c_q}) =  \binom{D-p}{q-p} \kappa^{(p)}(G) \,.
    \end{equation}
This is indeed true: for  $i_1 < \cdots < i_p \in \{c_1,\ldots,c_q\}$, the term $\kappa((G\vert_{c_1,\dots,c_q})\vert_{i_1, \dots,i_p})$ appears as many times as there are ways to complete $i_1, \dots , i_p$ to $q$ colors, that is, to choose $q-p$ colors among the remaining $D-p$. 
\end{proof}

A property of $\omega_p$ that will be needed is its invariance under $1$\textit{-dipole contractions} (this is known for $\omega_2$, see \eg Ref.~\cite{Gurau2011_2}).  

\begin{figure}[ht]
    \centering
    \includegraphics[height = 1.2cm]{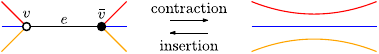}
    \caption{Edge insertion and contraction.}
    \label{fig:1dip}
\end{figure}

\begin{defi}\label{def:dipole}
If $G\in \cG_D$ and  $c \in \paa{1,\dots,D}$,  $G^{\hat{c}}$ denotes the graph without the edges of color $c$.
    \begin{itemize}
        \item A $1$-dipole is a single edge $e=(v,\bar v)$ such that  $v$ and $\bar{v}$ belong to different connected components of $G^{\hat{c}}$.
        \item An edge contraction  is the operation that removes an edge $e=(v,\bar v)$ and the vertices $v$ and $\bar{v}$, and connects the pending edges respecting the coloring (Fig.~\ref{fig:1dip}). The inverse operation is an edge insertion.
         \item A  $1$-dipole contraction (insertion) is the contraction (insertion) of a 1-dipole. 
    \end{itemize}
\end{defi}

\begin{lem} \label{lem:omega_p_1dip}
    For $2 \leq p \leq D-1$, the $p$-complete degree is invariant under $1$-dipole contractions and insertions.
\end{lem}

\begin{proof}
    Let $G \in \cG_D$ and $\tilde{G}$ resulting from the  contraction of a $1$-dipole $e$ of color $c$ in $G$. One has $k(G) - k(\tilde{G}) = 1.$
    Furthermore, since $e$ is a 1-dipole, $v$ and $\bar v$ belong to two different color $i,j$ faces of $G$ for any two $i,j$ different from $c$, so that in $\tilde G$ the edges of color $i,j$ belong to the same color $i,j$ face. The operation therefore, leaves the number of connected components unchanged: 
    \begin{equation}
        \label{eq:kappa-in-edge-contract}
        \kappa(G) = \kappa(\tilde{G})\,.
    \end{equation} The same is true for any subgraph $G\vert_{C}$ with $C\subset\{1,\ldots, D\}$, $c\in C$: $\kappa(G\vert_{C}) = \kappa(\tilde{G}\vert_{C})$. If, however, $c\notin C$,  then the edge-contraction reduces the number of connected components by one: $\kappa(G\vert_{C}) = \kappa(\tilde{G}\vert_{C}) + 1$.
    As a consequence:
    \begin{equation}
    \label{eq:kappap-in-edge-contract}
        \kappa^{(p)}(G) - \kappa^{(p)}(\tilde{G}) =   \binom{D-1}{p} \,.
    \end{equation}
    Compiling all the results, we have 
  $$
        \omega_p(\tilde{G}) = \binom{D-1}{p-1} \kappa(\tilde{G}) + \binom{D-1}{p}k(\tilde{G}) - \kappa^{(p)}(\tilde{G}) 
        = \binom{D-1}{p-1} \kappa(G) + \binom{D-1}{p}\pac{k(G) - 1} - \kappa^{(p)}(G) + \binom{D-1}{p} \,,
$$
    so that $\omega_p(G) = \omega_p(\tilde{G})$. 
\end{proof}

\subsection{Graphs of vanishing \texorpdfstring{$p$}{p}-complete degree}

We first focus on  $\omega_{D-1}$, defined as: 
\begin{equation}
\label{eq:omega_D-1}
\omega_{D-1}(G) = (D-1)\kappa(G) + k(G) - \kappa^{(D-1)}(G)\,.
\end{equation}

\begin{lem} \label{lem:omega_D-1}
    For any $D$-colored graph $G$, we have $\omega_{D-1}(G) \in\mathbb{N}$. Moreover, for $D \geq 3$, we have the equivalence 
    \begin{equation}
    \label{eq:vanish-omegaD-1-vs-degree}
        \omega_{D-1}(G) = 0  \quad \Longleftrightarrow \quad \omega_2(G) = 0 \,.
    \end{equation}
\end{lem}

\begin{proof}
    We first prove that $\omega_{D-1}(G) \geq 0$ (it is an integer by definition). Starting from a graph $G$, we choose a 1-dipole and contract it, and do so inductively until there are no more 1-dipoles, and we call $H$ the resulting graph. At each step,   both $k(G) - k(H)$ and $\kappa^{(D-1)}(G) - \kappa^{(D-1)}(H)$ grow by one (Eq.~\eqref{eq:kappap-in-edge-contract}), so that 
    \begin{equation}
    \label{eq:k-vs-kappa_D-1}
        k(G) - k(H) = \kappa^{(D-1)}(G) - \kappa^{(D-1)}(H)\,.
    \end{equation}
    However, since there is no 1-dipole in $H$, removing the edges of any color leaves the number of connected components unchanged, that is, for any $c$, $\kappa(H^{\hat c})=\kappa(H)=\kappa(G)$, where the second equality follows from Eq.~\eqref{eq:kappa-in-edge-contract}. As a consequence:
    
    	\begin{equation}
		\kappa^{(D-1)}(H) = D\kappa(G) \,. 
	\end{equation}
	Combining this with Eq.~\eqref{eq:k-vs-kappa_D-1}, we may express $\omega_{D-1}$ given in Eq.~\eqref{eq:omega_D-1} as 
	\begin{equation}
	\label{eq:omega_D-1-in-terms-of-H}
	    \omega_{D-1}(G) = k(H) - \kappa(G) = k(H) - \kappa(H)\ge 0\,.
	\end{equation}

    \ 
    
    We now proceed to showing Eq.~\eqref{eq:vanish-omegaD-1-vs-degree}. It is trivial for $D=3$, so we assume that $D>3$. From Eq.~\eqref{eq:recall-omegq-pq-diff} and Lem.~\ref{lem:omega_p,q}:
    \begin{equation}
    \omega_2^{(D-1)}(G)= (D-2) (\omega_2(G) - \omega_{D-1}(G))= \sum_{c=1}^D \omega_2(G^{\hat c})\,
    \end{equation}
    and it is well-known (see Refs.~\cite{Gurau:2011aq,Gurau2011_1,Gurau2011_2}) that the Gurau-degree $\omega_2$ is always non-negative, so that $\omega_2^{(D-1)}(G)\ge 0$. Therefore: 
    \begin{equation}
        \omega_2(G) \ge  \omega_{D-1}(G)\ge 0\,,
    \end{equation}
    so that $ \omega_2(G)=0 \Rightarrow \omega_{D-1}(G)=0$. 
    
    Reciprocally and with the notations above, from Eq.~\eqref{eq:omega_D-1-in-terms-of-H}, one has that $\omega_{D-1}(G) = \omega_{D-1}(H)$. Therefore,  $\omega_{D-1}(G)=0$ implies that $k(H)=\kappa(H)$, that is, $H$ is a collection of colored graphs with two vertices per connected component, so that $\omega_{2}(H)=0$. The graph $G$ is obtained from $H$ by a sequence of 1-dipole insertions, and since $\omega_2$ remains the same under such operations (Lem.~\ref{lem:omega_p_1dip} and   Ref.~\cite{Gurau2011_2}), we conclude that $\omega_{2}(G)=0$. 
\end{proof}

\begin{rem}
\label{rm:graphs-of-positive-omega-D-1}
Fixing $p>0$, from the proof of Lem.~\ref{lem:omega_D-1}, one can generate all the connected graphs $G$  with $\omega_{D-1}(G)=p$ starting from the 1-dipole-free colored graphs $H$ with $k(H)= p + 1$ and inductively inserting 1-dipoles. For instance, for $p=1$, one should start with the different cyclic graphs (Sec.~\ref{sss:necklace}) with $\abs{A}, \abs{\bar A} >1$ and insert 1-dipoles, and so on.
\end{rem}

\paragraph{Proof of Thm.~\ref{th:GenGurauDegPos}.}We first show that for any $G\in \cG_D$, $D\ge 3$, and any $p\in\{2,\ldots,D-1\}$, 
\begin{equation}
    \label{eq:omegap_is_non-neg}
    \omega_p(G) \in \mathbb{N}\,.  
\end{equation}

For $D=2$, $\omega_2=0$, identically. For $D=3$, it is known that $\omega_2 \in \bb{N}$ (see Refs.~\cite{Gurau:2011aq,Gurau2011_1,Gurau2011_2}).    
We fix $D\ge 4$ and $2 \leq p \leq D-1$, and assume that we have already shown that $\omega_p(G)\ge 0$ for any $G\in\cG_{D'}$, $D'<D$. From Lem.~\ref{lem:omega_p,q}, it is therefore true that
\begin{equation}
   \label{eq:omegapD-1_is_non-neg}
   \omega_p^{(D-1)}(G) = \sum_{c=1}^D \omega_p(G^{\hat c})\geq 0 \,.
\end{equation}
Furthermore, combining this with  Eq.~\eqref{eq:recall-omegq-pq} and Lem.~\ref{lem:omega_D-1}, we obtained the desired result
\begin{equation}
\label{eq:omega-p_ito_D-1_etc}
    \omega_p(G) = \frac{1}{D-1}\binom{D-1}{p-1} \omega_{D-1}(G)  + \frac{1}{D-p}\omega_p^{(D-1)}(G) \geq 0 \,.
\end{equation}
The proposition then follows by induction on $D$.

Finally, the second statement of Thm.~\ref{th:GenGurauDegPos} amounts to showing that for any $D\ge 3$, any $G\in\cG_D$, and  any $p \in \paa{2,\dots,D-1}$, 
\begin{equation}
    \label{eq:vanish-omegap-vs-degree}
	\omega_p(G) = 0 \qquad \Longleftrightarrow \qquad \omega_2(G) = 0 \,.
\end{equation}

Combining  Eq.~\eqref{eq:recall-omegq-pq-diff} (formulated as in Eq.~\eqref{eq:zeta_pq}), Lem.~\ref{lem:omega_p,q}, and Eq.~\eqref{eq:omegap_is_non-neg}, one has for any $1\le p < q\le D$:
	\begin{equation}
	\frac 1 {\cal{I}_p} \omega_p(G) \ge  \frac 1 {\cal{I}_q} \omega_q(G) \ge 0\,.
	\end{equation}
In particular, $\omega_2(G)=0 \Rightarrow \omega_q(G)=0$. Reciprocally, from Eq.~\eqref{eq:omega-p_ito_D-1_etc}, Eq.~\eqref{eq:omegap_is_non-neg} and Eq.~\eqref{eq:omegapD-1_is_non-neg}, if $\omega_p(G)=0$  then $\omega_{D-1}(G)=0$, and from Lem.~\ref{lem:omega_D-1},  $\omega_2(G)=0$. This concludes the proof of Eq.~\eqref{eq:vanish-omegap-vs-degree} and therefore of Thm.~\ref{th:GenGurauDegPos}. \qed

\section{Distinction power of trace-invariants considered in the literature}
\label{sec:table-in-appendix}

The tables below summarize for $D=3$ whether the trace-invariants listed in Sec.~\ref{sub:inv-from-lit} discriminate different HT states listed in Sec.~\ref{ss:discrimination} or not. We use the following notation: {\cmark} means that the distinction condition is satisfied, {\xmark} means that it is not satisfied,  and {\cmark/\xmark} means that there are examples of invariants or states for which the condition is satisfied, and others for which it is not.

\begin{table}[H]
    \begin{center}
        \begin{tabular}{ |>{\centering\arraybackslash}m{3.7cm}||c|c|c|c|c|c|c|c||c|c|c| } \hline
            \multirow{2}{*}{$D=3$} & \multicolumn{8}{c||}{$\GHZ$} & \multicolumn{3}{c|}{$\GHZ^{\frac 1 2}$} \\ \cline{2-12}
             & $\GHZ\vert_{c_1 c_2}$ & $\GHZ\vert_{c_1 c_3}$ & $\GHZ\vert_{c_2 c_3}$ &  $\Phi_{c_1}$ & $\Phi_{c_2}$ & $\Phi_{c_3}$  & $\phi_2$ & $\varphi$ & $\Phi_{c_1}$ & $\Phi_{c_2}$ & $\Phi_{c_3}$ \\ \hline \hline
            cyclic for $(c_1,c_2)$ vs $c_3$
            & \cmark & \multicolumn{2}{c|}{\xmark}       & \multicolumn{2}{c|}{\cmark} & \xmark    & \xmark     & \xmark & \xmark & \xmark       & \cmark\\ \hline
            $c_1$-melonic, non cyclic
                             & \multicolumn{2}{c|}{\cmark} & \xmark       & \multicolumn{3}{c|}{\cmark}     & \xmark     & \xmark & \xmark & \cmark       & \cmark \\ \hline
            other melonic 
                             & \multicolumn{3}{c|}{\cmark}                         & \multicolumn{3}{c|}{\cmark}     & \xmark       & \xmark & \multicolumn{3}{c|}{\cmark}\\ \hline\hline
            planar non-melonic\tablefootnote{Including $\RM_{2n}^{(c_1)}$ for $n>1$ and $\RE_{m,n}^{(c_1)}$ for $n>1$ and $m\ge 2$.}   
                             & \multicolumn{3}{c|}{\cmark}                         & \multicolumn{3}{c|}{\cmark}   & \xmark        & \cmark & \multicolumn{3}{c|}{\cmark}\\ \hline\hline
            $\PT_{2n+1}^{(c_1)}$, $n>0$  
                             & \multicolumn{3}{c|}{\xmark}                         & \multicolumn{3}{c|}{\xmark}     & \cmark       & \cmark  & \multicolumn{3}{c|}{\cmark}\\ \hline
            $\PT_{2n}^{(c_1)}$, $n>1$  
                             & \multicolumn{2}{c|}{\xmark} & \cmark       &\xmark  & \multicolumn{2}{c|}{\cmark}     & \cmark     & \cmark  &  \multicolumn{3}{c|}{\cmark} \\ \hline
            $\JRM_k^{\vec{i}}$, $k>3$ $^{\ref{foot:JRM0}}$
                             & \cmark/\xmark $^{\ref{foot:JRM}}$ & \cmark & \cmark/\xmark $^{\ref{foot:JRM}}$                        & \cmark & \cmark/\xmark $^{\ref{foot:JRM}}$ & \cmark   & \cmark        & \cmark  & \multicolumn{3}{c|}{\cmark}\\ \hline\hline
            $\ME_{n}^3$, $n>2$ 
                             & \multicolumn{3}{c|}{\cmark}                         & \multicolumn{3}{c|}{\cmark}   & \cmark        & \cmark  & \multicolumn{3}{c|}{\cmark}\\ \hline
        \end{tabular}
    \end{center}
\hspace{0.3cm}
    \begin{center}
        \begin{tabular}{ |>{\centering\arraybackslash}m{4cm}||c|c|c| } \hline
            $D=3$ & $\GHZ\vert_{c_1 c_2}$ vs $\GHZ\vert_{c_1 c_3}$ & $\GHZ\vert_{c_1 c_2}$ vs $\GHZ\vert_{c_2 c_3}$ &  $\GHZ\vert_{c_1 c_3}$ vs $\GHZ\vert_{c_2 c_3}$\\ \hline \hline
            cyclic for $(c_1,c_2)$ vs $c_3$ 
                            & \cmark & \cmark  & \xmark \\ \hline
            $\RM_{2n}^{(c_1)}$, $n>2$
                            & \xmark & \cmark  & \cmark \\ \hline
            $\RE_{m,n}^{(c_1)}$, $n>2$, $m \ge 2$
                            & \xmark & \cmark  & \cmark \\ \hline
            $\PT_{2n+1}^{(c_1)}$, $n>0$  
                             & \xmark & \xmark  & \xmark \\ \hline
            $\PT_{2n}^{(c_1)}$, $n>1$  
                             & \xmark & \cmark  & \cmark \\ \hline
            $\ME_{n}^3$, $n>1$ 
                             & \xmark & \xmark  & \xmark \\ \hline
        \end{tabular}
    \end{center}
\end{table}   

The table below provides the values taken by the different combinatorial quantities introduced in Sec.~\ref{ss:discrimination} for the trace-invariants listed in Sec.~\ref{sub:inv-from-lit} for which a simple answer exists.

In the first table, we consider $\{c_1, c_2, c_3\} = \{1,2,3\}$. In the second table, we let $c,c_1\in\{1,\ldots, D\}$ and $B_1, B_2, B_3, B, B'\subset \{1,\ldots, D\}$ are such that $c\in B_1$, $B_1,B_2,B_3$ form a partition of $\{1,\ldots, D\}$, $p_c=\lvert B_c\rvert$ and $p_1 \le \frac D 2$,  $B \cap B_1 \neq \emptyset$ and $B \cap \bar{B_1} \neq \emptyset$, while $B'\subset B_1$. The tripartite coarse-grained invariants are considered for blocks $B_1,B_2,B_3$ replacing the colors 1, 2, 3. We also set $\alpha_{p,p_1, p_2}= \binom{p_1+p_2-1}{p} - \binom{p_1}{p} - \binom{p_2}{p}$, as well as  $\alpha_{p,p_1, p_2, p_3}= \binom{p_1+p_2+p_3-1}{p} - \binom{p_1}{p} - \binom{p_2}{p} - \binom{p_3}{p}$, $\alpha'_{p,p_1,p_2}= \binom{p_1+p_2}{p} - \binom{p_1}{p} - \binom{p_2}{p}$, and  
$\beta_{p,q,p_1, p_2}=\binom{p_1+p_2-p}{q-p}\alpha_{p,p_1, p_2} - \binom{q-1}{p-1}\alpha_{q,p_1,p_2}$.

\begin{table}[H]
    \begin{center}
    \scalebox{0.75}{
        \begin{tabular}{ |c||c|c|c|c|c|c|c|c|c|c| } \hline
        $D=3$            & $\kappa - k$ & $K_{c_1c_2}$ & $K_{c_1 c_3}$ &   $K_{c_2c_3}$ &  $\frac{1}{2}F_{c_1} - \kappa$ & $\frac{1}{2}F_{c_3} - \kappa$ & $\Omega_{c_1}$ & $\Omega_{c_3}$ & $\omega_2 = 2g_\tau$ & $\Delta$ \\ \hline\hline
        cyclic   for $(c_1,c_2)$ vs $c_3$, $k>1$ 
            &$1-k$     &  $1-k$  &\multicolumn{2}{c|}{$0$} &   $\frac {k-1} 2$ &   $0$    & $0$ & $k-1$ & $0$     & $0$ \\ \hline
        $\RM_{2n}^{(c_1)}$, $n>1$  
                             & $1-2n$   &      \multicolumn{2}{c|}{$n-1$}  &   $1$  & $n-1$ &      $\frac n 2$                    & $1$& $n-1$  & $0$       & $1$ \\ \hline
        $\RE_{m,n}^{(c_1)}$ $n>1,m\ge 2$
                             & $1-mn$    &    \multicolumn{2}{c|}{$n-1$}  &   $n(m - 2) + 1$ & $n-1$&    $\frac{n(m-1)}{2}$                     & $1 + n(m - 2)$& $n-1$  & $0$      & $1$ \\ \hline
        $\PT_{2n+1}^{(c_1)}$, $n>0$  
                             & $-2n$  &      \multicolumn{3}{c|}{$0$}& \multicolumn{2}{c|}{$0$}                          & \multicolumn{2}{c|}{$2n$}   & $2n$      & $0$ \\ \hline
        $\PT_{2n}^{(c_1)}$, $n>1$  
                             & $1-2n$  &    \multicolumn{2}{c|}{$0$} &   $1$ &  $0$ &   $\frac 1 2$                     & $2n-1$  &  $2(n-1)$ & $2(n-1)$       & $0$ \\ \hline
        $\ME_{n}^3$, $n>2$ 
                             & $1 - n^2$   &      \multicolumn{3}{c|}{$n-1$} & \multicolumn{2}{c|}{$n-1$}                     & \multicolumn{2}{c|}{$(n-1)^2$}     & $(n-1)(n-2)$        & ?‘$\frac{n(n-1)}{2}$? \\ \hline
        \end{tabular}
        }
    \end{center}
\end{table}
\begin{table}[H]
    \begin{center}
        \scalebox{0.8}{
        \begin{tabular}{ |>{\centering\arraybackslash}m{2.7cm}||c|c|c|c|c| } \hline
            $D>3$            &  $K_{B'}$ &   $K_{B}$ &  $ \frac{1}{D-1}F_c-  \kappa$  &   $\Omega_{c}$  & $\omega_p$ \\ \hline \hline
            cyclic for $B_1$ vs $\bar{B_1}$ 
                             &$ k-1$&  $0$     &$(k-1)\frac{p_1-1}{D-1}$& $(k-1)(D-1-p_1)$                       & $(k-1)\alpha_{p,p_1, p_2}$     \\ \hline
            $\RM_{2n}^{(B_1)}$  
                            &\multicolumn{2}{c|}{Eq.~\eqref{eq:K-for-block-RM}}    & $n \frac{D+p_1 - 2 }{D-1} - 1$  & $1+n(D-p_1-2)$    & $  (2n-1)\alpha_{p,p_1,p_2, p_3} - (n-1)(\alpha'_{p, p_1, p_2} + \alpha'_{p, p_1, p_3}) - \alpha'_{p, p_2, p_3}$       \\ \hline 
            $\PT_{2n+1}^{(B_1)}$ 
                             &\multicolumn{2}{c|}{Eq.~\eqref{eq:K-for-block-PT}}    & $2n\frac{p_1-1}{D-1}$                      & $2n(D-1-p_1)$    & $2n \alpha_{p,p_1, p_2, p_3}$       \\ \hline 
            $\PT_{2n}^{(B_1)}$ 
                              &\multicolumn{2}{c|}{Eq.~\eqref{eq:K-for-block-PT}}     & $\frac{(2n-1)(p_1-1) +p_3}{D-1}$   & $(2n-1)(D-1-p_1) - p_3$    & $ (2n-1)\alpha_{p,p_1, p_2, p_3} - \alpha'_{p,p_2, p_3}$        \\  \hline
            $\ME_{n}^D$, $n>2$ 
                             & \multicolumn{2}{c|}{$n^{D-\abs{B}} - 1$}        &$n^{D-2} - 1$& $1+ n^{D-1} (D-2) - n^{D-2}(D-1)$                      &  $\cal{I}_p\left[1 - n^{D-1} + \frac{D}{p}(n^{p-1} - 1)n^{D-p}\right] $    \\ \hline
            $\RME_{m,n}^{(c_1)}$ $n>1,m\ge 2$
                             & \multicolumn{2}{c|}{Eq.~\eqref{eq:facesRME}}        &$> 0$& $> 0$                          & $\binom{D-1}{p-1} (1 - n^{D-p}) + 2\binom{D-1}{p} (n^{p-1} - 1) n^{D-p-1}$    \\ \hline
        \end{tabular}
        }
    \end{center}
    \begin{center}
        \scalebox{0.8}{
        \begin{tabular}{ |>{\centering\arraybackslash}m{2.7cm}||c|c|c|c| } \hline
            $D>3$             & $\omega_p^{(q)}$ & $g_\tau$ & $\Delta$ \\ \hline \hline
            cyclic for $B_1$ vs $\bar{B_1}$ 
                             & $(k-1)\beta_{p,q,p_1, p_2}$ \tablefootnote{For $D=4$, the non-melonic cyclic graph satisfies $\omega_2^{(3)} = 0$.}       & $0$\tablefootnote{\label{foot:exists-jacket}This is only true for the appropriate choice of $\tau$.} & $(k-1)\frac{p_1(p_1-1)}{2}$ \\ \hline
            $\RM_{2n}^{(B_1)}$  
                            & $\propto\ (2n-\mathrm{const})$ & $0^{\ref{foot:exists-jacket}}$ & Eq.~\eqref{eq:Delta-block-RM}\\ \hline 
            $\PT_{2n+1}^{(B_1)}$ 
                             & $\propto\ 2n$ & $2n^{\ref{foot:exists-jacket}}$ & Eq.~\eqref{eq:Delta-block-PTodd}\\ \hline 
            $\PT_{2n}^{(B_1)}$ 
                              & $\propto\ (2n-\mathrm{const})$ & $2(n-1)^{\ref{foot:exists-jacket}}$ & Eq.~\eqref{eq:Delta-block-PTeven} \\  \hline
            $\ME_{n}^D$, $n>2$ 
                             &  $\cal{I}_p \binom{D-p}{q-p}
      \frac{D}{pq} \pac{(q-p)n^{D-1} - q n^{D-p} + p n^{D-q}}$        & $1 - n^{D-1} + \frac{D}{2}(n-1)n^{D-2}$ & ?‘$ (n-1)n^{D-2}\frac{(D-1)(D-2)}{4}$? \\ \hline
            $\RME_{m,n}^{(c_1)}$ $n>1,m\ge 2$
                             &  
                             Eq.~\eqref{eq:omegapq-RME}     
     & $1 - n^{D-2} + (D-2)(n - 1)n^{D-3}$ & ?‘$\frac{(D-1)(D-2)}{2}n^{D-3}$? \\ \hline
        \end{tabular}
        }
    \end{center}
\end{table}

\begin{equation}
\label{eq:K-for-block-RM}
    K_{B''}(\RM_{2n}^{(B_1)}) = \left\{
    \begin{array}{ll}
        2n -1 \,, \rm{ if } B'' \subset B_1 \rm{ , } B'' \subset B_2 \rm{ or } B'' \subset B_3 \,, \\
        n - 1 \,, \rm{ if } B'' \subset B_2 \cup B_1 \rm{ with } B'' \cap B_2 \neq \emptyset \,, \, B'' \cap B_1 \neq \emptyset \,, \textrm{ or } 2\leftrightarrow 3 \\
        1 \,, \rm{ if } B'' \subset B_2 \cup B_3 \rm{ with } B'' \cap B_2 \neq \emptyset \,, \, B'' \cap B_3 \neq \emptyset \,, \\
        0 \,, \rm{ otherwise}\,.
    \end{array}
    \right.
\end{equation}

\begin{equation}
\label{eq:Delta-block-RM}
    \Delta(\RM_{2n}^{(B_1)}) = \left\{
    \begin{array}{ll} 
\binom{p_2}{2}(2n-2) + \binom{p_1}{2}n + p_1 p_2 (n-1) \textrm{ if } \abs{B_3} > \frac D 2\, \textrm{ or }2\leftrightarrow 3, \\
    \pac{\binom{p_2}{2} + \binom{p_3}{2}}n + p_2 p_3 \textrm{ otherwise}.
        \end{array}
    \right.
\end{equation}

\begin{equation}
\label{eq:K-for-block-PT}
    K_{B''}(\PT_k^{(B_1)}) = \left\{
    \begin{array}{ll}
        k - 1 \,, \rm{ if } B'' \subset B_1 \rm{ , } B'' \subset B_2 \rm{ or } B'' \subset B_3 \,, \\
        1 \,, \rm{ for } k \rm{ even, and if } B'' \subset B_1 \cup B_2 \rm{ with } B'' \not\subset B_1 \rm{ and } B'' \not\subset B_2 \,, \\
        0 \,, \rm{ otherwise}\,.
    \end{array}
    \right.
\end{equation}

\begin{equation}
\label{eq:Delta-block-PTodd}
    \Delta(\PT_{2n+1}^{(B_1)}) = \left\{
    \begin{array}{ll} 
(k-1)\pa{\binom {p_i}{2} + \binom {p_j}{2}  + \frac{p_i p_j}{2}} \textrm{ if } \abs{B_k} > \frac D 2\, \textrm{ for }\{i,j,k\}=\{1,2,3\}, \\
    \frac{k-1}{2} \sum_{i=1}^3\binom{p_i}{2} \textrm{ otherwise}.
        \end{array}
    \right.
\end{equation}

\begin{equation}
\label{eq:Delta-block-PTeven}
    \Delta(\PT_{2n}^{(B_1)}) = \left\{
    \begin{array}{lll} 
(k-2) \pa{\binom {p_1}{2} + \binom {p_2}{2}  + \frac{p_1 p_2}{2}} + \binom{p_1}{2}\textrm{ if } \abs{B_3} > \frac D 2\,  \textrm{ or }2\leftrightarrow 3, \\
(k-1) \pa{\binom {p_2}{2} + \binom {p_3}{2}  + \frac{p_2 p_3}{2}} + \frac{p_2p_3}{2}\textrm{ if } \abs{B_1} > \frac D 2\,, \\
    \frac{k-2}{2} \sum_{i=1}^3\binom{p_i}{2} + \binom{p_1}{2}\textrm{ otherwise}.
        \end{array}
    \right.
\end{equation}

\begin{equation} \label{eq:facesRME}
    K_{B''}(\RME_{m,n}^{(c_1)}) = \left\{
    \begin{array}{ll}
        2n^{D-1 - \abs{B''}} + n^{D-2}(m-2) -1 \,, \rm{ if } c_1 \notin B'' \,, \\
        n^{D-\abs{B''}}-1 \,, \rm{ if } c_1 \in B'' \,. 
    \end{array}
    \right.
\end{equation}

\begin{equation}
\label{eq:omegapq-RME}
 \begin{split}
     &\omega_p^{(q)}(\RME_{m,n}^{(c_1)})= {\cal{I}_p \binom{D-p}{q-p}}  \frac{p}{D} \Bigl[ \binom{D-1}{q-1} (n^{D-q}-n^{D-p}) \\
     &\hspace{3.5cm}+ 2 \binom{D-1}{q} \pac{\frac{q}{p}\frac{D-p}{D-q}(n^{p-1}-1)n^{D-p-1} - (n^{q-1}-1)n^{D-q-1}}\Bigr] 
      \end{split}  
\end{equation}

\section{Proof of Thm.~\ref{th:TreeDegComp}} \label{A:DeltaAB}

This section is subdivided in two parts. First, we prove a result regarding the disjoint union and the flip operation for graphs having tree-like dominant graphs and regarding the vertex contraction operation for graphs having \textit{only} tree-like dominant graphs. Secondly, we prove the statement of Thm.~\ref{th:TreeDegComp} for compatible graphs or graphs with bounds on the degree of compatibility.

\begin{defi}
    Let $D \geq 3$, $A,B \in \cG_D$, $e_A$ (resp.~$e_B$) be an edge of color $c \in \paa{1,\dots,D}$ of $A$ (resp.~$B$), and $v_A$, $v_B$ be vertices of distinct colors of respectively $A$ and $B$. For $\cal{O} \in \paa{\sqcup, \flip_{(e_A,e_B)},\op_{(v_A,v_B)}}$, we define the following sets:
    \begin{equation}
        \MD{}(A) \, \cal{O}\,\MD{}(B) \eqdef \paa{\widehat G \in \cG_{D+1}(A \, \cal{O}\, B) \, \vert \, \widehat G = \widehat A \, \cal{O}\, \widehat B\,, \rm{ for } (\widehat A, \widehat B) \in \MD{}(A) \times \MD{}(B)} \,,
    \end{equation}
    where the set $\MD{}$ was defined Eq.~\eqref{eq:MD} and Eq.~\eqref{eq:MD_2}.
\end{defi}

\begin{rem}
    We should emphasis that, in what follows, we will not treat the case $D=2$ which is well-known from the literature to be compatible graphs that factorize at large-$N$, and whose set $\MD{}(C_k)$ is given by non-crossing permutations (see \eg Refs.~\cite{NicaSpeicher2006,Gurau2025} and definition of non-crossing permutations in Sec.~\ref{subsec:GraphStructure}, paragraph ``The $c$-degree $\Omega_c$.''). The combinatorial factor is then: $\mu_{C_k}(\ket{\varphi}) = \mathrm{Cat}_k$.
\end{rem}

\paragraph{Objective of the proofs.} Let $D\geq 3$, $A,B \in \cG_D$, $e_A$ and $e_B$ be edges of a given color of respectively $A$ and $B$ and $v_A$ and $v_B$ be vertices of distinct colors of $A$ and $B$. If, for $\cal{O} \in \paa{\sqcup, \flip_{(e_A,e_B)},\op_{(v_A,v_B)}}$, one proves the following:
\begin{equation} \label{eq:toproveFLIP_OP}
    \MD{}(A  \, \cal{O} \, B) = \MD{}(A) \, \cal{O} \, \MD{}(B) \,,
\end{equation}
then, the multiplicative property of the $\mu$'s follows directly, \ie from Eqs.~\eqref{eq:MD} and \eqref{eq:mu_vs_chi}, one has
\begin{equation}
    \mu_{A\,\cal{O}\,B}(\ket{\varphi}) = \sum_{\widehat G \in \MD{}(A  \, \cal{O} \, B)} \Xi(\widehat G) = \sum_{\widehat A \in \MD{}(A)} \sum_{\widehat B \in \MD{}(B)} \Xi(\widehat A) \cdot \Xi(\widehat B) = \mu_{A}(\ket{\varphi}) \mu_{B}(\ket{\varphi}) \,.
\end{equation}

Regarding the additivity of the degree of compatibility, let us assume Eq.~\eqref{eq:toproveFLIP_OP}. Since $F(A \sqcup B) = F(A) + F(B)$, $k(A \sqcup B) = k(A) + k(B)$ and $F_0(\widehat A \sqcup \widehat B) = F_0(\widehat A) + F_0(\widehat B)$, Eq.~\eqref{eq:Delta_0} implies that
\begin{equation}
    \Delta(A \sqcup B) = \Delta_0(\widehat A \sqcup \widehat B) = \Delta_0(\widehat A) + \Delta_0(\widehat B) = \Delta(A) + \Delta(B) \,.
\end{equation}
Moreover, replacing the edges in indices by the given color, one sees that $F(A \flip_c B) = F(A) + F(B) - (D-1)$, $k(A \flip_c B) = k(A) + k(B)$ and $F_0(\widehat A \flip_c \widehat B) = F_0(\widehat A) + F_0(\widehat B) - 1$, which implies by Eq.~\eqref{eq:Delta_0} that
\begin{equation}
    \Delta(A \flip_c B) = \Delta_0(\widehat A \flip_c \widehat B) = \Delta_0(\widehat A) + \Delta_0(\widehat B) = \Delta(A) + \Delta(B) \,.
\end{equation}
The same goes for the vertex contraction operation, since, by forgetting the vertices in indices, one has $F(A \op B) = F(A) + F(B) - D(D-1)/2$, $k(A \op B) = k(A) + k(B) - 1$, and $F_0(\widehat A \op \widehat B) = F_0(\widehat A) + F_0(\widehat B) - D$, which leads to:
\begin{equation}
    \Delta(A \op B) = \Delta_0(\widehat A \op \widehat B) = \Delta_0(\widehat A) + \Delta_0(\widehat B) = \Delta(A) + \Delta(B) \,.
\end{equation}
So, the main point of the following section will be to prove, in various contexts, Eq.~\eqref{eq:toproveFLIP_OP} from which the desired results of Thm.~\ref{th:TreeDegComp} follow.

\subsection{On graphs having tree-like dominant graphs or only tree-like dominant graphs} \label{Ass:flip_op}

Graphs having tree-like dominant graphs exhibit, by the use of Ref.~\cite{Factorization2026}, the following property regarding binary operations.
\begin{prop} \label{prop:tree_union_flip}
    Let $D \geq 3$, $c \in \paa{1,\dots,D}$, $A,B \in \cG_D$ and $e_A$ (resp.~$e_B$) be an edge of color $c$ in $A$ (resp.~$B$). If $A \sqcup B$ has tree-like dominant graphs, then:
    \begin{equation}
        \forall \cal{O} \in \paa{\sqcup, \flip_{(e_A,e_B)}}\,, \quad \MD{}(A \, \cal{O} \, B) = \MD{}(A) \, \cal{O} \, \MD{}(B)\,. 
    \end{equation}
\end{prop}

\begin{proof}
    Assume that $\widehat G \in \cG_{D+1}(A \sqcup B)$ is not connecting $A$ and $B$ (but it can link connected components in $A$ or in $B$), then, there exists $\widehat A \in \cG_{D+1}(A)$ and $\widehat B \in \cG_{D+1}(B)$ such that $\widehat G = \widehat A \sqcup \widehat B$ and for which $F_0(\widehat G) = F_0(\widehat A) + F_0(\widehat B)$. However, if $\widehat G \in \MD{}(A \sqcup B)$ then, one should maximize the number $F_0(\widehat G)$ leading to $(\widehat A, \widehat B) \in \MD{}(A) \times \MD{}(B)$ and
    \begin{equation}
        \Delta(G) = \Delta(A) + \Delta(B) \,.
    \end{equation}
    On the contrary, if $\widehat G \in \cG_{D+1}(A \sqcup B)$ is connecting $A$ and $B$ (but not necessarily connecting all connected components of $A \sqcup B$) and $A \sqcup B$ has tree-like dominant graphs, then, from Ref.~\cite{Factorization2026}, for all $\widehat A \in \MD{}(A)$ and $\widehat B \in \MD{}(B)$ we have the following inequality 
    \begin{equation}
       F_0(\widehat G) \leq F_0(\widehat A) + F_0(\widehat B) - D  \quad \rm{ meaning that } \quad \Delta_0(\widehat G) \geq \Delta(A) + \Delta(B) + \frac{D(D-1)}{2} > \Delta(A) + \Delta(B)\,.
    \end{equation}
    The latter proves that if $\widehat G \in \MD{}(A \sqcup B)$ then $\widehat G$ cannot connect $A$ and $B$, \ie
    \begin{equation}
        \MD{}(A \sqcup B) = \MD{}(A) \sqcup \MD{}(B)\,.
    \end{equation}
    
    Regarding the flip operation, let $e_A$ (resp.~$e_B$) be an edge of color $c$ in $A$ (resp.~$B$). Let us write $G = A \flip_{(e_A,e_B)} B$ with $A$ and $B$ connected, since, from what we just proved, one can add connected components after the flip operation. From any $\widehat G \in \cG_{D+1}(G)$ one can construct a $(D+1)$-colored graph $\widehat H \in \cG_{D+1}(A \sqcup B)$ by unflipping the edges of color $c$ previously flipped to recover the edges $e_A$ and $e_B$. \\
    Assume first that $\widehat H$ is not connecting $A$ and $B$ (but it can link connected components in $A$ or in $B$), then, there exists $\widehat A \in \cG_{D+1}(A)$ and $\widehat B \in \cG_{D+1}(B)$ such that $\widehat H = \widehat A \sqcup \widehat B$. The latter means that $F_0(\widehat G) = F_0(\widehat H) - 1$, \ie 
    \begin{equation}
        \Delta_0(\widehat G) = \Delta_0(\widehat H) = \Delta_0(\widehat A) + \Delta_0(\widehat B) \,.
    \end{equation}
    Minimizing $\Delta_0(\widehat G)$ leads to $\widehat A \in \MD{}(A)$ and $\widehat B \in \MD{}(B)$ as well as $\Delta(G) = \Delta(A) + \Delta(B)$. \\
    On the contrary, if $\widehat H \in \cG_{D+1}(A \sqcup B)$ is connecting $A$ and $B$ (but not necessarily connecting all connected components of $A \sqcup B$) and $A \sqcup B$ has tree-like dominant graphs, then, from Ref.~\cite{Factorization2026}, for all $\widehat A \in \MD{}(A)$ and $\widehat B \in \MD{}(B)$ we have
    \begin{equation}
        F_0(\widehat H) \leq F_0(\widehat A) + F_0(\widehat B) - D \,.
    \end{equation}
    Generally, one has $F_0(\widehat G) = F_0(\widehat H) \pm 1$. It implies that the degrees of compatibility are related by:
    \begin{equation}
        \Delta_0(\widehat G) = \Delta_0(\widehat H) - \frac{D-1}{2} (1 \pm 1) \geq \Delta(A) + \Delta(B) + \frac{(D-1)(D-2)}{2} > \Delta(A) + \Delta(B) \,,
    \end{equation}
    meaning that if $\widehat G \in \MD{}(G)$ then $\widehat H$ cannot link $A$ and $B$. Hence, we have
    \begin{equation}
        \MD{}(A \flip_{(e_A,e_B)} B) = \MD{}(A) \flip_{(e_A,e_B)} \MD{}(B)\,.
    \end{equation}
\end{proof}

If one is interested in the vertex contraction operation, then one should assume graphs having only tree-like dominant graphs. Indeed, one has the following result:

\begin{prop} \label{prop:Tree_only}
    Let $D \geq 3$, $A,B \in \cG_D$ and $v_A$, $v_B$ be vertices of distinct colors in $A$ and $B$. If $A \sqcup B$ has only tree-like dominant graphs, then:
    \begin{equation}
        \MD{}(A \op_{(v_A,v_B)} B) = \MD{}(A) \op_{(v_A,v_B)} \MD{}(B)\,. 
    \end{equation}
\end{prop}

\begin{proof}
    From Thm.~\ref{prop:tree_union_flip}, we can restrict ourselves to the connected case, \ie $A$ and $B$ are connected $D$-colored graphs. Let $(v_A,v_B)$ be vertices of distinct colors of respectively $A$ and $B$. We denote by $G = A \op_{(v_A,v_B)} B$. For any $\widehat G \in \cG_{D+1}(G)$ we will refer to $\widehat H \in \cG_{D+1}^{\conn}(A\sqcup B)$ as the $(D+1)$-colored graph obtained from $\widehat G$ by recovering the vertices previously contracted (\ie $v_A$ and $v_B$) and by adding an edge $e_\rm{L}$ of color $0$ related $v_A$ to $v_B$. A direct computation yields $F_0(\widehat G) = F_0(\widehat H)$, thus leading to 
    \begin{equation}
        \Delta_0(\widehat G) =  \Delta_{0}(\widehat H) - \frac{D(D-1)}{2}\,.
    \end{equation}
    It is then equivalent to minimize $\Delta_0(\widehat G)$ for $\widehat G \in \cG_{D+1}(G)$ or $\Delta_{0}(\widehat H)$ for $\widehat H \in \cG_{D+1}^{\conn}(A \sqcup B)$ (with $v_A$ linked to $v_B$ by an edge of color $0$). However, if $A \sqcup B$ has only tree-like dominant graphs, then all connected $(D+1)$-colored graph  $\widehat H$ minimizing $\Delta_{0}(\widehat H)$ while having an edges of color $0$ related $A$ to $B$, are given by 
    \begin{equation}
        \widehat H = \widehat{A} \flip_{(e,e')} \widehat B \,,
    \end{equation}
    where $(\widehat A,\widehat B) \in \MD{}(A) \times \MD{}(B)$ and $e$ (resp.~$e'$) is an edge of color $0$ in $\widehat A$ (resp.~$\widehat B$) attached to $v_A$ (resp.~$v_B$). For that matter, contracted the vertices $v_A$ and $v_B$ leads to the following characterization of the set $\MD{}$:
    \begin{equation}
        \forall \widehat{G} \in \MD{}(A \op_{(v_A,v_B)} B)\,,\quad \exists (\widehat A,\widehat B) \in \MD{}(A) \times \MD{}(B)\,,\quad \rm{such that}\quad \widehat G = \widehat A \op_{(v_A,v_B)} \widehat B\,.
    \end{equation}
    Saying differently, $\MD{}(A \op_{(v_A,v_B)} B) = \MD{}(A) \op_{(v_A,v_B)} \MD{}(B)$. 
\end{proof}

\subsection{Results on graphs with bounded degrees of compatibility}

For the union operation and the flip operation, a generic result can be stated as long as the degree of compatibility of $A$ and $B$ is upper bounded.
\begin{prop} \label{prop:UnionFlipAdd}
    Let $D \geq 3$, $A,B \in \cG_D$ and $(e_A,e_B)$ be vertices of $A$ and $B$. 
    \begin{enumerate}
        \item If either $\Delta(A) + \Delta(B) < D(D-1)/2$ or $\Delta(A \sqcup B) < D(D-1)/2$ then:
        \begin{equation}
            \Delta(A\sqcup  B) = \Delta(A) + \Delta(B)\quad \rm{and} \quad \MD{}(A \sqcup B) = \MD{}(A) \sqcup \MD{}(B)\,.
        \end{equation}
        \item If either $\Delta(A) + \Delta(B) < (D-1)(D-2)/2$ or $\Delta(A \flip_{(e_A,e_B)} B) < (D-1)(D-2)/2$ then:
        \begin{equation}
            \Delta(A\flip_{(e_A,e_B)} B) = \Delta(A) + \Delta(B)\quad \rm{and} \quad \MD{}(A \flip_{(e_A,e_B)} B) = \MD{}(A) \flip_{(e_A,e_B)} \MD{}(B)\,.
        \end{equation}
    \end{enumerate}
\end{prop}

\begin{proof}
    Point 1. of Prop.~\ref{prop:UnionFlipAdd} was proved in Ref.~\cite{Factorization2026}. 
    
    Define $G = A \flip_{(e_A,e_B)} B$. Let $(\widehat A,\widehat B) \in \MD{}(A) \times \MD{}(B)$. Since, we have 
    \begin{equation}
        \widehat A \flip_{(e_A,e_B)} \widehat B \in \cG_{D+1}(A \flip_{(e_A,e_B)} B)\,,
    \end{equation} 
    we have the inequality: $\Delta(G) \leq \Delta(A) + \Delta(B)$. Hence, if $\Delta(A) + \Delta(B) < (D-1)(D-2)/2$ so does $\Delta(G)$. 
    
    Let us then assume that $\Delta(G) < (D-1)(D-2)/2$. For $\widehat G \in \cG_{D+1}(G)$, we define $\widehat H \in \cG_{D+1}(A \sqcup B)$ to be the $(D+1)$-colored graph obtained by restoring on $\widehat G$ the edges $(e_A,e_B)$ lost by the flip operation. Due to the bipartite structure of $G$, we have $F_0(\widehat G) = F_0(\widehat H) \pm 1$, which yields the following expression:
    \begin{equation}
        \Delta_0(\widehat G) = \Delta_0(\widehat H) - \frac{D-1}{2} \pac{1 \pm 1} \geq \Delta_0(\widehat H) - (D-1) \,.
    \end{equation}
    Applying the following lower bound shown in Ref.~\cite{Factorization2026} 
    \begin{equation}
        \forall \widehat H \in \cG_{D+1}(A \sqcup B) \,, \quad \Delta_0(\widehat H) \geq \frac{D(D-1)}{2}(\kappa(A \sqcup B) - \kappa(\widehat H)) \,,
    \end{equation}
    provides a lower bound on $\Delta_0(\widehat G)$, namely
    \begin{equation}
        \Delta_0(\widehat G) \geq \frac{D-1}{2}\pac{D(\kappa(A \sqcup B) - \kappa(\widehat H)) - 2} \,.
    \end{equation}
    Hence, if $\widehat G \in \MD{}(G)$ then using the assumption on the upper bound of $\Delta(G)$, one has that
    \begin{equation}
        \frac{D-1}{2}\pac{D(\kappa(A \sqcup B) - \kappa(\widehat H)) - 2} \leq \Delta( G) < \frac{(D-1)(D-2)}{2} \,,
    \end{equation}
    which leads to a contradiction whenever $\kappa(\widehat H) < \kappa(A \sqcup B)$. As a conclusion, $\widehat H$ is disconnected, \ie $\widehat G$ is written as a flip of color $c$ (the color of $e_A$ and $e_B$) between $(\widehat A, \widehat B) \in \cG_{D+1}(A) \times \cG_{D+1}(B)$. Maximizing the faces with color $0$ in $\widehat G$ leads to: $(\widehat A, \widehat B) \in \MD{}(A) \times \MD{}(B)$.
\end{proof}

For the vertex contraction operation, we only have a result regarding compatible graphs under given conditions. 

\begin{prop} \label{prop:comp}
    Let $D \geq 3$, $A,B \in \cG_D$, and $(v_A,v_B)$ be vertices of distinct colors of respectively $A$ and $B$. If $A$ and $B$ are both compatible $D$-colored graphs and $v_A$ or $v_B$ or both belong to at least one face of size four or less, then: 
    \begin{equation}
        \Delta(A \op_{(v_A,v_B)} B) = 0\,, \qquad \rm{and} \qquad \MD{}(A \op_{(v_A,v_B)}B) = \MD{}(A) \op_{(v_A,v_B)} \MD{}(B)\,.
    \end{equation}
\end{prop}

The proof requires the following lemma. 
\begin{lem} \label{lem:Local}
    Let $D \geq 3$. If we assume that: 
    \begin{enumerate}
        \item $A$ and $B$ are compatible connected $D$-colored graphs.
        \item There exists $(v_A,v_B)$, vertices of respectively $A$ and $B$ of distinct colors such that one or both belongs to, at least one face of size at most four.
    \end{enumerate}
    then, all $(D+1)$-colored graphs $\widehat G$ of $\cG_{D+1}^{\conn}(A \sqcup B)$ containing an edge of color $0$ linking $v_A$ to $v_B$, and satisfying:
    \begin{equation}
        \Delta_0(\widehat G) = \frac{D(D-1)}{2} \,,
    \end{equation}
    are given by:
    \begin{equation}
        \widehat G = \widehat A \flip_{(e_A^0,e_B^0)} \widehat B \,,
    \end{equation}
    where $(\widehat{A},\widehat B) \in \MD{}(A)  \times \MD{}(B)$ and $e_A^0$ (resp.~$e_B^0$) is an edge of color $0$ of $\widehat A$ (resp.~$\widehat B$) such that $e_A^0$ is attached to $v_A$ (resp.~$e_B^0$ is attached to $v_B$).
\end{lem}

\begin{proof}
    We suppose points 1. and 2. of Lem.~\ref{lem:Local}. Let $\widehat G$ be a $(D+1)$-colored graph of $\cG_{D+1}^{\conn}(A \sqcup B)$ containing an edge of color $0$ linking $v_A$ to $v_B$, and satisfying:
    \begin{equation}
        \Delta_0(\widehat G) = \frac{D(D-1)}{2} \,.
    \end{equation}
    Given a label of the vertices, we let $(\sigma_1,\dots,\sigma_D)$ be the permutations describing $G$ and $\nu$ the permutation describing the edges color $0$ of $\widehat G$. By point 2., we denote by $f_A$ (resp.~$f_B$) the smallest face whose $v_A$ (resp.~$v_B$) belongs to. By assumption, $\min \paa{\abs{f_A},\abs{f_B}} \leq 4$ and, without loss of generality, say that $\min \paa{\abs{f_A},\abs{f_B}} = \abs{f_A}$. 
    
    Since $A$ and $B$ as well as $\widehat G$ are connected, it means that there are no vanishing Gromov products. Indeed, recall that for all pairs of distinct colors $(i,j) \in \paa{1,\dots,D}$, we have
    \begin{equation}
        \GP{\sigma_i}{\sigma_j}{\nu} = g(\widehat G\vert_{\paa{0,i,j}}) + \kappa(A\sqcup B \vert_{\paa{i,j}}) - \kappa(\widehat G \vert_{\paa{0,i,j}}) \,,
    \end{equation}
    where, $\kappa(A\sqcup B \vert_{\paa{i,j}}) - \kappa(\widehat G \vert_{\paa{0,i,j}}) \geq 1$ due to the presence of the edge between $v_A$ and $v_B$ in $\widehat G$ denoted by $e_\rm{L}$. 
    
    For that matter, the fact that $\Delta_0(\widehat G) = D(D-1)/2$ implies -- under the assumptions that $\widehat G$ is connected -- that $\widehat G$ is maximizing the number $F_0(\widehat G)$ since all Gromov products must be one (\ie minimal in this context). 
    
    Moreover, since we must have $\kappa(A\sqcup B \vert_{\paa{i,j}}) - \kappa(\widehat G \vert_{\paa{0,i,j}}) = 1$ for all pairs of distinct colors, it means that the edges of color $0$ linking $A$ to $B$ can only link vertices from the face $f_A$ to vertices from the face $f_B$. What is more, the latter implies that at most $\abs{f_A}$ edges of color $0$ are connecting $A$ to $B$. 
    
    We now make a case distinction on the size of $f_A$ as shown in Fig.~\ref{fig:CaseDist}.
    
    \begin{figure}[ht]
        \centering
        \includegraphics[width = \textwidth]{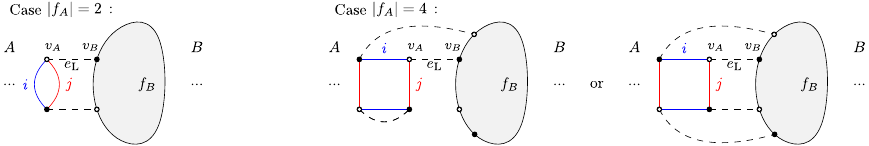}
        \caption{Case distinction considered in the proof of Lem.~\ref{lem:Local}.}
        \label{fig:CaseDist}
    \end{figure}

    \begin{itemize}
        \item Case $\abs{f_A} = 2$: Since $\widehat G$ is connected by the use of $e_\rm{L} = (v_A,v_B)$, it means that there exists another edge of color $0$ connecting the other vertex of $f_A$ to a vertex of $f_B$. Hence, there exists $(\widehat A, \widehat B) \in \cG_{D+1}(A) \times \cG_{D+1}(B)$ and edges of color $0$ related to $v_A$ and $v_B$, \ie the edges $(e_A^{0},e_B^{0})$ of respectively $\widehat A$ and $\widehat B$, such that:
        \begin{equation}
            \widehat G = \widehat A \flip_{(e_A^{0},e_B^{0})} \widehat B \qquad \rm{and} \qquad \Delta_0(\widehat G) = \Delta_0(\widehat A) + \Delta_0(\widehat B) + \frac{D(D-1)}{2} \,.
        \end{equation}
        Since $\Delta_0(\widehat G) = D(D-1)/2$, we must require: $(\widehat A, \widehat B) \in \MD{}(A) \times \MD{}(B)$.
        \item Case $\abs{f_A} = 4$: Since $\widehat G$ is connected by the use of $e_\rm{L} = (v_A,v_B)$, it means that there exists one or three more edges of color $0$ connecting the other vertices of $f_A$ to the vertices of $f_B$. If there is one other edge of color $0$, we can use the argument of ``Case $\abs{f_A} = 2$''.
        
        We will prove that, in addition to $e_\rm{L}$, there cannot be three other edges connecting the vertices of $f_A$ to the vertices of $f_B$. Indeed, note $i,j$ the colors of $f_A$. We can divide the set of colors $\paa{1,\dots,D}$ in two sets: $\sC_i$ and $\sC_j$. A color $c$ in $\sC_i$ (resp.~$\sC_j$) have a face of colors $(0,c)$ that are attached to the vertices related by one edge of $f_A$ of color $i$ (resp.~$j$) and another face of color $(0,c)$ attached to the other two vertices of $f_A$ related by the other edge of color $i$ (resp.~$j$). An example is given in Fig.~\ref{fig:Ex_C_i}. Trivially, one has $i \in \sC_i$ and $j\in \sC_j$. 
        
        \begin{figure}[ht]
            \centering
            \includegraphics[width = .7\textwidth]{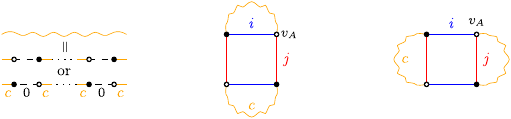}
            \caption{Left: a graphical representation of an alternation of colors $0$ and $c$ using a wavy segment of color $c$. Middle: a representation of a color $c \in \sC_i$. Right: a representation of a color $c \in \sC_j$.}
            \label{fig:Ex_C_i}
        \end{figure}
        
        Now, we cut the four edges of color $0$ linking $A$ to $B$. There are exactly two ways to glue them respecting the bipartite structure of the graph such that the new $(D+1)$-colored graph is non-connected, \ie of the form $\widehat G_\sqcup \eqdef \widehat A \sqcup \widehat B$ for $(\widehat A, \widehat B) \in \cG_{D+1}(A) \times \cG_{D+1}(B)$. Hence, since one cut-edge of color $0$ is folded down parallel to edges $i$ in $f_A$ we denote this cut: \textit{cut i}. The other cut is then denoted by \textit{cut j}. Fig.~\ref{fig:cuts} exemplified the two possible cuts.
        
        \begin{figure}[ht]
            \centering
            \includegraphics[width = .9\textwidth]{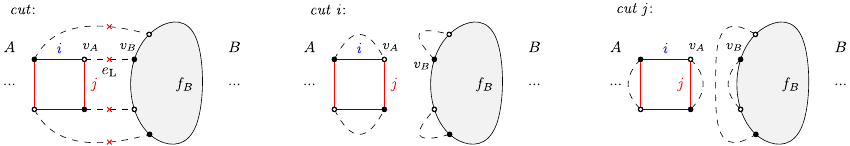}
            \caption{Left: the cuts are represented by red crosses. Middle: \textit{cut i} has been performed. Right: \textit{cut j} has been performed.}
            \label{fig:cuts}
        \end{figure}
        
        The cuts divide both sets $\sC_i$ and $\sC_j$ into two. Indeed, for $c \in \sC_i$, we say that $c\in \sC_i^\I$ (resp.~$\sC_i^\II$) if, restricted the edges of color $(0,c)$ in $\widehat G$, the number of faces attached to vertices of $f_A$ is one (resp.~two). Fig~\ref{fig:types} illustrates such sets, and straightforwardly, we have the following partition:
        \begin{equation}
            \paa{1,\dots,D} = \sC_i^\I \cup \sC_i^\II \cup \sC_j^\I \cup \sC_j^\II \,.
        \end{equation}
        where, $\sC_i^\I,\,\sC_i^\II,\, \sC_j^\I$ and $\sC_j^\II$ are two-by-two disjoint sets.
        
        \begin{figure}[ht]
            \centering
            \includegraphics[width = \textwidth]{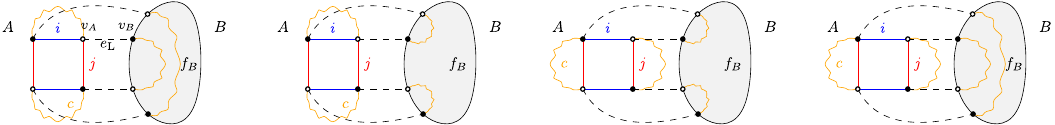}
            \caption{From left to right: representation of colors in, respectively, $\sC_i^\I$, $\sC_i^\II$, $\sC_j^\I$, and $\sC_j^\II$.}
            \label{fig:types}
        \end{figure}
        
        For the \textit{cut i}, we have:
        \begin{equation}
            F_{0c}(\widehat G) - F_{0c}(\widehat G_\sqcup) =  
            \begin{cases}
                    -2 \; \rm{ if }\, c \in \sC_i^\I \cup \sC_i^\II \cup \sC_j^\I \\
                    0 \; \rm{ otherwise}.
            \end{cases}
        \end{equation}
        Regarding the \textit{cut j}, one find:
        \begin{equation}
            F_{0c}(\widehat G) - F_{0c}(\widehat G_\sqcup) =  
            \begin{cases}
                    -2 \; \rm{ if }\, c \in \sC_j^\I \cup \sC_j^\II \cup \sC_i^\I \\
                    0 \; \rm{ otherwise}.
            \end{cases}
        \end{equation}
        Assume the \textit{cut i} is performed. We distinguish the permutation $\nu_{\widehat G}$ from $\nu_{\widehat G_\sqcup}$ where they respectively describe the edges of color $0$ before and after the cut. Hence, for $(c,c') \in \paa{1,\dots,D}$ two distinct colors, one can compute the following difference of Gromov product:
        \begin{align}
            \GP{\sigma_c}{\sigma_{c'}}{\nu_{\widehat G}} - \GP{\sigma_c}{\sigma_{c'}}{\nu_{\widehat G_\sqcup}} &= \frac{1}{2}\pac{F_{0c}(\widehat G_\sqcup) - F_{0c}(\widehat G) + F_{0c'}(\widehat G_\sqcup) - F_{0c'}(\widehat G)} \\
            &= 
            \begin{cases}
                    0 \; \rm{ if }\, (c,c') \in \sC_j^\II \times \sC_j^\II\\
                    1 \; \rm{ if }\, (c,c') \in \pa{\sC_i^\I \cup \sC_i^\II \cup \sC_j^\I} \times \sC_j^\II \\
                    2 \; \rm{ if }\, (c,c') \in \pa{\sC_i^\I \cup \sC_i^\II \cup \sC_j^\I}^{\times 2} \,. 
            \end{cases}
        \end{align}
        However, bear in mind that $\widehat G$ is such that $\Delta_0(\widehat G) = D(D-1)/2$, \ie it requires -- by the positivity of a Gromov product -- the inequality
        \begin{equation}
            \GP{\sigma_c}{\sigma_{c'}}{\nu_{\widehat G}} - \GP{\sigma_c}{\sigma_{c'}}{\nu_{\widehat G_\sqcup}} \leq 1 \,,
        \end{equation}
        \ie there cannot be two colors in the sets $\sC_i^\I \cup \sC_i^\II \cup \sC_j^\I$. Since, $i \in \sC_i = \sC_i^\I \cup \sC_i^\II$, we must have:
        \begin{equation} \label{eq:cuti}
            \abs{\sC_i} = 1 \,, \qquad \rm{and} \qquad \abs{\sC_j^\II} = D-1 \,.
        \end{equation}
        
        Nevertheless, there are no reasons why the \textit{cut j} cannot be performed (\ie namely one restores the edges after \textit{cut i} and performs \textit{cut j}). The latter leads to:
        \begin{equation} \label{eq:cutj}
            \abs{\sC_j} = 1 \,, \qquad \rm{and} \qquad \abs{\sC_i^\II} = D-1 \,.
        \end{equation}
        Finally, since $D \geq 3$, both Eq.~\eqref{eq:cuti} and Eq.~\eqref{eq:cutj} cannot simultaneously be true, leading to a contradiction. Hence, if $\widehat G$ is connected and minimizes the degree of compatibility, \ie $\Delta_0(\widehat G) = D(D-1)/2$, there cannot be four edges of color $0$ linking vertices of $f_A$ to vertices of $f_B$.
    \end{itemize}
\end{proof}

\begin{figure}[ht]
    \centering
    \includegraphics[height = 3.5cm]{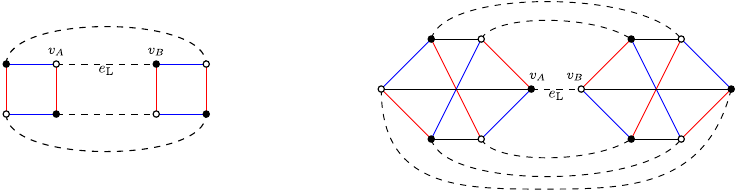}
    \caption{Left: a $3$-colored graph $\widehat G \in \cG_3(C_2 \sqcup C_2)$, achieving $\Delta_0(\widehat{G}) = 1$ and being constructed as two flips of graphs $\widehat C_2 \in \MD{}(C_2)$. Right: $4$-colored graph $\widehat G \in \cG_4^{\conn}(\PT_3 \sqcup \PT_3)$, realizing $\Delta_0(\widehat G) = 3$ without being constructed as one flip of $\widehat{\PT}_3 \in \MD{}(\PT_3)$.}
    \label{fig:CEflip}
\end{figure}

We remark that the previous proof is not valid for $D=2$ and for faces of size greater than or equal to six. Indeed, Fig.~\ref{fig:CEflip} (on the left) shows an example of two copies of $C_2$ for which four edges linking $A$ to $B$ is possible while achieving $\Delta_0(\widehat{C_2 \sqcup C_2}) = 1$. What is more, Fig.~\ref{fig:CEflip} (on the right) highlights an example of two copies of $\PT_3$ such that the $4$-colored graph presented satisfies a degree of compatibility of three while not being constructed as a flip between two copies of $\widehat{\PT}_3 \in \MD{}(\PT_3)$ but, instead, of three flips. Having a minimal face of size six, the latter example does not take part in the conditions of Lem.~\ref{lem:Local}.

We are now able to prove point 4. of Prop.~\ref{prop:comp}.
\begin{proof}
    Let $A$ and $B$ be two compatible connected\footnote{We restrict to connected graph since we can add the connected subgraphs of $A$ or $B$ by the use point 1. of Prop.~\ref{prop:comp} (see Ref.~\cite{Factorization2026}).} graphs and consider two vertices $v_A$ and $v_B$ such that we consider $G = A \op_{(v_A,v_B)} B$. We let $\widehat G \in \cG_{D+1}(G)$ and define $\widehat H \in \cG_{D+1}(A \sqcup B)$ to be the $(D+1)$-colored graph obtained by restoring the vertices $v_A$ and $v_B$ lost by the vertex-contraction operation and by adding an edge of color $0$ bridging $v_A$ to $v_B$ and noted $e_\rm{L}$. Since $F_0(\widehat G) = F_0(\widehat H)$, $k(\widehat G^{\widehat 0}) = k(\widehat H^{\widehat 0}) - 1 $ and $F(\widehat G^{\widehat 0}) = F(\widehat H^{\widehat 0}) - D(D-1)/2$, Eq.~\eqref{eq:Delta_0} yields: 
    \begin{equation}
        \Delta_0(\widehat G) = \Delta_0(\widehat H) - \frac{D(D-1)}{2} \,.
    \end{equation}
    However, since $\widehat A \op_{(v_A,v_B)} \widehat B \in \cG_{D+1}(G)$, we have $\Delta(G) \leq \Delta(A) + \Delta(B) = 0$. Hence, if $\widehat G \in \MD{}(G)$, we then have:
    \begin{equation}
        \Delta_0(\widehat H) = \frac{D(D-1)}{2} \,,
    \end{equation}
    with $\widehat H \in \cG_{D+1}^{\conn}(A \sqcup B)$, $\widehat H$ contains an edge of color $0$ such that $e_\rm{L} = (v_A, v_B)$ and $v_A$, $v_B$ or both belongs to at least one face of size at most four. For that matter, we can apply Lem.~\ref{lem:Local}.
    
    By Lem.~\ref{lem:Local}, for every $\widehat G \in \MD{}(G)$, there exists $(\widehat A , \widehat B) \in \MD{}(A) \times \MD{}(B)$ and edges $e_A^0$, $e_B^0$ attached respectively to $v_A$ and $v_B$ such that 
    \begin{equation}
        \widehat H = \widehat A \flip_{(e_A^0,e_B^0)} \widehat B \,.
    \end{equation}
    Hence, applying the vertex-contraction, one removes $e_\rm{L}$ and obtains that:
    \begin{equation}
        \widehat G = \widehat A \op_{(v_A,v_B)} \widehat B \,.
    \end{equation}
\end{proof}

\begin{figure}[ht]
    \centering
    \includegraphics[height = 4cm]{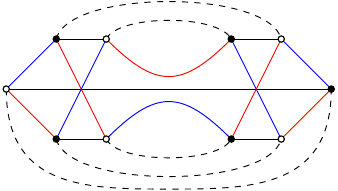}
    \caption{Example of a $4$-colored graph in $\MD{}(A \op B)$ that cannot be written as vertex contraction operation.}
    \label{fig:CEAoB}
\end{figure}

As a counterexample the contraction between any vertices of the connected compatible graphs $\PT_3$ and $\PT_3$ satisfies
\begin{equation}
    \MD{}(\PT_3 \op \PT_3) \supset \MD{}(\PT_3) \op \MD{}(\PT_3) \,,
\end{equation}
as exemplified by Fig.~\ref{fig:CEAoB}.

\newpage

\addcontentsline{toc}{section}{References}
\printbibliography

\end{document}

%% file: Packages.tex
\usepackage{graphicx} 
\usepackage[left=2cm,right=2cm,top=2cm,bottom=2cm]{geometry} 
\usepackage{xcolor} 
\usepackage{enumitem}
\usepackage{pifont}  
\usepackage{comment} 
\usepackage{float} 
\usepackage{multicol} 

\usepackage{soul}

\usepackage{mathtools} 
\usepackage{esvect} 
\usepackage{stackrel,amssymb} 
\usepackage{amsmath} 
\usepackage{amsthm} 
\usepackage{mathrsfs} 
\usepackage{cancel} 
\usepackage{mathabx} 
\usepackage{stackengine} 
\usepackage{tablefootnote}

\usepackage{pgfplots}
\pgfplotsset{compat=1.18}

\usepackage{slashbox} 
\usepackage{diagbox} 
\usepackage{multirow} 

\usepackage{authblk}

\usepackage{subcaption}
\usepackage{graphicx}      
\usepackage{eso-pic}       
\graphicspath{ {./images/} }

 \usepackage[
     backend=biber,
     style=numeric-comp,
     sorting=none,
   ]{biblatex}
\usepackage{ifthen}
\usepackage{stackengine} 
\usepackage{scalerel}

\usepackage{listings} 

\definecolor{codegreen}{rgb}{0,0.6,0}
\definecolor{codegray}{rgb}{0.5,0.5,0.5}
\definecolor{codepurple}{rgb}{0.58,0,0.82}
\definecolor{backcolour}{rgb}{0.95,0.95,0.92}

\lstdefinestyle{mystyle}{
    backgroundcolor=\color{backcolour},   
    commentstyle=\color{codegreen},
    keywordstyle=\color{magenta},
    numberstyle=\tiny\color{codegray},
    stringstyle=\color{codepurple},
    basicstyle=\ttfamily\footnotesize,
    breakatwhitespace=false,         
    breaklines=true,                 
    captionpos=b,                    
    keepspaces=true,                 
    numbers=left,                    
    numbersep=5pt,                  
    showspaces=false,                
    showstringspaces=false,
    showtabs=false,                  
    tabsize=2
}

\usepackage{hyperref}
\hypersetup{
    colorlinks=true,
    linkcolor=black,
    citecolor=black,      
    urlcolor=black,
    pdftitle={Tensorial invariants for the classification of multipartite entanglement},
    pdfpagemode=FullScreen,
}

\usepackage{lscape}
\usepackage{graphicx}

%% file: Commandes.tex
\makeatletter 

\newcommand*\sNeg[2][0mu]{\Neginternal{#1}{\snegslash}{#2}}

\newcommand*\Neginternal[3]{\mathpalette\Neg@{{#1}{#2}{#3}}}
\newcommand*\Neg@[2]{\Neg@@{#1}#2}
\newcommand*\Neg@@[4]{%
  \mathrel{\ooalign{%
    $\m@th#1#4$\cr
    \hidewidth$\m@th#3{#1}\mkern\muexpr#2*2$\hidewidth\cr
  }}%
}

\newcommand*\negslash[1]{\m@th#1\not\mathrel{\phantom{=}}}
\newcommand*\snegslash[1]{\rotatebox[origin=c]{60}{$\m@th#1-$}}
\newcommand*\ssnegslash[1]{\rotatebox[origin=c]{60}{$\m@th#1{\dabar@}\mkern-7mu{\dabar@}$}}
\newcommand*\sssnegslash[1]{\rotatebox[origin=c]{60}{$\m@th#1\dabar@$}}
\makeatother  

\newcommand{\notimplies}{\sNeg{\Longrightarrow}}

\newcommand{\lat}[1]{\textsl{#1}}
\newcommand{\apriori}{\lat{a~priori} }
\newcommand{\ie}{\lat{i.e.}~}

\newcommand{\etc}{\lat{etc.} }

\newcommand{\eg}{\lat{e.g.}~}
 
\renewcommand{\rm}[1]{\text{#1}}

\newcommand{\e}[1]{\rm{\textup{e}}^{#1}}           

\renewcommand{\d}{\rm{\textup{d}}} 		      	

\newcommand{\pa}[1]{\left( #1 \right)}
\newcommand{\pac}[1]{\left[ #1 \right]}
\newcommand{\paa}[1]{\left\{ #1 \right\}}

\DeclareMathOperator{\tr}{Tr}          

\DeclareMathOperator{\id}{id}

\renewcommand{\vec}[1]{\vv{#1}} 				
\renewcommand{\tilde}[1]{\widetilde{#1}}

\DeclareMathOperator{\flip}{\ast}

\DeclareMathOperator{\op}{\circ}

\newcommand{\ot}{\otimes}
\DeclareMathOperator{\End}{End}
\DeclareMathOperator{\red}{Supp}

\newcommand{\xmark}{\ding{55}}
\newcommand{\cmark}{\ding{51}}

\DeclareMathOperator{\spec}{Spec}
\newcommand{\norm}[1]{\left\lVert#1\right\rVert}

\newcommand{\bra}[1]{\left\langle #1 \right\vert} 			
\newcommand{\ket}[1]{\left\vert #1 \right\rangle} 			
\newcommand{\mean}[1]{\left\langle #1 \right\rangle}

\newcommand{\abs}[1]{\left\lvert #1 \right\rvert}
\newcommand{\innprod}[2]{\left\langle #1 \mid #2 \right\rangle}
\renewcommand{\bar}[1]{\overline{#1}}

\newcommand\eqdef{\mathrel{\overset{\makebox[0pt]{\mbox{\normalfont\tiny\sffamily def}}}{=}}}

\newcommand{\bb}[1]{\mathbb{#1}}

\newcommand{\Prob}[1]{\bb{P}\pa{#1}}
\newcommand{\GP}[3]{\pa{#1 \vert #2}_{#3}}
\renewcommand{\H}{\mathcal{H}}

\newcommand{\MD}[1]{\cal{M}}

\newcommand{\K}{\mathrm{K}}
\newcommand{\ME}{\mathrm{ME}}
\newcommand{\LN}{\mathrm{LN}}
\newcommand{\PTE}{\mathrm{PTE}}
\newcommand{\RE}{\mathrm{RE}}
\newcommand{\RM}{\mathrm{RM}}
\newcommand{\JRM}{\mathrm{JRM}}
\newcommand{\RME}{\mathrm{RME}}
\newcommand{\PT}{\mathrm{PT}}
\newcommand{\GHZ}{\mathrm{GHZ}}
\newcommand{\Bell}{\mathrm{Bell}}

\renewcommand{\S}{\cal{S}}

\theoremstyle{plain}
\newtheorem{theo}{Theorem}[section]
\newtheorem{prop}[theo]{Proposition}
\newtheorem{lem}[theo]{Lemma}
\newtheorem{rem}[theo]{Remark}
\newtheorem{ex}[theo]{Example}
\newtheorem{defi}[theo]{Definition}
\newtheorem{cor}[theo]{Corollary}

\renewcommand{\cal}[1]{\mathcal{#1}}

\definecolor{cadmiumred}{rgb}{0.89, 0.0, 0.13}
\newboolean{ifprintcolor}	

\ifthenelse{\boolean{ifprintcolor}}{}{}

\newcommand{\nlongrightarrow}{%
  \mathrel{%
    \tikz[baseline=-0.5ex]{%
      \draw[->] (0,0) -- (.75,0);        
      \draw[line width=0.4pt] (0.2,-0.1) -- (0.55,0.1); 
    }%
  }%
}

\newcommand{\fS}{\mathfrak{S}}
\newcommand{\sF}{\mathsf{F}}
\newcommand{\sM}{\mathsf{M}}
\newcommand{\U}{\mathsf{U}}
\newcommand{\LU}{\mathsf{LU}}
\newcommand{\LO}{\mathsf{LO}}
\newcommand{\LOCC}{\mathsf{LOCC}}
\newcommand{\LOSR}{\mathsf{LOSR}}

\newcommand{\cO}{\mathcal{O}}
\newcommand{\cI}{\mathcal{I}}
\newcommand{\cL}{\mathcal{L}}

\newcommand{\sC}{\mathscr{C}}
\newcommand{\I}{%
  \textup{\uppercase\expandafter{\romannumeral 1}}%
}
\newcommand{\II}{%
  \textup{\uppercase\expandafter{\romannumeral 2}}%
}
\newcommand{\cE}{\mathcal{E}}
\newcommand{\simLU}{\underset{\LU}{\sim}}
\newcommand{\toLO}{\underset{\LO}{\longrightarrow}}
\newcommand{\nottoLO}{\underset{\LO}{\nlongrightarrow}}
\newcommand{\toLOCC}{\underset{\LOCC}{\longrightarrow}}
\DeclareMathOperator{\iin}{in}
\DeclareMathOperator{\oout}{out}
\DeclareMathOperator{\im}{Im}
\DeclareMathOperator{\Span}{Span}
\DeclareMathOperator{\inv}{inv}
\DeclareMathOperator{\triv}{triv}
\DeclareMathOperator{\conn}{c}
\DeclareMathOperator{\aux}{aux}
\newcommand{\cA}{\mathcal{A}}
\newcommand\restr[2]{{
  \left.\kern-\nulldelimiterspace 
  #1 
  \vphantom{\big|} 
  \right|_{#2} 
  }}
\def\extd{\mathrm {d}}
\newcommand{\cG}{\mathcal{G}}
\newcommand{\pos}{\mathcal{P}}